\documentclass[twoside,english,aps,superscriptaddress,nobibnotes,nofootinbib]{revtex4-1}
\pdfoutput=1
\usepackage[latin9]{inputenc}
\usepackage{color}
\usepackage{babel}
\usepackage{verbatim}
\usepackage{cancel}
\usepackage{amsmath}
\usepackage{amssymb}
\usepackage{esint}
\usepackage[unicode=true,
 bookmarks=true,bookmarksnumbered=true,bookmarksopen=false,
 breaklinks=false,pdfborder={0 0 0},backref=false,colorlinks=true]
 {hyperref}
\hypersetup{pdftitle={The Minimal Modal Interpretation of Quantum Theory},
 pdfauthor={Jacob A. Barandes, David Kagan},
 pdfsubject={Physics},
 pdfkeywords={physics, quantum mechanics, interpretations}}

\makeatletter

\providecommand{\tabularnewline}{\\}

\@ifundefined{textcolor}{}
{%
 \definecolor{BLACK}{gray}{0}
 \definecolor{WHITE}{gray}{1}
 \definecolor{RED}{rgb}{1,0,0}
 \definecolor{GREEN}{rgb}{0,1,0}
 \definecolor{BLUE}{rgb}{0,0,1}
 \definecolor{CYAN}{cmyk}{1,0,0,0}
 \definecolor{MAGENTA}{cmyk}{0,1,0,0}
 \definecolor{YELLOW}{cmyk}{0,0,1,0}
}

\usepackage{url}
\usepackage{graphicx}
\usepackage{tikz}
\usetikzlibrary{matrix,arrows,decorations.pathmorphing,decorations.pathreplacing}

\makeatother

\begin{document}

\title{The Minimal Modal Interpretation of Quantum Theory}

\author{Jacob A. Barandes}

\email{barandes@physics.harvard.edu}

\affiliation{Jefferson Physical Laboratory, Harvard University, Cambridge, MA
02138}

\author{David Kagan}

\email{dkagan@umassd.edu}

\affiliation{Department of Physics, University of Massachusetts Dartmouth, North
Dartmouth, MA 02747}

\date{\today}
\begin{abstract}
We introduce a realist, unextravagant interpretation of quantum theory
that builds on the existing physical structure of the theory and allows
experiments to have definite outcomes, but leaves the theory's basic
dynamical content essentially intact. Much as classical systems have
specific states that evolve along definite trajectories through configuration
spaces, the traditional formulation of quantum theory asserts that
closed quantum systems have specific states that evolve unitarily
along definite trajectories through Hilbert spaces, and our interpretation
extends this intuitive picture of states and Hilbert-space trajectories
to the case of open quantum systems as well. We provide independent
justification for the partial-trace operation for density matrices,
reformulate wave-function collapse in terms of an underlying interpolating
dynamics, derive the Born rule from deeper principles, resolve several
open questions regarding ontological stability and dynamics, address
a number of familiar no-go theorems, and argue that our interpretation
is ultimately compatible with Lorentz invariance. Along the way, we
also investigate a number of unexplored features of quantum theory,
including an interesting geometrical structure---which we call subsystem space---that
we believe merits further study. We include an appendix that briefly
reviews the traditional Copenhagen interpretation and the measurement
problem of quantum theory, as well as the instrumentalist approach
and a collection of foundational theorems not otherwise discussed
in the main text.
\end{abstract}
\maketitle

\noindent \begin{center}

\global\long\def\hyphen{\mbox{-}}
\global\long\def\colon{\mbox{:}}

\global\long\def\difficult{\dagger}
\global\long\def\anc{*}

\global\long\def\exampleclose{\blacklozenge}
\global\long\def\solutionclose{\blacksquare}
\global\long\def\stepclose{\checkmark}

\global\long\def\fakespace{}

\global\long\def\subarrayleftleft#1#2{\begin{subarray}{l}
 #1\\
#2 
\end{subarray}}

\global\long\def\quote#1{``#1"}

\global\long\def\eqsbrace#1{\left.#1\qquad\qquad\qquad\right\}  }

\global\long\def\presup#1#2{\vphantom{#1}^{#2}#1}
\global\long\def\presub#1#2{\vphantom{#1}_{#2}#1}

\global\long\def\tight#1{\!\negthickspace\negthickspace#1\negthickspace\negthickspace\!}
\global\long\def\const{\mathrm{const}}
\global\long\def\degrees{\circ}

\global\long\def\units#1{\mathrm{#1}}
\global\long\def\hasunits#1{\left[#1\right]}

\global\long\def\from{\leftarrow}
\global\long\def\xto#1#2{\xrightarrow[#1]{#2}}
\global\long\def\xfrom#1#2{\xleftarrow[#1]{#2}}
\global\long\def\mapsfrom{\from}
\global\long\def\xmapsto#1#2{\overset{{\scriptstyle #2}}{\underset{{\scriptstyle #1}}{\mapsto}}}

\global\long\def\impliedby{\Longleftarrow}
\global\long\def\exchange{\leftrightarrow}

\global\long\def\binaryand{\mathrm{\ and\ }}
\global\long\def\binaryor{\mathrm{\ or\ }}
\global\long\def\binaryis{\mathrm{\ is\ }}
\global\long\def\binaryfor{\mathrm{\ for\ }}
\global\long\def\binaryfrom{\mathrm{\ from\ }}
\global\long\def\binarytext#1{\mathrm{\ #1\ }}

\global\long\def\whichfor{\mathrm{for\ }}
\global\long\def\whichif{\mathrm{if\ }}
\global\long\def\whichotherwise{\mathrm{otherwise}}
\global\long\def\whichtext#1{\mathrm{#1\ }}
\global\long\def\also{\mathrm{also}}

\global\long\def\given{\vert}

\global\long\def\QED{\mathrm{QED}}

\global\long\def\mapping#1#2#3{#1:#2\to#3}
\global\long\def\composition{\circ}

\global\long\def\set#1{\left\{  #1\right\}  }
\global\long\def\setindexed#1#2{\left\{  #1\right\}  _{#2}}

\global\long\def\setbuild#1#2{\left\{  \left.\!#1\right|#2\right\}  }
\global\long\def\complem{\mathrm{c}}

\global\long\def\union{\cup}
\global\long\def\intersection{\cap}
\global\long\def\cartesianprod{\times}
\global\long\def\disjointunion{\sqcup}

\global\long\def\isomorphic{\cong}

\global\long\def\sgn{\mathrm{sgn\,}}
\global\long\def\absval#1{\left|#1\right|}

\global\long\def\average#1{\left\langle #1\right\rangle }
\global\long\def\saverage#1{\left\langle #1\right\rangle }
\global\long\def\taverage#1{\overline{#1}}
\global\long\def\mean#1{\overline{#1}}

\global\long\def\evaluate#1{\left.#1\right|}

\global\long\def\notpropto{\not\propto}
\global\long\def\floor#1{\left\lfloor #1\right\rfloor }
\global\long\def\ceil#1{\left\lceil #1\right\rceil }

\global\long\def\integration#1#2{#1#2\;}
\global\long\def\differential#1{#1\;}

\global\long\def\csch{\mathrm{csch\,}}
\global\long\def\arccot{\mathrm{arccot\,}}
\global\long\def\oforder#1{\mathcal{O}\left(#1\right)}

\global\long\def\convolution{\circ}

\global\long\def\conj{\ast}
\global\long\def\re{\mathrm{Re\,}}
\global\long\def\im{\mathrm{Im\,}}

\global\long\def\Disc{\mathrm{Disc\,}}
\global\long\def\Res{\mathrm{Res\,}}

\global\long\def\Span{\mathrm{Span\,}}
\global\long\def\Tr{\mathrm{Tr\,}}
\global\long\def\tr{\mathrm{tr\,}}

\global\long\def\transp{\mathrm{T}}
\global\long\def\adj{\dagger}

\global\long\def\hasrep{\rightarrowtail}
\global\long\def\repfor{\leftarrowtail}

\global\long\def\xhasrep#1#2{\overset{#2}{\underset{#1}{\rightarrowtail}}}
\global\long\def\xrepfor#1#2{\overset{#2}{\underset{#1}{\leftarrowtail}}}

\global\long\def\diag#1{\mathrm{diag}\left(#1\right)}

\global\long\def\directsum{\oplus}
\global\long\def\directprod{\otimes}
\global\long\def\tensorprod{\otimes}

\global\long\def\boundary{\partial}

\global\long\def\notparallel{\not\parallel}
\global\long\def\notperp{\not\perp}

\global\long\def\svec#1{\vec{#1}}
\global\long\def\norm#1{\absval{#1}}
\global\long\def\stensor#1{\overleftrightarrow{#1}}
\global\long\def\dotprod{\cdot}
\global\long\def\crossprod{\times}

\global\long\def\del{\svec{\nabla}}
\global\long\def\dalemb{\square}

\global\long\def\gradient{\svec{\nabla}}
\global\long\def\divergence{\svec{\nabla}\cdot}
\global\long\def\curl{\svec{\nabla}\times}
\global\long\def\laplacian{\svec{\nabla}^{2}}

\global\long\def\gradientprime{\svec{\nabla}^{\prime}}
\global\long\def\divergenceprime{\svec{\nabla}^{\prime}\cdot}
\global\long\def\curlprime{\svec{\nabla}^{\prime}\times}
\global\long\def\laplacianprime{\svec{\nabla}^{\prime2}}

\global\long\def\homeomorphic{\cong}
\global\long\def\diffeomorphic{\cong}

\global\long\def\dual{\ast}
\global\long\def\extderiv{\mathrm{d}}
\global\long\def\hodge{\mathrm{\star}}

\global\long\def\hodgeprod#1#2{\left(#1,#2\right)_{\star}}

\global\long\def\liederiv#1#2{\pounds_{#2}#1}

\global\long\def\liebracket#1#2{\left[#1,#2\right]}

\global\long\def\connection{P}
\global\long\def\cov{\nabla}

\global\long\def\tud#1#2#3{#1_{\phantom{#2}#3}^{#2}}

\global\long\def\tdu#1#2#3{#1_{#2}^{\phantom{#2}#3}}

\global\long\def\tudu#1#2#3#4{#1_{\phantom{#2}#3}^{#2\phantom{#3}#4}}

\global\long\def\tdud#1#2#3#4{#1_{#2\phantom{#3}#4}^{\phantom{#2}#3}}

\global\long\def\tudud#1#2#3#4#5{#1_{\phantom{#2}#3\phantom{#4}#5}^{#2\phantom{#3}#4}}

\global\long\def\tdudu#1#2#3#4#5{#1_{#2\phantom{#3}#4}^{\phantom{#2}#3\phantom{#4}#5}}

\global\long\def\Probability#1{\mathrm{Prob}\left(#1\right)}
\global\long\def\Amplitude#1{\mathrm{Amp}\left(#1\right)}

\global\long\def\expectval#1{\left\langle #1\right\rangle }
\global\long\def\op#1{\hat{#1}}

\global\long\def\bra#1{\left\langle #1\right|}
\global\long\def\ket#1{\left|#1\right\rangle }
\global\long\def\braket#1#2{\left\langle \left.\!#1\right|#2\right\rangle }

\global\long\def\dbbra#1{\left\langle #1\right\Vert }
\global\long\def\dbket#1{\left\Vert #1\right\rangle }

\global\long\def\inprod#1#2{\left\langle #1,#2\right\rangle }
\global\long\def\normket#1{\left\Vert #1\right\Vert }

\global\long\def\comm#1#2{\left[#1,#2\right]}
\global\long\def\acomm#1#2{\left\{  #1,#2\right\}  }

\global\long\def\pbrack#1#2{\left\{  #1,#2\right\}  _{\mathrm{PB}}}

\global\long\def\timeorder{\mathrm{T}}

\par\end{center}

\makeatletter
\def\l@subsection#1#2{}
\def\l@subsubsection#1#2{}
\makeatother

\tableofcontents{}

\section{Introduction\label{sec:Introduction}}

\subsection{Why Do We Need a New Interpretation?\label{sub:Why-Do-We-Need-a-New-Interpretation}}

\subsubsection{The Copenhagen Interpretation\label{subsub:The-Copenhagen-Interpretation}}

Any mathematical-physical theory like quantum theory%
\footnote{We use the term ``quantum theory'' here in its broadest sense of
referring to the general theoretical framework consisting of Hilbert
spaces and state vectors that encompasses models as diverse as nonrelativistic
point particles and quantum field theories. We are not referring specifically
to the nonrelativistic models of quantum-mechanical point particles
that dominated the subject in its early days.%
} requires an interpretation, by which we mean some asserted connection
with the real world. The traditional Copenhagen interpretation, with
its axiomatic Born rule for computing empirical outcome probabilities
and its notion of wave-function collapse for establishing the persistence
of measurement outcomes, works quite well in most practical circumstances.%
\footnote{See Appendix~\ref{sub:The-Copenhagen-Interpretation-and-the-Measurement-Problem}
for a detailed definition of the Copenhagen interpretation, as well
as a description of the famous measurement problem of quantum theory
and a systematic classification of attempts to solve it according
to the various prominent interpretations of the theory.%
} At least according to some surveys \cite{Tegmark:1998iqmmwmw,SchlosshauerKoflerZeilinger:2013sfatqm},
the Copenhagen interpretation is still the most popular interpretation
today.

\begin{figure}
\begin{centering}
\begin{center}
\definecolor{cffffff}{RGB}{255,255,255}

\begin{tikzpicture}[y=0.60pt,x=0.60pt,yscale=-1, inner sep=0pt, outer sep=0pt]
\begin{scope}[shift={(-47.98725,-39.68933)}]
  \begin{scope}[cm={{0.53212,0.0,0.0,0.53212,(-88.27664,-60.44546)}}]
    \path[shift={(-12.5874,170.91883)},draw=black,miter limit=4.00,line
      width=0.800pt] (400.0000,237.3622) .. controls (400.0000,250.3803) and
      (389.4467,260.9336) .. (376.4286,260.9336) .. controls (363.4104,260.9336) and
      (352.8571,250.3803) .. (352.8571,237.3622) .. controls (352.8571,224.3440) and
      (363.4104,213.7908) .. (376.4286,213.7908) .. controls (389.4467,213.7908) and
      (400.0000,224.3440) .. (400.0000,237.3622) -- cycle;
    \path[draw=black,line join=miter,line cap=butt,line width=0.800pt]
      (351.0280,388.1869) .. controls (357.1764,401.3189) and (358.3277,414.4508) ..
      (351.0280,427.5828);
    \path[draw=black,line join=miter,line cap=butt,line width=0.800pt]
      (376.2826,388.6920) .. controls (370.1342,401.8239) and (368.9829,414.9559) ..
      (376.2826,428.0879);
    \path[draw=black,line join=miter,line cap=butt,line width=0.800pt]
      (350.5229,393.7427) -- (355.5737,390.4597);
    \path[draw=black,line join=miter,line cap=butt,line width=0.800pt]
      (351.7856,398.5410) -- (357.0889,395.2579);
    \path[draw=black,line join=miter,line cap=butt,line width=0.800pt]
      (352.5433,402.0765) -- (358.3516,401.5714);
    \path[draw=black,line join=miter,line cap=butt,line width=0.800pt]
      (353.5534,406.6222) -- (358.3516,407.3798);
    \path[draw=black,line join=miter,line cap=butt,line width=0.800pt]
      (353.0483,411.1678) -- (357.8465,413.1882);
    \path[draw=black,line join=miter,line cap=butt,line width=0.800pt]
      (352.2907,416.7237) -- (357.0889,417.9864);
    \path[draw=black,line join=miter,line cap=butt,line width=0.800pt]
      (350.7755,420.7643) -- (356.0788,422.7846);
    \path[draw=black,line join=miter,line cap=butt,line width=0.800pt]
      (349.5128,424.2998) -- (354.8161,427.0778);
    \path[draw=black,line join=miter,line cap=butt,line width=0.800pt]
      (372.9988,390.4597) -- (377.5445,393.4902);
    \path[draw=black,line join=miter,line cap=butt,line width=0.800pt]
      (371.7362,395.0054) -- (376.2818,396.7732);
    \path[draw=black,line join=miter,line cap=butt,line width=0.800pt]
      (370.2209,398.5410) -- (375.0191,400.3087);
    \path[draw=black,line join=miter,line cap=butt,line width=0.800pt]
      (369.2108,401.8239) -- (373.5039,403.5917);
    \path[draw=black,line join=miter,line cap=butt,line width=0.800pt]
      (368.7057,407.1272) -- (373.7565,407.1272);
    \path[draw=black,line join=miter,line cap=butt,line width=0.800pt]
      (369.2108,411.1678) -- (374.5141,410.4102);
    \path[draw=black,line join=miter,line cap=butt,line width=0.800pt]
      (368.9582,416.9762) -- (374.5141,415.9661);
    \path[draw=black,line join=miter,line cap=butt,line width=0.800pt]
      (370.4735,420.7643) -- (375.0191,419.5016);
    \path[draw=black,line join=miter,line cap=butt,line width=0.800pt]
      (372.4938,424.2998) -- (376.0293,422.2795);
    \path[draw=black,line join=miter,line cap=butt,line width=0.800pt]
      (373.5039,427.3303) -- (377.2920,424.8049);
    \path[draw=black,line join=miter,line cap=butt,line width=0.800pt]
      (321.2285,400.8138) .. controls (287.3022,407.3572) and (269.0644,425.6670) ..
      (256.5788,448.2910);
    \path[draw=black,line join=miter,line cap=butt,line width=0.800pt]
      (336.3808,424.0473) .. controls (307.7466,431.3340) and (292.4566,447.5169) ..
      (279.8123,465.4635);
    \path[draw=black,line join=miter,line cap=butt,line width=0.800pt]
      (357.5940,444.2504) .. controls (332.8225,452.8230) and (317.4292,467.0225) ..
      (306.0762,483.6463);
  \end{scope}
  \begin{scope}[cm={{0.53212,0.0,0.0,0.53212,(82.36707,6.48496)}}]
    \path[draw=black,line join=miter,line cap=butt,line width=0.800pt]
      (366.6854,201.8137) -- (366.6854,170.4990);
    \path[draw=black,line join=miter,line cap=butt,line width=0.800pt]
      (394.9696,207.8747) -- (413.1524,181.6107);
    \path[draw=black,line join=miter,line cap=butt,line width=0.800pt]
      (408.1016,230.0980) -- (435.3757,214.9457);
    \path[draw=black,line join=miter,line cap=butt,line width=0.800pt]
      (410.1219,249.2909) -- (437.3961,248.2807);
    \path[draw=black,line join=miter,line cap=butt,line width=0.800pt]
      (401.0306,277.5752) -- (420.2235,289.6970);
    \path[draw=black,line join=miter,line cap=butt,line width=0.800pt]
      (386.8275,290.7072) -- (395.9189,311.9204);
    \path[draw=black,line join=miter,line cap=butt,line width=0.800pt]
      (366.6448,293.7376) -- (366.6448,321.0117);
    \path[draw=black,line join=miter,line cap=butt,line width=0.800pt]
      (338.4011,292.7275) -- (321.2285,308.8899);
    \path[draw=black,line join=miter,line cap=butt,line width=0.800pt]
      (323.2488,271.5143) -- (302.0356,278.5853);
    \path[draw=black,line join=miter,line cap=butt,line width=0.800pt]
      (317.1879,242.2198) -- (288.9036,231.1082);
    \path[draw=black,line join=miter,line cap=butt,line width=0.800pt]
      (327.2894,213.9356) -- (307.0864,195.7528);
    \path[draw=black,line join=miter,line cap=butt,line width=0.800pt]
      (346.4823,208.8848) -- (329.3097,180.6005);
    \path[shift={(111.11678,102.02541)},draw=black,miter limit=4.00,line
      width=0.800pt] (426.2844,163.4280) .. controls (426.2844,196.3436) and
      (349.8523,223.0269) .. (255.5686,223.0269) .. controls (161.2849,223.0269) and
      (84.8528,196.3436) .. (84.8528,163.4280) .. controls (84.8528,130.5123) and
      (161.2849,103.8290) .. (255.5686,103.8290) .. controls (349.8523,103.8290) and
      (426.2844,130.5123) .. (426.2844,163.4280) -- cycle;
    \path[shift={(104.04571,77.78175)},draw=black,miter limit=4.00,line
      width=0.800pt] (458.6093,190.7021) .. controls (458.6093,230.3124) and
      (371.0968,262.4229) .. (263.1447,262.4229) .. controls (155.1927,262.4229) and
      (67.6802,230.3124) .. (67.6802,190.7021) .. controls (67.6802,151.0917) and
      (155.1927,118.9812) .. (263.1447,118.9812) .. controls (371.0968,118.9812) and
      (458.6093,151.0917) .. (458.6093,190.7021) -- cycle;
    \path[shift={(104.04571,77.78175)},draw=black,miter limit=4.00,line
      width=0.800pt] (483.8631,193.2275) .. controls (483.8631,237.5799) and
      (385.7225,273.5346) .. (264.6600,273.5346) .. controls (143.5974,273.5346) and
      (45.4569,237.5799) .. (45.4569,193.2275) .. controls (45.4569,148.8750) and
      (143.5974,112.9203) .. (264.6600,112.9203) .. controls (385.7225,112.9203) and
      (483.8631,148.8750) .. (483.8631,193.2275) -- cycle;
    \path[shift={(104.04571,77.78175)},draw=black,miter limit=4.00,line
      width=0.800pt] (342.4417,183.6310) .. controls (342.4417,200.3678) and
      (304.9041,213.9356) .. (258.5991,213.9356) .. controls (212.2940,213.9356) and
      (174.7564,200.3677) .. (174.7564,183.6310) .. controls (174.7564,166.8942) and
      (212.2940,153.3264) .. (258.5991,153.3264) .. controls (304.9041,153.3264) and
      (342.4417,166.8942) .. (342.4417,183.6310) -- cycle;
    \path[shift={(97.98479,75.76144)},draw=black,miter limit=4.00,line
      width=0.800pt] (366.6854,186.6615) .. controls (366.6854,207.3034) and
      (320.7809,224.0371) .. (264.1549,224.0371) .. controls (207.5288,224.0371) and
      (161.6244,207.3034) .. (161.6244,186.6615) .. controls (161.6244,166.0195) and
      (207.5289,149.2858) .. (264.1549,149.2858) .. controls (320.7809,149.2858) and
      (366.6854,166.0195) .. (366.6854,186.6615) -- cycle;
    \path[shift={(201.02036,92.93403)},draw=black,fill=cffffff,miter limit=4.00,line
      width=0.800pt] (201.0204,155.3467) .. controls (201.0204,175.4308) and
      (184.7390,191.7122) .. (164.6549,191.7122) .. controls (144.5708,191.7122) and
      (128.2894,175.4308) .. (128.2894,155.3467) .. controls (128.2894,135.2626) and
      (144.5708,118.9812) .. (164.6549,118.9812) .. controls (184.7390,118.9812) and
      (201.0204,135.2626) .. (201.0204,155.3467) -- cycle;
    \path[shift={(110.10663,81.82236)},draw=black,fill=cffffff,miter limit=4.00,line
      width=0.800pt] (434.3656,185.6513) .. controls (434.3656,190.6723) and
      (430.2953,194.7427) .. (425.2742,194.7427) .. controls (420.2532,194.7427) and
      (416.1829,190.6723) .. (416.1829,185.6513) .. controls (416.1829,180.6303) and
      (420.2532,176.5599) .. (425.2742,176.5599) .. controls (430.2953,176.5599) and
      (434.3656,180.6303) .. (434.3656,185.6513) -- cycle;
    \path[shift={(104.04571,77.78175)},draw=black,fill=cffffff,miter limit=4.00,line
      width=0.800pt] (355.5737,255.8569) .. controls (355.5737,259.4832) and
      (352.6340,262.4229) .. (349.0077,262.4229) .. controls (345.3814,262.4229) and
      (342.4417,259.4832) .. (342.4417,255.8569) .. controls (342.4417,252.2306) and
      (345.3814,249.2909) .. (349.0077,249.2909) .. controls (352.6340,249.2909) and
      (355.5737,252.2306) .. (355.5737,255.8569) -- cycle;
    \path[shift={(104.04571,77.78175)},draw=black,fill=cffffff,miter limit=4.00,line
      width=0.800pt] (57.5787,210.4000) .. controls (57.5787,214.5842) and
      (54.1867,217.9762) .. (50.0026,217.9762) .. controls (45.8184,217.9762) and
      (42.4264,214.5842) .. (42.4264,210.4000) .. controls (42.4264,206.2158) and
      (45.8184,202.8239) .. (50.0026,202.8239) .. controls (54.1867,202.8239) and
      (57.5787,206.2158) .. (57.5787,210.4000) -- cycle;
    \path[shift={(106.06602,76.77159)},draw=black,fill=cffffff,miter limit=4.00,line
      width=0.800pt] (159.6041,193.7325) .. controls (159.6041,196.5220) and
      (157.3428,198.7833) .. (154.5533,198.7833) .. controls (151.7639,198.7833) and
      (149.5026,196.5220) .. (149.5026,193.7325) .. controls (149.5026,190.9430) and
      (151.7639,188.6817) .. (154.5533,188.6817) .. controls (157.3428,188.6817) and
      (159.6041,190.9430) .. (159.6041,193.7325) -- cycle;
    \path[shift={(104.04571,77.78175)},draw=black,fill=cffffff,miter limit=4.00,line
      width=0.800pt] (181.8275,187.6716) .. controls (181.8275,190.4611) and
      (179.5662,192.7224) .. (176.7767,192.7224) .. controls (173.9872,192.7224) and
      (171.7259,190.4611) .. (171.7259,187.6716) .. controls (171.7259,184.8821) and
      (173.9872,182.6208) .. (176.7767,182.6208) .. controls (179.5662,182.6208) and
      (181.8275,184.8821) .. (181.8275,187.6716) -- cycle;
  \end{scope}
  \begin{scope}[cm={{0.53212,0.0,0.0,0.53212,(-72.15105,33.08298)}}]
    \path[cm={{1.00135,0.0,0.0,0.52346,(45.00199,57.46502)}},draw=black,line
      join=miter,line cap=butt,line width=0.800pt] (327.2894,49.2807) .. controls
      (332.8606,49.1288) and (330.2429,56.7156) .. (327.5420,58.5404) .. controls
      (320.2226,63.4855) and (311.2492,56.8977) .. (308.7700,49.7858) .. controls
      (304.3351,37.0642) and (314.3658,24.4320) .. (326.5318,21.5015) .. controls
      (344.3859,17.2009) and (361.1068,31.0650) .. (364.3284,48.2706) .. controls
      (368.6222,71.2029) and (350.7973,92.1283) .. (328.5521,95.5794) .. controls
      (300.5587,99.9221) and (275.3809,78.0820) .. (271.7310,50.7959) .. controls
      (267.3106,17.7487) and (293.1937,-11.7062) .. (325.5217,-15.5374) .. controls
      (363.6191,-20.0524) and (397.3656,9.8899) .. (401.3673,47.2604) .. controls
      (405.9874,90.4062) and (371.9757,128.4534) .. (329.5622,132.6183) .. controls
      (281.3694,137.3507) and (239.0151,99.2627) .. (234.6921,51.8061) .. controls
      (229.8424,-1.4332) and (272.0115,-48.0991) .. (324.5115,-52.5763) .. controls
      (382.7966,-57.5470) and (433.7774,-11.2934) .. (438.4062,46.2503);
    \path[draw=black,line join=miter,line cap=butt,line width=0.800pt]
      (470.7311,73.5244) -- (462.6499,112.9203);
    \path[draw=black,line join=miter,line cap=butt,line width=0.800pt]
      (476.7920,35.1386) -- (499.0154,71.5041);
    \path[draw=black,line join=miter,line cap=butt,line width=0.800pt]
      (371.7361,19.9863) -- (409.1118,12.9152);
    \path[draw=black,line join=miter,line cap=butt,line width=0.800pt]
      (267.6904,41.1995) -- (295.9747,13.9254);
    \path[draw=black,line join=miter,line cap=butt,line width=0.800pt]
      (261.6295,102.8188) -- (288.9036,122.0117);
    \path[draw=black,line join=miter,line cap=butt,line width=0.800pt]
      (360.6245,134.1335) -- (413.1524,141.2046);
    \path[draw=black,line join=miter,line cap=butt,line width=0.800pt]
      (381.8377,91.7071) -- (401.0306,72.5142);
    \path[draw=black,line join=miter,line cap=butt,line width=0.800pt]
      (342.4417,62.4127) -- (320.2184,81.6056);
    \path[draw=black,fill=cffffff,miter limit=4.00,line width=0.800pt]
      (399.0103,45.7452) .. controls (399.0103,47.6978) and (397.4273,49.2807) ..
      (395.4747,49.2807) .. controls (393.5221,49.2807) and (391.9392,47.6978) ..
      (391.9392,45.7452) .. controls (391.9392,43.7925) and (393.5221,42.2096) ..
      (395.4747,42.2096) .. controls (397.4273,42.2096) and (399.0103,43.7925) ..
      (399.0103,45.7452) -- cycle;
    \path[shift={(37.88071,8.5863)},draw=black,fill=cffffff,miter limit=4.00,line
      width=0.800pt] (399.0103,45.7452) .. controls (399.0103,47.6978) and
      (397.4273,49.2807) .. (395.4747,49.2807) .. controls (393.5221,49.2807) and
      (391.9392,47.6978) .. (391.9392,45.7452) .. controls (391.9392,43.7925) and
      (393.5221,42.2096) .. (395.4747,42.2096) .. controls (397.4273,42.2096) and
      (399.0103,43.7925) .. (399.0103,45.7452) -- cycle;
    \path[shift={(23.73858,18.68782)},draw=black,fill=cffffff,miter limit=4.00,line
      width=0.800pt] (399.0103,45.7452) .. controls (399.0103,47.6978) and
      (397.4273,49.2807) .. (395.4747,49.2807) .. controls (393.5221,49.2807) and
      (391.9392,47.6978) .. (391.9392,45.7452) .. controls (391.9392,43.7925) and
      (393.5221,42.2096) .. (395.4747,42.2096) .. controls (397.4273,42.2096) and
      (399.0103,43.7925) .. (399.0103,45.7452) -- cycle;
    \path[shift={(-9.59646,32.82996)},draw=black,fill=cffffff,miter limit=4.00,line
      width=0.800pt] (399.0103,45.7452) .. controls (399.0103,47.6978) and
      (397.4273,49.2807) .. (395.4747,49.2807) .. controls (393.5221,49.2807) and
      (391.9392,47.6978) .. (391.9392,45.7452) .. controls (391.9392,43.7925) and
      (393.5221,42.2096) .. (395.4747,42.2096) .. controls (397.4273,42.2096) and
      (399.0103,43.7925) .. (399.0103,45.7452) -- cycle;
    \path[shift={(-37.88073,7.57614)},draw=black,fill=cffffff,miter limit=4.00,line
      width=0.800pt] (399.0103,45.7452) .. controls (399.0103,47.6978) and
      (397.4273,49.2807) .. (395.4747,49.2807) .. controls (393.5221,49.2807) and
      (391.9392,47.6978) .. (391.9392,45.7452) .. controls (391.9392,43.7925) and
      (393.5221,42.2096) .. (395.4747,42.2096) .. controls (397.4273,42.2096) and
      (399.0103,43.7925) .. (399.0103,45.7452) -- cycle;
    \path[shift={(-64.1447,34.85026)},draw=black,fill=cffffff,miter limit=4.00,line
      width=0.800pt] (399.0103,45.7452) .. controls (399.0103,47.6978) and
      (397.4273,49.2807) .. (395.4747,49.2807) .. controls (393.5221,49.2807) and
      (391.9392,47.6978) .. (391.9392,45.7452) .. controls (391.9392,43.7925) and
      (393.5221,42.2096) .. (395.4747,42.2096) .. controls (397.4273,42.2096) and
      (399.0103,43.7925) .. (399.0103,45.7452) -- cycle;
    \path[shift={(-24.74875,56.06347)},draw=black,fill=cffffff,miter limit=4.00,line
      width=0.800pt] (399.0103,45.7452) .. controls (399.0103,47.6978) and
      (397.4273,49.2807) .. (395.4747,49.2807) .. controls (393.5221,49.2807) and
      (391.9392,47.6978) .. (391.9392,45.7452) .. controls (391.9392,43.7925) and
      (393.5221,42.2096) .. (395.4747,42.2096) .. controls (397.4273,42.2096) and
      (399.0103,43.7925) .. (399.0103,45.7452) -- cycle;
    \path[shift={(27.77918,50.00255)},draw=black,fill=cffffff,miter limit=4.00,line
      width=0.800pt] (399.0103,45.7452) .. controls (399.0103,47.6978) and
      (397.4273,49.2807) .. (395.4747,49.2807) .. controls (393.5221,49.2807) and
      (391.9392,47.6978) .. (391.9392,45.7452) .. controls (391.9392,43.7925) and
      (393.5221,42.2096) .. (395.4747,42.2096) .. controls (397.4273,42.2096) and
      (399.0103,43.7925) .. (399.0103,45.7452) -- cycle;
    \path[shift={(46.97208,37.88072)},draw=black,fill=cffffff,miter limit=4.00,line
      width=0.800pt] (399.0103,45.7452) .. controls (399.0103,47.6978) and
      (397.4273,49.2807) .. (395.4747,49.2807) .. controls (393.5221,49.2807) and
      (391.9392,47.6978) .. (391.9392,45.7452) .. controls (391.9392,43.7925) and
      (393.5221,42.2096) .. (395.4747,42.2096) .. controls (397.4273,42.2096) and
      (399.0103,43.7925) .. (399.0103,45.7452) -- cycle;
    \path[shift={(-43.94165,65.15484)},draw=black,fill=cffffff,miter limit=4.00,line
      width=0.800pt] (399.0103,45.7452) .. controls (399.0103,47.6978) and
      (397.4273,49.2807) .. (395.4747,49.2807) .. controls (393.5221,49.2807) and
      (391.9392,47.6978) .. (391.9392,45.7452) .. controls (391.9392,43.7925) and
      (393.5221,42.2096) .. (395.4747,42.2096) .. controls (397.4273,42.2096) and
      (399.0103,43.7925) .. (399.0103,45.7452) -- cycle;
    \path[shift={(-90.40866,33.84011)},draw=black,fill=cffffff,miter limit=4.00,line
      width=0.800pt] (399.0103,45.7452) .. controls (399.0103,47.6978) and
      (397.4273,49.2807) .. (395.4747,49.2807) .. controls (393.5221,49.2807) and
      (391.9392,47.6978) .. (391.9392,45.7452) .. controls (391.9392,43.7925) and
      (393.5221,42.2096) .. (395.4747,42.2096) .. controls (397.4273,42.2096) and
      (399.0103,43.7925) .. (399.0103,45.7452) -- cycle;
    \path[shift={(-62.12439,-2.52538)},draw=black,fill=cffffff,miter limit=4.00,line
      width=0.800pt] (399.0103,45.7452) .. controls (399.0103,47.6978) and
      (397.4273,49.2807) .. (395.4747,49.2807) .. controls (393.5221,49.2807) and
      (391.9392,47.6978) .. (391.9392,45.7452) .. controls (391.9392,43.7925) and
      (393.5221,42.2096) .. (395.4747,42.2096) .. controls (397.4273,42.2096) and
      (399.0103,43.7925) .. (399.0103,45.7452) -- cycle;
    \path[shift={(-17.67768,-1.51523)},draw=black,fill=cffffff,miter limit=4.00,line
      width=0.800pt] (399.0103,45.7452) .. controls (399.0103,47.6978) and
      (397.4273,49.2807) .. (395.4747,49.2807) .. controls (393.5221,49.2807) and
      (391.9392,47.6978) .. (391.9392,45.7452) .. controls (391.9392,43.7925) and
      (393.5221,42.2096) .. (395.4747,42.2096) .. controls (397.4273,42.2096) and
      (399.0103,43.7925) .. (399.0103,45.7452) -- cycle;
    \path[shift={(-2.02031,-4.04061)},draw=black,fill=cffffff,miter limit=4.00,line
      width=0.800pt] (400.0204,85.6462) .. controls (400.0204,93.4567) and
      (389.8445,99.7883) .. (377.2920,99.7883) .. controls (364.7394,99.7883) and
      (354.5635,93.4567) .. (354.5635,85.6462) .. controls (354.5635,77.8357) and
      (364.7394,71.5041) .. (377.2920,71.5041) .. controls (389.8445,71.5041) and
      (400.0204,77.8357) .. (400.0204,85.6462) -- cycle;
    \path[shift={(-53.03302,48.9924)},draw=black,fill=cffffff,miter limit=4.00,line
      width=0.800pt] (399.0103,45.7452) .. controls (399.0103,47.6978) and
      (397.4273,49.2807) .. (395.4747,49.2807) .. controls (393.5221,49.2807) and
      (391.9392,47.6978) .. (391.9392,45.7452) .. controls (391.9392,43.7925) and
      (393.5221,42.2096) .. (395.4747,42.2096) .. controls (397.4273,42.2096) and
      (399.0103,43.7925) .. (399.0103,45.7452) -- cycle;
    \path[shift={(-86.36805,59.09392)},draw=black,fill=cffffff,miter limit=4.00,line
      width=0.800pt] (399.0103,45.7452) .. controls (399.0103,47.6978) and
      (397.4273,49.2807) .. (395.4747,49.2807) .. controls (393.5221,49.2807) and
      (391.9392,47.6978) .. (391.9392,45.7452) .. controls (391.9392,43.7925) and
      (393.5221,42.2096) .. (395.4747,42.2096) .. controls (397.4273,42.2096) and
      (399.0103,43.7925) .. (399.0103,45.7452) -- cycle;
    \path[shift={(19.69796,76.26652)},draw=black,fill=cffffff,miter limit=4.00,line
      width=0.800pt] (399.0103,45.7452) .. controls (399.0103,47.6978) and
      (397.4273,49.2807) .. (395.4747,49.2807) .. controls (393.5221,49.2807) and
      (391.9392,47.6978) .. (391.9392,45.7452) .. controls (391.9392,43.7925) and
      (393.5221,42.2096) .. (395.4747,42.2096) .. controls (397.4273,42.2096) and
      (399.0103,43.7925) .. (399.0103,45.7452) -- cycle;
    \path[shift={(62.12437,76.26652)},draw=black,fill=cffffff,miter limit=4.00,line
      width=0.800pt] (399.0103,45.7452) .. controls (399.0103,47.6978) and
      (397.4273,49.2807) .. (395.4747,49.2807) .. controls (393.5221,49.2807) and
      (391.9392,47.6978) .. (391.9392,45.7452) .. controls (391.9392,43.7925) and
      (393.5221,42.2096) .. (395.4747,42.2096) .. controls (397.4273,42.2096) and
      (399.0103,43.7925) .. (399.0103,45.7452) -- cycle;
    \path[shift={(72.2259,-24.74874)},draw=black,fill=cffffff,miter limit=4.00,line
      width=0.800pt] (399.0103,45.7452) .. controls (399.0103,47.6978) and
      (397.4273,49.2807) .. (395.4747,49.2807) .. controls (393.5221,49.2807) and
      (391.9392,47.6978) .. (391.9392,45.7452) .. controls (391.9392,43.7925) and
      (393.5221,42.2096) .. (395.4747,42.2096) .. controls (397.4273,42.2096) and
      (399.0103,43.7925) .. (399.0103,45.7452) -- cycle;
    \path[shift={(-53.03302,-22.72843)},draw=black,fill=cffffff,miter
      limit=4.00,line width=0.800pt] (399.0103,45.7452) .. controls
      (399.0103,47.6978) and (397.4273,49.2807) .. (395.4747,49.2807) .. controls
      (393.5221,49.2807) and (391.9392,47.6978) .. (391.9392,45.7452) .. controls
      (391.9392,43.7925) and (393.5221,42.2096) .. (395.4747,42.2096) .. controls
      (397.4273,42.2096) and (399.0103,43.7925) .. (399.0103,45.7452) -- cycle;
    \path[shift={(-137.88583,-0.50508)},draw=black,fill=cffffff,miter
      limit=4.00,line width=0.800pt] (399.0103,45.7452) .. controls
      (399.0103,47.6978) and (397.4273,49.2807) .. (395.4747,49.2807) .. controls
      (393.5221,49.2807) and (391.9392,47.6978) .. (391.9392,45.7452) .. controls
      (391.9392,43.7925) and (393.5221,42.2096) .. (395.4747,42.2096) .. controls
      (397.4273,42.2096) and (399.0103,43.7925) .. (399.0103,45.7452) -- cycle;
    \path[shift={(-124.75385,48.9924)},draw=black,fill=cffffff,miter limit=4.00,line
      width=0.800pt] (399.0103,45.7452) .. controls (399.0103,47.6978) and
      (397.4273,49.2807) .. (395.4747,49.2807) .. controls (393.5221,49.2807) and
      (391.9392,47.6978) .. (391.9392,45.7452) .. controls (391.9392,43.7925) and
      (393.5221,42.2096) .. (395.4747,42.2096) .. controls (397.4273,42.2096) and
      (399.0103,43.7925) .. (399.0103,45.7452) -- cycle;
    \path[shift={(113.64215,81.31728)},draw=black,fill=cffffff,miter limit=4.00,line
      width=0.800pt] (399.0103,45.7452) .. controls (399.0103,47.6978) and
      (397.4273,49.2807) .. (395.4747,49.2807) .. controls (393.5221,49.2807) and
      (391.9392,47.6978) .. (391.9392,45.7452) .. controls (391.9392,43.7925) and
      (393.5221,42.2096) .. (395.4747,42.2096) .. controls (397.4273,42.2096) and
      (399.0103,43.7925) .. (399.0103,45.7452) -- cycle;
    \path[shift={(102.53047,45.96194)},draw=black,fill=cffffff,miter limit=4.00,line
      width=0.800pt] (399.0103,45.7452) .. controls (399.0103,47.6978) and
      (397.4273,49.2807) .. (395.4747,49.2807) .. controls (393.5221,49.2807) and
      (391.9392,47.6978) .. (391.9392,45.7452) .. controls (391.9392,43.7925) and
      (393.5221,42.2096) .. (395.4747,42.2096) .. controls (397.4273,42.2096) and
      (399.0103,43.7925) .. (399.0103,45.7452) -- cycle;
    \path[shift={(-60.10409,95.45942)},draw=black,fill=cffffff,miter limit=4.00,line
      width=0.800pt] (399.0103,45.7452) .. controls (399.0103,47.6978) and
      (397.4273,49.2807) .. (395.4747,49.2807) .. controls (393.5221,49.2807) and
      (391.9392,47.6978) .. (391.9392,45.7452) .. controls (391.9392,43.7925) and
      (393.5221,42.2096) .. (395.4747,42.2096) .. controls (397.4273,42.2096) and
      (399.0103,43.7925) .. (399.0103,45.7452) -- cycle;
  \end{scope}
  \path[draw=black,line join=miter,line cap=butt,line width=0.480pt]
    (48.2873,210.5448) .. controls (74.9594,206.5036) and (101.6314,206.5036) ..
    (128.3034,210.5448) .. controls (180.5027,222.8713) and (224.6837,211.1432) ..
    (272.5747,210.5448) .. controls (302.3680,202.4576) and (331.0775,204.1241) ..
    (359.2587,210.5448) .. controls (378.5233,219.8884) and (391.8369,213.3626) ..
    (406.5409,210.5448);
  \path[draw=black,line join=miter,line cap=butt,line width=0.480pt]
    (48.2873,236.0045) .. controls (62.1468,231.6454) and (77.3497,229.1671) ..
    (99.2066,236.0045) .. controls (122.0048,245.3293) and (150.3676,244.4520) ..
    (182.8598,236.0045) .. controls (225.7239,236.0591) and (276.2779,251.4932) ..
    (311.3704,236.0045) .. controls (352.9980,247.3235) and (375.0874,236.3131) ..
    (406.5409,236.0045);
  \path[fill=black] (190.05081,234.54469) node[above right] (text4098) {Heisenberg
    Cut};
  \path[fill=black] (61.619301,222.46349) node[above right] (text4102) {?};
  \path[fill=black] (91.593994,227.66312) node[above right] (text4102-2) {?};
  \path[fill=black] (134.0204,231.7037) node[above right] (text4102-7) {?};
  \path[fill=black] (170.38589,229.6631) node[above right] (text4102-21) {?};
  \path[fill=black] (305.74631,224.63266) node[above right] (text4102-8) {?};
  \path[fill=black] (336.05087,229.68343) node[above right] (text4102-5) {?};
  \path[fill=black] (358.27426,226.65292) node[above right] (text4102-82) {?};
  \path[fill=black] (379.48743,229.6834) node[above right] (text4102-0) {?};
  \begin{scope}[cm={{0.51568,0.3069,-0.3069,0.51568,(204.20183,-46.49847)}}]
    \path[draw=black,miter limit=4.00,line width=0.800pt] (291.9341,454.3519) ..
      controls (291.9341,458.2571) and (288.7683,461.4229) .. (284.8630,461.4229) ..
      controls (280.9578,461.4229) and (277.7919,458.2571) .. (277.7919,454.3519) ..
      controls (277.7919,450.4466) and (280.9578,447.2808) .. (284.8630,447.2808) ..
      controls (288.7683,447.2808) and (291.9341,450.4466) .. (291.9341,454.3519) --
      cycle;
    \path[shift={(-1.01015,-4.0)},draw=black,miter limit=4.00,line width=0.800pt]
      (311.1270,458.8976) .. controls (311.1270,463.6396) and (299.8205,467.4839) ..
      (285.8732,467.4839) .. controls (271.9259,467.4839) and (260.6194,463.6396) ..
      (260.6194,458.8976) .. controls (260.6194,454.1555) and (271.9259,450.3113) ..
      (285.8732,450.3113) .. controls (299.8205,450.3113) and (311.1270,454.1555) ..
      (311.1270,458.8976) -- cycle;
    \path[cm={{0.79622,0.60501,-0.60501,0.79622,(335.89171,-82.97669)}},draw=black,miter
      limit=4.00,line width=0.800pt] (311.1270,458.8976) .. controls
      (311.1270,463.6396) and (299.8205,467.4839) .. (285.8732,467.4839) .. controls
      (271.9259,467.4839) and (260.6194,463.6396) .. (260.6194,458.8976) .. controls
      (260.6194,454.1555) and (271.9259,450.3113) .. (285.8732,450.3113) .. controls
      (299.8205,450.3113) and (311.1270,454.1555) .. (311.1270,458.8976) -- cycle;
    \path[cm={{0.83229,-0.55435,0.55435,0.83229,(-206.44342,230.8909)}},draw=black,miter
      limit=4.00,line width=0.800pt] (311.1270,458.8976) .. controls
      (311.1270,463.6396) and (299.8205,467.4839) .. (285.8732,467.4839) .. controls
      (271.9259,467.4839) and (260.6194,463.6396) .. (260.6194,458.8976) .. controls
      (260.6194,454.1555) and (271.9259,450.3113) .. (285.8732,450.3113) .. controls
      (299.8205,450.3113) and (311.1270,454.1555) .. (311.1270,458.8976) -- cycle;
  \end{scope}
  \begin{scope}[cm={{0.53951,-0.26275,0.26275,0.53951,(66.11385,96.63684)}}]
    \path[draw=black,miter limit=4.00,line width=0.800pt] (291.9341,454.3519) ..
      controls (291.9341,458.2571) and (288.7683,461.4229) .. (284.8630,461.4229) ..
      controls (280.9578,461.4229) and (277.7919,458.2571) .. (277.7919,454.3519) ..
      controls (277.7919,450.4466) and (280.9578,447.2808) .. (284.8630,447.2808) ..
      controls (288.7683,447.2808) and (291.9341,450.4466) .. (291.9341,454.3519) --
      cycle;
    \path[shift={(-1.01015,-4.0)},draw=black,miter limit=4.00,line width=0.800pt]
      (311.1270,458.8976) .. controls (311.1270,463.6396) and (299.8205,467.4839) ..
      (285.8732,467.4839) .. controls (271.9259,467.4839) and (260.6194,463.6396) ..
      (260.6194,458.8976) .. controls (260.6194,454.1555) and (271.9259,450.3113) ..
      (285.8732,450.3113) .. controls (299.8205,450.3113) and (311.1270,454.1555) ..
      (311.1270,458.8976) -- cycle;
    \path[cm={{0.79622,0.60501,-0.60501,0.79622,(335.89171,-82.97669)}},draw=black,miter
      limit=4.00,line width=0.800pt] (311.1270,458.8976) .. controls
      (311.1270,463.6396) and (299.8205,467.4839) .. (285.8732,467.4839) .. controls
      (271.9259,467.4839) and (260.6194,463.6396) .. (260.6194,458.8976) .. controls
      (260.6194,454.1555) and (271.9259,450.3113) .. (285.8732,450.3113) .. controls
      (299.8205,450.3113) and (311.1270,454.1555) .. (311.1270,458.8976) -- cycle;
    \path[cm={{0.83229,-0.55435,0.55435,0.83229,(-206.44342,230.8909)}},draw=black,miter
      limit=4.00,line width=0.800pt] (311.1270,458.8976) .. controls
      (311.1270,463.6396) and (299.8205,467.4839) .. (285.8732,467.4839) .. controls
      (271.9259,467.4839) and (260.6194,463.6396) .. (260.6194,458.8976) .. controls
      (260.6194,454.1555) and (271.9259,450.3113) .. (285.8732,450.3113) .. controls
      (299.8205,450.3113) and (311.1270,454.1555) .. (311.1270,458.8976) -- cycle;
  \end{scope}
  \begin{scope}[cm={{0.60009,0.0,0.0,0.60009,(-31.12251,-9.03108)}}]
    \path[draw=black,miter limit=4.00,line width=0.800pt] (291.9341,454.3519) ..
      controls (291.9341,458.2571) and (288.7683,461.4229) .. (284.8630,461.4229) ..
      controls (280.9578,461.4229) and (277.7919,458.2571) .. (277.7919,454.3519) ..
      controls (277.7919,450.4466) and (280.9578,447.2808) .. (284.8630,447.2808) ..
      controls (288.7683,447.2808) and (291.9341,450.4466) .. (291.9341,454.3519) --
      cycle;
    \path[shift={(-1.01015,-4.0)},draw=black,miter limit=4.00,line width=0.800pt]
      (311.1270,458.8976) .. controls (311.1270,463.6396) and (299.8205,467.4839) ..
      (285.8732,467.4839) .. controls (271.9259,467.4839) and (260.6194,463.6396) ..
      (260.6194,458.8976) .. controls (260.6194,454.1555) and (271.9259,450.3113) ..
      (285.8732,450.3113) .. controls (299.8205,450.3113) and (311.1270,454.1555) ..
      (311.1270,458.8976) -- cycle;
    \path[cm={{0.79622,0.60501,-0.60501,0.79622,(335.89171,-82.97669)}},draw=black,miter
      limit=4.00,line width=0.800pt] (311.1270,458.8976) .. controls
      (311.1270,463.6396) and (299.8205,467.4839) .. (285.8732,467.4839) .. controls
      (271.9259,467.4839) and (260.6194,463.6396) .. (260.6194,458.8976) .. controls
      (260.6194,454.1555) and (271.9259,450.3113) .. (285.8732,450.3113) .. controls
      (299.8205,450.3113) and (311.1270,454.1555) .. (311.1270,458.8976) -- cycle;
    \path[cm={{0.83229,-0.55435,0.55435,0.83229,(-206.44342,230.8909)}},draw=black,miter
      limit=4.00,line width=0.800pt] (311.1270,458.8976) .. controls
      (311.1270,463.6396) and (299.8205,467.4839) .. (285.8732,467.4839) .. controls
      (271.9259,467.4839) and (260.6194,463.6396) .. (260.6194,458.8976) .. controls
      (260.6194,454.1555) and (271.9259,450.3113) .. (285.8732,450.3113) .. controls
      (299.8205,450.3113) and (311.1270,454.1555) .. (311.1270,458.8976) -- cycle;
  \end{scope}
  \begin{scope}[cm={{0.60009,0.0,0.0,0.60009,(85.46847,-28.43072)}}]
    \path[draw=black,line join=miter,line cap=butt,line width=0.800pt]
      (339.4113,533.1438) -- (385.8783,568.4991) -- (342.4417,600.8240);
    \path[draw=black,line join=miter,line cap=butt,line width=0.800pt]
      (384.8681,567.4890) .. controls (391.1085,570.3478) and (398.0143,574.8703) ..
      (399.0691,564.7655) .. controls (402.4175,553.8822) and (409.1672,560.0058) ..
      (415.0319,561.7041) .. controls (419.8901,569.6212) and (424.7484,564.1063) ..
      (429.6066,558.9090) .. controls (430.4308,548.3951) and (437.7022,553.9993) ..
      (443.6208,556.2213) .. controls (446.7361,568.4291) and (452.9224,559.1400) ..
      (458.6093,553.3468) -- (488.9138,510.9204);
    \path[draw=black,line join=miter,line cap=butt,line width=0.800pt]
      (458.6093,554.3570) -- (486.8935,592.7428);
  \end{scope}
  \begin{scope}[cm={{0.60009,0.0,0.0,0.60009,(55.56152,-29.64127)}}]
    \path[draw=black,line join=miter,line cap=butt,line width=0.800pt]
      (82.8325,529.1032) -- (116.1675,562.4382) .. controls (119.5725,567.4921) and
      (127.1823,572.5459) .. (115.3252,577.5998) -- (105.5563,588.7861) --
      (114.1472,598.8037) -- (78.7919,620.0169);
    \path[draw=black,line join=miter,line cap=butt,line width=0.800pt]
      (113.1371,597.7935) -- (154.5533,625.0677);
    \path[draw=black,line join=miter,line cap=butt,line width=0.800pt]
      (116.1675,562.4382) -- (167.6853,536.1742);
  \end{scope}
  \path[cm={{0.97255,0.23269,-0.23269,0.97255,(0.0,0.0)}},fill=black]
    (125.96268,237.67122) node[above right] (text4403) {$\Psi$};
  \path[cm={{0.97053,-0.24098,0.24098,0.97053,(0.0,0.0)}},fill=black]
    (191.92426,329.3291) node[above right] (text4403-4) {$\Psi$};
  \path[cm={{0.99133,-0.13138,0.13138,0.99133,(0.0,0.0)}},fill=black]
    (151.6591,340.97943) node[above right] (text4403-7) {$\Psi$};
\end{scope}

\end{tikzpicture}
\par\end{center}
\par\end{centering}

\caption{\label{fig:HeisenbergCut}The Heisenberg cut.}
\end{figure}
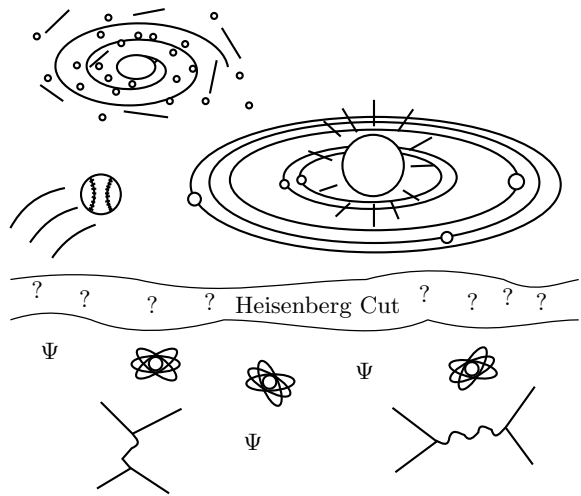

Unfortunately, the Copenhagen interpretation also suffers from a number
of serious drawbacks. Most significantly, the definition of a measurement
according to the Copenhagen interpretation relies on a questionable
demarcation, known as the Heisenberg cut (\emph{Heisenbergscher Schnitt})
\cite{vonNeumann:1932mgdq,Landsman:2007bcq}, between the large classical
systems that carry out measurements and the small quantum systems
that they measure; this ill-defined Heisenberg cut has never been
identified in any experiment to date and must be put into the interpretation
by hand. (See Figure~\ref{fig:HeisenbergCut}.) An associated issue
is the interpretation's assumption of wave-function collapse---known
more formally as the Von Neumann-Lüders projection postulate \cite{vonNeumann:1955mfqm,Luders:1950udzddm}---by
which we refer to the supposed instantaneous, discontinuous change
in a quantum system immediately following a measurement by a classical
system, in stark contrast to the smooth time evolution that governs
dynamically closed systems.

The Copenhagen interpretation is also unclear as to the ultimate meaning
of the state vector of a system: Does a system's state vector merely
represent the experimenter's knowledge, is it some sort of objective
probability distribution,%
\footnote{Recent work \cite{PuseyBarrettRudolph:2012rqs,BarrettCavalcantiLalMaroney:2013rqs,ColbeckRenner:2013swfudups}
casts considerable doubt on assertions that state vectors are nothing
more than probability distributions over more fundamental ingredients
of reality.%
} or is it an irreducible ingredient of reality like the state of a
classical system? For that matter, what constitutes an observer,
and can we meaningfully talk about the state of an observer within
the formalism of quantum theory? Given that no realistic system is
ever perfectly free of quantum entanglements with other systems, and
thus no realistic system can ever truly be assigned a specific state
vector in the first place, what becomes of the popular depiction of
quantum theory in which every particle is supposedly described by
a specific wave function propagating in three-dimensional space? The
Copenhagen interpretation leads to additional trouble when trying
to make sense of thought experiments like Schrödinger's cat, Wigner's friend,
and the quantum Zeno paradox.%
\footnote{We will discuss all of these thought experiments in Section~\ref{sub:Paradoxes-of-Quantum-Theory-Revisited}.%
}

\subsubsection{An Ideal Interpretation\label{subsub:An-Ideal-Interpretation}}

Physicists and philosophers have expended much effort over many decades
on the search for an alternative interpretation of quantum theory
that could resolve these problems. Ideally, such an interpretation
would eliminate the need for an \emph{ad hoc} Heisenberg cut, thereby
demoting measurements to an ordinary kind of interaction and allowing
quantum theory to be a complete theory that seamlessly encompasses
\emph{all} systems in Nature, including observers as physical systems
with quantum states of their own. Moreover, an acceptable interpretation
should fundamentally (even if not always \emph{superficially}) be
consistent with all experimental data and other reliably known features
of Nature, including relativity, and should be general enough to accommodate
the large variety of both presently known and hypothetical physical
systems. Such an interpretation should also address the key no-go
theorems developed over the years by researchers working on the foundations
of quantum theory, should not depend on concepts or quantities whose
definitions require a physically unrealistic measure-zero level of
sharpness, and should be insensitive to potentially unknowable features
of reality, such as whether we can sensibly define ``the universe
as a whole'' as being a closed or open system.

\subsubsection{Instrumentalism\label{subsub:Instrumentalism}}

In principle, an alternative to this project is always available:
One could instead simply insist upon instrumentalism---known in some
quarters as the ``shut-up-and-calculate approach'' \cite{Mermin:1989wwp,Mermin:2004cfhst}---meaning
that one should regard the mathematical formalism of quantum theory
merely as a calculational recipe or algorithm for predicting measurement
results and empirical outcome probabilities obtained by the kinds
of physical systems (agents) that can self-consistently act as observers,
and, furthermore, that one should refuse to make definitive metaphysical
claims about any underlying reality.%
\footnote{``Whereof one cannot speak, thereof one must be silent.'' \cite{Wittgenstein:1921tlp}%
} We provide a more extensive description of the instrumentalist approach
in Appendix~\ref{subsub:The-Instrumentalist-Approach}.

On the other hand, if the physics community had uniformly accepted
instrumentalism from the very beginning of the history of quantum
theory, then we might have missed out on the many important spin-offs
from the search for a better interpretation: Decoherence, quantum
information, quantum computing, quantum cryptography, and the black-hole
information paradox are just a few of the far-reaching ideas that
ultimately owe their origin to people thinking seriously about the
meaning of quantum theory.

\subsection{Our Interpretation\label{sub:Our-Interpretation}}

In this paper and in a brief companion letter \cite{BarandesKagan:2014nmmiqt},
we present a realist interpretation of quantum theory that hews closely
to the basic structure of the theory in its widely accepted current
form. Our primary goal is to move beyond instrumentalism and describe
an actual reality that lies behind the mathematical formalism of quantum
theory. We also intend to provide new hope to those who find themselves
disappointed with the pace of progress on making sense of the theory's
foundations \cite{Weinberg:2013qasw,Weinberg:2012loqm}.

Our interpretation is fully quantum in nature. However, for purposes
of motivation, consider the basic theoretical structure of classical
physics: A classical system has a specific state that evolves in time
through the system's configuration space according to some dynamical
rule that may or may not be stochastic, and this dynamical rule exists
whether or not the system's state lies beneath a nontrivial evolving
probability distribution on the system's configuration space; moreover,
the dynamical rule for the system's underlying state is consistent
with the overall evolution of the system's probability distribution,
in the sense that if we consider a probabilistic ensemble over the
system's initial underlying state and apply the dynamical rule to
each underlying state in the ensemble, then we correctly obtain the
overall evolution of the system's probability distribution as a whole.

In particular, note the \emph{insufficiency} of specifying a dynamical
rule \emph{solely} for the evolution of the system's overall probability
distribution but \emph{not} for the system's underlying state itself,
because then the system's underlying state would be free to fluctuate
wildly and discontinuously between macroscopically distinct configurations.
For example, even if a classical system's probability distribution
describes constant probabilities $p_{1}$ and $p_{2}$ for the system
to be in macroscopically distinct states $q_{1}$ or $q_{2}$, there
would be nothing preventing the system's state from hopping discontinuously
between $q_{1}$ and $q_{2}$ with respective frequency ratios $p_{1}$
and $p_{2}$ over arbitrarily short time intervals. Essentially, by
imposing a dynamical rule on the system's underlying state, we can
provide a ``smoothness condition'' for the system's physical configuration
over time and thus eliminate these kinds of instabilities.

In quantum theory, a system that is \emph{exactly} closed and that
is \emph{exactly} in a pure state (both conditions that are unphysical
idealizations) evolves along a well-defined trajectory through the
system's Hilbert space according to a well-known dynamical rule, namely,
the Schrödinger equation. However, in traditional formulations of
quantum theory, an open quantum system that must be described by a
density matrix due to entanglements with other systems---a so-called
improper mixture---does not have a specific underlying state vector,
let alone a Hilbert-space trajectory or a dynamical rule governing
the time evolution of such an underlying state vector and consistent
with the overall evolution of the system's density matrix. It is a
chief goal of our interpretation of quantum theory to provide these
missing ingredients.

\subsection{Conceptual Summary\label{sub:Conceptual-Summary}}

We present a technical summary of our interpretation in Section~\ref{sub:Summary}.
In short, for a quantum system in an improper mixture, our interpretation
identifies the \emph{eigenstates} of the system's density matrix with
the \emph{possible states} of the system in reality and identifies
the \emph{eigenvalues} of that density matrix with the \emph{probabilities}
that one of those possible states is \emph{actually} occupied,%
\footnote{In the language of \cite{Weinberg:2012loqm}, our interpretation
therefore comports ``with the idea that the state of a physical system
is described by a vector in {[}a{]} Hilbert space rather than by numerical
values of the positions and momenta of all the particles in the system,''
and is not an interpretation ``with no description of physical states
at all.'' That is, our interpretation is not ``only an algorithm
for calculating probabilities.''%
} and introduces just enough minimal structure beyond this simple picture
to provide a dynamical rule for underlying state vectors as they evolve
along Hilbert-space trajectories and to evade criticisms made in the
past regarding similar interpretations. This minimal additional structure
consists of a single additional class of conditional probabilities
amounting essentially to a series of smoothness conditions that kinematically
relate the states of parent systems to the states of their subsystems,
as well as dynamically relate the states of a single system to each
other over time.

\subsection{Comparison with Other Interpretations of Quantum Theory\label{sub:Comparison-with-Other-Interpretations-of-Quantum-Theory}}

Our interpretation, which builds on the work of many others, is general,
model-independent, and encompasses relativistic systems, but is also
conservative and unextravagant: It includes only metaphysical objects
that are either already a standard part of quantum theory or that
have counterparts in classical physics. We do not posit the existence
of exotic ``many worlds'' \cite{Everett:1957rsfqm,Wheeler:1957aersfqt,DeWitt:1970qmr,EverettDeWittGraham:1973mwiqm,Everett:1973tuwf,Deutsch:1985qtupt,Deutsch:1999qtpd,Wallace:2002wei,Wallace:2003erddapei,BrownWallace:2004smpdbble},
physical ``pilot waves'' \cite{deBroglie:1930iswm,Bohm:1952siqtthvi,Bohm:1952siqtthvii,BohmHiley:1993uu},
or any fundamental GRW-type dynamical-collapse or spontaneous-localization
modifications to quantum theory \cite{GhirardiRiminiWeber:1986udmms,Penrose:1989enm,Pearle:1989csdsvrwsl,BassiGhirardi:2003drm,Weinberg:2011csv,AdlerBassi:2009qte}.
Indeed, our interpretation leaves the widely accepted mathematical
structure of quantum theory essentially intact.%
\footnote{Moreover, our interpretation does not introduce any new violations
of time-reversal symmetry, and gives no fundamental role to  relative
states \cite{Everett:1957rsfqm}; a cosmic multiverse or self-locating
uncertainty \cite{AguirreTegmark:2011biuciqm}; coarse-grained histories
or decoherence functionals \cite{Hartle:1991scgnqm,Hartle:1993sqmqms,GellMannHartle:1993ceqs,GellMannHartle:2013acgesdqr};
decision theory \cite{Deutsch:1999qtpd,Wallace:2002qpdtr,Wallace:2009fpbrdta};
Dutch-book coherence, SIC-POVMs, or \emph{urgleichungs} \cite{CavesFuchsSchack:2002qpbp,CavesFuchsSchack:2007spqc,FuchsSchack:2009qbc,Fuchs:2010qbp,Fuchs:2010qbpqb,Fuchs:2011qbrqss,ApplebyEricssonFuchs:2011pqbss,Fuchs:2012bcrpqd,Fuchs:2012iqb,FuchsSchack:2013qbcnnv};
circular frequentist arguments involving unphysical ``limits'' of
infinitely many copies of measurement devices \cite{FarhiGoldstoneGutmann:1989hpaqm,CavesSchack:2005pfodniqpp,BuniyHsuZee:2006dopqm,AguirreTegmark:2011biuciqm};
infinite imaginary ensembles \cite{Ballentine:1970siqm,Ballentine:1998qmmd};
quantum reference systems or perspectivalism \cite{Bene:1997qrsnfqm,Bene:1997qpdnvpl,Bene:1997onqsms,Bene:2001qrsrlqm,Bene:2001rmirqft,BeneDieks:2002pvmiqm};
relational or non-global quantum states \cite{Rovelli:1996rqm,SmerlakRovelli:2007repr,BerkovitzHemmo:2005miqmrr,Hollowood:2013eciqm,Hollowood:2013nmqm};
many-minds states \cite{AlbertLoewer:1988imwi}; mirror states \cite{Hollowood:2013eciqm,Hollowood:2013nmqm};
faux-Boolean algebras \cite{Dickson:1995fbacm,Dickson:1995fbacpd,Vermaas:1999puqm};
``atomic'' subsystems \cite{BacciagaluppiDickson:1997gqtm,BacciagaluppiDickson:1999dmi};
algebraic quantum field theory \cite{EarmanRuetsche:2005rimi}; secret
classical superdeterminism or fundamental information loss \cite{tHooft:2005dbqm,tHooft:2007mbdqm,tHooft:2011hwfccwvse,tHooft:2014sfqm};
cellular automata \cite{tHooft:2012dds,tHooft:2014caiqm}; classical
matrix degrees of freedom or trace dynamics \cite{Adler:2002sdguimmpqm,Adler:2013igtdiga};
or discrete Hilbert spaces or appeals to unknown Planck-scale physics
\cite{BuniyHsuZee:2005hsd,BuniyHsuZee:2006dopqm}.%
} At the same time, we will argue that our interpretation is ultimately
compatible with Lorentz invariance and is nonlocal only in the mild
sense familiar from the framework of classical gauge theories.

Furthermore, we make no assumptions about as-yet-unknown aspects of
reality, such as the fundamental discreteness or continuity of time
or the dimensionality of the ultimate Hilbert space of Nature. Nor
does our interpretation rely in any crucial way upon the existence
of a well-defined maximal parent system that encompasses all other
systems and is dynamically closed in the sense of having a so-called
cosmic pure state or universal wave function\emph{ }that \emph{precisely}
obeys the Schrödinger equation; by contrast, this sort of cosmic assumption
is a necessarily \emph{exact} ingredient in the traditional formulations
of the de Broglie-Bohm pilot-wave interpretation \cite{deBroglie:1930iswm,Bohm:1952siqtthvi,Bohm:1952siqtthvii,BohmHiley:1993uu}
and the Everett-DeWitt many-worlds interpretation \cite{Everett:1957rsfqm,Wheeler:1957aersfqt,DeWitt:1970qmr,EverettDeWittGraham:1973mwiqm,Everett:1973tuwf,Deutsch:1985qtupt,Deutsch:1999qtpd,Wallace:2002wei,Wallace:2003erddapei,BrownWallace:2004smpdbble}.
(Given their stature among interpretations of quantum theory, we will
have more to say about the de Broglie-Bohm interpretation in Section~\ref{subsub:The-de-Broglie-Bohm-Pilot-Wave-Interpretation-of-Quantum-Theory}
and the Everett-DeWitt interpretation in Sections~\ref{subsub:The-Everett-DeWitt-Many-Worlds-Interpretation-of-Quantum-Theory}
and \ref{subsub:Nonlocality in the Everett-DeWitt Many-Worlds Interpretation}.)

Indeed, by considering merely the \emph{possibility} that our observable
universe is but a small region of an eternally inflating cosmos of
indeterminate spatial size and age \cite{Linde:1986eci,Linde:1986eesrciu,GarrigaVilenkin:1998ru,Guth:2000iei,GarrigaVilenkin:2001mwo,GuthKaiser:2005iei,Susskind:2005cl,Vilenkin:2006mwo,Guth:2007eii,GarrigaGuthVilenkin:2007eibcpm,Anninos:2012dsm},
it becomes clear that the idea of a biggest closed system (``the
universe as a whole'') may not generally be a sensible or empirically
verifiable concept to begin with, let alone an axiom on which a robust
interpretation of quantum theory can safely rely. Our interpretation
certainly allows for the existence of a biggest closed system, but
is also fully able to accommodate the alternative circumstance that
if we were to imagine gradually enlarging our scope to parent systems
of increasing physical size, then we might well find that the hierarchical
sequence never terminates at any maximal, dynamically closed system,
but may instead lead to an unending ``Russian-doll'' succession
of ever-more-open parent systems.%
\footnote{One might try to argue that one can always \emph{formally} define
a closed maximal parent system just to be ``the system containing
all systems.'' Whatever logicians might say about such a construction,
we run into the more prosaic issue that if we cannot construct this
closed maximal parent system via a well-defined succession of parent
systems of incrementally increasing size, then it becomes unclear
mathematically how we can generally define any human-scale system
as a subsystem of the maximal parent system and thereby define the
partial-trace operation, to be described in detail in Section~\ref{subsub:The-Partial-Trace-Operation}.
Furthermore, if our observable cosmic region is indeed an open system,
then its own time evolution may not be exactly linear, in which case
it is far from obvious that we can safely and rigorously embed that
open-system time evolution into the hypothetical unitary dynamics
of any conceivable closed parent system.%
}

\subsection{Outline of this Paper\label{sub:Outline-of-this-Paper}}

In Section~\ref{sec:Preliminary-Concepts}, we lay down the conceptual
groundwork for our interpretation and review features of classical
physics whose quantum counterparts will play an important role. In
Section~\ref{sec:The-Minimal-Modal-Interpretation}, we define our
interpretation of quantum theory in precise detail and compare it
to some of the other prominent interpretations, as well as identify
an important nontrivial geometrical structure, herein called subsystem space,
that has been lurking in quantum theory all along.

Next, in Section~\ref{sec:The-Measurement-Process}, we describe
how our interpretation makes sense of the measurement process, provide
a first-principles derivation of the Born rule for computing empirical
outcome probabilities, and discuss possible corrections to the naïve
Born rule that are otherwise invisible in the traditional Copenhagen
interpretation according to which, in contrast to our own interpretation,
the Born rule is taken axiomatically to be an exact statement about
reality. We also revisit several familiar ``paradoxes'' of quantum
theory, including Schrödinger's cat, Wigner's friend, and the quantum Zeno paradox.

In Section~\ref{sec:Lorentz-Invariance-and-Locality}, we study issues
of locality and Lorentz invariance and consider several well-known
thought experiments and no-go theorems. Additionally, we show that
our interpretation evades claims \cite{DicksonClifton:1998limi,Myrvold:2002mir,EarmanRuetsche:2005rimi,BerkovitzHemmo:2005miqmrr,Myrvold:2009cc}
that interpretations similar to our own are incompatible with Lorentz
invariance and necessarily depend on the existence of a ``preferred''
inertial reference frame. We also address the question of nonlocality
more generally, and argue that the picture of quantum theory that
emerges from our interpretation is no more nonlocal than are classical
gauge theories.

We conclude in Section~\ref{sec:Conclusion}, which contains a concise
summary of our interpretation as well as a discussion of falsifiability,
future research directions, and relevant philosophical issues. In
our \hyperlink{hypertarget:appendix}{appendix}, we present a brief
summary of the Copenhagen interpretation and the measurement problem
of quantum theory---including a systematic analysis of the various
ways that the prominent interpretations have attempted to solve the
measurement problem---as well as a description of the instrumentalist
approach to quantum theory and a summary of several important foundational
theorems not covered in the main text.

\section{Preliminary Concepts\label{sec:Preliminary-Concepts}}

\subsection{Ontology and Epistemology\label{sub:Ontology-and-Epistemology}}

Our aim is to present our interpretation of quantum theory in a language
familiar to physicists. We make an allowance, however, for a small
amount of widely used, model-independent philosophical terminology
that will turn out to be very helpful.

For our purposes, we will use the term ``ontology'' (adjective
``ontic'') to refer to a state of being or objective physical existence---that
is, the way things, including ourselves as physical observers, \emph{really
are}.%
\footnote{Bell coined the term ``beables'' (that is, ``be-ables'') to refer
to ontic states of \emph{being} or \emph{reality}, as opposed to the
mere ``observables'' (``observe-ables'') that appear in experiments.
As Bell wrote \cite{Bell:2004suqm}, ``The beables of the theory
are those elements which might correspond to elements of reality,
to things which exist. Their existence does not depend on `observation.'\char`\"{}
Today, the term ``beables'' is often used in the quantum foundations
community to refer more specifically to a \emph{fixed} and \emph{unchanging}
set of elements of reality, such as the coordinate basis in the original
nonrelativistic formulation of the de Broglie-Bohm interpretation
of quantum theory.%
} This language is crucial for being able to talk about \emph{realist}
interpretations of quantum theory such as our own.

Meanwhile, we will use the term ``epistemology'' (adjective ``epistemic'')
to refer to knowledge or information regarding a particular ontic
piece of reality. We can further subdivide epistemology into subjective
and objective parts: The subjective component refers to information
that a particular observer possesses about an ontic piece of reality,
whereas the objective component refers to the information an ontic
piece of reality reveals about itself to the rest of the world beyond
it---meaning the \emph{most complete} information that any observer
could possibly possess about that piece of reality.

All successful scientific theories and models make predictions about
things we can directly or indirectly observe, but until we attach
an interpretation, we can't really talk of an underlying ontology
or its associated epistemology. To say that one's interpretation of
a mathematical theory of physics adds an ontology and an associated
epistemology to the theory is to say that one is establishing some
sort of connection between, on the one hand, the mathematical objects
of the theory, and, on the other hand, the ontic states of things
as they really exist as well as what epistemic information those ontic
ingredients of reality make known about themselves and can be known
by observers.

We regard an interpretation of a mathematical theory of physics as
being realist if it asserts an underlying ontology of some kind,
and anti-realist if it asserts otherwise. We call an interpretation
agnostic if it refrains from making any definitive claims about an
ontology, either whether one exists at all or merely whether we can
say anything specific about it. In the sense of these definitions,
instrumentalism is agnostic, whereas the interpretation we introduce
in this paper is realist.

\subsection{Classical Theories\label{sub:Classical-Theories}}

Before laying out our interpretation of quantum theory in detail,
we consider the salient features of the classical story, presented
as generally as possible and in a manner intended to clarify the conceptual
parallels and distinctions with the key ingredients that we'll need
in the quantum case.

\subsubsection{Classical Kinematics and Ontology}

\begin{figure}
\begin{centering}
\begin{center}
\begin{tikzpicture}[y=0.80pt, x=0.8pt,yscale=-1, inner sep=0pt, outer sep=0pt]
\begin{scope}[shift={(-128.70036,-26.10314)}]
  \begin{scope}[cm={{0.0,-0.54041,1.0,0.0,(114.19047,275.35636)}}]
    \path[draw=black,line join=miter,line cap=butt,line width=0.800pt]
      (219.2857,48.0765) .. controls (223.5714,6.6479) and (320.7143,21.6479) ..
      (388.5714,51.6479) .. controls (456.4286,81.6479) and (486.4286,133.7908) ..
      (433.5714,162.3622) .. controls (380.7143,190.9336) and (423.5714,183.0765) ..
      (436.4286,236.6479) .. controls (449.2857,290.2193) and (372.8571,360.9336) ..
      (272.1429,328.0765) .. controls (171.4286,295.2193) and (132.1429,244.5050) ..
      (180.0000,198.7907) .. controls (202.0789,177.7005) and (193.7594,173.0296) ..
      (193.6902,168.0848) .. controls (193.6094,162.3115) and (196.9930,156.1650) ..
      (154.2857,123.0765) .. controls (75.0000,61.6479) and (271.4286,120.9336) ..
      (219.2857,48.0765) -- cycle;
  \end{scope}
  \path[cm={{0.70053,0.0,0.0,0.70053,(11.14695,-12.70093)}},draw=black,fill=black,miter
    limit=4.00,line width=0.800pt]
    (346.4823,150.8010)arc(-0.058:180.058:2.525)arc(-180.058:0.058:2.525) --
    cycle;
  \path[cm={{0.70053,0.0,0.0,0.70053,(11.50077,-5.2707)}},draw=black,fill=black,miter
    limit=4.00,line width=0.800pt]
    (346.4823,150.8010)arc(-0.058:180.058:2.525)arc(-180.058:0.058:2.525) --
    cycle;
  \path[cm={{0.70053,0.0,0.0,0.70053,(3.62358,-13.91118)}},draw=black,fill=black,miter
    limit=4.00,line width=0.800pt]
    (346.4823,150.8010)arc(-0.058:180.058:2.525)arc(-180.058:0.058:2.525) --
    cycle;
  \path[cm={{0.70053,0.0,0.0,0.70053,(3.71672,-20.13117)}},draw=black,fill=black,miter
    limit=4.00,line width=0.800pt]
    (346.4823,150.8010)arc(-0.058:180.058:2.525)arc(-180.058:0.058:2.525) --
    cycle;
  \path[cm={{0.70053,0.0,0.0,0.70053,(15.74662,-23.66937)}},draw=black,fill=black,miter
    limit=4.00,line width=0.800pt]
    (346.4823,150.8010)arc(-0.058:180.058:2.525)arc(-180.058:0.058:2.525) --
    cycle;
  \path[cm={{0.70053,0.0,0.0,0.70053,(23.53067,-13.7624)}},draw=black,fill=black,miter
    limit=4.00,line width=0.800pt]
    (346.4823,150.8010)arc(-0.058:180.058:2.525)arc(-180.058:0.058:2.525) --
    cycle;
  \path[cm={{0.70053,0.0,0.0,0.70053,(25.6536,-5.97834)}},draw=black,fill=black,miter
    limit=4.00,line width=0.800pt]
    (346.4823,150.8010)arc(-0.058:180.058:2.525)arc(-180.058:0.058:2.525) --
    cycle;
  \path[cm={{0.70053,0.0,0.0,0.70053,(16.45426,6.7592)}},draw=black,fill=black,miter
    limit=4.00,line width=0.800pt]
    (346.4823,150.8010)arc(-0.058:180.058:2.525)arc(-180.058:0.058:2.525) --
    cycle;
  \path[cm={{0.70053,0.0,0.0,0.70053,(1.5938,7.46684)}},draw=black,fill=black,miter
    limit=4.00,line width=0.800pt]
    (346.4823,150.8010)arc(-0.058:180.058:2.525)arc(-180.058:0.058:2.525) --
    cycle;
  \path[cm={{0.70053,0.0,0.0,0.70053,(-9.72846,-5.2707)}},draw=black,fill=black,miter
    limit=4.00,line width=0.800pt]
    (346.4823,150.8010)arc(-0.058:180.058:2.525)arc(-180.058:0.058:2.525) --
    cycle;
  \path[cm={{0.70053,0.0,0.0,0.70053,(-11.14374,-22.96173)}},draw=black,fill=black,miter
    limit=4.00,line width=0.800pt]
    (346.4823,150.8010)arc(-0.058:180.058:2.525)arc(-180.058:0.058:2.525) --
    cycle;
  \path[cm={{0.70053,0.0,0.0,0.70053,(0.17852,-34.99163)}},draw=black,fill=black,miter
    limit=4.00,line width=0.800pt]
    (346.4823,150.8010)arc(-0.058:180.058:2.525)arc(-180.058:0.058:2.525) --
    cycle;
  \path[cm={{0.70053,0.0,0.0,0.70053,(23.53067,-24.54514)}},draw=black,fill=black,miter
    limit=4.00,line width=0.800pt]
    (346.4823,150.8010)arc(-0.058:180.058:2.525)arc(-180.058:0.058:2.525) --
    cycle;
  \path[cm={{0.70053,0.0,0.0,0.70053,(18.63402,-30.522)}},draw=black,fill=black,miter
    limit=4.00,line width=0.800pt]
    (346.4823,150.8010)arc(-0.058:180.058:2.525)arc(-180.058:0.058:2.525) --
    cycle;
  \path[cm={{0.70053,0.0,0.0,0.70053,(31.85543,-29.27366)}},draw=black,fill=black,miter
    limit=4.00,line width=0.800pt]
    (346.4823,150.8010)arc(-0.058:180.058:2.525)arc(-180.058:0.058:2.525) --
    cycle;
  \path[cm={{0.70053,0.0,0.0,0.70053,(38.39114,6.05156)}},draw=black,fill=black,miter
    limit=4.00,line width=0.800pt]
    (346.4823,150.8010)arc(-0.058:180.058:2.525)arc(-180.058:0.058:2.525) --
    cycle;
  \path[cm={{0.70053,0.0,0.0,0.70053,(22.82303,14.54325)}},draw=black,fill=black,miter
    limit=4.00,line width=0.800pt]
    (346.4823,150.8010)arc(-0.058:180.058:2.525)arc(-180.058:0.058:2.525) --
    cycle;
  \path[cm={{0.70053,0.0,0.0,0.70053,(30.60709,-21.54644)}},draw=black,fill=black,miter
    limit=4.00,line width=0.800pt]
    (346.4823,150.8010)arc(-0.058:180.058:2.525)arc(-180.058:0.058:2.525) --
    cycle;
  \path[cm={{0.70053,0.0,0.0,0.70053,(36.26821,-10.22419)}},draw=black,fill=black,miter
    limit=4.00,line width=0.800pt]
    (346.4823,150.8010)arc(-0.058:180.058:2.525)arc(-180.058:0.058:2.525) --
    cycle;
  \path[cm={{0.70053,0.0,0.0,0.70053,(30.60709,-0.31721)}},draw=black,fill=black,miter
    limit=4.00,line width=0.800pt]
    (346.4823,150.8010)arc(-0.058:180.058:2.525)arc(-180.058:0.058:2.525) --
    cycle;
  \path[cm={{0.70053,0.0,0.0,0.70053,(19.99247,0.39043)}},draw=black,fill=black,miter
    limit=4.00,line width=0.800pt]
    (346.4823,150.8010)arc(-0.058:180.058:2.525)arc(-180.058:0.058:2.525) --
    cycle;
  \path[cm={{0.70053,0.0,0.0,0.70053,(17.1619,-13.7624)}},draw=black,fill=black,miter
    limit=4.00,line width=0.800pt]
    (346.4823,150.8010)arc(-0.058:180.058:2.525)arc(-180.058:0.058:2.525) --
    cycle;
  \path[cm={{0.70053,0.0,0.0,0.70053,(9.41534,-21.11823)}},draw=black,fill=black,miter
    limit=4.00,line width=0.800pt]
    (346.4823,150.8010)arc(-0.058:180.058:2.525)arc(-180.058:0.058:2.525) --
    cycle;
  \path[cm={{0.70053,0.0,0.0,0.70053,(7.77629,-30.69013)}},draw=black,fill=black,miter
    limit=4.00,line width=0.800pt]
    (346.4823,150.8010)arc(-0.058:180.058:2.525)arc(-180.058:0.058:2.525) --
    cycle;
  \path[cm={{0.70053,0.0,0.0,0.70053,(-1.38555,-28.19464)}},draw=black,fill=black,miter
    limit=4.00,line width=0.800pt]
    (346.4823,150.8010)arc(-0.058:180.058:2.525)arc(-180.058:0.058:2.525) --
    cycle;
  \path[cm={{0.70053,0.0,0.0,0.70053,(-6.8979,-18.71588)}},draw=black,fill=black,miter
    limit=4.00,line width=0.800pt]
    (346.4823,150.8010)arc(-0.058:180.058:2.525)arc(-180.058:0.058:2.525) --
    cycle;
  \path[cm={{0.70053,0.0,0.0,0.70053,(0.17852,-8.10126)}},draw=black,fill=black,miter
    limit=4.00,line width=0.800pt]
    (346.4823,150.8010)arc(-0.058:180.058:2.525)arc(-180.058:0.058:2.525) --
    cycle;
  \path[cm={{0.70053,0.0,0.0,0.70053,(8.67021,-1.02485)}},draw=black,fill=black,miter
    limit=4.00,line width=0.800pt]
    (346.4823,150.8010)arc(-0.058:180.058:2.525)arc(-180.058:0.058:2.525) --
    cycle;
  \path[->,>=latex,draw=black,line join=miter,line cap=butt,line width=0.560pt]
    (291.4127,137.7485) .. controls (324.7858,122.0030) and (359.4753,126.0028) ..
    (394.7283,138.4562);
  \path[cm={{0.70053,0.0,0.0,0.70053,(46.70396,34.08356)}},draw=black,fill=black,miter
    limit=4.00,line width=0.800pt]
    (346.4823,150.8010)arc(-0.058:180.058:2.525)arc(-180.058:0.058:2.525) --
    cycle;
  \path[cm={{0.70053,0.0,0.0,0.70053,(59.60844,28.28269)}},draw=black,fill=black,miter
    limit=4.00,line width=0.800pt]
    (346.4823,150.8010)arc(-0.058:180.058:2.525)arc(-180.058:0.058:2.525) --
    cycle;
  \path[cm={{0.70053,0.0,0.0,0.70053,(71.28392,24.59569)}},draw=black,fill=black,miter
    limit=4.00,line width=0.800pt]
    (346.4823,150.8010)arc(-0.058:180.058:2.525)arc(-180.058:0.058:2.525) --
    cycle;
  \path[cm={{0.70053,0.0,0.0,0.70053,(89.1044,21.5232)}},draw=black,fill=black,miter
    limit=4.00,line width=0.800pt]
    (346.4823,150.8010)arc(-0.058:180.058:2.525)arc(-180.058:0.058:2.525) --
    cycle;
  \path[cm={{0.70053,0.0,0.0,0.70053,(104.46687,21.5232)}},draw=black,fill=black,miter
    limit=4.00,line width=0.800pt]
    (346.4823,150.8010)arc(-0.058:180.058:2.525)arc(-180.058:0.058:2.525) --
    cycle;
  \path[cm={{0.70053,0.0,0.0,0.70053,(120.44385,23.36669)}},draw=black,fill=black,miter
    limit=4.00,line width=0.800pt]
    (346.4823,150.8010)arc(-0.058:180.058:2.525)arc(-180.058:0.058:2.525) --
    cycle;
  \path[cm={{0.70053,0.0,0.0,0.70053,(135.19183,27.05369)}},draw=black,fill=black,miter
    limit=4.00,line width=0.800pt]
    (346.4823,150.8010)arc(-0.058:180.058:2.525)arc(-180.058:0.058:2.525) --
    cycle;
  \path[cm={{0.70053,0.0,0.0,0.70053,(158.54279,33.81318)}},draw=black,fill=black,miter
    limit=4.00,line width=0.800pt]
    (346.4823,150.8010)arc(-0.058:180.058:2.525)arc(-180.058:0.058:2.525) --
    cycle;
  \path[fill=black] (199.21518,58.356892) node[above right] (text6299)
    {$\mathcal{C}$};
  \path[fill=black] (297.70227,159.38274) node[above right] (text6299-1)
    {trajectory $q\left(t\right)$};
  \path[->,>=latex,draw=black,line join=miter,line cap=butt,line width=0.560pt]
    (207.7867,108.1667) .. controls (209.9880,99.0819) and (214.6292,91.4611) ..
    (224.9926,87.2737);
  \path[fill=black] (204.31883,124.09956) node[above right] (text6299-9) {$q$};
  \path[->,>=latex,draw=black,line join=miter,line cap=butt,line width=0.560pt]
    (215.7751,109.3957) -- (226.8361,102.6362);
  \path[->,>=latex,draw=black,line join=miter,line cap=butt,line width=0.560pt]
    (215.1606,120.4566) .. controls (223.3764,119.7733) and (231.6459,119.1707) ..
    (236.6681,113.6972);
\end{scope}

\end{tikzpicture}
\par\end{center}
\par\end{centering}

\caption{\label{fig:ClassicalConfigurationSpace}A schematic picture of a classical
configuration space $\mathcal{C}$, with examples of allowed ontic
states $q$ and an example of an ontic-state trajectory $q\left(t\right)$.}
\end{figure}
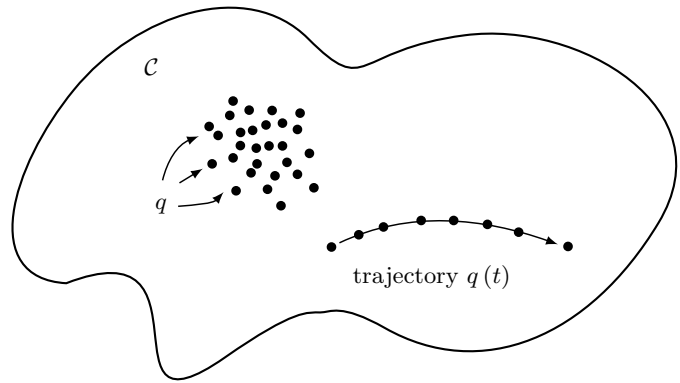

The kinematical structure of a classical theory is conceptually straightforward
and admits an intuitively simple ontology: An ontic state of a classical
system, meaning the state of the system as it could truly exist in
reality, corresponds at each instant in time $t$ to some element
$q$ of a configuration space $\mathcal{C}$ whose elements are by
definition \emph{mutually exclusive} possibilities. A full sequence
$q\left(t\right)$ of such elements over time $t$ constitutes an
ontic-state trajectory for the system. (See Figure~\ref{fig:ClassicalConfigurationSpace}.)

\subsubsection{Classical Epistemology}

From the mutual exclusivity of all the system's allowed ontic states,
we also naturally obtain the associated epistemology of classical
physics: In the language of probability theory, this mutual exclusivity
allows us to regard the system's configuration space as being a \emph{single}
sample space  $\mathcal{C}=\Omega$, meaning that if we don't know
the system's \emph{actual} ontic state precisely, then we are free
to describe the system in terms of an epistemic state defined to
be a probability distribution $\mapping p{\mathcal{C}}{\left[0,1\right]}$
on $\mathcal{C}=\Omega$, where $p\left(q\right)\in\left[0,1\right]$
 is the epistemic probability that the actual ontic state is truly
$q$ and definitely not any other allowed ontic state $q^{\prime}\ne q$.%
\footnote{We do not attempt to wade into the centuries-old philosophical debate
regarding the ultimate meaning of probability in terms of anything
more fundamental, as encapsulated wonderfully by Russell in 1929 (quoted
in \cite{Jauch:1974qpc}): ``Probability is the most important concept
in modern science, especially as nobody has the slightest notion what
it means.''  Interestingly, this quotation parallels Feynman's famous
quip in 1961 that ``I think I can safely say that nobody understands
quantum mechanics.'' \cite{Feynman:1965cpl} For our purposes, we
treat probability as a primitive, irreducible concept, just as we
treat concepts like ontology and epistemology.%
} Equivalently, we can regard the system's epistemic state $p$ as
the collection $\setindexed{\left(p\left(q\right),q\right)}{q\in\mathcal{C}}$
of ordered pairs that each identify one of the system's \emph{possible}
ontic states $q\in\mathcal{C}$ together with the corresponding probability
$p\left(q\right)\in\left[0,1\right]$ with which $q$ is the system's
\emph{actual} ontic state. In analogy with ontic-state trajectories
$q\left(t\right)$, we will sometimes say that a full sequence $p\left(t\right)$
of a system's epistemic states over time constitutes an epistemic-state trajectory
for the system.

Note that epistemic states and epistemic probabilities describing
kinematical configurations in classical physics are always ultimately
subjective, in the sense that they depend on the particular observer
in question and are nontrivial only due to prosaic reasons of ignorance
on the part of that observer. For instance, the system's ontic state
might be determined by a hidden random number generator or by the
roll of an unseen die, or the observer's recording precision might
be limited by technological constraints.

An optimal observer has no subjective uncertainty and knows the system's
present ontic state precisely; in this idealized case, the probability
distribution $p\left(q\right)$ singles out that one ontic state with
unit probability and we say that the system's epistemic state is pure.
In the more general and realistic case in which the observer's knowledge
is incomplete and the probability distribution $p\left(q\right)$
is nontrivial, we say that the system's epistemic state is mixed.

\subsubsection{Formal Epistemic States\label{subsub:Formal-Epistemic-States}}

The subjectivity of classical epistemic states means that, if we wish,
we could work with \emph{formal} epistemic probability distributions
over possibilities that are \emph{not} mutually exclusive. As an example,
suppose that we fill a large container with marbles, $72\%$ of which
are known to come from a box containing only red marbles and $28\%$
of which are known to come from a box containing marbles that are
either red or blue according to an unknown red:blue ratio. If we we
single out one marble from the container at random, then we can formally
describe the epistemic state of the marble in terms of the non-exclusive
possible statements $x=\quote{\mathrm{red}}$ and $y=\quote{\mathrm{red\ or\ blue}}$
as $\set{\left(0.72,x\right),\left(0.28,y\right)}$. Similar reasoning
would apply when describing the epistemic state of a dispenser that
stochastically releases marbles in the state $x$ at a frequency of
72\% and in the state $y$ at a frequency of 28\%.

However, to avoid ambiguities in our discussion ahead---especially
when employing the notion of entropy to quantify our level of uncertainty---we
will generally assume that all our classical epistemic states always
encode \emph{logically rigorous} probability distributions involving
only \emph{mutually exclusive} possibilities.

\subsubsection{Classical Surprisal and Entropy}

Given a system's epistemic state $p$, the smooth function 
\begin{equation}
\log\frac{1}{p\left(q\right)}\in\left[0,\infty\right],\label{eq:DefSurprisalFn}
\end{equation}
 called the surprisal \cite{Tribus:1961tt}, captures our ``surprise''
at learning that the system's actual ontic state happens to be $q$.
Our ``average level of surprise,'' also called the (Shannon) entropy
of the epistemic state $p$, is defined by 
\begin{equation}
S\equiv\expectval{\log\frac{1}{p}}=-\sum_{q}p\left(q\right)\log p\left(q\right)\in\left[0,\log\left(\#\mathrm{states\ in\ }\mathcal{C}\right)\right]\label{eq:DefClassicalEntropy}
\end{equation}
 and provides an overall measure of how much we currently do \emph{not}
yet know about the system's ontology \cite{Shannon:1948mtc,ShannonWeaver:1949mtc,Jaynes:1957itsm,Jaynes:1957itsmii}.
That is, if we think of an epistemic state $p$ as characterizing
our information about a system's underlying ontic state, then the
entropy $S$ defined in \eqref{eq:DefClassicalEntropy} provides us
with a quantitative measure of how much information we \emph{lack}
and that is therefore still hiding in the system.

Indeed, in the limit of a pure epistemic state, meaning that we know
the actual ontic state of the system with certainty and thus $p\left(q\right)\to1$
for a single value of $q$ and $p\left(q^{\prime}\right)\to0$ for
all $q^{\prime}\ne q$, the entropy goes to zero, $S\to0$. On the
other hand, if we know nothing about the actual ontic state of the
system, so that all the probabilities $p\left(q\right)$ become essentially
equal, then the entropy approaches its maximal value $S\to\log\left(\#\mathrm{states\ in\ }\mathcal{C}\right)$.

\subsubsection{Classical Measurements and Signals}

Combining classical kinematics with the definition of entropy in \eqref{eq:DefClassicalEntropy}
as a total measure of the information that we lack about a system's
underlying ontic state, we can give a precise meaning to the notions
of classical measurements, signals, and information transfer.

\begin{figure}
\begin{centering}
\begin{tabular}{ccc}
\definecolor{ce1e1e1}{RGB}{225,225,225}
\definecolor{cffffff}{RGB}{255,255,255}

\begin{tikzpicture}[y=0.80pt,x=0.80pt,yscale=-1, inner sep=0pt, outer sep=0pt]
\begin{scope}[shift={(-43.03757,-49.82106)}]
  \path[draw=black,fill=ce1e1e1,miter limit=4.00,line width=0.587pt,rounded
    corners=0.0000cm] (43.4047,50.1881) rectangle (121.0482,118.9161);
  \path[draw=black,fill=cffffff,miter limit=4.00,line width=0.800pt,rounded
    corners=0.0000cm] (49.4975,55.3213) rectangle (116.1675,112.9000);
  \path[fill=black] (76.615807,74.483757) node[above right] (text5199) {$Q$};
  \path[draw=black,fill=ce1e1e1,miter limit=4.00,line width=0.651pt,rounded
    corners=0.0000cm] (221.7193,52.2182) rectangle (296.5049,119.0741);
  \path[draw=black,fill=ce1e1e1,line join=miter,line cap=butt,line width=0.800pt]
    (221.8021,77.5650) -- (199.5889,52.3112) -- (199.5889,118.9812) --
    (221.8123,96.7579) -- cycle;
  \path[draw=black,fill=cffffff,miter limit=4.00,line width=0.800pt,rounded
    corners=0.0000cm] (226.0559,56.8569) rectangle (292.7260,114.4355);
  \path[fill=black] (74.705872,88.846024) node[above right] (text5199-8) {$p_Q$};
  \path[fill=black] (62.35231,105.19958) node[above right] (text5199-8-8)
    {$S_Q>0$};
  \path[fill=black] (252.5278,72.316536) node[above right] (text5199-5) {$A$};
  \path[fill=black] (248.61786,88.678802) node[above right] (text5199-8-2)
    {$``\emptyset"$};
  \path[fill=black] (239.26431,105.03236) node[above right] (text5199-8-8-3)
    {$S_A=0$};
\end{scope}

\end{tikzpicture} & ~~~~~~~~~~ & 
\definecolor{ce1e1e1}{RGB}{225,225,225}
\definecolor{cffffff}{RGB}{255,255,255}

\begin{tikzpicture}[y=0.80pt,x=0.80pt,yscale=-1, inner sep=0pt, outer sep=0pt]
\begin{scope}[shift={(-43.03757,-49.82106)}]
  \path[draw=black,fill=ce1e1e1,miter limit=4.00,line width=0.587pt,rounded
    corners=0.0000cm] (43.4047,50.1881) rectangle (121.0482,118.9161);
  \path[draw=black,fill=cffffff,miter limit=4.00,line width=0.800pt,rounded
    corners=0.0000cm] (49.4975,55.3213) rectangle (116.1675,112.9000);
  \path[fill=black] (76.615807,74.483757) node[above right] (text5199) {$Q$};
  \path[draw=black,fill=ce1e1e1,miter limit=4.00,line width=0.651pt,rounded
    corners=0.0000cm] (149.7193,52.2182) rectangle (224.5049,119.0741);
  \path[draw=black,fill=ce1e1e1,line join=miter,line cap=butt,line width=0.800pt]
    (149.8021,77.5650) -- (127.5889,52.3112) -- (127.5889,118.9812) --
    (149.8123,96.7579) -- cycle;
  \path[draw=black,fill=cffffff,miter limit=4.00,line width=0.800pt,rounded
    corners=0.0000cm] (154.0559,56.8569) rectangle (220.7260,114.4355);
  \path[fill=black] (74.705872,88.846024) node[above right] (text5199-8) {$p_Q$};
  \path[fill=black] (62.35231,105.19958) node[above right] (text5199-8-8)
    {$S_Q>0$};
  \path[fill=black] (180.5278,72.316536) node[above right] (text5199-5) {$A$};
  \path[fill=black] (180.61786,88.678802) node[above right] (text5199-8-2)
    {$p_A^\prime$};
  \path[fill=black] (163.26431,105.03236) node[above right] (text5199-8-8-3)
    {$S_A^\prime=S_Q$};
  \path[->,>=latex,shift={(43.03757,49.82106)},draw=black,line join=miter,line cap=butt,line
    width=0.800pt] (221.6780,36.0726) -- (189.5046,36.0726);
\end{scope}

\end{tikzpicture}\tabularnewline
(a) &  & (b)\tabularnewline
\end{tabular}
\par\end{centering}

\caption{\label{fig:ClassicalMeasurement}The subject system $Q$ and the measurement
apparatus $A$ (a) before the measurement and (b) after the measurement.}
\end{figure}
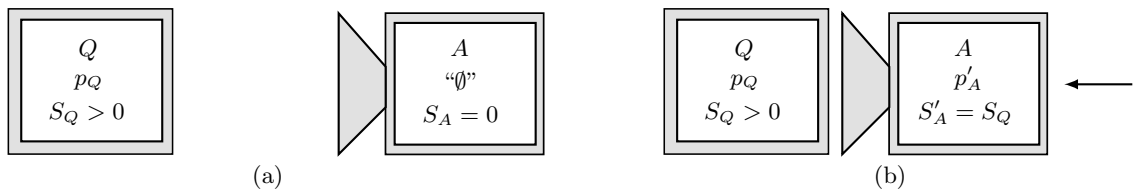

Suppose that we wish to study a system $Q$ whose epistemic state
is $p_{Q}$---that is, consisting of individual probabilities $p_{Q}\left(q\right)$
for each of the system's possible ontic states $q$---and whose entropy
$S_{Q}>0$ represents the amount of information that we currently
lack about the system's actual ontic state. We proceed by sending
over a measurement apparatus $A$ in an initially known ontic state
$\quote{\emptyset}$, by which we mean that the dial on the apparatus
initially reads ``empty'' and the initial entropy of the apparatus
is $S_{A}=0$. (See Figure~\ref{fig:ClassicalMeasurement}a.)

When the apparatus $A$ comes into local contact with the subject
system $Q$ and performs a measurement to determine the state of $Q$,
the state of the measurement apparatus becomes correlated with the
state of the subject system; for example, if the actual ontic state
of the subject system $Q$ happens to be $q$, then the ontic state
of the apparatus $A$ evolves from $\quote{\emptyset}$ to $\quote q$.
As a consequence, the measurement apparatus, which is still far away
from us, develops a nontrivial epistemic state $p_{A}^{\prime}$ identical
to that of the subject system---meaning that $p_{A}^{\prime}\left(\quote q\right)=p_{Q}\left(q\right)$---and
thus the apparatus develops a nonzero final correlational entropy
$S_{A}^{\prime}=S_{Q}$. (See Figure~\ref{fig:ClassicalMeasurement}b.)

Correlational entropy of this kind therefore indicates that there
has been the transmission of a signal---that is, the exchange of
\emph{observable} information, which, as we know, is constrained by
the vacuum speed of light $c$. In the present example, this signal
is communicated from the subject system $Q$ to the measurement apparatus
$A$. We therefore also obtain a criterion for declaring that \emph{no}
signal has passed between two systems, namely, when their respective
epistemic states evolve independently of one another and thus neither
system develops any correlational entropy.

With these concepts of information, signaling, and correlational entropy
in hand, we can obtain a measure for the precision of a system's information
in terms of its correlational entropy. Consider a system---say, a
measurement apparatus---that measures some previously unknown numerical
property of another system to a precision of $n$ bits. Then the correlational
entropy of the apparatus grows by at least an amount $\Delta S\sim\log n$,
and so the minimum measurement error of the apparatus satisfies the
error-entropy bound 
\begin{equation}
\mathrm{minimum\ error}\sim1/n\sim e^{-\Delta S}\geq e^{-S},\label{eq:MinimumMeasurementErrorEntropyBound}
\end{equation}
 where $S\geq\Delta S$ is the \emph{total} final entropy \eqref{eq:DefClassicalEntropy}
of the apparatus. Hence, the precision with which a measurement apparatus
can specify any numerical quantity is bounded from below by $\exp\left(-S\right)$.
We will see that this same sort of error crops up in the analogous
case of quantum measurements when we study them in Section~\ref{sub:Measurements-and-Decoherence}.%
\footnote{\label{fn:BlackHoleInfoParadox}Curiously, $\sim\exp\left(-S\right)$
effects also seem to play an important role in the famous black-hole information paradox
\cite{Hawking:1975pcbh,Hawking:1976bpgc,Preskill:1992,Giddings:1995bhip,Giddings2006:bhiun},
because the mixed-state final density matrix computed semiclassically
appears to differ by $\sim\exp\left(-S\right)$ off-diagonal entries
\cite{BalasubramanianCzech:2011qairbh,IizukaKabat:2013omihw} from
that of the pure-state final density matrix state that would be required
by information conservation. See also footnote~\ref{fn:BekensteinBound}.%
}

\subsubsection{Classical Dynamics\label{subsub:Classical-Dynamics}}

A classical system has well-defined dynamics if final ontic states
can be predicted (probabilistically at least) based on given data
characterizing initial ontic states. More precisely, a system has
dynamics if for any choice of final time $t^{\prime}$ there exists
a rule for picking prescribed initial times $t_{1}\geq t_{2}\geq\cdots$
and there exists a mapping $p\left(\cdot;t^{\prime}\given\left(\cdot;t_{1}\right),\left(\cdot;t_{2}\right),\dotsc\right)$
that takes \emph{arbitrary} initial ontic states $q_{1}$ at $t_{1}$,
$q_{2}$ at $t_{2}$, $\dotsc$ and yields corresponding conditional probabilities
$p\left(q^{\prime};t^{\prime}\given\left(q_{1};t_{1}\right),\left(q_{2};t_{2}\right),\dotsc\right)\in\left[0,1\right]$
that the system's ontic state at $t^{\prime}$ is $q^{\prime}$: 
\begin{equation}
p\left(\cdot;t^{\prime}\given\left(\cdot;t_{1}\right),\left(\cdot;t_{2}\right),\dotsc\right)\colon\ \underbrace{\left(q_{1};t_{1}\right),\left(q_{2};t_{2}\right),\dotsc}_{\mathrm{initial\ data}},\ \underbrace{\left(q^{\prime};t^{\prime}\right)}_{\mathrm{final\ data}}\mapsto\underbrace{p\left(q^{\prime};t^{\prime}\given\left(q_{1};t_{1}\right),\left(q_{2};t_{2}\right),\dotsc\right)}_{\mathrm{conditional\ probabilities}}.\label{eq:ClassicalOnticLevelDynMapGeneral}
\end{equation}
 The initial times $t_{1},t_{2},\dotsc$ that are necessary to define
the mapping could, in principle, be only infinitesimally separated,
and their total number, which could be infinite, is called the order
of the dynamics. (As an example, Markovian dynamics, to be defined
shortly, requires initial data at only a \emph{single} initial time,
and is therefore first order.)

Just as was the case for the \emph{epistemic} probabilities characterizing
an observer's knowledge of a given system's ontic state at a single
moment in time, the \emph{conditional} probabilities that define a
classical system's dynamics can be subjective in the sense that they
are merely a matter of the observer's prosaic ignorance regarding
details of the given system. However, these conditional probabilities
can also (or instead) be objective in the sense that they cannot
be trivialized based solely on knowledge about the given system itself;
for example, nontrivial conditional probabilities may arise from the
observer's ignorance about the properties or dynamics of a larger
environment in which the given system resides---an environment that
could, for instance, be infinitely big and old---or they may emerge
in an effective sense from chaos, to be described later on.

We call the dynamics deterministic in the idealized case in which
the conditional probabilities $p\left(q^{\prime};t^{\prime}\given\left(q_{1};t_{1}\right),\left(q_{2};t_{2}\right),\dotsc\right)$
specify a particular ontic state $q^{\prime}$ at $t^{\prime}$ with
unit probability for each choice of initial data $\left(q_{1};t_{1}\right),\left(q_{2};t_{2}\right),\dotsc$.%
\footnote{We address the definition and status of ``superdeterminism,'' a
stronger notion of determinism, in Section~\ref{subsub:The-Status-of-Superdeterminism}.%
} Otherwise, whether subjective, objective, or a combination of both,
we say that the dynamics is stochastic.

By assumption, the conditional probabilities appearing in \eqref{eq:ClassicalOnticLevelDynMapGeneral}
are determined \emph{solely} by the initial ontic data $\left(q_{1};t_{1}\right),\left(q_{2};t_{2}\right),\dotsc$
together with a specification of the final ontic state $q^{\prime}$
at $t^{\prime}$. In particular, the conditional probabilities are
required to be independent of the system's evolving epistemic state
$p$, and thus $p\left(\cdot;t^{\prime}\given\left(\cdot;t_{1}\right),\left(\cdot;t_{2}\right),\dotsc\right)$
naturally lifts to a (multi-)linear dynamical mapping relating the
system's evolving epistemic state at each of the initial times $t_{1}\geq t_{2}\geq\cdots$
to the system's epistemic state at the final time $t^{\prime}$, 
\begin{equation}
p\left(\cdot;t^{\prime}\given\left(\cdot;t_{1}\right),\left(\cdot;t_{2}\right),\dotsc\right)\colon\ \underbrace{p\left(q^{\prime};t^{\prime}\right)}_{\substack{\mathrm{epistemic}\\
\mathrm{state\ at\ }t^{\prime}
}
}=\sum_{q_{1},q_{2},\dotsc}\underbrace{p\left(q^{\prime};t^{\prime}\given\left(q_{1};t_{1}\right),\left(q_{2};t_{2}\right),\dotsc\right)}_{\substack{\mathrm{conditional\ probabilities}\\
\mathrm{(independent\ of\ epistemic\ states)}
}
}\underbrace{p\left(q_{1};t_{1}\right)p\left(q_{2};t_{2}\right)\dotsm}_{\substack{\mathrm{epistemic\ states}\\
\mathrm{at\ }t_{1},t_{2},\dotsc
}
},\label{eq:ClassicalMultilinearEpistemicLevelDynMapGeneral}
\end{equation}
 which we can think of as a kind of dynamical Bayesian propagation formula.

If a system has well-defined dynamics \eqref{eq:ClassicalOnticLevelDynMapGeneral},
then we see from \eqref{eq:ClassicalMultilinearEpistemicLevelDynMapGeneral}
that the dynamics actually operates at \emph{two} levels---namely,
at the level of ontic states and at the level of epistemic states---and
consistency requires that these two levels of dynamics must related
by $p\left(\cdot;t^{\prime}\given\left(\cdot;t_{1}\right),\left(\cdot;t_{2}\right),\dotsc\right)$.
Indeed, we can regard $p\left(\cdot;t^{\prime}\given\left(\cdot;t_{1}\right),\left(\cdot;t_{2}\right),\dotsc\right)$
equivalently as describing the ontic-level dynamics in the sense of
a mapping \eqref{eq:ClassicalOnticLevelDynMapGeneral} from initial
ontic states to the conditional probabilities for final ontic states,
or as describing epistemic-level dynamics in the sense of a multilinear
mapping \eqref{eq:ClassicalMultilinearEpistemicLevelDynMapGeneral}
between initial and final epistemic states. This equivalence means
that not only does the existence of ontic-level dynamics naturally
define epistemic-level dynamics, but any given epistemic-level dynamics
also determines the system's ontic-level dynamics because we can just
read off the ontic-level dynamical mapping \eqref{eq:ClassicalOnticLevelDynMapGeneral}
from the matrix elements \eqref{eq:ClassicalMultilinearEpistemicLevelDynMapGeneral}
of the multilinear dynamical mapping.

Although it's possible to conceive of an alternative world in which
dynamics and trajectories are specified \emph{only} at the epistemic
level and \emph{do not} determine ontic-level dynamics or trajectories,
it's important to recognize the \emph{inadequacy} of such a description:
If all we knew were the rules dictating how epistemic states $p$
evolve in time, then there would be nothing to ensure that a system's
underlying ontic state should evolve in a sensible manner and avoid
fluctuating wildly between macroscopically distinct configurations.
Quantum theory, as traditionally formulated, finds itself in just
such a predicament, because a quantum system's dynamics (when it exists)
is specified only in terms of density matrices and not directly as
a multilinear dynamical mapping \eqref{eq:ClassicalMultilinearEpistemicLevelDynMapGeneral}
acting on epistemic states themselves. As part of our interpretation
of quantum theory, we fill in this missing ingredient by providing
an explicit translation of density-matrix dynamics into dynamical
mappings for ontic and epistemic states, and in such a way that we
avoid macroscopic instabilities (such as eigenstate swaps, to be
defined in Section~\ref{subsub:Near-Degeneracies-and-Eigenstate-Swaps})
that have presented problems for other interpretations.

Returning to our classical story, the assumed independence of the
conditional probabilities $p\left(q^{\prime};t^{\prime}\given\left(q_{1};t_{1}\right),\dotsc\right)$
from the system's evolving epistemic state $p$ also implies that
any conditional probabilities depending on ontic states at \emph{additional}
initial times $\tilde{t}_{1},\tilde{t}_{2},\dotsc\notin\set{t_{1},t_{2},\dotsc}$
must (if they exist) be equal to $p\left(q^{\prime};t^{\prime}\given\left(q_{1};t_{1}\right),\dotsc\right)$,
\begin{equation}
p\left(q^{\prime};t^{\prime}\given\left(q_{1};t_{1}\right),\dotsc,\left(\tilde{q}_{1},\tilde{t}_{1}\right),\dotsc\right)=p\left(q^{\prime};t^{\prime}\given\left(q_{1};t_{1}\right),\dotsc\right),\label{eq:TrivilizationHigherCondProbs}
\end{equation}
 because otherwise the dynamical Bayesian propagation formula 
\[
p\left(q^{\prime};t^{\prime}\given\left(q_{1};t_{1}\right),\dotsc\right)=\sum_{\tilde{q}_{1},\dotsc}p\left(q^{\prime};t^{\prime}\given\left(q_{1};t_{1}\right),\dotsc,\left(\tilde{q}_{1},\tilde{t}_{1}\right),\dotsc\right)p\left(\tilde{q}_{1};\tilde{t}_{1}\right)\dotsm
\]
 would imply that $p\left(q^{\prime};t^{\prime}\given\left(q_{1};t_{1}\right),\dotsc\right)$
has a disallowed dependence on the epistemic state of the system at
the additional times $\tilde{t}_{1},\tilde{t}_{2}\dotsc$. As an example,
in the case of a system whose dynamics is Markovian \cite{Markov:1954ta},
meaning first order, the mapping $p\left(\cdot;t^{\prime}\given\cdot;t\right)$
defined in \eqref{eq:ClassicalOnticLevelDynMapGeneral} requires the
specification of just a \emph{single} initial time $t$, and any additional
mappings \eqref{eq:ClassicalOnticLevelDynMapGeneral} involving multiple
initial times $t\geq t_{1}\geq t_{2}\geq$ must be equal to the single-initial-time
mapping $p\left(\cdot;t^{\prime}\given\cdot;t\right)$: 
\begin{equation}
p\left(\cdot;t^{\prime}\given\left(\cdot;t\right),\left(\cdot;t_{1}\right),\left(\cdot;t_{2}\right),\dotsc\right)=p\left(\cdot;t^{\prime}\given\cdot;t\right).\label{eq:ClassicalMarkovDynMap}
\end{equation}
 In other words, a system whose dynamics is Markovian retains no ``memory''
of its ontic states prior to its most recent ontic state, apart from
whatever memory is directly encoded in that most recent ontic state.

We call a system closed in the idealized case in which the system
does not interact with or exchange information with its environment.
Unless we allow for literal destruction of information, closed classical
systems are always governed \emph{fundamentally} by deterministic
dynamics. (Subtleties arise for chaotic systems, to be discussed shortly.)
More realistically, however, systems are never exactly closed and
are instead said to be open, in which case information can ``leak
out.'' The existence of dynamics for an open system is a delicate
question because conditional probabilities inherited from an enclosing
parent system may contain a nontrivial dependence on the parent system's
own epistemic state, but any such open-system dynamics, if it exists,
is generically stochastic.

\subsubsection{Classical Systems with Continuous Configuration Spaces}

Many familiar classical systems have \emph{continuous} configuration
spaces that we can regard as manifolds of some dimension $N\geq1$.
When working with a classical system of this kind, we can parameterize
the ontic states $q=\left(q_{1},\dotsc,q_{N}\right)$ of the configuration
space $\mathcal{C}$ in terms of $N$ continuously valued degrees of freedom
$q_{\alpha}$ ($\alpha=1,\dotsc,N$), which play the role of coordinates
for the configuration-space manifold. It is then more natural to describe
the system's epistemic states in terms of probability \emph{densities}
$\rho\left(q\right)\equiv\rho\left(q_{1},\dotsc.q_{N}\right)\equiv dp\left(q\right)/d^{N}q$
rather than in terms of probabilities $p\left(q\right)$ \emph{per
se}.

If we can treat time as being a continuous parameter in our description
of a classical system having a continuous configuration space, then,
purely at the level of kinematics, the ontic states $q\left(t\right)=\left(q_{1}\left(t\right),\dotsc,q_{N}\left(t\right)\right)$
of the system at infinitesimally separated times $t,$ $t+dt$, $t+2dt$,
$\dotsc$ are independent kinematical variables, and thus the instantaneous coordinate values
$q_{\alpha}\left(t\right)$, instantaneous velocities $\dot{q}_{\alpha}\left(t\right)\equiv dq_{\alpha}\left(t\right)/dt\equiv\left(q_{\alpha}\left(t+dt\right)-q_{\alpha}\left(t\right)\right)/dt$,
instantaneous accelerations $\ddot{q}_{\alpha}\left(t\right)\equiv d^{2}q_{\alpha}\left(t\right)/dt^{2}$,
and so forth, are all likewise independent kinematical variables.
We are therefore free to extend the notion of the system's epistemic
state to describe the joint probabilities $p\left(q,\dot{q},\ddot{q},\dotsc\right)$
that the system's ontic state is $q\equiv\left(q_{1},\dotsc,q_{N}\right)$,
that its instantaneous velocities have values given by $\dot{q}\equiv\left(\dot{q}_{1},\dotsc,\dot{q}_{N}\right)$,
and so forth.

This generalization is particularly convenient when examining a system
whose dynamical mapping \eqref{eq:ClassicalOnticLevelDynMapGeneral}
is \emph{second order} in time, meaning that it involves initial ontic
states $q$ at a \emph{pair} of infinitesimally separated times $t$
and $t+dt$. In that case, we can equivalently express the dynamics
in terms of a mapping involving the system's ontic state $q=\left(q_{1},\dotsc,q_{N}\right)$
and its instantaneous velocities $\dot{q}=\left(\dot{q}_{1},\dotsc,\dot{q}_{N}\right)$
at the \emph{single} initial time $t$. Hence, if we now think of
the ordered pair $\left(q,\dot{q}\right)$ as describing a point in
a generalized kind of configuration space (mathematically speaking,
the tangent bundle of the configuration-space manifold), then we can
treat the dynamics as though it were Markovian---that is, depending
only on data at a \emph{single} initial time. When the dynamics is,
moreover, deterministic, and we can express the dynamical mapping
from initial data $\left(q,\dot{q};t\right)$ to final data $\left(q^{\prime},\dot{q}^{\prime};t^{\prime}\right)$
as a collection of equations, known as the system's equations of motion,
then we enter the familiar terrain of textbook classical physics,
with its language of Lagrangians $L\left(q,\dot{q}\right)$, action functionals
$S\left[q\right]\equiv\integration{\int}{dt}L\left(q\left(t\right),\dot{q}\left(t\right);t\right)$,
canonical momenta $p_{\alpha}$, phase spaces $\left(q,p\right)$
(mathematically speaking, the \emph{cotangent} bundle of the system's
configuration-space manifold), Hamiltonians $H\left(q,p\right)$,
and Poisson brackets 
\begin{equation}
\pbrack fg\equiv\sum_{\alpha}\left[\frac{\partial f}{\partial q_{\alpha}}\frac{\partial g}{\partial p_{\alpha}}-\frac{\partial g}{\partial q_{\alpha}}\frac{\partial f}{\partial p_{\alpha}}\right].\label{eq:PBracks}
\end{equation}

For classical systems of this type, we can formulate the second-order
equations of motion as the so-called canonical equations of motion
\begin{equation}
\dot{q}_{\alpha}=\frac{\partial H}{\partial p_{\alpha}}=-\pbrack H{q_{\alpha}},\qquad\dot{p}_{\alpha}=-\frac{\partial H}{\partial q_{\alpha}}=-\pbrack H{p_{\alpha}},\label{eq:ClassicalCanEqMot}
\end{equation}
 and, for consistency, the system's generalized epistemic state $\rho\left(q,p\right)$
on its phase space must then satisfy the classical Liouville equation
\begin{equation}
\frac{\partial\rho}{\partial t}=\pbrack H{\rho}.\label{eq:ClassicalLiouvilleEq}
\end{equation}
 A familiar corollary is the Liouville theorem: The \emph{total}
time derivative of the probability distribution $\rho\left(q\left(t\right),p\left(t\right);t\right)$
vanishes on any trajectory that solves the equations of motion, 
\begin{equation}
\frac{d\rho}{dt}=0,\label{eq:ClassicalIncompressProbDistr}
\end{equation}
 meaning that the ensemble of phase-space trajectories described by
$\rho\left(q\left(t\right),p\left(t\right);t\right)$ behaves like
an incompressible fluid.

\subsubsection{Chaos}

Classical systems with continuous configurations spaces can exhibit
an important kind of dynamics known as chaos. While chaotic dynamics
is \emph{formally} deterministic according to the classification scheme
described earlier---initial ontic states are mapped to unique final
ontic states with unit probability---the predicted values of final
ontic states are exponentially sensitive to minute changes in the
initial data, and thus the decimal precision needed to describe the
system's final epistemic state is exponentially greater than the decimal
precision needed to specify the system's initial data. Indeed, our
relative predictive power, which we could define as being the ratio
of our output precision to our input precision, typically goes to
zero as we try to approach the idealized limit of infinitely sharp
input precision. Because any realistic recording device is limited
to a fixed and finite decimal precision, chaotic dynamics is therefore
\emph{effectively} stochastic even if the system of interest is closed
off from its larger environment.

\subsection{Quantum Theories\label{sub:Quantum-Theories}}

\subsubsection{Quantum Kinematics\label{subsub:Quantum-Kinematics}}

The picture our interpretation paints for quantum theory is surprisingly
similar to the classical story, the key kinematical difference being
that a quantum system's configuration space is ``too large'' to
be a \emph{single} sample space of mutually exclusive states. More
precisely, our interpretation asserts that the configuration space
of a quantum system corresponds (up to meaningless overall normalization
factors) to a \emph{vector space} $\mathcal{H}$,%
\footnote{It is not a goal of this paper to provide deeper principles underlying
this fact, which we take on as an axiom; for a detailed discussion
of the complex vector-space structure of Hilbert spaces in quantum
theory, and attempts to derive it from more basic principles and generalize
it in new directions, see Section~\ref{subsub:Understanding-and-Generalizing-the-Hilbert-Space-Structure-of-Quantum-Theory}.
It's interesting to note the historical parallel between the way that
quantum theory modifies classical ontology by replacing classical
configuration spaces (even those that are discrete) with smoothly
interpolating continuous vector spaces, and, similarly, the way that
probability theory itself modified epistemology centuries ago by replacing
binary true/false assignments with a smoothly interpolating continuum
running from false ($0$) to true ($1$).%
} called the system's Hilbert space, for which the notion of mutual
exclusivity between states is defined by the vanishing of their inner
product, much in the same way that nondegenerate eigenstates of the
Hermitian operators representing observables in the traditional formulation
of quantum theory are always orthogonal.

As a consequence, every quantum system exhibits a \emph{continuous
infinity} of distinct sample spaces, each corresponding to a particular
orthonormal basis for the system's Hilbert space. It follows that
an epistemic state makes \emph{logically rigorous} sense only as a
probability distribution over an orthonormal basis, although, just
as we saw in Section~\ref{subsub:Formal-Epistemic-States} for the
classical case, it is sometimes useful to consider \emph{formal} subjective
epistemic states over possibilities that are not mutually exclusive.

\begin{figure}
\begin{centering}
\begin{center}
\definecolor{ce1e1e1}{RGB}{225,225,225}

\begin{tikzpicture}[y=0.80pt, x=0.8pt,yscale=-1, inner sep=0pt, outer sep=0pt]
\begin{scope}[shift={(-128.70036,-26.10314)}]
  \path[fill=black] (148.57146,184.50504) node[above right] (text3819)
    {$\left\{\Psi\right\}_i$};
  \path[fill=black] (422.41739,185.57451) node[above right] (text3819-2)
    {$\Psi^\prime$};
  \begin{scope}[cm={{0.0,-0.54041,1.0,0.0,(114.19047,275.35636)}}]
    \path[draw=black,fill=ce1e1e1,miter limit=4.00,line width=0.800pt]
      (321.4286,260.5765) .. controls (321.4286,280.1037) and (302.4007,295.9336) ..
      (278.9286,295.9336) .. controls (255.4565,295.9336) and (236.4286,280.1037) ..
      (236.4286,260.5765) .. controls (236.4286,241.0493) and (255.4565,225.2193) ..
      (278.9286,225.2193) .. controls (302.4007,225.2193) and (321.4286,241.0493) ..
      (321.4286,260.5765) -- cycle;
    \path[draw=black,line join=miter,line cap=butt,line width=0.800pt]
      (219.2857,48.0765) .. controls (223.5714,6.6479) and (320.7143,21.6479) ..
      (388.5714,51.6479) .. controls (456.4286,81.6479) and (486.4286,133.7908) ..
      (433.5714,162.3622) .. controls (380.7143,190.9336) and (423.5714,183.0765) ..
      (436.4286,236.6479) .. controls (449.2857,290.2193) and (372.8571,360.9336) ..
      (272.1429,328.0765) .. controls (171.4286,295.2193) and (132.1429,244.5050) ..
      (180.0000,198.7907) .. controls (202.0789,177.7005) and (193.7594,173.0296) ..
      (193.6902,168.0848) .. controls (193.6094,162.3115) and (196.9930,156.1650) ..
      (154.2857,123.0765) .. controls (75.0000,61.6479) and (271.4286,120.9336) ..
      (219.2857,48.0765) -- cycle;
    \path[draw=black,line join=miter,line cap=butt,line width=0.800pt]
      (305.0000,25.9336) .. controls (277.8571,68.0765) and (227.1429,73.7908) ..
      (227.1429,73.7908);
    \path[draw=black,line join=miter,line cap=butt,line width=0.800pt]
      (207.1429,87.3622) .. controls (232.8571,115.9336) and (159.2857,127.3622) ..
      (159.2857,127.3622);
    \path[draw=black,line join=miter,line cap=butt,line width=0.800pt]
      (277.8571,53.0765) .. controls (271.4286,97.3622) and (212.1429,128.0765) ..
      (212.1429,128.0765);
    \path[draw=black,line join=miter,line cap=butt,line width=0.800pt]
      (250.7143,99.5050) .. controls (281.4286,157.3622) and (193.5714,180.9336) ..
      (193.5714,180.9336);
    \path[draw=black,line join=miter,line cap=butt,line width=0.800pt]
      (161.4286,250.2193) .. controls (245.0000,202.3622) and (240.0000,156.6479) ..
      (240.0000,156.6479);
    \path[draw=black,line join=miter,line cap=butt,line width=0.800pt]
      (215.7143,208.0765) .. controls (211.4286,258.0765) and (186.4286,284.5050) ..
      (186.4286,284.5050);
    \path[draw=black,line join=miter,line cap=butt,line width=0.800pt]
      (265.0000,83.0765) .. controls (337.1429,83.0765) and (332.8571,33.0765) ..
      (332.8571,33.0765);
    \path[draw=black,line join=miter,line cap=butt,line width=0.800pt]
      (316.4286,70.2193) .. controls (304.2857,133.0765) and (257.1429,126.6479) ..
      (257.1429,126.6479);
    \path[draw=black,line join=miter,line cap=butt,line width=0.800pt]
      (227.1429,193.7908) .. controls (293.5714,204.5050) and (285.7143,120.9336) ..
      (285.7143,120.9336);
    \path[draw=black,line join=miter,line cap=butt,line width=0.800pt]
      (397.1429,55.2193) .. controls (400.7143,121.6479) and (326.4286,55.2193) ..
      (326.4286,55.2193);
    \path[draw=black,line join=miter,line cap=butt,line width=0.800pt]
      (297.8571,112.3622) .. controls (370.7143,125.2193) and (361.4286,79.5050) ..
      (361.4286,79.5050);
    \path[draw=black,line join=miter,line cap=butt,line width=0.800pt]
      (280.0000,164.5050) .. controls (351.4286,163.7908) and (337.1429,112.3622) ..
      (337.1429,112.3622);
    \path[draw=black,line join=miter,line cap=butt,line width=0.800pt]
      (233.5714,312.3622) .. controls (282.8571,275.9336) and (325.7143,335.9336) ..
      (325.7143,335.9336);
    \path[draw=black,line join=miter,line cap=butt,line width=0.800pt]
      (290.7143,306.6479) .. controls (352.1429,252.3622) and (392.8571,315.9336) ..
      (392.8571,315.9336);
    \path[draw=black,line join=miter,line cap=butt,line width=0.800pt]
      (325.0000,151.6479) .. controls (402.1429,134.5050) and (358.5714,96.6479) ..
      (358.5714,96.6479);
    \path[draw=black,line join=miter,line cap=butt,line width=0.800pt]
      (370.0000,121.6479) .. controls (415.0000,121.6479) and (418.5714,68.7908) ..
      (418.5714,68.7908);
    \path[draw=black,line join=miter,line cap=butt,line width=0.800pt]
      (400.7143,108.7908) .. controls (427.8571,131.6479) and (458.5714,134.5050) ..
      (458.5714,134.5050);
    \path[draw=black,line join=miter,line cap=butt,line width=0.800pt]
      (414.2857,173.7908) .. controls (380.0000,156.6479) and (378.5714,120.2193) ..
      (378.5714,120.2193);
    \path[draw=black,line join=miter,line cap=butt,line width=0.800pt]
      (333.5714,149.5050) .. controls (362.1429,189.5050) and (397.8571,163.0765) ..
      (397.8571,163.0765);
    \path[draw=black,line join=miter,line cap=butt,line width=0.800pt]
      (427.8571,218.0765) .. controls (375.0000,220.2193) and (365.7143,171.6479) ..
      (365.7143,171.6479);
    \path[draw=black,line join=miter,line cap=butt,line width=0.800pt]
      (394.2857,210.2193) .. controls (358.5714,244.5050) and (423.5714,287.3622) ..
      (423.5714,287.3622);
    \path[draw=black,line join=miter,line cap=butt,line width=0.800pt]
      (355.7143,288.0765) .. controls (355.7143,249.5050) and (383.5714,237.3622) ..
      (383.5714,237.3622);
    \path[draw=black,line join=miter,line cap=butt,line width=0.800pt]
      (265.7143,300.2193) .. controls (277.8571,243.7907) and (361.4286,259.5050) ..
      (361.4286,259.5050);
    \path[draw=black,line join=miter,line cap=butt,line width=0.800pt]
      (209.2857,243.0765) .. controls (258.5714,264.5050) and (259.2857,300.2193) ..
      (259.2857,300.2193);
    \path[draw=black,line join=miter,line cap=butt,line width=0.800pt]
      (235.7143,260.9336) .. controls (307.1429,220.2193) and (297.8571,263.0765) ..
      (297.8571,263.0765);
    \path[draw=black,line join=miter,line cap=butt,line width=0.800pt]
      (270.7143,180.2193) .. controls (297.1429,202.3622) and (262.1429,248.7908) ..
      (262.1429,248.7908);
    \path[draw=black,line join=miter,line cap=butt,line width=0.800pt]
      (280.7143,204.5050) .. controls (340.0000,204.5050) and (364.2857,254.5050) ..
      (364.2857,254.5050);
    \path[draw=black,line join=miter,line cap=butt,line width=0.800pt]
      (341.4286,158.7908) .. controls (335.0000,190.2193) and (300.7143,205.9336) ..
      (300.7143,205.9336);
    \path[draw=black,line join=miter,line cap=butt,line width=0.800pt]
      (375.7143,194.5050) .. controls (350.7143,195.9336) and (342.1429,226.6479) ..
      (342.1429,226.6479);
    \path[draw=black,line join=miter,line cap=butt,line width=0.800pt]
      (182.1429,121.6479) .. controls (225.0000,124.5050) and (244.2857,152.3622) ..
      (244.2857,152.3622);
    \path[->,>=latex,draw=black,line join=miter,line cap=butt,line width=0.800pt]
      (181.4286,302.3622) .. controls (202.8571,260.2193) and (225.0000,264.5050) ..
      (265.0000,264.5050);
    \path[->,>=latex,draw=black,line join=miter,line cap=butt,line width=0.800pt]
      (197.1429,26.6479) .. controls (239.2857,10.9336) and (256.4286,32.3622) ..
      (259.4286,34.3622);
    \path[->,>=latex,draw=black,line join=miter,line cap=butt,line width=0.800pt]
      (164.2857,61.6479) .. controls (142.1429,89.5050) and (162.7143,100.9336) ..
      (165.7143,105.9336);
    \path[->,>=latex,draw=black,line join=miter,line cap=butt,line width=0.800pt]
      (191.4286,61.6479) .. controls (193.5714,85.9336) and (229.3143,88.5050) ..
      (234.7143,89.5050);
    \path[->,>=latex,draw=black,line join=miter,line cap=butt,line width=0.800pt]
      (208.5714,42.3622) .. controls (257.8571,42.3622) and (289.0000,50.7908) ..
      (289.0000,50.7908);
  \end{scope}
\end{scope}

\end{tikzpicture}
\par\end{center}
\par\end{centering}

\caption{\label{fig:QuantumConfigSpaceVennDiagram}A Venn diagram representing
a quantum system's configuration space, with an orthonormal basis
for the system's Hilbert space corresponding to a partitioning $\set{\Psi_{i}}_{i}$
by mutually exclusive ontic states. An ontic state $\Psi^{\prime}$
not orthogonal to all the members of the orthonormal basis is also
displayed, and is not mutually exclusive with all the ontic states
in $\set{\Psi_{i}}_{i}$.}
\end{figure}
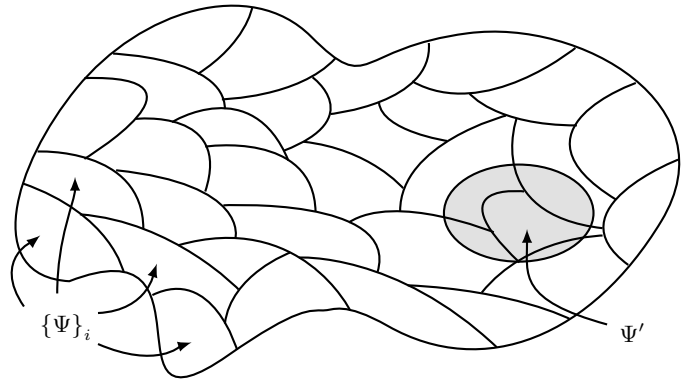

Just as a helpful visual analogy, we can loosely imagine all of a
system's allowed ontic states as corresponding to finite-size regions
collectively covering an abstract Venn diagram. Then any orthonormal
basis for the system's Hilbert space corresponds to a partition of
the Venn diagram into disjoint regions, whereas any two non-orthogonal
ontic states describe overlapping regions. (See Figure~\ref{fig:QuantumConfigSpaceVennDiagram}.)
Keep in mind, however, that this picture is entirely metaphorical:
We regard quantum states as irreducible sample-space elements, and
thus we are not \emph{literally} identifying the Venn diagram's \emph{individual
points} as elements of the system's sample space.

This new feature of quantum theory opens up a possibility not available
to classical systems, namely, that a system's sample space is fundamentally
contextual, changing over time from one orthonormal basis to another---that
is, from one sample space to another sample space incompatible with
the first but that nonetheless lies in the \emph{same} configuration
space---as the system interacts with other systems, such as measurement
devices or the system's larger environment. Hence, our interpretation
of quantum theory builds in the notion of quantum contextuality \cite{Spekkens:2005cptum,Spekkens2008:ncenn,Spekkens:2013wannumqt}
from the very beginning, in close keeping with the Kochen-Specker
theorem \cite{KochenSpecker:1967phvqm} described in Appendix~\ref{sub:Foundational-Theorems}.
Many interpretations of quantum theory, including our own, take advantage
of this general feature of the theory in order to resolve the measurement
problem through the well-known quantum phenomenon of decoherence \cite{Bohm:1951qt,JoosZeh:1985ecptiwe,Joos:2003dacwqt,Zurek:2003dtqc,Schlosshauer:2005dmpiqm,SchlosshauerCamilleri:2008qct,BreuerPetruccione:2002toqs}.

\subsubsection{Density Matrices}

As we will review in Section~\ref{sub:The-Correspondence-Between-Objective-Epistemic-States-and-Density-Matrices},
the formalism of quantum theory naturally furnishes a mathematical
object $\op{\rho}$, called a density matrix \cite{Landau:1927dpwm,vonNeumann:1927wadq,vonNeumann:1932mgdq},
that our interpretation places in a central role for defining a system's
epistemic state and its underlying ontology, especially in the context
of relating parent systems with their subsystems in the presence of
quantum entanglement. A density matrix, which we regard as being related
to but conceptually distinct from the system's epistemic state, is
a Hermitian operator on the system's Hilbert space whose eigenvalues
are all nonnegative and sum to unity.

\begin{figure}
\begin{centering}
\begin{center}
\scalebox{1.5}{
\definecolor{ce1e1e1}{RGB}{225,225,225}

\begin{tikzpicture}[y=0.80pt, x=0.8pt,yscale=-1, inner sep=0pt, outer sep=0pt]
\begin{scope}[shift={(-41.15005,-13.21949)}]
  \path[draw=black,fill=ce1e1e1,line join=miter,miter limit=4.00,line width=0.761pt,rounded
    corners=0.2000cm] (165,51) rectangle (182,70);
  \path[fill=black] (92.06411,87.079956) node[above right] (text5199-52)
    {$\hat{\rho}$};
  \path[->,>=latex,draw=black,line join=miter,line cap=butt,line width=0.924pt]
    (108.6989,83.3966) -- (143.0093,83.3966);
  \path[->,>=latex,draw=black,line join=miter,line cap=butt,line width=0.925pt]
    (109.0520,75) -- (142.3517,63.2528);
  \path[->,>=latex,draw=black,line join=miter,line cap=butt,line width=1.015pt]
    (107.9561,67.1956) -- (141.6060,42.5951);
  \path[fill=black] (120.57199,103.81787) node[above right] (text5199-52-2)
    {$\vdots$};
  \path[fill=black] (150.37277,46.663536) node[above right] (text5199-52-1) {$p_1,
    \Psi_1$};
  \path[fill=black] (150.37251,66.615074) node[above right] (text5199-52-1-9)
    {$p_2, \Psi_2$};
  \path[fill=black] (150.37251,87.630531) node[above right] (text5199-52-1-9-5)
    {$p_3, \Psi_3$};
  \path[->,>=latex,draw=black,line join=miter,line cap=butt,line width=0.924pt]
    (140,120) -- (152,95);
  \path[->,>=latex,draw=black,line join=miter,line cap=butt,line width=0.924pt]
    (172,120) -- (172,95);
  \path[->,>=latex,draw=black,line join=miter,line cap=butt,line width=0.924pt]
    (223,60) -- (192,60);
\path[fill=black] (100,145) node[above right] (text5199-52-2-0) {$\begin{subarray}{l}\text{epistemic} \\ \text{probabilities} \\ \text{(eigenvalues)}\end{subarray}$};
\path[fill=black] (170,143) node[above right] (text5199-52-2-0) {$\begin{subarray}{l}\text{possible} \\ \text{ontic states} \\ \text{(eigenstates)}\end{subarray}$};
\path[fill=black] (230,65) node[above right] (text5199-52-2-0) {$\begin{subarray}{l}\text{actual} \\ \text{ontic state} \end{subarray}$};
\end{scope}

\end{tikzpicture}
}
\par\end{center}
\par\end{centering}

\caption{\label{fig:DensityMatrixFulcrum}A schematic depiction of our postulated
relationship between a system's density matrix $\op{\rho}$ and its
associated epistemic state $\set{\left(p_{1},\Psi_{1}\right),\left(p_{2},\Psi_{2}\right),\left(p_{3},\Psi_{3}\right),\dotsc}$,
the latter consisting of epistemic probabilities $p_{1},p_{2},p_{3},\dotsc$
(the eigenvalues of the density matrix) and possible ontic states
$\Psi_{1},\Psi_{2},\Psi_{3},\dotsc$ (represented by the eigenstates
of the density matrix), where one of those \emph{possible} ontic states
(in this example, $\Psi_{2}$) is the system's \emph{actual} ontic
state.}
\end{figure}
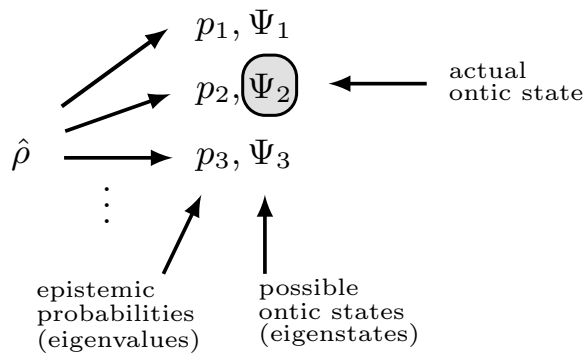

For the case of a density matrix that arises entirely due to external
quantum entanglements---that is, for a system in a totally improper mixture---our
interpretation identifies the eigenvalues of the density matrix as
collectively describing a probability distribution---the system's
epistemic state---over a set of possible ontic states that are represented
by the eigenstates of the density matrix, where one of those \emph{possible}
ontic states is the system's \emph{actual} ontic state. (See Figure~\ref{fig:DensityMatrixFulcrum}.)
In the very rough sense in which observables in the traditional formulation
of quantum theory correspond to Hermitian operators whose eigenvalues
represent measurement outcomes and whose eigenstates represent their
associated state vectors, our interpretation therefore identifies
a system's density matrix as the Hermitian operator corresponding
to the system's ``probability observable.''

\subsubsection{Quantum Dynamics}

As is well known, density matrices describing \emph{closed} systems
evolve in time according to a \emph{quantum} version of the Liouville
equation, known as the Von Neumann equation, which bears a striking
resemblance%
\footnote{This resemblance becomes even closer after an appropriate change to
complex phase-space variables, as we discuss in Section~\ref{subsub:Coherent-States}.%
} to the classical Liouville equation \eqref{eq:ClassicalLiouvilleEq}
and is given by 
\begin{equation}
\frac{\partial\op{\rho}}{\partial t}=-\frac{i}{\hbar}\comm{\op H}{\op{\rho}}.\label{eq:QuantumLiouvilleEq}
\end{equation}
 This equation for the time evolution of density matrices has been
generalized \cite{Lindblad:1976gqds,BreuerPetruccione:2002toqs,Joos:2003dacwqt,SchumacherWestmoreland:2010qpsi}
to cases involving open systems losing information to their larger
environments, but a missing ingredient has been a general specification
of the dynamics that governs the system at the level of its \emph{ontic
and epistemic states}.

Our interpretation supplies this key ingredient by translating the
evolution law governing density matrices into a Markovian dynamical
mapping in the sense of \eqref{eq:ClassicalOnticLevelDynMapGeneral}
that satisfies the key consistency requirement \eqref{eq:ClassicalMultilinearEpistemicLevelDynMapGeneral}
relating the dynamics of ontic states with the dynamics of epistemic
states. This mapping belongs to a new class of conditional probabilities
that both dynamically relate the ontic states of a single quantum
system at initial and final times and also define the kinematical
relationship between the ontic states of parent systems with those
of their subsystems. In addition, the mapping serves as a kind of
smoothness condition that sews together ontic states in a sensible
manner and avoids some of the ontic-level instabilities that arise
in other interpretations of quantum mechanics.

\section{The Minimal Modal Interpretation\label{sec:The-Minimal-Modal-Interpretation}}

\subsection{Basic Ingredients\label{sub:Basic-Ingredients}}

Having presented the motivation, conceptual foundations, and rough
outlines of our interpretation in the preceding sections, we now commence
our precise exposition. The axiomatic content of our interpretation,
which we summarize in detail in Section~\ref{sub:Summary}, is intended
to be minimal and consists of 
\begin{enumerate}
\item definitions provided in this section of what we mean by ontic and
epistemic states in quantum theory, as well as a dichotomy between
subjective (``classical'') and objective epistemic states;
\item a postulated relationship detailed in Section~\ref{sub:The-Correspondence-Between-Objective-Epistemic-States-and-Density-Matrices}
between objective epistemic states and density matrices, and, as a
consequence, a relationship also between ontic states and density
matrices;
\item a rule (the partial-trace operation) described in Section~\ref{sub:Parent-Systems,-Subsystems,-and-the-Partial-Trace-Operation}
for calculating density matrices for subsystems given density matrices
for their parent systems, and, as an immediate corollary, also a rule
for relating the objective epistemic states of parent systems with
those of their subsystems;
\item and, finally, a general formula (defining a class of quantum conditional probabilities)
in Section~\ref{sub:Quantum-Conditional-Probabilities} both for
\emph{kinematically} relating the ontic states of parent systems with
those of their subsystems, as well as for \emph{dynamically} relating
ontic states to each other over time and objective epistemic states
to each other over time.
\end{enumerate}
Note that apart from the references to density matrices, all these
rules and metaphysical entities have necessary (if often implicit)
counterparts in the description of classical systems detailed in Section~\ref{sub:Classical-Theories}:
For classical systems as well as for quantum systems, we must define
what we mean by ontic and epistemic states, establish relationships
between epistemic states of parent systems and subsystems as well
as between ontic states of parent systems and subsystems (the latter
relationship being trivial in the classical case, but crucial for
making sense of measurements in the quantum case), and dynamical relationships
between ontic states over time and epistemic states over time.

There are two chief differences between the classical and quantum
cases, the first difference being that in the quantum case, we need
to invoke density matrices, ultimately because quantum entanglement
leads to the existence of fundamentally objective epistemic states
and makes density matrices a crucial link between parent systems and
their subsystems. The second difference is that we can't afford to
take as many of the other ingredients for granted in the quantum case
as we could in the classical case---quantum theory forces us to be
much more explicit about our assumptions.

We claim that all the familiar features of quantum theory follow as
consequences when we take all these basic principles their logical
conclusions, including the Born rule for computing empirical outcome
probabilities and also consistency with the various no-go theorems,
as we will discuss starting in Section~\ref{sec:The-Measurement-Process}.
We treat several additional no-go theorems in Appendix~\ref{sub:Foundational-Theorems}.

\subsubsection{Ontic States}

Quantum physics first enters our story with the condition that the
distinct ontic states $\Psi_{i}$ that make up our system's configuration
space are each represented by a particular unit vector $\ket{\Psi_{i}}$
(up to overall phase) in a complex inner-product vector space $\mathcal{H}$,
called the system's Hilbert space: 
\begin{equation}
\boxed{\Psi_{i}\exchange\ket{\Psi_{i}}\in\mathcal{H}\ \left(\mathrm{up\ to\ overall\ phase}\right).}\label{eq:OntStateVecCorresp}
\end{equation}
 That is, we identify the system's configuration space as the quotient
$\mathcal{H}/\sim$, where $\sim$ represents equivalence of vectors
in the Hilbert space $\mathcal{H}$ up to overall normalization and
phase. Note that we regard \eqref{eq:OntStateVecCorresp} as merely
a \emph{correspondence} between an ontic object and a mathematical
object, and so we will not use the terms ``ontic state'' and ``state
vector'' synonymously in this paper, in contrast to the practice
of some authors.

\subsubsection{Epistemic States\label{subsub:Epistemic-States}}

As part of the definition of our interpretation, and in close parallel
with the classical case presented in Section~\ref{sub:Classical-Theories},
we postulate that a quantum system's evolving epistemic state consists
at each moment in time $t$ of a collection $\setindexed{\left(p_{i},\Psi_{i}\right)}i$
of ordered pairs that for each $i$ identify one of the system's \emph{possible}
ontic states $\Psi_{i}=\Psi_{i}\left(t\right)$ together with the
corresponding epistemic probability $p_{i}=p_{i}\left(t\right)\in\left[0,1\right]$
for which $\Psi_{i}$ is the system's \emph{actual} ontic state at
the current time $t$.%
\footnote{Throughout this paper, we will always work in the so-called Schrödinger picture,
in which a system's time evolution is carried by states and not by
operators. Indeed, because most of the systems that we consider in
this paper aren't dynamically closed and thus don't evolve according
to unitary dynamics---unitary dynamics usually being a good approximation
only for systems that are microscopic and therefore easy to isolate
from their environments---none of the other familiar pictures (such
as the Heisenberg picture) will generally be well-defined anyway.%
} As in the classical case, an epistemic state is called pure in the
idealized case in which its probability distribution is trivial---that
is, when one epistemic probability $p_{i}$ is equal to unity and
all the others vanish---and is called mixed when the probability
distribution is nontrivial.

Note that a system's epistemic probabilities $p_{i}$ at a particular
moment in time should not generally be confused with the empirical outcome probabilities
that are used to quantify the results of experiments and that are
computed via the Born rule. Whereas epistemic probabilities describe
a system's \emph{present} state of affairs, we will ultimately show
that empirical outcome probabilities can be identified with a measurement
device's predicted \emph{future} epistemic probabilities conditioned
on the hypothetical assumption that the device will have performed
a particular measurement on a given subject system.

\subsubsection{Modal Interpretations and Minimalism}

Our use of the modifiers ``possible and ``actual,'' together known
formally as modalities, identifies our interpretation of quantum
theory as belonging to the general class of modal interpretations
originally introduced by Krips in 1969 \cite{Krips:1969tpqm,Krips:1975siqt,Krips:1987mqt}
and then independently developed by van Fraassen (whose early formulations
involved the fusion of modal logic \cite{LewisLangford:1932sl,CocchiarellaFreund2008:mliss,Williamson2013:mlm}
with quantum logic \cite{BirkhoffVonNeumann:1936lqm,Gibbins:1987pplql}),
Dieks, Vermaas, and others \cite{vanFraassen:1972faps,Cartwright:1974vfmmqm,vanFraassen:1991qmmv,Bub:1992qmwpp,VermaasDieks:1995miqmgdo,BacciagaluppiDickson:1997gqtm,BacciagaluppiDickson:1999dmi,Vermaas:1999puqm,LombardiCastagnino:2008mhiqm,ArdenghiCastagninoLombardi:2011mhiqmco,LombardiArdenghFortiniNarvaja:2011fqmdi}.

The modal interpretations are now understood to encompass a very large
set of interpretations of quantum theory, including most interpretations
that fall between the ``many worlds'' of the Everett-DeWitt approach
and the ``no worlds'' of the instrumentalist approaches.  Generally
speaking, in a modal interpretation, one singles out some preferred
basis for each system's Hilbert space and then regards the elements
of that basis as the system's \emph{possible} ontic states---one of
which is the system's \emph{actual} ontic state---much in keeping
with how we think conceptually about classical probability distributions.
For example, as we will explain more fully in Section~\ref{subsub:The-de-Broglie-Bohm-Pilot-Wave-Interpretation-of-Quantum-Theory},
and as emphasized in \cite{Vermaas:1999puqm}, the de Broglie-Bohm
pilot-wave interpretation can be regarded as a special kind of modal
interpretation in which the preferred basis is permanently fixed \emph{for
all} systems at a universal choice. Other modal interpretations, such
as our own, instead allow the preferred basis for a given system to
change---in our case by choosing the preferred basis to be the evolving
eigenbasis of that system's density matrix.  However, we claim that
no existing modal interpretation captures the one that we introduce
in this paper.

A central guiding principle of our interpretation is\emph{ minimalism}:
By intention, we make no changes to the way quantum theory should
be used in practice, and the only requirements we impose are those
that are absolutely mandated by the need for our notion of ontology
and epistemology to account for the observable predictions of quantum
theory---no more, no less. We are also metaphysically conservative
in the sense that we include only ingredients that are either already
a part of the standard formalism of quantum theory or that have counterparts
in classical physics. We therefore naturally call our interpretation
the \emph{minimal modal interpretation} of quantum theory.

Our reasons for this minimalism go beyond philosophically satisfying
notions of axiomatic simplicity and parsimony. Based on an abundance
of historical examples, we know that when trying to add an ontology
to quantum theory, including (even implicitly) more features than
is strictly necessary is not just a metaphysical extravagance, but
also usually leads to trouble. This trouble may take the form of unacceptable
ontological instabilities, conflicts with various no-go theorems,
an uncontrollable profusion of ``epicycles,'' or just an overwhelming
structural ornateness.

By contrast, we will show that our minimal modal interpretation evades
a recent no-go theorem \cite{DicksonClifton:1998limi,Myrvold:2002mir,EarmanRuetsche:2005rimi,BerkovitzHemmo:2005miqmrr,Myrvold:2009cc}
asserting that other members of the class of modal interpretations
are incompatible with Lorentz invariance at an ontological level.
We will also argue that the same minimalism makes it possible to provide
first-principles derivations of familiar aspects of quantum theory,
including the Born rule (and possible corrections to it) for computing
the empirical outcome probabilities that emerge from experiments.

\subsubsection{Hidden Variables and the Irreducibility of Ontic States\label{subsub:Hidden-Variables-and-the-Irreducibility-of-Ontic-States}}

To the extent that our interpretation of quantum theory involves hidden variables,
the actual ontic states underlying the epistemic states of systems
play that role. However, one could also argue that calling them hidden
variables is just an issue of semantics because they are on the same
metaphysical footing as both the \emph{traditional} notion of quantum
states as well as the actual ontic states of \emph{classical} systems.

In any event, it is important to note that our interpretation includes
\emph{no other} hidden variables: Just as in the classical case, we
regard ontic states as being \emph{irreducible} objects, and, in keeping
with this interpretation, we  \emph{do not} regard a system's ontic
state itself as being an epistemic probability distribution---much
less a ``pilot wave''---over a set of more basic hidden variables.
In a rough sense, our interpretation \emph{unifies} the de Broglie-Bohm
interpretation's pilot wave and hidden variables into a single ontological
entity that we call an ontic state.

In particular, we do not attach an epistemic probability interpretation
to the \emph{components} of a vector representing a system's ontic
state, nor do we assume \emph{a priori} the Born rule, which we will
ultimately \emph{derive} as a means of computing empirical outcome
probabilities. Otherwise, we would need to introduce an unnecessary
\emph{additional }level of probabilities into our interpretation and
thereby reduce its axiomatic parsimony and explanatory power.

Via the phenomenon of environmental decoherence, our interpretation
ensures that the evolving ontic state of a sufficiently macroscopic
system---with significant energy and in contact with a larger environment---is
highly likely to be represented by a temporal sequence of state vectors
that presumably approximate coherent states, as described in Section~\ref{subsub:Coherent-States},
and whose labels evolve in time according to recognizable semiclassical
equations of motion. For microscopic, isolated systems, by contrast,
we simply accept that the ontic state vector may not always have an
intuitively familiar classical description.

\subsubsection{Constraints from the Kolmogorov Axioms}

The logically rigid \emph{Kolmogorov axioms} \cite{Kolmogorov:1933ftp},
which are satisfied by any well-defined epistemic probability distribution,
require that our epistemic probabilities $p_{i}=p_{i}\left(t\right)$
at any moment in time $t$ can never sum to a value greater than unity.
In the idealized limit in which the system cannot decay, we can require
the stronger condition that the epistemic probabilities always sum
exactly to unity, $\sum_{i}p_{i}=1$. By contrast, for systems having
$\sum_{i}p_{i}<1$, we naturally interpret the discrepancy $\left(1-\sum_{i}p_{i}\right)\in\left[0,1\right]$
as the system's probability of no longer existing, as we will explain
in greater detail in Section~\ref{subsub:Imperfect-Tensor-Product-Factorizations-Truncated-Hilbert-Spaces-Approximate-Density-Matrices-and-Unstable-Systems}.

In keeping with the Kolmogorov axioms, we additionally require that
the system's corresponding possible ontic states $\Psi_{i}=\Psi_{i}\left(t\right)$
at any single moment in time $t$ are mutually exclusive. We translate
this requirement into quantum language as the condition that no member
$\ket{\Psi_{i}}$ of the system's set of possible ontic state vectors
at the time $t$ can be expressed nontrivially as superposition involving
any of the others; none is a \textquotedblleft{}blend\textquotedblright{}
involving any of the others. This condition is equivalent to requiring
that the system's possible ontic state vectors at the time $t$ must
all be mutually orthogonal, 
\begin{equation}
\braket{\Psi_{i}}{\Psi_{j}}=0\binaryfor i\ne j,\label{eq:MutualExclOrthog}
\end{equation}
 much in the familiar way that eigenstates corresponding to nondegenerate
eigenvalues of the Hermitian operators representing observables in
the traditional formulation of quantum theory are always orthogonal.
Hence, \emph{any} orthonormal basis for that Hilbert space defines
a distinct allowed sample space admitting logically rigorous probability
distributions. However, sets of non-orthogonal state vectors (such
as the system's Hilbert space \emph{as a whole}) do not, strictly
speaking, admit logically rigorous probability distributions, although,
in keeping with our discussion in Section~\ref{subsub:Formal-Epistemic-States},
it is occasionally useful to define \emph{formal} probability distributions
even in those cases.

\subsubsection{The de Broglie-Bohm Pilot-Wave Interpretation of Quantum Theory\label{subsub:The-de-Broglie-Bohm-Pilot-Wave-Interpretation-of-Quantum-Theory}}

In contrast to regarding \emph{any} orthonormal basis as a potentially
valid sample space, the so-called ``fixed'' modal interpretations
\cite{Bub:1992qmwpp,Vermaas:1999puqm}---which include the well-known
de Broglie-Bohm pilot-wave interpretation as a special case---require
that the sample space remains permanently fixed forever at some universally
preferred choice, which is then a matter for the interpretation to
try to justify once and for all. A generic state vector is then regarded
as both being an epistemic probability distribution over this fixed
sample space of ontic states and also as being an ontological entity
in its own right, namely, a physical ``pilot wave'' that guides
the evolution of the system's hidden ontic state through this fixed
sample space while being conceptually distinct from that hidden ontic
state.

In its original nonrelativistic formulation, the de Broglie-Bohm interpretation
\cite{deBroglie:1930iswm,Bohm:1952siqtthvi,Bohm:1952siqtthvii,BohmHiley:1993uu}
takes its fixed orthonormal basis to be the position eigenbasis of
point particles, and offers what might appear to be a straightforward
solution to questions concerning measurements in quantum theory \cite{Maudlin:1995wbtsmp}.
Unfortunately, in the \emph{relativistic} regime, spatial position
ceases to exist as an orthonormal basis: The Hilbert-space inner product
of two would-be position eigenstates is nonvanishing, although it
is exponentially suppressed in the particle's Compton wavelength $\lambda_{\mathrm{Compton}}=h/mc$
and therefore indeed goes to zero in the nonrelativistic limit $c\to\infty$
\cite{PeskinSchroeder:1995iqft}. (Preserving causality in the presence
of this spontaneous superluminal propagation famously necessitates
the existence of antiparticles \cite{PeskinSchroeder:1995iqft,Weinberg:1972gcpagtr,FeynmanWeinberg:1987eplp}.)

Attempts to recast the de Broglie-Bohm interpretation in a relativistic
context require giving up the elegance and axiomatic parsimony of
the interpretation's nonrelativistic formulation and involve replacing
the nonrelativistic position eigenbasis with bases that remain orthonormal
in the relativistic regime, such as the field-amplitude eigenbasis
for bosonic fields \cite{Bohm:1952siqtthvii,BohmHiley:1984pqbss}.
However, fermionic fields still represent a serious problem, because
their ``field amplitudes'' are inherently non-classical, anticommuting,
nilpotent Grassmann numbers; although Grassmann numbers provide a
convenient \emph{formal} device for expressing fermionic scattering
amplitudes in terms of Berezin path integrals \cite{Berezin:1966msq,PeskinSchroeder:1995iqft},
taking Grassmann numbers seriously as physical ingredients in quantum-mechanical
Hilbert spaces would lead to nonsense ``probabilities'' that are
not ordinary numbers. Trying instead to choose the fermion-number
eigenbasis \cite{Bell:2004suqm} brings back the question of ill-definiteness
of spatial position, leading some advocates to drop orthonormal bases
altogether in favor of POVMs \cite{Struvye:2012ozzpwaf}. Relativity
therefore implies that there's no safe choice of fixed orthonormal
basis to provide the de Broglie-Bohm interpretation with its foundation:
At best, there's no canonical choice of basis to fix once and for
all, and, at worst, there is no choice that's consistent or sensible.

Whether or not the de Broglie-Bohm interpretation's proponents ultimately
find a satisfactory fixed orthonormal basis to define their sample
space,%
\footnote{Even for a nonrelativistic system, it's not clear why the coordinate
basis should be favored in the de Broglie-Bohm interpretation over
an orthonormal basis approximating the system's far more classical-looking
coherent states \cite{Schrodinger:1926dsuvdmzm,Glauber:1963cisrf},
as defined in Section~\ref{subsub:Coherent-States}.%
} the interpretation still runs into other troubles as well, including
its inability to accommodate the non-classical changes of particle
spectrum that can arise in quantum field theories and the non-classical
changes in configuration space that can emerge from quantum dualities,
both of which play a central role in much of modern physics.%
\footnote{Non-classical changes in particle spectrum in quantum field theories
include prosaic examples like transitions from tachyonic particle
modes to massive radial ``Higgs'' modes and massless Nambu-Goldstone
modes after the spontaneous symmetry breaking of a continuous symmetry
\cite{Nambu:1960qpgits,Goldstone:1961ftss,GoldstoneSalamWeinberg:1962bs},
as well as more exotic examples like Skyrmions \cite{Skyrme:1961nlft,Skyrme:1961psqmf,PerringSkyrme:1962mufe,Skyrme:1962mufe}
and bosonization \cite{Coleman:1975qsgemtm,Mandelstam:1975soqsge}.
Prominent examples of dualities include generalizations of electric-magnetic
duality in certain supersymmetric gauge theories \cite{MontonenOlive:1977mmgp,Seiberg:1995emdsnagt},
holographic dualities like the AdS/CFT correspondence between gauge
theories and theories of quantum gravity in higher dimensions \cite{BrownHenneaux:1986cccrasetdg,Maldacena:1999lnlsfts,GubserKlebanovPolyakov:1998gtcncst,Witten:1998adsh,AharonyGubserMaldacenaOoguriOz:2000lnftstg,GuicaHartmanSongStrominger:2009kcftc},
and various dualities that play key roles in string theory \cite{GiveonPorratiRabinovici:1994tsdst,Sen:1994swcdfdst,StromingerYauZaslow:1996mstd}.
In particular, the AdS/CFT correspondence as well as string theory
suggest that spacetime can undergo radical changes in structure \cite{Witten:1993pn2ttd,AspinwallGreeneMorrison:1994mstd,Witten:1996ptmtft}
via intermediary non-geometric phases, and even perhaps that the ``space''
in ``spacetime'' is itself an emergent property of more primitive
ingredients \cite{Seiberg:2006es,ElShowkPapadodimas:2012eshcft}.%
}

The de Broglie-Bohm interpretation also suffers from a somewhat more
metaphysical problem: Because a pilot wave has an ontological existence
\emph{over and above} that of the hidden ontic state of its corresponding
physical system, we run into the well-known difficulty \cite{BrownWallace:2004smpdbble}
of making sense of the ontological status of all its many branches.
Indeed, the pilot wave's branches behave precisely as the ``many
worlds'' of the Everett-DeWitt interpretation of quantum theory in
all their complexity, despite the fact that only one of those branches
is supposedly ``occupied'' by the system's hidden ontic state and
the rest of the branches are ``empty worlds'' filled with ghostly
people living out presumably ghostly lives.

\subsubsection{Subjective Uncertainty and Proper Mixtures\label{subsub:Subjective-Uncertainty-and-Proper-Mixtures}}

As we explained in Section~\ref{sub:Classical-Theories}, we use
nontrivial epistemic states $\setindexed{\left(p\left(q\right),q\right)}q$
in classical physics in order to account for subjective uncertainty
about a classical system's actual ontic state at a given moment in
time. Similarly, one way that epistemic states $\setindexed{\left(p_{i},\Psi_{i}\right)}i$
can arise in quantum theory is when we have subjective uncertainty
about a quantum system's actual ontic state at a particular moment
in time, in which case we call the system's epistemic state a proper mixture.

Strictly speaking, the requirements of a logically rigorous probability
distribution require that the possible ontic states $\Psi_{i}$ that
make up a proper mixture must be mutually exclusive and hence correspond
to mutually orthogonal state vectors $\ket{\Psi_{i}}$ in accordance
with \eqref{eq:MutualExclOrthog}. However, after deriving the Born
rule later in this paper, we will describe in Section~\ref{sub:Subjective-Density-Matrices-and-Proper-Mixtures}
how to accommodate contexts in which it is useful to relax this mutual
exclusivity, just as we explained in Section~\ref{subsub:Formal-Epistemic-States}
how we could consider \emph{formal} probability distributions over
non-exclusive possibilities in classical physics.

There is no real controversy or dispute over the metaphysical meaning
of proper mixtures, at least in the sense that there is wide acceptance
for regarding a proper mixture as a prosaic, subjective probability
distribution over a hidden underlying ontic state. We will therefore
put proper mixtures entirely aside for most of this paper, returning
to them in Section~\ref{sub:Subjective-Density-Matrices-and-Proper-Mixtures}
only after deriving the Born rule.

\subsubsection{Objective Uncertainty and Improper Mixtures\label{subsub:Objective-Uncertainty-and-Improper-Mixtures}}

Quantum theory also features what our minimal modal interpretation
regards as an \emph{objective} kind of uncertainty that has no real
classical counterpart, and in this case we will \emph{always} insist
upon logically rigorous epistemic states involving mutually exclusive
possible ontic states. To see where this new form of uncertainty comes
from, and to make clear the importance of the relationship between
parent systems and subsystems in the context of our interpretation
of quantum theory, we need to step back for a moment and compare the
notions of parent systems and subsystems in classical physics and
in quantum physics.

Classically, a system $C$ with configuration space $\mathcal{C}_{C}$
is said to be a parent or composite system consisting of two subsystems
$A$ and $B$ with respective configuration spaces $\mathcal{C}_{A}$
and $\mathcal{C}_{B}$ if the configuration space of system $C$ is
expressible as the Cartesian product $\mathcal{C}_{C}=\mathcal{C}_{A}\cartesianprod\mathcal{C}_{B}=\setbuild{\left(a,b\right)}{a\in\mathcal{C}_{A},\ b\in\mathcal{C}_{B}}$,
meaning that each element $c=\left(a,b\right)$ of $\mathcal{C}_{C}$
is an ordered pair that identifies a specific element $a$ of $\mathcal{C}_{A}$
and a specific element $b$ of $\mathcal{C}_{B}$. We then naturally
denote the parent system $C$ by $A+B$, and, at least in the case
in which all the configuration spaces in question have finitely many
elements $N_{A},N_{B},N_{A+B}<\infty$, we have the simple relation
that $N_{A+B}=N_{A}N_{B}$. Note, of course, that either subsystem
$A$ or $B$ (or both) could well be a parent system to \emph{even
more elementary} subsystems.

The quantum case is much more subtle because the Cartesian product
of a pair of nontrivial Hilbert spaces is \emph{not} another Hilbert
space. We instead identify a given quantum system $C$ having a Hilbert
space $\mathcal{H}_{C}$ as being the parent system of two quantum
subsystems $A$ and $B$ with respective Hilbert spaces $\mathcal{H}_{A}$
and $\mathcal{H}_{B}$ if there exists an orthonormal basis for $\mathcal{H}_{C}$
whose elements are of the tensor-product form $\ket{a,b}=\ket a\tensorprod\ket b$,
where the sets of vectors $\ket a$ and $\ket b$ respectively constitute
orthonormal bases for $\mathcal{H}_{A}$ and $\mathcal{H}_{B}$. Then,
by construction, every vector in $\mathcal{H}_{C}$ consists of some
linear combination of the vectors $\ket{a,b}$. We express this fact
by writing the parent system's Hilbert space as the tensor product
$\mathcal{H}_{C}=\mathcal{H}_{A}\tensorprod\mathcal{H}_{B}$, and
we denote the parent system, as we did in the classical case, by $C=A+B$.
Observe that the dimensions of these various Hilbert spaces (assuming
they are all of finite dimension) then satisfy $\dim\mathcal{H}_{A+B}=\left(\dim\mathcal{H}_{A}\right)\left(\dim\mathcal{H}_{B}\right)$.

Given this background, consider a parent quantum system $A+B$ consisting
of two subsystems $A$ and $B$ and whose ontic state $\Psi_{A+B}$
is described by a so-called entangled state vector of the form 
\begin{equation}
\ket{\Psi_{A+B}}=\alpha\ket{\Psi_{A,1}}\ket{\Psi_{B,1}}+\beta\ket{\Psi_{A,2}}\ket{\Psi_{B,2}},\qquad\alpha,\beta\in\mathbb{C},\ \absval{\alpha}^{2}+\absval{\beta}^{2}=1.\label{eq:GenericTwoSubsysEntangledStateVec}
\end{equation}
 What is the ontic state of subsystem $A$? What is the ontic state
of subsystem $B$?

In the Copenhagen interpretation of quantum theory, these two questions
do not possess well-defined answers. Instrumentalist interpretations
do not regard even the questions themselves as being sensible or meaningful.

\subsubsection{The Everett-DeWitt Many-Worlds Interpretation of Quantum Theory\label{subsub:The-Everett-DeWitt-Many-Worlds-Interpretation-of-Quantum-Theory}}

By contrast, according to the Everett-DeWitt many-worlds interpretation
of quantum theory, \eqref{eq:GenericTwoSubsysEntangledStateVec} implies
that there exist two simultaneous ``worlds'' or ``realities''
or ``branches'': In one world, associated with a probability $\absval{\alpha}^{2}$,
the ontic state of $A$ is $\Psi_{A,1}$ and the ontic state of $B$
is $\Psi_{B,1}$, whereas in the other world, associated with a probability
$\absval{\beta}^{2}$, the ontic state of $A$ is $\Psi_{A,2}$ and
the ontic state of $B$ is $\Psi_{B,2}$.

Of course, because we can always choose from a \emph{continuously
infinite} set of different orthonormal bases for the Hilbert space
of the parent system $A+B$, as we discussed in Section~\ref{subsub:Quantum-Kinematics},
these two ``worlds'' are radically non-unique even for a fixed parent-system
state vector; moreover, the different possible world-bases are not
generally related to one another in a manner that can be conceptualized
classically, and preserving manifest locality in the many-worlds interpretation
generically requires switching from one world-basis to another as
a function of time. These issues, known collectively as the preferred-basis problem,
imply a breakdown in the popular portrayal of the many-worlds interpretation
as describing unfolding reality in terms of a well-defined forking
structure in the manner of Borges' \emph{Garden of Forking Paths}
\cite{Borges:1941gfp} or Lewis's modal realism \cite{Lewis:1973c,Lewis:1986opw},
and are unsolvable without introducing additional axiomatic ingredients
into the interpretation. In particular, without additional postulates,
one cannot evade the preferred-basis problem merely by appealing to
environment-induced decoherence, because the many-worlds interpretation
traditionally assumes that ``the universe as a whole''---which determines
the single branch-set shared by all systems in Nature---is described
by an always-pure state that never undergoes decoherence and therefore
doesn't possess a canonical preferred basis.%
\footnote{For an explicit discussion of these points in the context of the EPR-Bohm
thought experiment, including the issue of nonlocality, see Section~\ref{subsub:Nonlocality in the Everett-DeWitt Many-Worlds Interpretation}.
There are additional reasons to be suspicious of attempts to give
all the branches of the many-worlds interpretation an equal ontological
meaning, as, for example, Aaronson describes in the context of quantum
computing in \cite{Aaronson:2013sapp}. We also emphasize that attaching
a many-worlds ontology to mathematical vectors in the first place
is \emph{itself} a nontrivial axiom; indeed, the instrumentalist approach,
which we describe in Appendix~\ref{subsub:The-Instrumentalist-Approach},
does not postulate \emph{any} ontological status for state vectors.%
}

Another conundrum of the many-worlds interpretation is the difficulty
in making sense of the components $\alpha$ and $\beta$ in \eqref{eq:GenericTwoSubsysEntangledStateVec}
in terms of \emph{probabilities} when both worlds are \emph{deterministically}
and \emph{simultaneously} realized,%
\footnote{Maudlin eloquently captures this key shortcoming of the many-worlds
interpretation in ``Problem 2: The problem of statistics'' in \cite{Maudlin:1995tmp},
and \cite{Deutsch:1999qtpd,Wallace:2002qpdtr,Wallace:2009fpbrdta}
attempt to suppress the problem by burying it under the elaborate
 axiomatic apparatus of decision theory.%
} especially given the fundamental logical obstruction to deriving
probabilistic conclusions from deterministic assumptions and in light
of the preferred-basis problem and the continuously infinite non-uniqueness
of the choice of world-basis. Putting aside the preferred-basis problem,
a common approach \cite{FarhiGoldstoneGutmann:1989hpaqm,AguirreTegmark:2011biuciqm}
is to study a pure state defined in terms of a ``limit'' of many
identical copies of a given measurement set-up, and then argue that
``maverick branches''---meaning terms in the final-state superposition
whose outcome frequency ratios deviate significantly from the Born
rule---have arbitrarily small Hilbert-space amplitudes and thus can
safely be ignored. However, infinite limits are always rigorously
defined in terms of increasing but finite sequences, and for any finite
number of copies of a given measurement set-up, maverick branches
have nonzero amplitudes and outnumber branches with better-behaved
frequency ratios. Asserting that the smallness of their amplitudes
makes maverick branches ``unlikely'' therefore implicitly assumes
the very probability interpretation to be derived \cite{CavesSchack:2005pfodniqpp,BuniyHsuZee:2006dopqm},
and is akin to the sort of circular reasoning inherent in all attempts
to use the law of large numbers to turn frequentism into a rigorous
notion of probability.

Even if one could somehow add enough additional axioms to justify
interpreting state-vector components like $\alpha$ and $\beta$ in
\eqref{eq:GenericTwoSubsysEntangledStateVec} as \emph{instantaneous},\emph{
kinematical} probabilities, the traditional many-worlds interpretation
of quantum theory lacks an explicit model describing how the experiential
trajectory of an individual observer \emph{dynamically} unfolds from
moment to moment through the interpretation's multitudinous (and ill-defined)
branching worlds. That is, even if the many-worlds interpretation
admits a sequence of static probability distributions at individual
moments in time, it does not possess anything like the dynamical conditional
probabilities \eqref{eq:ClassicalOnticLevelDynMapGeneral} connecting
one moment in time of an observer's experiential trajectory to the
next moment in time.

In particular, without adding on significantly more assumptions and
metaphysical structure, the many-worlds interpretation is unable to
ensure the \emph{ontological stability} of such an observer's experiential
trajectory through time: Observers in the many-worlds interpretation
of quantum theory are vulnerable to radical macroscopic instabilities
in experience (and memory) that parallel the sorts of macroscopic
instabilities that are possible in a hypothetical version of classical
physics that lacks ontic-level dynamics, as we explained in Section~\ref{subsub:Classical-Dynamics}
shortly after \eqref{eq:ClassicalMultilinearEpistemicLevelDynMapGeneral}.
We will detail a concrete quantum analogue of such dynamical ontological
instabilities (eigenstate swaps) in Section~\ref{subsub:Near-Degeneracies-and-Eigenstate-Swaps},
and we will eliminate them in the context of our own interpretation
of quantum theory when we introduce a dynamical notion of quantum
conditional probabilities in Section~\ref{sub:Quantum-Conditional-Probabilities}.
Modal interpretations have been criticized in the past for lacking
any such dynamical smoothing conditions on ontic-state trajectories
to eliminate these sorts of metaphysical instabilities,%
\footnote{See ``Problem 3: The problem of effect'' in \cite{Maudlin:1995tmp}.%
} but to the extent that such criticisms of the traditional modal interpretations
are justified, the same criticisms apply equally to the many-worlds
interpretation as well.

Finally, because all realistic systems exhibit a nonzero degree of
quantum entanglement with other systems, the traditional many-worlds
interpretation depends crucially upon the existence of a maximal closed
system---again, ``the universe as a whole''---that admits a description
in terms of an \emph{exact} cosmic pure state. Indeed, the many-worlds
interpretation relies on the existence of this cosmic pure state in
order to identify the branches on which the superposed copies of various
subsystems reside, and with what associated probabilities. However,
as we discussed in Section~\ref{sub:Comparison-with-Other-Interpretations-of-Quantum-Theory},
there are reasons to be skeptical that any such maximal closed system
is physically guaranteed to exist and be well-defined, thus opening
up the real possibility that the many-worlds interpretation isn't
fundamentally well-defined either.

\subsubsection{Our Interpretation\label{subsub:Our-Interpretation}}

Answering our ontological questions about the state vector \eqref{eq:GenericTwoSubsysEntangledStateVec}
in the context of our own interpretation of quantum theory will require
that we first introduce a class of well-known mathematical objects,
called density matrices, that have no counterpart in classical physics.
For now, we will simply say that our uncertainty over each subsystem's
actual ontic state is captured by what we call an objective epistemic state
for that subsystem. A quantum epistemic state that includes at least
some objective uncertainty of this kind---possibly together with some
subjective uncertainty as well---is called an improper mixture. Turning
things around, we can then identify an objective epistemic state as
an improper mixture in the extremal case of zero subjective uncertainty.

In contrast to proper mixtures (that is, wholly subjective epistemic
states), improper mixtures do not have a widely accepted \emph{a priori}
meaning, and it is a central purpose of our minimal modal interpretation
of quantum theory to provide one, starting with the requirement that
an objective epistemic state must be a logically rigorous probability
distribution and therefore always involve mutually exclusive \eqref{eq:MutualExclOrthog}
possible ontic states. In Section~\ref{sub:The-Correspondence-Between-Objective-Epistemic-States-and-Density-Matrices},
we will define density matrices and relate them to objective epistemic
states, and then, once we have developed a prescription for relating
the density matrices of parent systems with those of their subsystems
in Section~\ref{sub:Parent-Systems,-Subsystems,-and-the-Partial-Trace-Operation},
we will finally have in our hands a precise means of resolving the
aforementioned conundrum about the ontic states of $A$ and $B$.

\subsection{The Correspondence Between Objective Epistemic States and Density
Matrices\label{sub:The-Correspondence-Between-Objective-Epistemic-States-and-Density-Matrices}}

Even if we happen to know a system's objective epistemic state $\setindexed{\left(p_{i}\left(t\right),\Psi_{i}\left(t\right)\right)}i$
at a given time $t$, we have not yet presented a framework for determining
the system's objective epistemic state at any other time $t^{\prime}\ne t$,
nor a prescription for relating ontic states to each other over time
or for relating objective epistemic states or ontic states between
parent systems and their subsystems. Our first step will be to define
a correspondence between the objective epistemic states of quantum
systems and a class of mathematical objects---density matrices---that
will serve as the fulcrum of this framework.

\subsubsection{Density Matrices}

Any objective epistemic state $\setindexed{\left(p_{i}\left(t\right),\Psi_{i}\left(t\right)\right)}i$
of the kind defined in Section~\ref{subsub:Epistemic-States}---meaning,
in particular, that it involves \emph{mutually exclusive} possible
ontic states $\Psi_{i}\left(t\right)$---can be identified uniquely%
\footnote{We address subtleties regarding degeneracies $p_{i}\left(t\right)=p_{j}\left(t\right)$
($i\ne j$) in Section~\ref{subsub:Probability-Crossings-and-Degeneracies}.%
} with a unit-trace, positive semi-definite matrix $\op{\rho}\left(t\right)$
known as the system's density matrix: 
\begin{equation}
\eqsbrace{\begin{aligned}\op{\rho}\left(t\right)=\sum_{i}p_{i}\left(t\right)\ket{\Psi_{i}\left(t\right)}\bra{\Psi_{i}\left(t\right)},\qquad\mathrm{with}\  & \op{\rho}\left(t\right)^{\adj}=\op{\rho}\left(t\right),\qquad\Tr\left[\op{\rho}\left(t\right)\right]=1,\\
 & 0\leq p_{i}\left(t\right)\leq1,\qquad\sum_{i}p_{i}\left(t\right)=1,\\
 & \braket{\Psi_{i}\left(t\right)}{\Psi_{j}\left(t\right)}=\delta_{ij}.
\end{aligned}
}\label{eq:DensMatrixTildes}
\end{equation}
 By ``identified'' here, we mean that there exists a precise correspondence%
\footnote{In the traditional language of the modal interpretations \cite{Vermaas:1999puqm},
this correspondence is termed a (core) property ascription or an
ontic ascription, our ontic states $\Psi_{i}$ are called (core) properties
or value states, and our objective epistemic states $\setindexed{\left(p_{i},\Psi_{i}\right)}i$
are called property sets, mathematical states, or dynamical states
\cite{VermaasDieks:1995miqmgdo}.%
} between the objective epistemic state $\setindexed{\left(p_{i}\left(t\right),\Psi_{i}\left(t\right)\right)}i$
of the system and the eigenvalue-eigenvector pairs of the density
matrix $\op{\rho}\left(t\right)$ at each moment in time $t$: 
\begin{equation}
\boxed{\setindexed{\left(p_{i}\left(t\right),\Psi_{i}\left(t\right)\right)}i\exchange\op{\rho}\left(t\right)=\sum_{i}p_{i}\left(t\right)\ket{\Psi_{i}\left(t\right)}\bra{\Psi_{i}\left(t\right)}.}\label{eq:EpStateDensMatrixCorresp}
\end{equation}
 Essentially, the correspondence \eqref{eq:EpStateDensMatrixCorresp}
provides a means of encoding a list of non-negative real numbers---the
system's epistemic probabilities---and a list of state vectors---representing
the system's possible ontic states---in a basis-independent manner,
namely, as the eigenvalue-eigenvector spectrum of the system's density
matrix.

Recalling the definition \eqref{eq:DefClassicalEntropy} of the entropy
of a classical system's epistemic state, the correspondence \eqref{eq:EpStateDensMatrixCorresp}
immediately implies that we can express the entropy $-\sum_{i}p_{i}\log p_{i}$
of a quantum objective epistemic state $\setindexed{\left(p_{i}\left(t\right),\Psi_{i}\left(t\right)\right)}i$
in terms of the corresponding density matrix $\op{\rho}$ according
to the basis-independent Von Neumann entropy formula 
\begin{equation}
S\equiv-\Tr\left[\op{\rho}\log\op{\rho}\right].\label{eq:DefVonNeumannEntropy}
\end{equation}
  For recent work attempting to derive important aspects of statistical
mechanics from fundamentally quantum reasoning, including various
fluctuation theorems and the second law of thermodynamics, see, for
example, \cite{PopescuShortWinter:2005fsmeisa,PopescuShortWinter:2006efsm,EspositoMukamel:2006ftqme,GoldsteinLebowitzTumulkaZanghi:2006ct,LindenPopescuShortWinter:2009,KawamotoHatano:2011tftnmoqs,LeggioNapoliBreuerMessina:2013ftqme,AndersGiovannetti:2013tftnmoqs}.

In light of the association \eqref{eq:EpStateDensMatrixCorresp} between
probabilities and matrices, together with linear-algebraic expressions
like the Von Neumann entropy formula \eqref{eq:DefVonNeumannEntropy},
quantum theory has been called ``a noncommutative generalization
of classical probability theory'' \cite{LeiferSpekkens:2011fqtcntbi}.
However, we regard \eqref{eq:EpStateDensMatrixCorresp} as merely
a \emph{correspondence} between an epistemic object and a mathematical
object, and so, just as we do not use the terms ``ontic state''
and ``state vector'' synonymously in this paper even though the
associated objects are related by \eqref{eq:OntStateVecCorresp},
we will not use the terms ``(objective) epistemic state'' and ``density
matrix'' synonymously either, again in contrast to the practice of
some authors.

The correspondence \eqref{eq:EpStateDensMatrixCorresp}, which lies
at the core of our minimal modal interpretation of quantum theory,
provides a natural context for emphasizing the following tautological
but crucial point:  

\begin{equation}
\boxed{\begin{subarray}{l}
\mbox{If our interpretation does not \emph{explicitly} identify a given}\\
\mbox{quantity as being a literal epistemic probability, then}\\
\mbox{that quantity is not---or at least not \emph{yet}---a literal}\\
\mbox{epistemic probability.}
\end{subarray}}\label{eq:TautStatementReEpProbs}
\end{equation}
 In particular, in keeping with our earlier comment in Section~\ref{subsub:Hidden-Variables-and-the-Irreducibility-of-Ontic-States}
that, insofar as our interpretation involves hidden variables, ontic
states play the role of those hidden variables, and also our comment
that we do not regard the state vectors representing them as being
epistemic probability distributions for anything else, we do not assume
the Born rule $\mathrm{Prob}\left(\dotsc\right)=\absval{\dotsc}^{2}$
\emph{a priori}, nor do we correspondingly interpret the absolute-value-squared
values of the complex components of state vectors as being \emph{literal}
probabilities for any of our hidden variables. According to our interpretation,
and in accordance with the basic correspondence \eqref{eq:EpStateDensMatrixCorresp}
above, we do not interpret the absolute-value-squared components of
state vectors as describing literal probabilities for our hidden variables
until some process---perhaps measurement- or environment-induced decoherence---turns
them into the eigenvalues of \emph{some} system's density matrix.
For these reasons, once we eventually do derive the Born rule in Section~\ref{subsub:Von-Neumann-Measurements},
we will refer to the absolute-value-squared components of pre-measurement
state vectors as empirical outcome probabilities, because it is only
after a measurement device performs a specific measurement that they
eventually become \emph{actualized} as true epistemic probabilities
for some system.

\subsubsection{Probability Crossings and Degeneracies\label{subsub:Probability-Crossings-and-Degeneracies}}

One frequently cited anomaly in the purportedly one-to-one relationship
\eqref{eq:EpStateDensMatrixCorresp} between objective epistemic states
and density matrices concerns the issue of eigenvalue degeneracies.
However, although one can easily picture a \emph{classical epistemic
state} evolving smoothly through a \emph{probability} crossing 
\begin{equation}
p_{i}\left(t_{c}\right)=p_{j}\left(t_{c}\right)\label{eq:ProbCrossing}
\end{equation}
 that occurs at some specific moment in time $t=t_{\mathrm{c}}$,
\emph{eigenvalue }crossings in \emph{density matrices} are arrangements
that require infinitely sharp---that is, measure-zero---fine-tuning
and thus never realistically occur: They are co-dimension-three events
in the abstract four-dimensional space of $2\times2$ density-matrix
blocks \cite{BacciagaluppiDonaldVermaas:1995cddpmi,Hollowood:2013cbr}
and would therefore require that the off-diagonal entries vanish \emph{precisely}
when the diagonal elements \emph{exactly} agree.

Just for purposes of illustration, consider the following \emph{static}
example: Although the two-particle, spin-singlet state vector 
\begin{equation}
\ket{\Psi_{\mathrm{EPR\hyphen Bohm}}}=\frac{1}{\sqrt{2}}\left(\ket{\uparrow\downarrow}-\ket{\downarrow\uparrow}\right)\label{eq:EPRPairStateVecPerfDegen}
\end{equation}
 familiar from the EPR-Bohm thought experiment (to be discussed in
detail in Section~\ref{sub:The-EPR-Bohm-Thought-Experiment-and-Bell's-Theorem})
would naïvely lead to degenerate $2\times2$ reduced density matrices
for each individual particle, setting up this state vector \emph{exactly}
would require unrealistically fine-tuning the total spin to $S_{\mathrm{tot},z}=0$
to infinite precision---a measure-zero state of affairs. In any realistic
scenario, there will unavoidably exist degeneracy-breaking deviations
among the components of state vectors like $\ket{\Psi_{\mathrm{EPR\hyphen Bohm}}}$,
so that, at best, the state vector actually takes the form 
\begin{equation}
\ket{\Psi_{\mathrm{EPR\hyphen Bohm}}}=\frac{1}{\sqrt{2}}\left(\left(1+\epsilon_{1}\right)\ket{\uparrow\downarrow}-\left(1+\epsilon_{2}\right)\ket{\downarrow\uparrow}+\epsilon_{3}\ket{\uparrow\uparrow}+\epsilon_{4}\ket{\downarrow\downarrow}\right),\quad\absval{\epsilon_{1}},\absval{\epsilon_{2}},\absval{\epsilon_{3}},\absval{\epsilon_{4}}\ll1.\label{eq:EPRPairStateVecDegenBreaking}
\end{equation}

Some have suggested \cite{Albert:1994qme} that the problem with density
matrices is how to interpret them when they exhibit \emph{exact} degeneracies,
which we have now seen are unphysical idealizations. However, as we
will explain next, the \emph{actual} danger for interpretations that
are based on a correspondence like \eqref{eq:EpStateDensMatrixCorresp}
(such as most modal interpretations) arises from the fact that density
matrices can never realistically \emph{have} exact degeneracies in
the first place.

\subsubsection{Near-Degeneracies and Eigenstate Swaps\label{subsub:Near-Degeneracies-and-Eigenstate-Swaps}}

The closest density-matrix counterpart to a probability crossing \eqref{eq:ProbCrossing}
is a near-degeneracy in which two probability eigenvalues reach a
point of closest approach 
\begin{equation}
\absval{p_{i}\left(t\right)-p_{j}\left(t\right)}\sim\rho_{0}\xi>0\label{eq:NearDegenProbClosestApproach}
\end{equation}
 at some specific moment in time $t_{0}$ and then turn around again,
while their associated orthogonal eigenstates $\ket{\Psi_{i}\left(t\right)}\perp\ket{\Psi_{j}\left(t\right)}$
undergo an ultra-fast eigenstate swap%
\footnote{Eigenstate swaps are called \emph{core property instabilities} or
\emph{fluctuations} in \cite{Vermaas:1999puqm} and \emph{crossovers}
in \cite{Hollowood:2013cbr}.%
} 
\begin{equation}
\eqsbrace{\begin{aligned}\ket{\Psi_{i}\left(t\right)} & \mapsto\ket{\Psi_{i}\left(t+\delta t_{\mathrm{swap}}\right)}\approx\ket{\Psi_{j}\left(t\right)}\perp\ket{\Psi_{i}\left(t\right)},\\
\ket{\Psi_{j}\left(t\right)} & \mapsto\ket{\Psi_{j}\left(t+\delta t_{\mathrm{swap}}\right)}\approx\ket{\Psi_{i}\left(t\right)}\perp\ket{\Psi_{j}\left(t\right)}
\end{aligned}
}\label{eq:EigenstateSwap}
\end{equation}
 over an ultra-short time scale of order 
\begin{equation}
\delta t_{\mathrm{swap}}\sim\rho_{0}\xi\tau.\label{eq:EigenstateSwapTimeScale}
\end{equation}
 The complex-valued dimensionless quantity 
\begin{equation}
\xi\sim\exp\left(-\#\mathrm{degrees\ of\ freedom}\right)\label{eq:NearDegenOffDiagParam}
\end{equation}
 has exponentially small magnitude in the total number of degrees
of freedom of the system itself and of all other systems that substantially
interact and entangle with it, and characterizes the size of the off-diagonal
elements in the relevant $2\times2$ density-matrix block in a basis
in which the diagonal elements become momentarily equal at precisely
$t=t_{0}$. The real-valued quantity $\rho_{0}$ is the would-be-degenerate
eigenvalue in the idealized but unphysical measure-zero case $\xi=0$.
Hence, near $t=t_{0}$, the relevant $2\times2$ density-matrix block
takes the approximate form 
\begin{equation}
\begin{pmatrix}\rho_{0}+\left(t-t_{0}\right)/\tau & \rho_{0}\xi\\
\rho_{0}\xi^{\conj} & \rho_{0}-\left(t-t_{0}\right)/\tau
\end{pmatrix}\subset\op{\rho}.\label{eq:NearDeg2x2Block}
\end{equation}
 Meanwhile, the real-valued quantity $\tau$, which has units of time,
is the characteristic time scale over which the probability eigenvalues
$p_{i}\left(t\right)$ and $p_{j}\left(t\right)$ would be changing
in the \emph{absence} of eigenstate-swap effects. To the approximate
extent that we can speak of well-defined energies $E$ for a system
that is open and thus whose dynamics is not strictly unitary, ordinary
time evolution corresponds roughly to the $\exp\left(-iEt/\hbar\right)$
relative phase factors familiar from elementary quantum theory, and
thereby implies that 
\begin{equation}
\tau\sim\hbar/E.\label{eq:OrdinaryDynamicalRate}
\end{equation}

Taken seriously, eigenstate swaps \eqref{eq:EigenstateSwap} could
conceivably cause large systems to undergo frequent but sudden fluctuations
between \emph{macroscopically distinct} ontic states over arbitrarily
short time scales, and thus have long been considered to be serious
ontological instabilities inherent in density-matrix-centered interpretations
of quantum theory.%
\footnote{As Vermaas writes on p. 133 of \cite{Vermaas:1999puqm} in critiquing
his own modal interpretation of quantum theory: ``If one {[}takes
eigenstate swaps seriously{]}, it follows that the set of eigenprojections
of a state that comes arbitrarily close to a degeneracy can change
maximally in an arbitrarily small time interval $t_{2}-t_{1}$. {[}...{]}
{[}T{]}he set of core properties can change rapidly, resulting in
an unstable property ascription during a finite time interval.''
Later, on p. 260, he writes: \char`\"{}{[}I{]}f a state has a spectral
resolution which is nearly degenerate, then an arbitrarily small change
of that state (by an interaction with the environment or by internal
dynamics) can maximally change the set of the possible core properties.
Perhaps this instability is one of the more serious defects of modal
interpretations because, firstly, it can have consequences for their
ability to solve the measurement problem. For even when a modal interpretation
ascribes readings to a pointer at a specific instant, a small fluctuation
of the state may mean that at the next instant the pointer possesses
properties which are radically different to readings.\char`\"{}%
} However, as we will explain in detail when we discuss the internal
dynamics of systems in our minimal modal interpretation of quantum
theory in Section~\ref{sub:Quantum-Conditional-Probabilities}, eigenstate
swaps fortunately turn out to be a mirage.

\subsubsection{The Fundamentally Unobservable Nature of Eigenstate Swaps}

As a first hint that we should not take eigenstate swaps \eqref{eq:EigenstateSwap}
seriously as physically real phenomena for macroscopic systems, notice
that a macroscopic system's eigenstate-swap time scale $\delta t_{\mathrm{swap}}$
in \eqref{eq:EigenstateSwapTimeScale} is \emph{parametrically} small
in the exponentially tiny quantity $\xi$ described in \eqref{eq:NearDegenOffDiagParam},
and is therefore always \emph{exponentially} smaller than the system's
ordinary characteristic time $\tau$ described in \eqref{eq:OrdinaryDynamicalRate}.
Thus, the hierarchical discrepancy between a macroscopic system's
characteristic time scale $\tau$ and its corresponding eigenstate-swap
time scale $\delta t_{\mathrm{swap}}$ actually \emph{gets worse}
if we could somehow arrange for the system to approach the idealized
limit of an exact degeneracy, because that would just make $\delta t_{\mathrm{swap}}\sim\xi$
even smaller. Similarly, if we were to attempt to hook the system
up to a macroscopic measuring device with the goal of trying to observe
the eigenstate swap experimentally---perhaps introducing a lot more
energy in order to achieve a very fine temporal measurement resolution---then
we \emph{would} decrease the overall system's ordinary characteristic
time scale $\tau$, but simultaneously we would \emph{vastly} decrease
$\xi$ and thereby end up pushing the eigenstate-swap time scale even
farther out of reach.

Interestingly, because a macroscopic quantum system's Hilbert space
has dimension 
\begin{equation}
\dim\mathcal{H}\sim\negthickspace\negthickspace\negthickspace\negthickspace\prod_{\substack{\mathrm{degrees}\\
\mathrm{of\ freedom}
}
}\negthickspace\negthickspace\negthickspace\negthickspace\left(\substack{\mathrm{range\ of\ each}\\
\mathrm{degree\ of\ freedom}
}
\right)\sim\exp\left(\#\mathrm{degrees\ of\ freedom}\right)\label{eq:DimHilbFromDegOfFreedom}
\end{equation}
 and its maximum entropy \eqref{eq:DefVonNeumannEntropy} goes as
\begin{equation}
S\sim-\sum_{\mathrm{basis}}\frac{1}{\dim\mathcal{H}}\log\frac{1}{\dim\mathcal{H}}\sim\log\dim\mathcal{H},\label{eq:MaxEntropy}
\end{equation}
 we see that the parameter $\xi$ from \eqref{eq:NearDegenOffDiagParam}
loosely corresponds to our error-entropy bound \eqref{eq:MinimumMeasurementErrorEntropyBound}:
\begin{equation}
\xi\sim\exp\left(-\#\mathrm{degrees\ of\ freedom}\right)\sim e^{-S}\sim\mathrm{minimum\ error}.\label{eq:NearDegenOffDiagParamFromError}
\end{equation}
 This result provides additional justification for regarding both
the distance of closest approach \eqref{eq:NearDegenProbClosestApproach}
between the near-degenerate probability eigenvalues, as well as the
eigenstate-swap time scale $\delta t_{\mathrm{swap}}$ in \eqref{eq:EigenstateSwapTimeScale},
as being unobservably small.

Eigenstate swaps for macroscopic systems therefore remain effectively
decoupled from the observable predictions of quantum theory. This
decoupling open up an important window of opportunity for trying to
smooth these kinds of instabilities out of existence altogether by
a suitable choice of internal dynamics for quantum ontic states, with
the added benefit of making the decoupling more manifest. We put forward
just such a proposal in Section~\ref{sub:Quantum-Conditional-Probabilities},
where we show explicitly that our choice ensures that a macroscopic
system's time-evolving actual ontic state essentially never undergoes
eigenstate swaps.

\subsection{Parent Systems, Subsystems, and the Partial-Trace Operation\label{sub:Parent-Systems,-Subsystems,-and-the-Partial-Trace-Operation}}

Up to now, our discussion has centered on the case in which we consider
just one particular system of interest. Of course, any study of quantum
theory must begin with some particular system, but there will generally
be other systems that we can consider simultaneously, and each may
naturally be interpreted as a parent system enclosing our original
system, or as a subsystem, or as an adjacent system, or as something
else entirely. In this section, we motivate and explain how our minimal
modal interpretation of quantum theory defines the relationship between
the objective epistemic states of parent systems and those of their
subsystems, especially in situations like \eqref{eq:GenericTwoSubsysEntangledStateVec}
that feature quantum entanglement.

\subsubsection{The Partial-Trace Operation\label{subsub:The-Partial-Trace-Operation}}

Recalling our discussion of classical parent systems and subsystems
in Section~\ref{subsub:Objective-Uncertainty-and-Improper-Mixtures},
and given an epistemic state $p_{A+B}$ for a classical composite
parent system $A+B$, we can naturally obtain reduced (or marginal)
epistemic states $p_{A}$ and $p_{B}$ for the respective subsystems
$A$ and $B$ by the familiar partial-sum (or marginalization) operation:
\begin{equation}
p_{A}\left(a\right)=\sum_{b}p_{A+B}\left(a,b\right),\qquad p_{B}\left(b\right)=\sum_{a}p_{A+B}\left(a,b\right).\label{eq:DefPartialSums}
\end{equation}
 Our next goal will be to motivate an analogous \emph{quantum} operation,
called the partial-trace operation, that relates the objective epistemic
states of parent systems with those of their subsystems, and we will
see that density matrices play a crucial role.

It is important to note that we will make no appeals to the Born rule
or Born-rule-based averages in justifying the definition of the partial-trace
operation. Indeed, we will ultimately find that we can \emph{derive}
the Born rule and its corollaries (as well as possible corrections
that are invisible in the traditional Copenhagen interpretation) from
the partial-trace operation's deeper principles of logical self-consistency
and its relationship with classical partial sums.

Central to our entire interpretation is the notion that every quantum
system has an ontology and objective epistemology defined through
a density matrix of its own.%
\footnote{In particular, it is important to remember that in our interpretation
of quantum theory, the density matrix or objective epistemic state
of any \emph{other} system, even that of a \emph{parent} system, does
not \emph{directly} define the ontology or objective epistemology
of a system of interest.%
} Given a quantum objective epistemic state for a parent system $W=A+B+C+D+\dotsb$
consisting of an arbitrary number of subsystems $A,B,C,D,\dotsc$
and associated to some density matrix $\op{\rho}_{W}$, we therefore
require a \emph{universal} prescription for \emph{nontrivially} assigning
unit-trace, positive semi-definite density matrices to all the various
possible definable subsystems $A$, $B$, $C$, $D$, $\dotsc$, $A+B$,
$A+C$, $A+D$, $\dotsc$, $A+B+C$, $A+B+D$, $\dotsc$, $A+B+C+D$,
$\dotsc$ in such a way that there is no dependence on our arbitrary
choice of orthonormal basis for each of the individual Hilbert spaces
of the subsystems, nor on the arbitrary order in which we could imagine
defining a descending sequence of density matrices for the subsystems.
We must, for instance, end up with the \emph{same} final density matrix
$\op{\rho}_{A}$ for subsystem $A$ whether we choose to define intermediate
density matrices according to the sequence $W\mapsto A+B+C\mapsto A+B\mapsto A$
or according to the sequence $W\mapsto A+B+D\mapsto A+D\mapsto A$.
Equivalently, any arbitrarily complicated diagram typified by the
following example must internally commute: 
\begin{equation}
\eqsbrace{\begin{array}{c}
\op{\rho}_{A+B+C}\\
\swarrow\qquad\qquad\searrow\\
\op{\rho}_{A+B},\ \op{\rho}_{C}\qquad\qquad\op{\rho}_{A},\ \op{\rho}_{B+C}\\
\searrow\qquad\qquad\swarrow\\
\op{\rho}_{A},\ \op{\rho}_{B},\ \op{\rho}_{C}.
\end{array}}\label{eq:SubsysDensMatrAltSeqDiag}
\end{equation}

Furthermore, our final result for any one subsystem's density matrix---say,
$\op{\rho}_{A}$, for subsystem $A$---cannot depend on our arbitrary
choice among the \emph{continuously infinite} different ways that
we could have defined the rest of the subsystems $B,C,\dotsc$. In
particular, we must ultimately get the \emph{same} answer $\op{\rho}_{A}$
whether we decompose $\mathcal{H}_{W}$ as $\mathcal{H}_{A}\tensorprod\mathcal{H}_{B}\tensorprod\mathcal{H}_{C}\tensorprod\dotsm$
and then define $\op{\rho}_{A}$ by the sequence $W\mapsto A+B\mapsto A$,
or whether we instead decompose $\mathcal{H}_{W}$ as $\mathcal{H}_{A}\tensorprod\mathcal{H}_{B^{\prime}}\tensorprod\mathcal{H}_{C^{\prime}}\tensorprod\dotsm$
for some different definitions $B^{\prime},C^{\prime},\dotsc$ of
the other subsystems and then define $\op{\rho}_{A}$ by the sequence
$W\mapsto A+B^{\prime}\mapsto A$. That is, the continuous infinity
of possible diagrams generalizing the following example must each
internally commute: 
\begin{equation}
\eqsbrace{\begin{array}{c}
\op{\rho}_{A+B+C}=\op{\rho}_{A+B^{\prime}+C^{\prime}}\\
\swarrow\qquad\qquad\searrow\\
\op{\rho}_{A+B}\qquad\qquad\op{\rho}_{A+B^{\prime}}\\
\searrow\qquad\swarrow\\
\op{\rho}_{A}.
\end{array}}\label{eq:SubsysDensMatrAltOtherRedefDiag}
\end{equation}

It is remarkable that a universal prescription meeting all these nontrivial
requirements of logical self-consistency exists at all, much less
that it turns out to be so straightforward. To define this prescription
explicitly and to ensure that it is indeed nontrivial and has the
correct behavior in the classical regime, we require furthermore that
it should reduce to classical partial sums \eqref{eq:DefPartialSums}
if the density matrix of a parent system $A+B$ happens to be diagonal
in the tensor-product basis $\ket{a,b}=\ket a\tensorprod\ket b$ with
probability eigenvalues $p_{A+B}\left(a,b\right)$---that is, if there
is no quantum entanglement in the sense of \eqref{eq:GenericTwoSubsysEntangledStateVec}
between the subsystems $A$ and $B$: 
\begin{equation}
\eqsbrace{\begin{gathered}\op{\rho}_{A+B}=\sum_{a,b}p_{A+B}\left(a,b\right)\ket a\tensorprod\ket b\bra a\tensorprod\bra b\\
\implies\op{\rho}_{A}=\sum_{a}\underbrace{\left(\sum_{b}p_{A+B}\left(a,b\right)\right)}_{p_{A}\left(a\right)}\ket a\bra a,\qquad\op{\rho}_{B}=\sum_{b}\underbrace{\left(\sum_{a}p_{A+B}\left(a,b\right)\right)}_{p_{B}\left(b\right)}\ket b\bra b.
\end{gathered}
}\label{eq:PartialTraceReducePartialSum}
\end{equation}
 Additionally, we require that the prescription should be linear
over the space of operators on the parent system's Hilbert space,
in keeping with the linearity property that holds for classical partial
sums over convex combinations $xp_{A+B}+yp_{A+B}^{\prime}$ ($x,y>0$,
$x+y=1$) of classical epistemic states $p_{A+B}$ and $p_{A+B}^{\prime}$.

The unique resulting partial-trace operation is well known and surprisingly
simple to describe. We begin by considering a parent system $C$ that
we can regard as consisting of just two subsystems $A$ and $B$ (where
either $A$ or $B$ could be parent systems encompassing subsystems
of their own), and then we write the density matrix $\op{\rho}_{A+B}$
for $C=A+B$ in the tensor-product basis $\ket a\tensorprod\ket b$,
which \emph{won't} generally be its diagonalizing basis, namely, if
$A$ and $B$ are entangled: 
\begin{equation}
\op{\rho}_{A+B}=\sum_{a,b,a^{\prime},b^{\prime}}\rho_{A+B}\left(\left(a,b\right),\left(a^{\prime},b^{\prime}\right)\right)\ket a\tensorprod\ket b\bra{a^{\prime}}\tensorprod\bra{b^{\prime}}.\label{eq:GeneralDensMatrixNondiag}
\end{equation}
 Linearity and consistency with partial sums in the special case 
\begin{equation}
\rho_{A+B}\left(\left(a,b\right),\left(a^{\prime},b^{\prime}\right)\right)=p_{A+B}\left(a,b\right)\label{eq:FactorizeDensMatrixElems}
\end{equation}
 then immediately imply that the partial trace down to either subsystem
$A$ or $B$ must be defined by formally ``turning around'' the
bras and kets of the \emph{other} subsystem and evaluating the resulting
inner products, where the result then defines the reduced density matrix
$\op{\rho}_{A}$ or $\op{\rho}_{B}$ of the subsystem $A$ or $B$,
respectively: 
\begin{equation}
\eqsbrace{\begin{aligned}\op{\rho}_{A}\equiv\Tr_{B}\left[\op{\rho}_{A+B}\right] & \equiv\sum_{a,b,a^{\prime},b^{\prime}}\rho_{A+B}\left(\left(a,b\right),\left(a^{\prime},b^{\prime}\right)\right)\ket a\bra{a^{\prime}}\underbrace{\braket b{b^{\prime}}}_{\delta\left(b,b^{\prime}\right)}\\
 & =\sum_{a,a^{\prime}}\left(\sum_{b}\rho_{A+B}\left(\left(a,b\right),\left(a^{\prime},b\right)\right)\right)\ket a\bra{a^{\prime}},\\
\op{\rho}_{B}\equiv\Tr_{A}\left[\op{\rho}_{A+B}\right] & \equiv\sum_{a,b,a^{\prime},b^{\prime}}\rho_{A+B}\left(\left(a,b\right),\left(a^{\prime},b^{\prime}\right)\right)\underbrace{\braket a{a^{\prime}}}_{\delta\left(a,a^{\prime}\right)}\ket b\bra{b^{\prime}}\\
 & =\sum_{b,b^{\prime}}\left(\sum_{a}\rho_{A+B}\left(\left(a,b\right),\left(a,b^{\prime}\right)\right)\right)\ket b\bra{b^{\prime}}.
\end{aligned}
}\label{eq:DefPartialTraces}
\end{equation}
 That is, the matrix elements of $\op{\rho}_{A}$ and $\op{\rho}_{B}$
are given in the respective orthonormal bases $\ket a$ for the Hilbert
space $\mathcal{H}_{A}$ of subsystem $A$ and $\ket b$ for the Hilbert
space $\mathcal{H}_{B}$ of subsystem $B$ by the natural matrix-generalizations
of classical partial sums \eqref{eq:DefPartialSums}: 
\begin{equation}
\rho_{A}\left(a,a^{\prime}\right)=\sum_{b}\rho_{A+B}\left(\left(a,b\right),\left(a^{\prime},b\right)\right),\qquad\rho_{B}\left(b,b^{\prime}\right)=\sum_{a}\rho_{A+B}\left(\left(a,b\right),\left(a,b^{\prime}\right)\right).\label{eq:PartialTracesMatrElems}
\end{equation}

\subsubsection{An Example}

With the formula \eqref{eq:DefPartialTraces} for partial traces in
hand, we are finally ready to answer the questions that we posed in
Section~\ref{subsub:Objective-Uncertainty-and-Improper-Mixtures}
regarding the respective ontic states of the two subsystems $A$ and
$B$ belonging to a composite parent system $A+B$ described by the
entangled state vector \eqref{eq:GenericTwoSubsysEntangledStateVec},
\[
\ket{\Psi_{A+B}}=\alpha\ket{\Psi_{A,1}}\ket{\Psi_{B,1}}+\beta\ket{\Psi_{A,2}}\ket{\Psi_{B,2}},\qquad\alpha,\beta\in\mathbb{C},\ \absval{\alpha}^{2}+\absval{\beta}^{2}=1,
\]
 where we assume just for simplicity that $\ket{\Psi_{A,1}}\perp\ket{\Psi_{A,2}}$
and $\ket{\Psi_{B,1}}\perp\ket{\Psi_{B,2}}$ in \eqref{eq:GenericTwoSubsysEntangledStateVec}.

Given that $A+B$ has the definite ontic state $\Psi_{A+B}$ defined
by \eqref{eq:GenericTwoSubsysEntangledStateVec}, the density matrix
of $A+B$ is 
\begin{equation}
\eqsbrace{\begin{aligned}\op{\rho}_{A+B} & =\ket{\Psi_{A+B}}\bra{\Psi_{A+B}}\\
 & =\left(\alpha\ket{\Psi_{A,1}}\ket{\Psi_{B,1}}+\beta\ket{\Psi_{A,2}}\ket{\Psi_{B,2}}\right)\left(\alpha^{\conj}\bra{\Psi_{A,1}}\bra{\Psi_{B,1}}+\beta^{\conj}\bra{\Psi_{A,2}}\bra{\Psi_{B,2}}\right),
\end{aligned}
}\label{eq:GenericTwoSubsysEntangDensMatr}
\end{equation}
 thereby implying from the partial-trace prescription \eqref{eq:DefPartialTraces}
that the respective reduced density matrices for $A$ and $B$ are
\begin{equation}
\eqsbrace{\begin{aligned}\op{\rho}_{A} & =\absval{\alpha}^{2}\ket{\Psi_{A,1}}\bra{\Psi_{A,1}}+\absval{\beta}^{2}\ket{\Psi_{A,2}}\bra{\Psi_{A,2}},\\
\op{\rho}_{B} & =\absval{\alpha}^{2}\ket{\Psi_{B,1}}\bra{\Psi_{B,1}}+\absval{\beta}^{2}\ket{\Psi_{B,2}}\bra{\Psi_{B,2}}.
\end{aligned}
}\label{eq:GenericTwoSubsysEntangRedDensMatr}
\end{equation}
 Hence, according to our interpretation of quantum theory, the possible
ontic states of subsystem $A$ are $\Psi_{A,1}$ and $\Psi_{A,2}$,
and the possible ontic states of subsystem $B$ are $\Psi_{B,1}$
and $\Psi_{B,2}$, with respective epistemic probabilities given by
\begin{equation}
\eqsbrace{\begin{gathered}p_{A}\left(\Psi_{A,1}\right)=\absval{\alpha}^{2},\qquad p_{A}\left(\Psi_{A,2}\right)=\absval{\beta}^{2},\\
p_{B}\left(\Psi_{B,1}\right)=\absval{\alpha}^{2},\qquad p_{B}\left(\Psi_{B,2}\right)=\absval{\beta}^{2}.
\end{gathered}
}\label{eq:GenericTwoSubsysRedProbs}
\end{equation}

\subsubsection{Correlation and Entanglement}

Given a parent quantum system $A+B$ consisting of two subsystems
$A$ and $B$, we say that $A$ and $B$ are correlated if the parent
system's density matrix $\op{\rho}_{A+B}$ is not a simple tensor
product of the reduced density matrices $\op{\rho}_{A}$ and $\op{\rho}_{B}$:
\begin{equation}
A\mathrm{\ and\ }B\mathrm{\ correlated}:\qquad\op{\rho}_{A+B}\ne\op{\rho}_{A}\tensorprod\op{\rho}_{B}.\label{eq:DefCorrSubsys}
\end{equation}
 Correlation is a property that exists even in classical physics and
corresponds to the failure of the epistemic probabilities for the
composite system $A+B$ to factorize: $p\left(a,b\right)\ne p\left(a\right)p\left(b\right)$.

But quantum systems are capable of an even stronger property known
as entanglement, which we first encountered in \eqref{eq:GenericTwoSubsysEntangledStateVec}
and now define more generally within the context of our interpretation
of quantum theory as describing the case in which the parent system's
density matrix $\op{\rho}_{A+B}$ is not diagonal in the tensor-product
basis $\ket{a,b}=\ket a\tensorprod\ket b$ corresponding to the two
subsystems: 
\begin{equation}
A\mathrm{\ and\ }B\mathrm{\ entangled}:\qquad\mathrm{eigenbasis\ of\ }\op{\rho}_{A+B}\ \mathrm{is\ not\ }\ket{a,b}=\ket a\tensorprod\ket b.\label{eq:DefEntangledSubsys}
\end{equation}
 It is precisely due to entanglement that the partial-trace operation
\eqref{eq:DefPartialTraces} does not generically reduce to simple
partial sums \eqref{eq:DefPartialSums}, and, indeed, is arguably
the very reason why we need density matrices in quantum theory in
the first place.%
\footnote{Note that the \emph{traditional} formulation of quantum theory---unlike
our own interpretation---does not single out the diagonalizing eigenbasis
of a density matrix as being \emph{fundamentally} preferred. Outside
of the idealized case of a quantum system belonging to a parent system
in an exactly pure state, entanglement \eqref{eq:DefEntangledSubsys}
therefore ceases to be a more than a purely formal property of density
matrices known as non-separability, and there exist various measures,
such as quantum discord \cite{HendersonVedral:2001cqtc,OllivierZurek2001:qdmqc},
for characterizing the quantumness of the resulting correlations.%
}

\subsubsection{Proper Mixtures and Improper Mixtures\label{subsub:Proper-Mixtures-and-Improper-Mixtures}}

Recall from their definition in Section~\ref{subsub:Subjective-Uncertainty-and-Proper-Mixtures}
that proper mixtures refer to subjective epistemic states, by which
we mean epistemic states that merely encode \emph{subjective} uncertainty
about a system's actual ontic state and make perfect sense even in
the context of classical physics. By contrast, improper mixtures,
as we defined them in Section~\ref{subsub:Our-Interpretation}, refer
to epistemic states that encode \emph{objective} uncertainty arising
from quantum entanglements with other systems, in addition to any
subjective uncertainty that may also be present.

In the traditional language of quantum theory, there exists a sharp
conceptual distinction between proper mixtures and improper mixtures,
whereas our minimal modal interpretation of quantum theory blurs that
distinction to a significant degree. Indeed, we regard both kinds
of mixtures as describing an epistemic state that hides the system's
actual ontic state.

However, looking back at our basic correspondence \eqref{eq:EpStateDensMatrixCorresp},
note that we have so far introduced density matrices \emph{solely}
to describe objective epistemic states---that is, \emph{fully objective}
improper mixtures. (We will extend the use of density matrices to
more general epistemic states in Section~\ref{sub:Subjective-Density-Matrices-and-Proper-Mixtures}
only after deriving the Born rule.) Furthermore, it is important
to keep in mind that \emph{it really matters} whether or not a system's
actual ontic state is hiding behind a density matrix describing a
truly nontrivial objective epistemic state: There can exist physical
differences between the case of a system whose density matrix happens
to be pure $\op{\rho}=\ket{\Psi}\bra{\Psi}$ and a system whose density
matrix is mixed but whose actual underlying ontic state nonetheless
happens to be $\Psi$.

For example, consider a composite system $A+B$ whose density matrix
has the \emph{pure} form 
\begin{equation}
\op{\rho}_{A+B}=\ket 1\ket{\Psi}\bra 1\bra{\Psi},\label{eq:PureStateExample}
\end{equation}
 where $1$ is an allowed state of $A$ and $\Psi$ is an allowed
state of $B$. We would naturally conclude not only that the parent
system $A+B$ has actual ontic state $\left(1,\Psi\right)$ with unit
epistemic probability, but, moreover, that the subsystems $A$ and
$B$ themselves have respective actual ontic states $1$ and $\Psi$
each with unit epistemic probability, for the simple reason that their
own reduced density matrices are respectively $\op{\rho}_{A}=\ket 1\bra 1$
and $\op{\rho}_{B}=\ket{\Psi}\bra{\Psi}$.

But now suppose instead that the parent system's density matrix has
the \emph{mixed} form 

\begin{equation}
\op{\rho}_{A+B}=p_{1}\ket 1\ket{\Psi}\bra 1\bra{\Psi}+p_{2}\ket 2\ket{\Phi}\bra 2\bra{\Phi},\qquad p_{1}+p_{2}=1,\label{eq:MixedStateExample}
\end{equation}
 where we assume that $\ket 1$ is orthogonal to $\ket 2$---that
is, $\braket 12=0$---but \emph{not necessarily} that $\ket{\Psi}$
is orthogonal to $\ket{\Phi}$. (Note, however, that $\ket 1\ket{\Psi}$
is nonetheless orthogonal to $\ket 2\ket{\Phi}$.) Remember that in
our interpretation of quantum theory, we determine a system's possible
ontic states and associated epistemic probabilities from the orthonormal
spectrum of the system's \emph{own} density matrix. Hence, although
it would be correct to say that the actual ontic state of the parent
system could be $\left(1,\Psi\right)$ with epistemic probability
$p_{1}$ or $\left(2,\Phi\right)$ with epistemic probability $p_{2}$,
it would be \emph{incorrect} to conclude that the actual ontic state
of the subsystem $B$ is $\Psi$ with epistemic probability $p_{1}$
or $\Phi$ with epistemic probability $p_{2}$, because $\braket{\Psi}{\Phi}\ne0$
means that $\Psi$ and $\Phi$ are not mutually exclusive possibilities
and therefore cannot correspond to the mutually orthogonal eigenstates
of the density matrix of subsystem $B$. As a further consequence,
it would also be incorrect to say that if the actual ontic state of
the parent system happens to be, say, $\left(1,\Psi\right)$, then
the actual ontic state of subsystem $B$ must be $\Psi$.

The reason for all this trouble is, unsurprisingly, entanglement \eqref{eq:DefEntangledSubsys}:
The diagonalizing basis for $\op{\rho}_{A+B}$ given in \eqref{eq:MixedStateExample}
is not the ``correct'' tensor-product orthonormal basis that corresponds
to our original tensor-product factorization $\mathcal{H}_{A+B}=\mathcal{H}_{A}\tensorprod\mathcal{H}_{B}$
of the parent system's Hilbert space $\mathcal{H}_{A+B}$ into the
Hilbert spaces $\mathcal{H}_{A}$ of subsystem $A$ and $\mathcal{H}_{B}$
of subsystem $B$. Of course, if the inner product $\braket{\Psi}{\Phi}$
is very small in magnitude, then we are welcome to choose a \emph{slightly
different} tensor-product decomposition $\mathcal{H}_{A+B}=\mathcal{H}_{A^{\prime}}\tensorprod\mathcal{H}_{B^{\prime}}$
in order to make the diagonalizing basis for $\op{\rho}_{A+B}$ the
tensor-product basis $\ket{a^{\prime},b^{\prime}}=\ket{a^{\prime}}\tensorprod\ket{b^{\prime}}$
corresponding to \emph{suitably redefined} subsystems $A^{\prime}$
and $B^{\prime}$. In that case, we could immediately read off from
the eigenvalues of $\op{\rho}_{A+B}$ the corresponding epistemic
probabilities for the subsystems $A^{\prime}$ and $B^{\prime}$,
and declare that if the actual ontic state of the composite system
happens to be, say, $\left(a^{\prime},b^{\prime}\right)$, then the
actual ontic states of $A^{\prime}$ and $B^{\prime}$ must respectively
be $a^{\prime}$ and $b^{\prime}$.

However, if we insist on working with our \emph{original} subsystems
$A$ and $B$, then it might seem that all we can conclude about subsystem
$B$ from the density matrix \eqref{eq:MixedStateExample} of the
parent system $A+B$ is that the reduced density matrix of $B$ is
given by the partial trace $\op{\rho}_{B}=\Tr_{A}\left[\op{\rho}_{A+B}\right]$,
as defined in \eqref{eq:DefPartialTraces}. In particular, the fact
that the parent system $A+B$ has a specific actual ontic state might
not appear to imply anything about the actual ontic state of $B$
alone.

There is a seemingly obvious connection between the actual ontic state
of the parent system $A+B=A^{\prime}+B^{\prime}$ and the actual ontic
state of $B^{\prime}$, but does the actual ontic state of the parent
system tell us anything about the actual ontic state of $B$, which
is presumably ``just a slightly different version'' of the same
subsystem as $B^{\prime}$? Without addressing this question, our
interpretation of quantum theory would seem to be woefully inadequate,
as our notion of an actual ontic state would be infinitely sensitive
to arbitrarily small (and thus observationally meaningless) redefinitions
of subsystems.%
\footnote{Indeed, a potential instability somewhat analogous to eigenstate swaps
\eqref{eq:EigenstateSwap} arises in the context of small changes
in the definition of a subsystem, as explored in \cite{BacciagaluppiDonaldVermaas:1995cddpmi,Donald:1998dcdpmi}.
As Vermaas writes on p. 134 of \cite{Vermaas:1999puqm}: ``A final
remark concerns yet another source for incorrect property ascriptions.
In this book I always assume that one can precisely identify the systems.
Consequently, one can also precisely identify the composites of these
systems. If, however, one proceeds the other way round and starts
with a set of composites, one has to answer the question of how exactly
to factor these composites into disjoint subsystems. In Bacciagaluppi,
Donald and Vermaas (1995, Example 7.3) it is proved that the property
ascription to a subsystem can depend with high sensitivity on the
precise identification of that subsystem.''%
} We will study and resolve this important issue in Section~\ref{sub:Quantum-Conditional-Probabilities}
when we explicitly define the relationship between the ontic states
of parent systems and the ontic states of their subsystems. In Section~\ref{sub:Subsystem-Spaces},
we will introduce the relatively unexplored notion of ``subsystem
spaces'' to describe the continuously infinite set of different ways
of defining different versions of a particular subsystem.

\subsubsection{Imperfect Tensor-Product Factorizations, Truncated Hilbert Spaces,
Approximate Density Matrices, and Unstable Systems\label{subsub:Imperfect-Tensor-Product-Factorizations-Truncated-Hilbert-Spaces-Approximate-Density-Matrices-and-Unstable-Systems}}

Given a system's Hilbert space, there will generally exist many possible
tensor-product factorizations defining valid subsystems. But, going
the other way, there may be cases in which an \emph{a priori} desired
choice of subsystem cannot be realized as an \emph{exact} tensor-product
factorization of a given parent system's Hilbert space.

For example, the Hilbert space of Nature is presently unknown, and
so we have no reason to believe that it admits a simple tensor-product
factorization $\mathcal{H}_{\mathrm{Nature}}=\mathcal{H}_{\mathrm{system}}\tensorprod\mathcal{H}_{\mathrm{other}}$
that includes the exact Hilbert space $\mathcal{H}_{\mathrm{system}}$
of any of the kinds of systems known today. Nonetheless, we profitably
employ such Hilbert spaces all the time. Indeed, even for the familiar
example of a quantum field theory, we know that the full system's
Hilbert-Fock space $\mathcal{H}_{\mathrm{Fock}}=\mathcal{H}_{0\ \mathrm{particles}}\directsum\mathcal{H}_{1\ \mathrm{particle}}\directsum\mathcal{H}_{2\ \mathrm{particles}}\directsum\dotsb$
does not neatly tensor-product-factorize as $\mathcal{H}_{\mathrm{1\ particle}}\tensorprod\mathcal{H}_{\mathrm{other}}$,
where $\mathcal{H}_{1\ \mathrm{particle}}$ is the Hilbert space of
a single particle of a certain species, and yet we successfully make
use of one-particle Hilbert spaces $\mathcal{H}_{1\ \mathrm{particle}}$
whenever we wish to study nonrelativistic or semi-relativistic physics.

We clearly need a prescription for obtaining \emph{approximate} subsystem
Hilbert spaces even when the full Hilbert space of the parent system
is either unknown or does not exactly permit the desired tensor-product
factorization. That prescription, which is implicit in all the aforementioned
examples, consists of first \emph{truncating} the full Hilbert space---or,
if the full Hilbert space is unknown, then regarding it as \emph{already}
being appropriately truncated---in order to make the desired tensor-product
factorization possible, and only then taking appropriate partial traces.

An immediate corollary is that the restricted density matrix on the
parent system's truncated Hilbert space, and thus the associated reduced
density matrix of the subsystem as well, are only \emph{approximate}
objects and, despite still being positive semi-definite, will no longer
have exactly unit trace, corresponding to the statement that their
probability eigenvalues no longer add up all the way to unity: $\sum_{i}p_{i}<1$.
The discrepancy $\left(1-\sum_{i}p_{i}\right)\in\left[0,1\right]$
arises from the absence of the states in which our subsystem does
not exist (or no longer exists)---the discrepancy is directly related
to our inability to capture the system's actual ontic state among
the possible ontic states $\Psi_{i}$ remaining in our description
of the system---and so we naturally interpret the discrepancy as representing
the probability that our system, which we now recognize as being unstable,
has actually decayed.

\subsection{Quantum Conditional Probabilities\label{sub:Quantum-Conditional-Probabilities}}

We are now ready to say more about the time evolution and dynamics
of ontic states and objective epistemic states, as well as address
lingering questions regarding eigenstate swaps \eqref{eq:EigenstateSwap}
that we first encountered in Section~\ref{subsub:Near-Degeneracies-and-Eigenstate-Swaps}
and parent-subsystem discrepancies arising from nontrivial objective
epistemic states that we first encountered in Section~\ref{subsub:Objective-Uncertainty-and-Improper-Mixtures}.

\subsubsection{Classical Dynamics and (Multi-)Linear Dynamical Mappings\label{subsub:Classical-Dynamics-and-Multi-Linear-Dynamical-Mappings}}

Recall from Section~\ref{subsub:Classical-Dynamics} that classical
systems with well-defined dynamics are those that, to an acceptable
level of approximation, possess an ontic-level dynamical mapping \eqref{eq:ClassicalOnticLevelDynMapGeneral}
that is independent of the system's epistemic state and naturally
lifts to a multilinear dynamical mapping \eqref{eq:ClassicalMultilinearEpistemicLevelDynMapGeneral}
relating initial and final epistemic states as well. In the special
case of Markovian dynamics \eqref{eq:ClassicalMarkovDynMap}---that
is, for a system whose dynamical mapping is first order, meaning that
it requires the input of initial data at only a \emph{single} initial
time---the ontic-level dynamics takes the general form 
\begin{equation}
p\left(\cdot;t^{\prime}\given\cdot;t\right)\colon\ \underbrace{\left(q;t\right)}_{\substack{\mathrm{initial}\\
\mathrm{data}
}
},\ \underbrace{\left(q^{\prime};t^{\prime}\right)}_{\substack{\mathrm{final}\\
\mathrm{data}
}
}\mapsto\underbrace{p\left(q^{\prime};t^{\prime}\given q;t\right)}_{\substack{\mathrm{conditional}\\
\mathrm{probabilities}
}
},\label{eq:ClassicalOnticLevelDynMapMarkov}
\end{equation}
 and the corresponding epistemic-level dynamics \eqref{eq:ClassicalMultilinearEpistemicLevelDynMapGeneral}
becomes a simple linear mapping 
\begin{equation}
p\left(\cdot;t^{\prime}\given\cdot;t\right)\colon\ \underbrace{p\left(q^{\prime};t^{\prime}\right)}_{\substack{\mathrm{epistemic}\\
\mathrm{state\ at\ }t^{\prime}
}
}=\sum_{q}\underbrace{p\left(q^{\prime};t^{\prime}\given q;t\right)}_{\substack{\mathrm{conditional}\\
\mathrm{probability}\\
\mathrm{(independent}\\
\mathrm{of\ epistemic}\\
\mathrm{states})
}
}\underbrace{p\left(q;t\right)}_{\substack{\mathrm{epistemic}\\
\mathrm{state\ at\ }t
}
},\label{eq:ClassicalLinearEpistemicLevelDynMapMarkov}
\end{equation}
 which we can regard as a kind of dynamical Bayesian propagation formula.

Consider now a classical system $Q$ with a configuration space $\mathcal{C}_{Q}=\setindexed qq$,
and suppose that $Q$ is an open subsystem of some larger classical
system $W=Q+E$, where $E$ is the environment of $Q$ inside $W$
and has configuration space $\mathcal{C}_{E}=\setindexed ee$ so that
the configuration space of $W$ is the Cartesian product $\mathcal{C}_{W}=\mathcal{C}_{Q}\cartesianprod\mathcal{C}_{E}=\setbuild{w=\left(q,e\right)}{q\in\mathcal{C}_{Q}\binaryand e\in\mathcal{C}_{E}}$.
Then even if the parent system $W$ as a whole has well-defined dynamics
in the sense of a linear mapping $p_{W}\left(\cdot;t^{\prime}\given\cdot;t\right)$
that is independent of the epistemic state of $W$ and that we assume
for simplicity is of the Markovian form \eqref{eq:ClassicalLinearEpistemicLevelDynMapMarkov},
the open subsystem $Q$ of $W$ will not necessarily have well-defined
dynamics of its own.

Indeed, by expressing the time evolution for the epistemic state of
$Q$ over the time interval from $t$ to $t^{\prime}$ as 
\begin{align}
p_{Q}\left(q^{\prime};t^{\prime}\right) & =\sum_{e^{\prime}}p_{W}\left(w^{\prime}=\left(q^{\prime},e^{\prime}\right);t^{\prime}\right)\nonumber \\
 & =\sum_{e^{\prime},w}p_{W}\left(w^{\prime}=\left(q^{\prime},e^{\prime}\right);t^{\prime}\given w=\left(q,e\right);t\right)p_{W}\left(w=\left(q,e\right);t\right)\nonumber \\
 & =\sum_{q}\left[\sum_{e^{\prime},e}p_{W}\left(w^{\prime}=\left(q^{\prime},e^{\prime}\right);t^{\prime}\given w=\left(q,e\right);t\right)\left(\frac{p_{W}\left(w=\left(q,e\right);t\right)}{p_{Q}\left(q;t\right)}\right)\right]p_{Q}\left(q;t\right),\label{eq:ClassicalNonlinearEpLevelDynEvExplicit}
\end{align}
 we see immediately that the failure of $Q$ to possess its own dynamics
is characterized by the complicated, \emph{nonlinear} dependence of
the ratio in parentheses (both its numerator and denominator) on the
epistemic state $p_{Q}\left(q;t\right)$ of $Q$. From Bayes' theorem,
we readily identify this ratio as being the conditional probability
that the environment $E$ is in the ontic state $e$ given that the
subsystem $Q$ is in the ontic state $q$ at the same time $t$: 
\begin{equation}
\frac{p_{W}\left(w=\left(q,e\right);t\right)}{p_{Q}\left(q;t\right)}=p_{E\given Q}\left(e;t\given q;t\right).\label{eq:ClassicalNonlinearDynRatioEqCondEnv}
\end{equation}

Notice that the open-subsystem evolution law \eqref{eq:ClassicalNonlinearEpLevelDynEvExplicit}
provides us with a \emph{coarse-grained} or \emph{effective} (albeit
nonlinear) notion of dynamics $p_{Q\subset W}\left(\cdot;t^{\prime}\given\cdot;t\right)$
for $Q$, 
\begin{equation}
p_{Q\subset W}\left(\cdot;t^{\prime}\given\cdot;t\right)\colon\ \underbrace{p_{Q}\left(q^{\prime};t^{\prime}\right)}_{\substack{\mathrm{epistemic}\\
\mathrm{state\ at\ }t^{\prime}
}
}=\sum_{q}\underbrace{p_{Q\subset W}\left(q^{\prime};t^{\prime}\given q;t\right)}_{\substack{\mathrm{conditional}\\
\mathrm{probability}\\
\mathrm{(has\ dependence}\\
\mathrm{on\ epistemic}\\
\mathrm{states})
}
}\underbrace{p_{Q}\left(q;t\right)}_{\substack{\mathrm{epistemic}\\
\mathrm{state\ at\ }t
}
},\label{eq:ClassicalNonlinearEpLevelDynMap}
\end{equation}
 where we have defined the coarse-grained or effective conditional
probabilities $p_{Q\subset W}\left(q^{\prime};t^{\prime}\given q;t\right)$
to be the factor appearing in brackets in \eqref{eq:ClassicalNonlinearEpLevelDynEvExplicit}---that
is, in accordance with Bayes' theorem, by multiplying each parent-system
conditional probability $p_{W}\left(w^{\prime};t^{\prime}\given w;t\right)$
by the epistemic probability $p_{W}\left(w;t\right)$ for $W$ at
the time $t$, marginalizing over the environment $E$ at both the
initial and final times, and then conditioning on $Q$ at the time
$t$: 
\begin{align}
p_{Q\subset W}\left(q^{\prime};t^{\prime}\given q;t\right) & =\frac{p_{Q\subset W}\left(q^{\prime};t\binaryand q;t\right)}{p_{Q}\left(q;t\right)}\nonumber \\
 & =\frac{1}{p_{Q}\left(q;t\right)}\sum_{e^{\prime},e}p_{W}\left(w^{\prime}=\left(q^{\prime},e^{\prime}\right);t^{\prime}\binaryand w=\left(q,e\right);t\right)\nonumber \\
 & =\sum_{e^{\prime},e}p_{W}\left(w^{\prime}=\left(q^{\prime},e^{\prime}\right);t^{\prime}\given w=\left(q,e\right);t\right)\underbrace{\left(\frac{p_{W}\left(w=\left(q,e\right);t\right)}{p_{Q}\left(q;t\right)}\right)}_{p_{E\given Q}\left(e;t\given q;t\right)}.\label{eq:ClassicalNonlinearEpLevelDynMapExplicit}
\end{align}

In the case in which the correlations between $Q$ and its environment
$E$ inside $W$ wash out over some short characteristic time scale
$\delta t_{Q}\ll t^{\prime}-t$---say, through irreversible thermal
radiation into outer space---the epistemic probabilities for $W$
approximately factorize, 
\begin{equation}
p_{W}\left(w=\left(q,e\right)\right)\approx p_{Q}\left(q\right)p_{E}\left(e\right),\label{eq:ClassicalCorrWashOut}
\end{equation}
 and so the ratio \eqref{eq:ClassicalNonlinearDynRatioEqCondEnv}
appearing in parentheses in the evolution equation \eqref{eq:ClassicalNonlinearEpLevelDynEvExplicit}
for $Q$ reduces to 
\begin{equation}
p_{E\given Q}\left(e;t\given q;t\right)=\frac{p_{W}\left(w=\left(q,e\right);t\right)}{p_{Q}\left(q;t\right)}=\frac{\cancel{p_{Q}\left(q;t\right)}p_{E}\left(e;t\right)}{\cancel{p_{Q}\left(q;t\right)}}=p_{E}\left(e;t\right).\label{eq:ClassicalCancelUncorrelated}
\end{equation}
 Hence, the dynamical mapping \eqref{eq:ClassicalNonlinearEpLevelDynMap}
for $Q$ now defines properly linear dynamics $p_{Q\subset W}\left(\cdot;t^{\prime}\given\cdot;t\right)=p_{Q}\left(\cdot;t^{\prime}\given\cdot;t\right)$
for $Q$ on its own, 
\begin{equation}
p_{Q}\left(\cdot;t^{\prime}\given\cdot;t\right)\colon\ \underbrace{p_{Q}\left(q^{\prime};t^{\prime}\right)}_{\substack{\mathrm{epistemic}\\
\mathrm{state\ at\ }t^{\prime}
}
}=\sum_{q}\underbrace{p_{Q}\left(q^{\prime};t^{\prime}\given q;t\right)}_{\substack{\mathrm{conditional}\\
\mathrm{probability}\\
\mathrm{(independent}\\
\mathrm{of\ epistemic}\\
\mathrm{states})
}
}\underbrace{p_{Q}\left(q;t\right)}_{\substack{\mathrm{epistemic}\\
\mathrm{state\ at\ }t
}
}\ \mathrm{for\ }t^{\prime}-t\gg\delta t_{Q},\label{eq:ClassicalNowLinEpLevelDynMap}
\end{equation}
 where \eqref{eq:ClassicalNonlinearEpLevelDynMapExplicit} has reduced
to 
\begin{equation}
p_{Q}\left(q^{\prime};t^{\prime}\given q;t\right)=\sum_{e^{\prime},e}p_{W}\left(w^{\prime}=\left(q^{\prime},e^{\prime}\right);t^{\prime}\given w=\left(q,e\right);t\right)p_{E}\left(e;t\right).\label{eq:ClassicalLinCondProbsAfterFac}
\end{equation}
 Nonetheless, even if the dynamics of the parent system $W$ is deterministic,
so that its own dynamical conditional probabilities $p_{W}\left(w^{\prime};t^{\prime}\given w;t\right)$
are trivial and relate each initial ontic state $w$ to a unique final
ontic state $w^{\prime}$ with unit probability, keep in mind that
the presence of the environment's instantaneous epistemic probabilities
$p_{E}\left(e;t\right)$ in the formula \eqref{eq:ClassicalLinCondProbsAfterFac}
for the conditional probabilities $p_{Q}\left(q^{\prime};t^{\prime}\given q;t\right)$
generally implies that the dynamics of $Q$ is stochastic.

Notice that the characteristic time scale $\delta t_{Q}$ sets a natural
short-time cutoff on the dynamics for our open subsystem $Q$. Although
$Q$ is a sensible system at the level of \emph{kinematics} over time
scales shorter than $\delta t_{Q}$, it does not possess truly well-defined
\emph{dynamics} on time scales shorter than $\delta t_{Q}$.

\subsubsection{Linear CPT Dynamical Mappings}

Interestingly, the classical linear dynamical mapping \eqref{eq:ClassicalLinearEpistemicLevelDynMapMarkov}
has a much-studied quantum counterpart that generalizes the ``deterministic''
unitary Schrödinger dynamics of closed quantum systems to a form of
linear \emph{stochastic} dynamics governing the time evolution of
a large class of open quantum systems.%
\footnote{See \cite{SudarshanMatthewsRau:1961sdqms,JordanSudarshan:1961dmdoqm,Choi:1975cplmcm}
for early work in this direction, and see \cite{SchumacherWestmoreland:2010qpsi}
for a modern pedagogical review.%
} This quantum-dynamical mapping consists of a linear function $\mathcal{E}^{t^{\prime}\from t}\left[\cdot\right]$
relating an open system's initial and final density matrices at respectively
initial and final times $t\leq t^{\prime}$: 
\begin{equation}
\op{\rho}\left(t\right)\mapsto\op{\rho}\left(t^{\prime}\right)=\mathcal{E}^{t^{\prime}\from t}\left[\op{\rho}\left(t\right)\right].\label{eq:LinDynMapOnDensMatr}
\end{equation}
 Notice that $\mathcal{E}^{t^{\prime}\from t}\left[\cdot\right]$
defines dynamics on \emph{density matrices}, as opposed to classical-type
dynamics \eqref{eq:ClassicalLinearEpistemicLevelDynMapMarkov} on
objective epistemic states directly. Functions like $\mathcal{E}^{t^{\prime}\from t}\left[\cdot\right]$
that map operators to operators are called superoperators.

Intuitively, if \eqref{eq:LinDynMapOnDensMatr} describes a mapping
taking in an initial density matrix and producing a final density
matrix, then it should, in particular, preserve the \emph{unit-trace}
of density matrices: 
\begin{equation}
\Tr\left[\op{\rho}\left(t\right)\right]=1\implies\Tr\left[\mathcal{E}^{t^{\prime}\from t}\left[\op{\rho}\left(t^{\prime}\right)\right]\right]=1.\label{eq:LinDynMapUnitTrPreserv}
\end{equation}
 The assumed linearity of $\mathcal{E}^{t^{\prime}\from t}\left[\cdot\right]$,
combined with its preservation \eqref{eq:LinDynMapUnitTrPreserv}
of the trace of unit-trace operators, means that it must in fact preserve
the traces of \emph{all} operators $\op O$, and thus must in general
be a trace-preserving (``T'' or ``TP'') mapping: 
\begin{equation}
\mathcal{E}^{t^{\prime}\from t}\left[\cdot\right]\mathrm{\ is\ T}\colon\qquad\Tr\left[\mathcal{E}^{t^{\prime}\from t}\left[\op O\right]\right]=\Tr\left[\op O\right]\ \mathrm{for\ all\ }\op O.\label{eq:LinDynMapT}
\end{equation}

Furthermore, \eqref{eq:LinDynMapOnDensMatr} should be a positive mapping,
meaning that it preserves the positive-semi-definiteness of density
matrices: 
\begin{equation}
\op{\rho}\left(t\right)\geq0\implies\mathcal{E}^{t^{\prime}\from t}\left[\op{\rho}\left(t^{\prime}\right)\right]\geq0.\label{eq:LinDynMapPos}
\end{equation}
 If we were to imagine introducing an arbitrary, causally disconnected
ancillary system with trivial dynamics determined by the identity
mapping $\mathrm{id}\left[\cdot\right]$ on operators, then it would
be reasonable to impose the (nontrivial) requirement that the resulting
composite dynamics $\mathcal{E}^{t^{\prime}\from t}\left[\cdot\right]\tensorprod\mathrm{id}\left[\cdot\right]$
governing the pair of systems should likewise be a positive mapping,
a condition on our original dynamical mapping $\mathcal{E}^{t^{\prime}\from t}\left[\cdot\right]$
called complete positivity (``CP''): 
\begin{equation}
\mathcal{E}^{t^{\prime}\from t}\left[\cdot\right]\mathrm{\ is\ CP}\colon\qquad\mathcal{E}^{t^{\prime}\from t}\left[\cdot\right]\tensorprod\mathrm{id}\left[\cdot\right]\ \mathrm{is\ a\ positive\ mapping}.\label{eq:LinDynMapCP}
\end{equation}

We are therefore led to the study of linear completely-positive-trace-preserving
(``CPT'')%
\footnote{Despite the unfortunate but conventional acronym, ``CPT'' here should
not be confused with the (C)harge-(P)arity-(T)ime-reversal transformations
that are familiar from particle physics; some authors use the acronym
``CPTP'' instead. %
} dynamical mappings of density matrices. In generalizing unitary
dynamics to linear CPT dynamics in this manner, as is necessary in
order to account for the crucial and non-reductive quantum relationships
between parent systems and their subsystems, note that we are \emph{not}
proposing any fundamental modification to the dynamics of quantum
theory, such as in GRW-type spontaneous-localization models \cite{GhirardiRiminiWeber:1986udmms,Penrose:1989enm,Pearle:1989csdsvrwsl,BassiGhirardi:2003drm,Weinberg:2011csv,AdlerBassi:2009qte},
but are simply accommodating the fact that generic mesoscopic and
macroscopic quantum systems are typically open to their environments
to some nonzero degree.%
\footnote{Given the generically local interactions found in realistic fundamental
physical models like quantum field theories and the resulting tendency
of decoherence to reduce open systems to states of relatively well-defined
spatial position \cite{JoosZeh:1985ecptiwe} automatically, proponents
of GRW-type spontaneous-localization models must justify why their
modifications to quantum theory aren't redundant \cite{Joos:1987cudmms}
and how we would experimentally distinguish those claimed modifications
from the more prosaic effects of decoherence.%
} Indeed, linear CPT dynamical mappings are widely used in quantum
chemistry as well as in quantum information science, in which they
are known as quantum operations; when specifically regarded as carriers
of quantum information, they are usually called\emph{ }quantum channels.%
\footnote{Starting from a simple measure of distinguishability between density
matrices that is non-increasing under linear CPT dynamics \cite{Ruskai:1994bss},
one can argue \cite{BreuerLainePiilo:2009mdnmbqpos,LainePiiloBreuer:2010mnmqp}
that linear CPT dynamics implies the absence of any backward flow
of information into the system from its environment. \cite{Buscemi:2013cppic}
strengthens this reasoning by proving that exact linear CPT dynamics
exists for a given quantum system if and only if the system's initial
correlations with its environment satisfy a quantum data-processing
inequality that prevents backward information flow.%
}

\subsubsection{Quantum Conditional Probabilities}

To motivate our defining formula for the generalized quantum counterpart
to the classical conditional probabilities appearing in \eqref{eq:ClassicalOnticLevelDynMapMarkov},
we begin by considering a collection of mutually disjoint quantum
systems $Q_{1},\dotsc,Q_{n}$ that we can identify as being subsystems
of some parent system $W=Q_{1}+\dotsb+Q_{n}$, where we include the
possibility that $Q_{2}=\cdots=Q_{n}=\emptyset$ are all trivial so
that $W=Q_{1}$. Suppose furthermore that the parent system $W$
has dynamics over a given time interval $\Delta t\equiv t^{\prime}-t\geq0$,
meaning that we can approximate the time evolution of the density
matrix $\op{\rho}_{W}$ for $W$ over the time interval $\Delta t$
by a linear CPT dynamical mapping \eqref{eq:LinDynMapOnDensMatr}:
\begin{equation}
\op{\rho}_{W}\left(t\right)\mapsto\op{\rho}_{W}\left(t^{\prime}\right)=\mathcal{E}_{W}^{t^{\prime}\from t}\left[\op{\rho}_{W}\left(t\right)\right].\label{eq:LinearCPTDynMapParentSys}
\end{equation}
 We can expand the density matrix $\op{\rho}_{W}\left(t\right)$ of
the parent system $W$ at the initial time $t$ in terms of its probability
eigenvalues $p_{W}\left(w;t\right)\equiv p_{W,w}\left(t\right)$ and
the projection operators (or eigenprojectors) $\op P_{W}\left(w;t\right)\equiv\ket{\Psi_{W}\left(w;t\right)}\bra{\Psi_{W}\left(w;t\right)}$
onto its orthonormal eigenbasis: 
\begin{equation}
\op{\rho}_{W}\left(t\right)=\sum_{w}p_{W}\left(w;t\right)\ket{\Psi_{W}\left(w;t\right)}\bra{\Psi_{W}\left(w;t\right)}=\sum_{w}p_{W}\left(w;t\right)\op P_{W}\left(w;t\right).\label{eq:SpectralDecompParentSysIntoProjOps}
\end{equation}
 Similarly, selecting subsystem $Q_{1}$ without loss of generality,
we can expand its (reduced) density matrix $\op{\rho}_{Q_{1}}\left(t^{\prime}\right)=\Tr_{Q_{2}+\dotsb+Q_{n}}\left[\op{\rho}_{W}\left(t^{\prime}\right)\right]$
at the final time $t^{\prime}$ in terms of its own probability eigenvalues
$p_{Q_{1}}\left(i_{1};t^{\prime}\right)\equiv p_{Q_{1},i_{1}}\left(t^{\prime}\right)$
and its own eigenprojectors $\op P_{Q_{1}}\left(i_{1};t^{\prime}\right)\equiv\ket{\Psi_{Q_{1}}\left(i_{1};t^{\prime}\right)}\bra{\Psi_{Q_{1}}\left(i_{1};t^{\prime}\right)}$:
\begin{equation}
\op{\rho}_{Q_{1}}\left(t^{\prime}\right)=\Tr_{Q_{2}+\dotsb+Q_{n}}\left[\op{\rho}_{W}\left(t^{\prime}\right)\right]=\sum_{i_{1}}p_{Q_{1}}\left(i_{1};t^{\prime}\right)\ket{\Psi_{Q_{1}}\left(i_{1};t^{\prime}\right)}\bra{\Psi_{Q_{1}}\left(i_{1};t^{\prime}\right)}=\sum_{i_{1}}p_{Q_{1}}\left(i_{1};t^{\prime}\right)\op P_{Q_{1}}\left(i_{1};t^{\prime}\right).\label{eq:SpectralDecompSubsysIntoProjOps}
\end{equation}
 We can also expand each of the identity operators $\op 1_{Q_{2}},\dotsc,\op 1_{Q_{n}}$
on the respective Hilbert spaces of the other mutually disjoint subsystem
$Q_{2},\dotsc,Q_{n}$ in terms of their own respective eigenprojectors
$\op P_{Q_{2}}\left(i_{2};t^{\prime}\right)$, $\dotsc$, $\op P_{Q_{n}}\left(i_{n};t^{\prime}\right)$,
\begin{equation}
\op 1_{Q_{2}}=\sum_{i_{2}}\op P_{Q_{2}}\left(i_{2};t^{\prime}\right),\ \dotsc,\ \op 1_{Q_{n}}=\sum_{i_{n}}\op P_{Q_{n}}\left(i_{n};t^{\prime}\right),\label{eq:SpectralDecompEnvIdOpIntoProjOps}
\end{equation}
 where we'll see that symmetry with $Q_{1}$ and consistency with
the linear CPT dynamical mapping \eqref{eq:LinearCPTDynMapParentSys}
for the parent system $W$ necessitates that these projection operators
are all evaluated at the \emph{same} final time $t^{\prime}$.

From the spectral decompositions \eqref{eq:SpectralDecompParentSysIntoProjOps},
\eqref{eq:SpectralDecompSubsysIntoProjOps}, and \eqref{eq:SpectralDecompEnvIdOpIntoProjOps},
together with our formula \eqref{eq:LinearCPTDynMapParentSys} for
the linear CPT dynamics of the parent system $W$, we can trivially
express the epistemic probability $p_{Q_{1}}\left(i_{1};t^{\prime}\right)\equiv p_{Q_{1},i_{1}}\left(t^{\prime}\right)$
for subsystem $Q_{1}$ to be in the ontic state $\Psi_{i_{1}}\left(i_{1};t^{\prime}\right)$
at the final time $t^{\prime}$ as 
\begin{align}
p_{Q_{1}}\left(i_{1};t^{\prime}\right) & =\Tr_{Q_{1}}\left[\op P_{Q_{1}}\left(i_{1};t^{\prime}\right)\op{\rho}_{Q_{1}}\left(t^{\prime}\right)\right]\nonumber \\
 & =\Tr_{W}\left[\left(\op P_{Q_{1}}\left(i_{1};t^{\prime}\right)\tensorprod\op 1_{Q_{2}}\tensorprod\dotsm\tensorprod\op 1_{Q_{n}}\right)\op{\rho}_{W}\left(t^{\prime}\right)\right]\nonumber \\
 & =\Tr_{W}\left[\left(\op P_{Q_{1}}\left(i_{1};t^{\prime}\right)\tensorprod\op 1_{Q_{2}}\tensorprod\dotsm\tensorprod\op 1_{Q_{n}}\right)\mathcal{E}_{W}^{t^{\prime}\from t}\left[\op{\rho}_{W}\left(t\right)\right]\right]\nonumber \\
 & =\sum_{w}\Tr_{W}\left[\left(\op P_{Q_{1}}\left(i_{1};t^{\prime}\right)\tensorprod\op 1_{Q_{2}}\tensorprod\dotsm\tensorprod\op 1_{Q_{n}}\right)\mathcal{E}_{W}^{t^{\prime}\from t}\left[\op P_{W}\left(w;t\right)\right]\right]p_{W}\left(w;t\right),\nonumber \\
 & =\negthickspace\negthickspace\negthickspace\negthickspace\sum_{i_{2},\dotsc,i_{n},w}\negthickspace\negthickspace\negthickspace\Tr_{W}\left[\left(\op P_{Q_{1}}\left(i_{1};t^{\prime}\right)\tensorprod\op P_{Q_{2}}\left(i_{2};t^{\prime}\right)\tensorprod\dotsm\tensorprod\op P_{Q_{n}}\left(i_{n};t^{\prime}\right)\right)\mathcal{E}_{W}^{t^{\prime}\from t}\left[\op P_{W}\left(w;t\right)\right]\right]p_{W}\left(w;t\right).\label{eq:MotivationForQuantCondProbs}
\end{align}

This last expression reduces to an intuitive Bayesian propagation
formula with marginalization, 
\begin{equation}
p_{Q_{1}}\left(i_{1};t^{\prime}\right)=\negthickspace\negthickspace\negthickspace\negthickspace\sum_{i_{2},\dotsc,i_{n},w}\negthickspace\negthickspace\negthickspace p_{Q_{1},\dotsc,Q_{n}\given W}\left(i_{1},\dotsc,i_{n};t^{\prime}\given w;t\right)p_{W}\left(w;t\right),\label{eq:MotivationForQuantCondProbsBayes}
\end{equation}
 provided that we interpret the trace over $W$ appearing in the summation
on $w$ in \eqref{eq:MotivationForQuantCondProbs} as the quantum conditional probability
$p_{Q_{1},\dotsc,Q_{n}\given W}\left(i_{1},\dotsc,i_{n};t^{\prime}\given w;t\right)$
for each subsystem $Q_{\alpha}$ to be in the ontic state $\Psi_{Q_{\alpha}}\left(i_{\alpha};t^{\prime}\right)\equiv\Psi_{Q_{\alpha},i_{\alpha}}\left(t^{\prime}\right)$
at time $t^{\prime}$ for $\alpha=1,\dotsc,n$ given that the parent
system $W$ was in the ontic state $\Psi_{W}\left(w;t\right)\equiv\Psi_{W,w}\left(t\right)$
at time $t$: 
\begin{equation}
\eqsbrace{\boxed{\begin{aligned}p_{Q_{1},\dotsc,Q_{n}\given W}\left(i_{1},\dotsc,i_{n};t^{\prime}\given w;t\right) & \equiv\Tr_{W}\left[\left(\op P_{Q_{1}}\left(i_{1};t^{\prime}\right)\tensorprod\dotsm\tensorprod\op P_{Q_{n}}\left(i_{n};t^{\prime}\right)\right)\mathcal{E}_{W}^{t^{\prime}\from t}\left[\op P_{W}\left(w;t\right)\right]\right]\\
 & \sim\Tr\left[\op P_{i_{1}}\left(t^{\prime}\right)\dotsm\op P_{i_{n}}\left(t^{\prime}\right)\mathcal{E}\left[\op P_{w}\left(t\right)\right]\right].
\end{aligned}
}}\label{eq:DefQuantCondProbGenFromTrParent}
\end{equation}
  As we'll see, these quantum conditional probabilities serve essentially
as smoothness conditions that sew together ontologies in a natural
way.

Note that our quantum conditional probabilities \eqref{eq:DefQuantCondProbGenFromTrParent}
are \emph{only} defined in terms of the projection operators onto
the orthonormal eigenstates of density matrices---that is, the eigenstates
representing the system's possible ontic states in accordance with
our general correspondence \eqref{eq:EpStateDensMatrixCorresp}---and
\emph{not} in terms of projection operators onto \emph{generic} state
vectors, in contrast to the Born rule for computing empirical outcome
probabilities. Furthermore, observe that the linear CPT dynamical
mapping $\mathcal{E}_{W}^{t^{\prime}\from t}\left[\cdot\right]$ in
our definition \eqref{eq:DefQuantCondProbGenFromTrParent} plays the
role of a parallel-transport superoperator that carries the parent-system
projection operator $\op P_{W}\left(w;t\right)$ from $t$ to $t^{\prime}$
before we compare it with the subsystem projection operators $\op P_{Q_{1}}\left(i_{1};t^{\prime}\right)$,
$\dotsc$, $\op P_{Q_{n}}\left(i_{n};t^{\prime}\right)$; the Born
rule involves the comparison of state vectors without any kind of
parallel transport.  Moreover, we emphasize that our definition \eqref{eq:DefQuantCondProbGenFromTrParent},
unlike the axiomatic wave-function collapse of the traditional Copenhagen
interpretation of quantum theory, does not introduce any new \emph{fundamental}
violations of time-reversal symmetry; indeed, given a system that
possesses well-defined \emph{reversed} dynamics expressible in terms
of a corresponding \emph{reversed} linear CPT dynamical mapping that
evolves the system's density matrix \emph{backward} in time, we are
free to use that reversed linear CPT dynamical mapping in our definition
\eqref{eq:DefQuantCondProbGenFromTrParent} of the quantum conditional
probabilities.

Observe also that the tensor-product operator $\op P_{Q_{1}}\left(i_{1};t^{\prime}\right)\tensorprod\dotsm\tensorprod\op P_{Q_{n}}\left(i_{n};t^{\prime}\right)$
and the time-evolved projection operator $\mathcal{E}_{W}^{t^{\prime}\from t}\left[\op P_{W}\left(w;t\right)\right]$
are both positive semi-definite operators. Hence, each $p_{Q_{1},\dotsc,Q_{n}\given W}\left(i_{1},\dotsc,i_{n};t^{\prime}\given w;t\right)$
defined according to \eqref{eq:DefQuantCondProbGenFromTrParent} consists
of a trace over a product of two positive semi-definite operators,
and is therefore guaranteed to be non-negative,%
\footnote{Proof: Let $A\geq0$ and $B\geq0$ be positive semi-definite matrices.
Then $\sqrt{A}$ and $\sqrt{B}$ exist and are likewise positive semi-definite,
and so, using the cyclic property of the trace, we have 
\[
\Tr\left[AB\right]=\Tr\left[\sqrt{A}\sqrt{A}\sqrt{B}\sqrt{B}\right]=\Tr\left[\sqrt{B}\sqrt{A}\sqrt{A}\sqrt{B}\right]=\Tr\left[\left(\sqrt{A}\sqrt{B}\right)^{\adj}\sqrt{A}\sqrt{B}\right]\geq0.\qquad\QED
\]
 Notice that this proof does \emph{not} generalize to traces over
products of more than two operators.%
} 
\begin{equation}
p_{Q_{1},\dotsc,Q_{n}\given W}\left(i_{1},\dotsc,i_{n};t^{\prime}\given w;t\right)\geq0.\label{eq:QuantCondProbsNonNeg}
\end{equation}
 From this starting point, we can show that the quantities $p_{Q_{1},\dotsc,Q_{n}\given W}\left(i_{1},\dotsc,i_{n};t^{\prime}\given w;t\right)$
miraculously satisfy several important and nontrivial requirements
that support their interpretation as a quantum class of conditional
probabilities.
\begin{enumerate}
\item From the completeness of the subsystem eigenprojectors $\op P_{Q_{1}}\left(i_{1};t^{\prime}\right)$,
$\dotsc$, $\op P_{Q_{n}}\left(i_{n};t^{\prime}\right)$, the unit-trace
of the parent-system eigenprojectors $\op P_{W}\left(w;t\right)$,
and the trace-preserving property of $\mathcal{E}_{W}^{t^{\prime}\from t}\left[\cdot\right]$,
the set of quantities $p_{Q_{1},\dotsc,Q_{n}\given W}\left(i_{1},\dotsc,i_{n};t^{\prime}\given w;t\right)$
for any fixed value of $w$ give unity when summed over the indices
$i_{1},\dotsc,i_{n}$ labeling all the possible ontic states of the
mutually disjoint partitioning subsystems $Q_{1},\dotsc,Q_{n}$: 
\begin{equation}
\sum_{i_{1},\dotsc,i_{n}}\negthickspace p_{Q_{1},\dotsc,Q_{n}\given W}\left(i_{1},\dotsc,i_{n};t^{\prime}\given w;t\right)=1.\label{eq:QuantCondProbsSumToUnity}
\end{equation}

\item Combining the properties \eqref{eq:QuantCondProbsNonNeg} and \eqref{eq:QuantCondProbsSumToUnity},
we see immediately that the quantities $p_{Q_{1},\dotsc,Q_{n}\given W}\left(i_{1},\dotsc,i_{n};t^{\prime}\given w;t\right)$
are each real numbers in the interval between $0$ and $1$: 
\begin{equation}
p_{Q_{1},\dotsc,Q_{n}\given W}\left(i_{1},\dotsc,i_{n};t^{\prime}\given w;t\right)\in\left[0,1\right].\label{eq:QuantCondProbsInUnitInterval}
\end{equation}

\item Recalling our derivation of \eqref{eq:MotivationForQuantCondProbs},
and from the linearity property of $\mathcal{E}_{W}^{t^{\prime}\from t}\left[\cdot\right]$
together with the spectral decomposition \eqref{eq:SpectralDecompParentSysIntoProjOps}
of the density matrix $\op{\rho}_{W}\left(t\right)$ of $W$ in terms
of its eigenprojectors $\op P_{W}\left(w;t\right)$ and the dynamical
equation \eqref{eq:LinearCPTDynMapParentSys} for the time evolution
of $\op{\rho}_{W}\left(t\right)$, we see that multiplying the quantities
$p_{Q_{1},\dotsc,Q_{n}\given W}\left(i_{1},\dotsc,i_{n};t^{\prime}\given w;t\right)$
by the epistemic probabilities $p_{W}\left(w;t\right)\equiv p_{W,w}\left(t\right)$
for $W$ at time $t$ and summing over $w$ and $i_{1},\dotsc,i_{n}$
\emph{except for one subsystem index} $i_{\alpha}$ gives the epistemic
probabilities $p_{Q_{\alpha}}\left(i_{\alpha};t^{\prime}\right)\equiv p_{Q_{\alpha},i_{\alpha}}\left(t^{\prime}\right)$
for $Q_{\alpha}$, in agreement with the rule \eqref{eq:MotivationForQuantCondProbsBayes}
for Bayesian propagation and marginalization: 
\begin{equation}
\sum_{i_{1},\dotsc\left(\mathrm{no\ }i_{\alpha}\right)\dotsc,i_{n},w}\negthickspace\negthickspace\negthickspace\negthickspace\negthickspace\negthickspace\negthickspace\negthickspace p_{Q_{1},\dotsc,Q_{n}\given W}\left(i_{1},\dotsc,i_{n};t^{\prime}\given w;t\right)p_{W}\left(w;t\right)=p_{Q_{\alpha}}\left(i_{\alpha};t^{\prime}\right).\label{eq:QuantCondProbsCompLaw}
\end{equation}

\item Due to the cyclic property of the trace, the quantities $p_{Q_{1},\dotsc,Q_{n}\given W}\left(i_{1},\dotsc,i_{n};t^{\prime}\given w;t\right)$
are manifestly invariant under arbitrary unitary transformations.
(Subtleties can arise for unitary transformations that involve time,
such as for Lorentz transformations, as we explain in Section~\ref{sub:The-Myrvold-No-Go-Theorem}.)
\item In the idealized case in which $W=Q_{1}\equiv Q$ is a closed system
undergoing ``deterministic'' unitary dynamics, so that 
\[
\op{\rho}_{Q}\left(t^{\prime}\right)=U_{Q}\left(t^{\prime}\from t\right)\op{\rho}_{Q}\left(t\right)U_{Q}^{\adj}\left(t^{\prime}\from t\right)
\]
 for some unitary time-evolution operator $U_{Q}\left(t^{\prime}\from t\right)$,
we have 
\[
\op P_{Q}\left(j;t^{\prime}\right)=U_{Q}\left(t^{\prime}\from t\right)\op P_{Q}\left(j;t\right)U_{Q}^{\adj}\left(t^{\prime}\from t\right)
\]
 and thus, as expected, the quantities $p_{Q\given Q}\left(i;t^{\prime}\given j;t\right)$
trivialize to the deterministic formula 
\begin{equation}
p_{Q\given Q}\left(i;t^{\prime}\given j;t\right)\equiv\Tr_{Q}\left[\op P_{Q}\left(i;t^{\prime}\right)\op P_{Q}\left(j;t^{\prime}\right)\right]=\delta_{ij}=\begin{cases}
1 & \whichfor i=j,\\
0 & \whichfor i\ne j.
\end{cases}\label{eq:QuantCondProbsTrivUnitary}
\end{equation}

\end{enumerate}
Considering the inflexibility of quantum theory and its famed disregard
for the sorts of familiar concepts favored by human beings, it's quite
remarkable that the theory contains such a large class of quantities
$p_{Q_{1},\dotsc,Q_{n}\given W}\left(i_{1},\dotsc,i_{n};t^{\prime}\given w;t\right)$
satisfying the properties \eqref{eq:QuantCondProbsNonNeg}-\eqref{eq:QuantCondProbsTrivUnitary}
of conditional probabilities. Essentially, our interpretation of quantum
theory takes this fact at face value by actually \emph{calling} these
quantities conditional probabilities.

\subsubsection{Probabilistic and Non-Probabilistic Uncertainty\label{subsub:Probabilistic-and-Non-Probabilistic-Uncertainty}}

However, one should keep in mind that quantum theory does appear to
limit what kinds of probabilities we can safely define. In particular,
our derivation \eqref{eq:MotivationForQuantCondProbs} does not extend
to \emph{non-disjoint} subsystems, nor to subsystems at \emph{multiple
distinct} final times, nor to multiple \emph{parent} systems. Hence,
certain kinds of hypothetical statements involving the ontic states
of our interpretation of quantum theory turn out not to admit generally
well-defined probabilities, despite lying behind a veil of uncertainty.

Physicists tend to use the terms ``uncertainty'' and ``probability''
almost synonymously, but the two concepts are distinct. Indeed, there
is no rigorous \emph{a priori} reason to believe as a general truth
about Nature that the frequency ratio of \emph{every} kind of repeatable
event should ``tend to'' some specific value in the limit of many
trials; the existence of such a limit is actually a highly nontrivial
and non-obvious constraint because we could easily imagine outcomes
instead occurring in a completely unpredictable way that never ``settles
down.'' Science would certainly be a far less successful predictive
enterprise if \emph{observable} phenomena did not generally obey such
a constraint, but it makes no difference to the predictive power of
science if \emph{hidden variables} do not always comply.

In fact, economists have known for many years that although certain
types of uncertainty, called probabilistic uncertainty or risk,
could safely be described in terms of probabilities, other kinds of
uncertainty, called non-probabilistic uncertainty, could not be.%
\footnote{In his light-hearted paper \cite{Aaronson:2013gqtm}, Aaronson refers
to the latter as ``Knightian'' uncertainty, in honor of economist
Frank Knight's seminal 1921 book \cite{Knight:1921rup} on the subject.%
} An example of this distinction from computer programming is the difference
between the possibility of two \emph{pseudo-random numbers} agreeing
and the possibility of a \emph{user} deciding to input two numbers
that agree. Further examples closer to physics are examined in \cite{SvetlichnyRedheadBrownButterfield:1988dbirejpd},
which elegantly explains the nonexistence of a joint probability distribution
$p\left(x,y,z,\dotsc\right)$ as the failure of its arguments $x,y,z,\dotsc$
to possess a well-defined joint limiting relative frequency, perhaps
due to the need for proper subsets of the arguments $x,y,z,\dotsc$
to satisfy their own mandated limiting relative frequencies.

We can present a more concrete example \cite{HessPhilipp:2004btcpwwi}
demonstrating that joint probabilities may not exist for certain sets
of random variables even in classical probability theory. Consider
a set of three classical random bits $X,Y,Z$ with individual probabilities
given by the chart 

\noindent \begin{center}
\begin{tabular}{|c|c|c|}
\hline 
 & $+$ & $-$\tabularnewline
\hline 
$p_{X}\left(x\right)$ & $1/2$ & $1/2$\tabularnewline
\hline 
$p_{Y}\left(y\right)$ & $1/2$ & $1/2$\tabularnewline
\hline 
$p_{Z}\left(z\right)$ & $1/2$ & $1/2$\tabularnewline
\hline 
\end{tabular}
\par\end{center}

\noindent and pairwise-joint probabilities 

\noindent \begin{center}
\begin{tabular}{|c|c|c|c|c|}
\hline 
 & $\left(+,+\right)$ & $\left(+,-\right)$ & $\left(-,+\right)$ & $\left(-,-\right)$\tabularnewline
\hline 
$p_{X,Y}\left(x,y\right)$ & $\frac{1}{4}\left(1+\frac{1}{\sqrt{2}}\right)$ & $\frac{1}{4}\left(1-\frac{1}{\sqrt{2}}\right)$ & $\frac{1}{4}\left(1-\frac{1}{\sqrt{2}}\right)$ & $\frac{1}{4}\left(1+\frac{1}{\sqrt{2}}\right)$\tabularnewline
\hline 
$p_{X,Z}\left(x,z\right)$ & $\frac{1}{4}\left(1+\frac{1}{\sqrt{2}}\right)$ & $\frac{1}{4}\left(1-\frac{1}{\sqrt{2}}\right)$ & $\frac{1}{4}\left(1-\frac{1}{\sqrt{2}}\right)$ & $\frac{1}{4}\left(1+\frac{1}{\sqrt{2}}\right)$\tabularnewline
\hline 
$p_{Y,Z}\left(y,z\right)$ & $\frac{1}{4}$ & $\frac{1}{4}$ & $\frac{1}{4}$ & $\frac{1}{4}$\tabularnewline
\hline 
\end{tabular}
\par\end{center}

\noindent satisfying the correct partial sums \eqref{eq:DefPartialSums}:
\begin{align*}
\sum_{y}p_{X,Y}\left(x,y\right) & =\sum_{z}p_{X,Z}\left(x,z\right)=\frac{1}{2}=p_{X}\left(x\right),\\
\sum_{x}p_{X,Y}\left(x,y\right) & =\sum_{z}p_{Y,Z}\left(y,z\right)=\frac{1}{2}=p_{Y}\left(y\right),\\
\sum_{x}p_{X,Z}\left(x,z\right) & =\sum_{y}p_{Y,Z}\left(y,z\right)=\frac{1}{2}=p_{Z}\left(z\right).
\end{align*}
 Then it is impossible to define a joint probability distribution
$p_{X,Y,Z}\left(x,y,z\right)$ for all three random bits $X,Y,Z$.%
\footnote{\noindent We can prove this claim by contradiction: Supposing to the
contrary that we could indeed assign joint probabilities to $\left(x,y,z\right)=\left(-,-,-\right)$
and $\left(x,y,z\right)=\left(-,-,+\right)$, we see that $p_{X,Y,Z}\left(-,-,-\right)\leq p_{Y,Z}\left(-,-\right)=1/4$
and $p_{X,Y,Z}\left(-,-,+\right)\leq p_{X,Z}\left(-,+\right)=\left(1/4\right)\left(1-1/\sqrt{2}\right)$,
but then the partial-sum formula \eqref{eq:DefPartialSums} breaks
down because $p_{X,Y,Z}\left(-,-,-\right)+p_{X,Y,Z}\left(-,-,+\right)\leq\left(1/4\right)\left(2-1/\sqrt{2}\right)<\left(1/4\right)\left(1+1/\sqrt{2}\right)=p_{X,Y}\left(-,-\right)$.
$\QED$%
}

\subsubsection{Hidden Ontic-Level Nonlocality\label{subsub:Hidden-Ontic-Level-Nonlocality}}

Note that the definition \eqref{eq:DefQuantCondProbGenFromTrParent}
of our quantum conditional probabilities is not manifestly local at
the ontic level, as we'll make clear explicitly when we discuss the
EPR-Bohm thought experiment in Section~\ref{sub:The-EPR-Bohm-Thought-Experiment-and-Bell's-Theorem}
in the context of our minimal modal interpretation of quantum theory.
However, the causal structure of special relativity only places constraints
on \emph{observable} signals, and our quantum conditional probabilities
are, by construction, compatible with standard density-matrix dynamics
and thus constrained by the no-communication theorem \cite{Hall:1987imnlqm,PeresTerno:2004qirt}
to disallow superluminal observable signals, as we will explain in
greater detail Section~\ref{sec:Lorentz-Invariance-and-Locality}.

The lack of manifest ontic-level locality in \eqref{eq:DefQuantCondProbGenFromTrParent}
is a feature, not a bug, because, as we'll see when we analyze the
EPR-Bohm thought experiment in Section~\ref{sub:The-EPR-Bohm-Thought-Experiment-and-Bell's-Theorem}
and the GHZ-Mermin thought experiment in Section~\ref{sub:The-GHZ-Mermin-Thought-Experiment},
there is no way to accommodate ontic hidden variables without hidden
ontic-level nonlocality.%
\footnote{See Section~\ref{subsub:Nonlocality in the Everett-DeWitt Many-Worlds Interpretation}
for a discussion of the status of locality in the Everett-DeWitt many-worlds
interpretation.%
} Hence, if our formula \eqref{eq:DefQuantCondProbGenFromTrParent}
\emph{didn't} allow for benign ontic-level nonlocality, then we would
clearly be doing something wrong.

\subsubsection{Kinematical Relationships Between Ontic States of Parent Systems
and Subsystems\label{subsub:Kinematical-Relationships-Between-Ontic-States-of-Parent-Systems-and-Subsystems}}

Setting $t^{\prime}=t$ and suppressing time from our notation for
clarity, the dynamical mapping $\mathcal{E}^{t^{\prime}\from t}\left[\cdot\right]$
drops out and the quantum conditional probabilities \eqref{eq:DefQuantCondProbGenFromTrParent}
yield an explicit instantaneous kinematical (and generically probabilistic)
relationship between the ontic states of the parent system $W=Q_{1}+\dotsb+Q_{n}$
and the ontic states of the partitioning collection of mutually disjoint
subsystems $Q_{1},\dotsc,Q_{n}$: 
\begin{equation}
\eqsbrace{\boxed{\begin{aligned}p_{Q_{1},\dotsc,Q_{n}\given W}\left(i_{1},\dotsc,i_{n}\given w\right) & \equiv\Tr_{W}\left[\left(\op P_{Q_{1}}\left(i_{1}\right)\tensorprod\dotsm\tensorprod\op P_{Q_{n}}\left(i_{n}\right)\right)\op P_{W}\left(w\right)\right]\\
 & =\bra{\Psi_{W,w}}\left(\ket{\Psi_{Q_{1},i_{1}}}\bra{\Psi_{Q_{1},i_{1}}}\tensorprod\dotsm\tensorprod\ket{\Psi_{Q_{n},i_{n}}}\bra{\Psi_{Q_{n},i_{n}}}\right)\ket{\Psi_{W,w}}.
\end{aligned}
}}\label{eq:QuantCondProbKinParentSub}
\end{equation}
 In particular, this result leads to a simpler version of our formula
\eqref{eq:MotivationForQuantCondProbsBayes} for Bayesian propagation
and marginalization: 
\begin{equation}
p_{Q_{1}}\left(i_{1}\right)=\negthickspace\negthickspace\negthickspace\sum_{i_{2},\dotsc,i_{n},w}\negthickspace\negthickspace\negthickspace p_{Q_{1},\dotsc,Q_{n}\given W}\left(i_{1},\dotsc,i_{n}\given w\right)p_{W}\left(w\right).\label{eq:QuantCondProbKinParentSubBayes}
\end{equation}
Notice how our quantum conditional probabilities play the role of
sewing together the ontologies of $W$ and $Q_{1},\dotsc,Q_{n}$.

In the simplest case, for which the actual ontic state $\ket{\Psi_{W,w}}=\ket{\Psi_{Q_{1},i_{1}}}\tensorprod\dotsm\tensorprod\ket{\Psi_{Q_{n},i_{n}}}$
of the parent system $W$ involves approximately no entanglement between
its mutually disjoint subsystems $Q_{1},\dotsc,Q_{n}$, we obtain
the classical-looking result $p_{Q_{1},\dotsc,Q_{n}\given W}\left(j_{1},\dotsc,j_{n}\given w=\left(i_{1},\dotsc,i_{n}\right)\right)=\delta_{j_{1}i_{1}}\dotsm\delta_{j_{n}i_{n}}$,
as expected. Because decoherence ensures that human-scale macroscopic
systems exhibit negligible quantum entanglement with one another,
we see immediately that the ontologies of typical macroscopic parent
systems and their macroscopic subsystems fit together in a classically
intuitive, reductionist manner. However, naïve reductionism generically
breaks down for microscopic systems like electrons, and thus demanding
classically intuitive relationships%
\footnote{Vermaas \cite{Vermaas:1999puqm} refers to the two logical directions
underlying these classically reductionist relationships as the property of composition
and the property of division.%
} between \emph{microscopic} parent systems and subsystems would mean
committing a fallacy of composition or division.%
\footnote{Maudlin implicitly makes this kind of error in criterion 1.A of his
three-part classification of interpretations of quantum theory in
\cite{Maudlin:1995tmp} when he assumes that the metaphysical completeness
of state vectors implies that the state vector of a \emph{parent system}
completely determines the state vectors of all its \emph{subsystems}
even for the case of microscopic systems. It's also worth mentioning
that Maudlin's criterion 1.B that state vectors always evolve according
to linear dynamical equations is somewhat misleading, because all
realistic systems are always at least slightly open and thus must
be described by density matrices that do not evolve according to exactly
linear dynamical equations.%
}

With the formula \eqref{eq:QuantCondProbKinParentSub} in hand, we
can easily generalize our earlier entangled example \eqref{eq:GenericTwoSubsysEntangDensMatr}
to the case in which the composite parent system $A+B$ itself has
a nontrivial density matrix 
\[
\op{\rho}_{A+B}=p_{\Psi}\ket{\Psi_{A+B}}\bra{\Psi_{A+B}}+p_{\Phi}\ket{\Phi_{A+B}}\bra{\Phi_{A+B}},\qquad p_{\Psi},p_{\Phi}\in\left[0,1\right],\ p_{\Psi}+p_{\Phi}=1,
\]
 where 
\[
\begin{aligned}\ket{\Psi_{A+B}} & =\alpha\ket{\Psi_{A,1}}\ket{\Psi_{B,1}}+\beta\ket{\Psi_{A,2}}\ket{\Psi_{B,2}},\qquad\alpha,\beta\in\mathbb{C},\ \absval{\alpha}^{2}+\absval{\beta}^{2}=1,\\
\ket{\Phi_{A+B}} & =\gamma\ket{\Psi_{A,3}}\ket{\Psi_{B,3}}+\delta\ket{\Psi_{A,4}}\ket{\Psi_{B,4}},\qquad\gamma,\delta\in\mathbb{C},\ \absval{\gamma}^{2}+\absval{\delta}^{2}=1
\end{aligned}
\]
 and where we assume for simplicity that the state vectors $\ket{\Psi_{A,i}}$
for subsystem $A$ are all mutually orthogonal and likewise that the
state vectors $\ket{\Psi_{B,i}}$ for subsystem $B$ are all mutually
orthogonal. The partial-trace prescription \eqref{eq:DefPartialTraces}
then yields the reduced density matrix of $A$, 
\[
\begin{aligned}\op{\rho}_{A} & =p_{\Psi}\absval{\alpha}^{2}\ket{\Psi_{A,1}}\bra{\Psi_{A,1}}+p_{\Psi}\absval{\beta}^{2}\ket{\Psi_{A,1}}\bra{\Psi_{A,1}}\\
 & \quad+p_{\Phi}\absval{\gamma}^{2}\ket{\Psi_{A,3}}\bra{\Psi_{A,3}}+p_{\Phi}\absval{\delta}^{2}\ket{\Psi_{A,4}}\bra{\Psi_{A,4}},
\end{aligned}
\]
 with a similar formula for $B$, and the instantaneous kinematical
relationship \eqref{eq:QuantCondProbKinParentSub} yields the conditional
probabilities 
\[
\begin{aligned} & p\left(\Psi_{A,1}\given\Psi_{A+B}\right)=p_{\Psi}\absval{\alpha}^{2}, & \qquad & p\left(\Psi_{A,2}\given\Psi_{A+B}\right)=p_{\Psi}\absval{\beta}^{2},\\
 & p\left(\Psi_{A,3}\given\Psi_{A+B}\right)=0, & \qquad & p\left(\Psi_{A,4}\given\Psi_{A+B}\right)=0,\\
 & p\left(\Psi_{A,1}\given\Phi_{A+B}\right)=0, & \qquad & p\left(\Psi_{A,2}\given\Phi_{A+B}\right)=0,\\
 & p\left(\Psi_{A,3}\given\Phi_{A+B}\right)=p_{\Phi}\absval{\gamma}^{2}, & \qquad & p\left(\Psi_{A,4}\given\Phi_{A+B}\right)=p_{\Phi}\absval{\delta}^{2},
\end{aligned}
\]
 again with similar formulas for $B$.

The formula \eqref{eq:QuantCondProbKinParentSub} also allows us to
resolve an issue that we brought up in our discussion surrounding
\eqref{eq:MixedStateExample} in Section~\ref{subsub:Proper-Mixtures-and-Improper-Mixtures},
where we saw that the existence of entanglement between a pair of
systems $A$ and $B$ seemed to prevent us from relating the ontic
state of subsystem $B$ to a given ontic state of its parent system
$A+B$ even if there existed a \emph{slight redefinition} of our subsystem
decomposition $A+B=A^{\prime}+B^{\prime}$ for which the parent system's
ontic state had the simple non-entangled form $\left(a^{\prime},b^{\prime}\right)$.
In that case, the quantum conditional probability $p_{B\given A+B}\left(b\given w=\left(a^{\prime},b^{\prime}\right)\right)\approx p_{B^{\prime}\given A^{\prime}+B^{\prime}}\left(b^{\prime}\given w=\left(a^{\prime},b^{\prime}\right)\right)=1$
would be very close to unity, thereby smoothing out the supposed discrepancy
and, furthermore, eliminating the need to define subsystem $B$ with
measure-zero sharpness.

\subsubsection{Quantum Dynamics of Open Subsystems}

Our discussion of dynamics in quantum theory begins with the assumption
of a system $W$ having well-defined quantum dynamics to an acceptable
level of approximation over a given time interval $\Delta t\equiv t^{\prime}-t\geq0$,
an assumption that we take to mean that $W$ approximately admits
a well-defined linear CPT dynamical mapping $\mathcal{E}_{W}^{t^{\prime}\from t}\left[\cdot\right]$
in the sense of \eqref{eq:LinearCPTDynMapParentSys}. Note that we
include the possibility that $\mathcal{E}_{W}^{t^{\prime}\from t}\left[\cdot\right]$
is a unitary mapping, as would be appropriate for the case in which
$W$ is a closed system and a good approximation for systems that
are sufficiently microscopic and therefore relatively easy to isolate
from their environments.

If we can regard this system as being a composite parent system $W=Q+E$
consisting of a subsystem $Q$ and its larger environment $E$ inside
$W$, then there is no guarantee that $Q$ itself has well-defined
dynamics of its own. Indeed, the reduced density matrix of $Q$ at
the final time $t^{\prime}$ is generally given by 
\begin{equation}
\op{\rho}_{Q}\left(t^{\prime}\right)=\Tr_{E}\left[\op{\rho}_{W}\left(t^{\prime}\right)\right]=\Tr_{E}\left[\mathcal{E}_{W}^{t^{\prime}\from t}\left[\op{\rho}_{W}\left(t\right)\right]\right],\label{eq:SubsysDensMatrGenericTimeEv}
\end{equation}
 which is not generically linear (or even analytic) in the reduced
density matrix $\op{\rho}_{Q}\left(t\right)$ at the initial time
$t$ because of the complicated way that $\op{\rho}_{Q}\left(t\right)$
is related to $\op{\rho}_{W}\left(t\right)$ through the partial trace
\eqref{eq:DefPartialTraces}.

We encountered a similar issue in Section~\ref{subsub:Classical-Dynamics-and-Multi-Linear-Dynamical-Mappings}
when we discussed the general relationship between classical parent-system
and subsystem dynamics, where we saw in \eqref{eq:ClassicalNonlinearEpLevelDynEvExplicit}
that a classical open subsystem's dynamical mapping is not generically
linear. Mimicking our approach in that discussion, we begin by rewriting
the subsystem evolution equation \eqref{eq:SubsysDensMatrGenericTimeEv}
as 
\begin{equation}
\op{\rho}_{Q}\left(t^{\prime}\right)=\Tr_{E}\left[\mathcal{E}_{W}^{t^{\prime}\from t}\left[\mathcal{A}_{Q\subset W}^{t}\left[\op{\rho}_{Q}\left(t\right)\right]\right]\right],\label{eq:SubsysDensMatrGenericTimeEvFromAssnMap}
\end{equation}
 where we have introduced the nonlinear, manifestly positive-definite
(but not generally trace-preserving) mapping 
\begin{equation}
\mathcal{A}_{Q\subset W}^{t}\left[\cdot\right]\equiv\op{\rho}_{W}^{1/2}\left(t\right)\left(\op{\rho}_{Q}^{-1/2}\left(t\right)\tensorprod\op 1_{E}\right)\left(\left(\cdot\right)\tensorprod\op 1_{E}\right)\left(\op{\rho}_{Q}^{-1/2}\left(t\right)\tensorprod\op 1_{E}\right)\op{\rho}_{W}^{1/2}\left(t\right),\label{eq:DefNonlinearAssgnMap}
\end{equation}
which is known as an assignment mapping. The alternative version
\eqref{eq:SubsysDensMatrGenericTimeEvFromAssnMap} of our original
subsystem evolution equation \eqref{eq:SubsysDensMatrGenericTimeEv}
for $Q$ is the natural noncommutative quantum counterpart to the
classical formula \eqref{eq:ClassicalNonlinearEpLevelDynEvExplicit},
and motivates introducing a coarse-grained or effective version
$p_{Q\subset W}\left(\cdot;t^{\prime}\given\cdot;t\right)$ of our
original quantum conditional probabilities \eqref{eq:DefQuantCondProbGenFromTrParent}
for $Q_{1}\equiv Q$ inside $W$: 
\begin{equation}
p_{Q\subset W}\left(j;t^{\prime}\given i;t\right)\equiv\Tr_{W}\left[\left(\op P_{Q}\left(j;t^{\prime}\right)\tensorprod\op 1_{E}\right)\mathcal{E}_{W}^{t^{\prime}\from t}\left[\mathcal{A}_{Q\subset W}^{t}\left[\op P_{Q}\left(i;t\right)\right]\right]\right].\label{eq:CoarseGrainedOpenSubsysDynMap}
\end{equation}

Again in parallel with the classical case \eqref{eq:ClassicalCorrWashOut},
if initial correlations between $Q$ and its environment $E$ approximately
wash out 
\begin{equation}
\op{\rho}_{W}\approx\op{\rho}_{Q}\tensorprod\op{\rho}_{E}\label{eq:QuantumCorrWashOut}
\end{equation}
 over some short characteristic time scale 
\begin{equation}
\delta t_{Q}\ll t^{\prime}-t,\label{eq:QuantCharacWashoutTimescale}
\end{equation}
 then 
\[
\op{\rho}_{W}^{1/2}\left(t\right)\left(\op{\rho}_{Q}^{-1/2}\left(t\right)\tensorprod\op 1_{E}\right)=\op 1_{Q}\tensorprod\op{\rho}_{E}^{1/2}\left(t\right)
\]
 and thus the assignment mapping \eqref{eq:DefNonlinearAssgnMap}
trivializes, 
\[
\mathcal{A}_{Q\subset W}^{t}\left[\op{\rho}_{Q}\left(t\right)\right]=\op{\rho}_{Q}\left(t\right)\tensorprod\op 1_{E}.
\]
 It follows as an immediate consequence that over time scales $t^{\prime}-t\gg\delta t_{Q}$,
the subsystem evolution equation \eqref{eq:SubsysDensMatrGenericTimeEvFromAssnMap}
reduces to linear CPT dynamics for $Q$ on its own, 
\begin{equation}
\op{\rho}_{Q}\left(t^{\prime}\right)=\Tr_{E}\left[\mathcal{E}_{W}^{t^{\prime}\from t}\left[\op{\rho}_{Q}\left(t\right)\tensorprod\op 1_{E}\right]\right]=\sum_{\alpha}\op E_{\alpha}^{t^{\prime}\from t}\op{\rho}_{Q}\left(t\right)\op E_{\alpha}^{t^{\prime}\from t\adj},\qquad\sum_{\alpha}\op E_{\alpha}^{t^{\prime}\from t\adj}\op E_{\alpha}^{t^{\prime}\from t}=\op 1_{Q},\label{eq:KrausRepDynMap}
\end{equation}
 where $\setindexed{\op E_{\alpha}^{t^{\prime}\from t}}{\alpha}$
is the usual set of Kraus operators for the dynamics \cite{Kraus:1971gscqt}.
Indeed, this line of reasoning, together with the additional assumptions
that the linear CPT density-matrix dynamics for $Q$ is Markovian
and homogeneous in time, precisely leads to the well-known and highly
effective Lindblad equation \cite{Lindblad:1976gqds}, which can
be expressed in its diagonal form as \cite{BreuerPetruccione:2002toqs,Joos:2003dacwqt,SchumacherWestmoreland:2010qpsi}
\begin{equation}
\frac{\partial\op{\rho}_{Q}}{\partial t}=-\frac{i}{\hbar}\comm{\op H}{\op{\rho}_{Q}}+\sum_{k=1}^{N^{2}-1}\gamma_{k}\left(\op A_{k}\op{\rho}_{Q}\op A_{k}^{\adj}-\frac{1}{2}\op A_{k}^{\adj}\op A_{k}\op{\rho}_{Q}-\frac{1}{2}\op{\rho}_{Q}\op A_{k}^{\adj}\op A_{k}\right)\label{eq:LindbladEq}
\end{equation}
 for a suitable collection of parameters $\gamma_{k}$ and operators
$\op H,\op A_{k}$. Letting $\mathcal{E}_{Q}^{t^{\prime}\from t}$
denote this reduced linear CPT dynamics for $Q$ alone, 
\begin{equation}
\op{\rho}_{Q}\left(t\right)\mapsto\op{\rho}_{Q}\left(t^{\prime}\right)=\mathcal{E}_{Q}^{t^{\prime}\from t}\left[\op{\rho}_{Q}\left(t\right)\right],\label{eq:SubsysDensMatrExactCPTTimeEv}
\end{equation}
the coarse-grained conditional probabilities \eqref{eq:CoarseGrainedOpenSubsysDynMap}
now coincide with our \emph{exact} general definition \eqref{eq:DefQuantCondProbGenFromTrParent}
for $W=Q_{1}\equiv Q$: 
\begin{equation}
p_{Q\subset W}\left(j;t^{\prime}\given i;t\right)\to p_{Q}\left(j;t^{\prime}\given i;t\right)\equiv\Tr_{Q}\left[\op P_{Q}\left(j;t^{\prime}\right)\mathcal{E}_{Q}^{t^{\prime}\from t}\left[\op P_{Q}\left(i;t\right)\right]\right]\sim\Tr\left[\op P_{j}\left(t^{\prime}\right)\mathcal{E}\left[\op P_{i}\left(t\right)\right]\right]\ \mathrm{for\ }t^{\prime}-t\gg\delta t_{Q}.\label{eq:ExactOpenSubsysDynMapAfterTimeCoarseGrain}
\end{equation}
 In other words, by coarse-graining \emph{in time} on the scale $\delta t_{Q}$
over which correlations between $Q$ and its environment $E$ wash
out, we no longer need to coarse grain in the sense \eqref{eq:CoarseGrainedOpenSubsysDynMap}
of explicitly referring to the parent system $W=Q+E$.

As we remarked in the classical case, the characteristic time scale
$\delta t_{Q}$ determines a natural short-time cutoff on the existence
of dynamics for our open subsystem $Q$. Notice also that our interpretation
of quantum theory can fully accommodate the possibility that the parent
system $W$ likewise has a nonzero characteristic time scale $\delta t_{W}\ne0$
that sets the cutoff on the existence of its own dynamics as well;
for all we know, there is no maximal parent system in Nature for which
this temporal cutoff scale exactly vanishes. That is, it may well
be that \emph{all} linear CPT dynamics is ultimately only an approximate
notion, although the same might well be true for density matrices
themselves, as we explained in the context of discussing truncated
Hilbert spaces in Section~\ref{subsub:Imperfect-Tensor-Product-Factorizations-Truncated-Hilbert-Spaces-Approximate-Density-Matrices-and-Unstable-Systems}.

\subsubsection{Dynamical Relationships Between Ontic States Over Time and Objective
Epistemic States Over Time}

Whether the dynamical quantum conditional probabilities $p_{Q}\left(j;t^{\prime}\given i;t\right)$
are exact 
\begin{equation}
\boxed{p_{Q}\left(j;t^{\prime}\given i;t\right)\equiv\Tr_{Q}\left[\op P_{Q}\left(j;t^{\prime}\right)\mathcal{E}_{Q}^{t^{\prime}\from t}\left[\op P_{Q}\left(i;t\right)\right]\right]\sim\Tr\left[\op P_{j}\left(t^{\prime}\right)\mathcal{E}\left[\op P_{i}\left(t\right)\right]\right]}\label{eq:ExactDynQuantCondProbs}
\end{equation}
 or coarse-grained either in the sense \eqref{eq:CoarseGrainedOpenSubsysDynMap}
of making explicit reference to a parent system or in the sense \eqref{eq:ExactOpenSubsysDynMapAfterTimeCoarseGrain}
of existing only over time scales exceeding some nonzero temporal
cutoff $\delta t_{Q}$ as introduced in \eqref{eq:QuantCharacWashoutTimescale},
they manifestly define dynamics for the ontic states of $Q$ in a
manner that parallels the classical case \eqref{eq:ClassicalOnticLevelDynMapMarkov}:%
\footnote{These dynamical quantum conditional probabilities also supply an important
ingredient that is missing from the traditional modal interpretations
and that is identified in ``Problem 3: The problem of effect'' in
\cite{Maudlin:1995tmp}, namely, the lack of a ``detailed dynamics
for the value {[}ontic{]} states.''%
} 
\begin{equation}
p_{Q}\left(\cdot;t^{\prime}\given\cdot;t\right)\colon\ \underbrace{\left(i;t\right)}_{\substack{\mathrm{initial}\\
\mathrm{data}
}
},\ \underbrace{\left(j;t^{\prime}\right)}_{\substack{\mathrm{final}\\
\mathrm{data}
}
}\mapsto\underbrace{p_{Q}\left(j;t^{\prime}\given i;t\right)}_{\substack{\mathrm{conditional}\\
\mathrm{probabilities}
}
}.\label{eq:QuantOnticLevelDynMapMarkov}
\end{equation}
 The dynamical quantum conditional probabilities $p_{Q}\left(j;t^{\prime}\given i;t\right)$
also provide an explicit dictionary that translates between the manifestly
quantum linear CPT dynamics \eqref{eq:SubsysDensMatrExactCPTTimeEv}
of density matrices and the linear dynamical mapping of objective
epistemic states familiar from the equation \eqref{eq:ClassicalLinearEpistemicLevelDynMapMarkov}
for classical epistemic states: 
\begin{equation}
p_{Q}\left(\cdot;t^{\prime}\given\cdot;t\right)\colon\ \underbrace{p_{Q}\left(j;t^{\prime}\right)}_{\substack{\mathrm{epistemic}\\
\mathrm{state\ at\ }t^{\prime}
}
}=\sum_{i}\underbrace{p_{Q}\left(j;t^{\prime}\given i;t\right)}_{\substack{\mathrm{conditional}\\
\mathrm{probability}\\
\mathrm{(independent}\\
\mathrm{of\ epistemic}\\
\mathrm{states})
}
}\underbrace{p_{Q}\left(i;t\right)}_{\substack{\mathrm{epistemic}\\
\mathrm{state\ at\ }t
}
}.\label{eq:QuantumNowLinEpLevelDynMap}
\end{equation}

\subsubsection{Entanglement and Imprecision in Coarse-Grained Quantum Conditional
Probabilities}

When working over time intervals so small that we \emph{cannot} assume
approximate factorization of the density matrix $\op{\rho}_{W}\not\approx\op{\rho}_{Q}\tensorprod\op{\rho}_{E}$
of the parent system $W=Q+E$, we claim that quantum entanglement
\eqref{eq:DefEntangledSubsys}---and not merely classical correlation
\eqref{eq:DefCorrSubsys}---between $Q$ and its environment $E$
inside $W$ determines the size of our imprecision in using the coarse-grained
quantum conditional probabilities $p_{Q\subset W}\left(j;t^{\prime}\given i;t\right)$
defined in \eqref{eq:CoarseGrainedOpenSubsysDynMap}.

As evidence in support of this claim, suppose that the dynamics for
the parent system $W$ over the time interval $t^{\prime}-t$ exactly
decouples into dynamics for subsystem $Q$ alone and dynamics for
the environment $E$ alone, with no interactions between the two subsystems,
so that the linear CPT dynamical mapping \eqref{eq:LinearCPTDynMapParentSys}
factorizes: 
\begin{equation}
\mathcal{E}_{W}^{t^{\prime}\from t}\left[\cdot\right]=\left(\mathcal{E}_{Q}^{t^{\prime}\from t}\tensorprod\mathcal{E}_{E}^{t^{\prime}\from t}\right)\left[\cdot\right].\label{eq:DecoupleLinearCPTDynMap}
\end{equation}
 (This circumstance includes the trivial case $t^{\prime}=t$ in which
$\mathcal{E}_{W}^{t^{\prime}\from t}\left[\cdot\right]$ is just the
identity mapping, as well as the case in which all these dynamical
mappings describe unitary time evolution.) If $Q$ and $E$ are only
classically correlated with each other at the initial time $t$ but
are not entangled in the sense of \eqref{eq:DefEntangledSubsys},
then the density matrix $\op{\rho}_{W}\left(t\right)$ of the parent
system takes the form 
\begin{equation}
\op{\rho}_{W}\left(t\right)=\sum_{i,e}p_{W}\left(\left(i,e\right);t\right)\op P_{Q}\left(i;t\right)\tensorprod\op P_{E}\left(e;t\right),\label{eq:ParentDensMatrCorrNoEntang}
\end{equation}
 and a simple calculation shows that the coarse-grained quantum conditional
probabilities $p_{Q\subset W}\left(j;t^{\prime}\given i;t\right)$
defined in \eqref{eq:CoarseGrainedOpenSubsysDynMap} reduce to the
expected \emph{exact} quantum conditional probabilities $p_{Q}\left(j;t^{\prime}\given i;t\right)$
defined in \eqref{eq:ExactDynQuantCondProbs} for $Q$ alone: 
\begin{equation}
p_{Q\subset W}\left(j;t^{\prime}\given i;t\right)=\Tr_{Q}\left[\op P_{Q}\left(j;t^{\prime}\right)\mathcal{E}_{Q}^{t^{\prime}\from t}\left[\op P_{Q}\left(i;t\right)\right]\right]\equiv p_{Q}\left(j;t^{\prime}\given i;t\right).\label{eq:CoarseGrainedQuantCondProbsReduceToExactForNoEntang}
\end{equation}
 Within the scope of our assumption of decoupled dynamics \eqref{eq:DecoupleLinearCPTDynMap},
deviations from \eqref{eq:CoarseGrainedQuantCondProbsReduceToExactForNoEntang}
can therefore arise only if $Q$ is entangled with its environment
$E$. We interpret this result as implying that for general parent-system
dynamics not necessarily factorizing \eqref{eq:DecoupleLinearCPTDynMap}
into independent dynamics for $Q$ and $E$, we should only ever trust
the validity of coarse-grained conditional probabilities \eqref{eq:CoarseGrainedOpenSubsysDynMap}
up to an intrinsic error---a new kind of ``uncertainty principle''---determined
by the amount of entanglement between $Q$ and $E$.

\subsubsection{Eigenstate Swaps}

Recall that an eigenstate swap \eqref{eq:EigenstateSwap}, as we have
defined it in Section~\ref{subsub:Near-Degeneracies-and-Eigenstate-Swaps},
describes an exchange between two orthogonal density-matrix eigenstates
over a time scale $\delta t_{\mathrm{swap}}$ that is exponentially
small in the total number of degrees of freedom of both the system
itself and of all other systems that substantially interact and entangle
with it, in keeping with \eqref{eq:EigenstateSwapTimeScale}.

It is now a simple matter to explain why actual ontic states avoid
eigenstate swaps: If an eigenstate swap in the system's density matrix
takes place from $t$ to $t+\delta t_{\mathrm{swap}}$, where $\delta t_{\mathrm{swap}}$
is assumed to be larger than the minimal time scale $\delta t_{Q}$
over which the dynamics of our system $Q$ actually exists, then,
in accordance with \eqref{eq:ExactOpenSubsysDynMapAfterTimeCoarseGrain},
there is approximately zero conditional probability $p_{Q}\left(i;t+\delta t_{\mathrm{swap}}\given i;t\right)\sim\absval{\braket{\Psi_{Q,i}\left(t+\delta t_{\mathrm{swap}}\right)}{\Psi_{Q,i}\left(t\right)}}^{2}\approx0$
for the system's ontic state to be $\Psi_{Q,i}\left(t+\delta t_{\mathrm{swap}}\right)\approx\perp\Psi_{Q,i}\left(t\right)$
at the time $t+\delta t_{\mathrm{swap}}$ given that it was $\Psi_{Q,i}\left(t\right)$
at the earlier time $t$.%
\footnote{Observe that the operation $\mathcal{E}_{Q}^{t^{\prime}\from t}\left[\op P_{Q}\left(i;t\right)\right]$
appearing in \eqref{eq:ExactOpenSubsysDynMapAfterTimeCoarseGrain}
evolves ontic states as though they were not part of density matrices,
and so is blind to the eigenstate swap. Note also that over long time
scales $t^{\prime}-t\gg\delta t_{\mathrm{swap}}>\delta t_{Q}$, there
can be appreciable conditional probabilities for the final ontic state
of a system to end up being nearly orthogonal to its initial ontic
state. That is, the present arguments eliminate only \emph{ultra-fast}
swaps to orthogonal ontic states, but allow for transitions that take
place over reasonably long time intervals.%
} Indeed, the probability that the system's actual ontic state will
instead be $\Psi_{Q,j}\left(t+\delta t_{\mathrm{swap}}\right)\approx\Psi_{Q,i}\left(t\right)$
is approximately unity. Notice again the smoothing role played by
our quantum conditional probabilities---they sew together the evolving
ontology of $Q$ in a manner that avoids eigenstate-swap instabilities.

These results may seem surprising given that $\Psi_{Q,i}\left(t\right)$
is \emph{nearly certain} to become $\Psi_{Q,i}\left(t^{\prime}\right)$
over \emph{even shorter} time scales $t^{\prime}-t\ll\delta t_{\mathrm{swap}}$,
but it is important to keep in mind that our objective quantum conditional
probabilities do not naïvely compose: We cannot generically express
$p_{Q}\left(i;t+\delta t_{\mathrm{swap}}\given i;t\right)$, defined
in accordance with \eqref{eq:ExactOpenSubsysDynMapAfterTimeCoarseGrain},
as a sum of products of quantum conditional probabilities of the form
$p_{Q}\left(i_{m};t_{m}\given i_{m-1};t_{m-1}\right)$ over a sequence
of many tiny intermediate time intervals $t_{m}-t_{m-1}\ll\delta t_{\mathrm{swap}}$
going from $t$ to $t+\delta t_{\mathrm{swap}}$.

\subsubsection{Ergodicity Breaking, Classical States, and the Emergence of Statistical
Mechanics}

On short time scales, our dynamical quantum conditional probabilities
\eqref{eq:ExactDynQuantCondProbs} generically allow a macroscopic
system's ontic-level trajectory to explore ergodically a large set
of ontic states differing in the values of only a few of the system's
degrees of freedom, while (super-)exponentially suppressing transitions
to ontic states that differ in the values of large numbers of degrees
of freedom. Hence, \eqref{eq:ExactDynQuantCondProbs} leads emergently
to a partitioning of the macroscopic system's overall Hilbert space
into distinct ergodic components that do not mix over short time
intervals and that we can therefore identify as the system's classical states,
much in the way that a ferromagnet undergoes a phase transition from
a single disordered ergodic component to a collection of distinct
ordered ergodic components as we cool the system below its Curie temperature.
For a treatment of ergodicity breaking and the related emergence of
statistical mechanics from quantum considerations, see \cite{PopescuShortWinter:2005fsmeisa,PopescuShortWinter:2006efsm,LindenPopescuShortWinter:2009,Hollowood:2013eciqm}.%
\footnote{In particular, ergodicity breaking is crucial for explaining the complications
that the authors of \cite{LindenPopescuShortWinter:2009} encounter
with their assumption that an open subsystem's final equilibrium state
should satisfy ``subsystem state independence''---that is, that
the open subsystem's final equilibrium state should be independent
of the subsystem's initial state. This initial-state sensitivity also
plays an important role in obtaining a robust resolution of the measurement
problem for realistic measurement devices, an issue that arose in
discussions with the author of \cite{Hollowood:2013eciqm} in the
context of a similar modal interpretation of quantum theory developed
concurrently with our own modal interpretation; we compare these two
modal interpretations in Section~\ref{sub:Comparison-with-the-Hollowood-Modal-Interpretation}.%
}

\subsubsection{Connections to Other Work}

In a spirit reminiscent of our minimal modal interpretation of quantum
theory, a series of approaches \cite{GisinPercival:1992qsdmaos,Carmichael:1993osaqo}
to studying open quantum systems involve ``unraveling'' the dynamical
equation for a given subsystem's reduced density matrix as an ensemble
average over a collection of stochastically evolving state vectors.
Such techniques have been applied to quantum optics \cite{DalibardCastinMolmer:1992wfadpqo,Carmichael:1993osaqo},
entanglement \cite{CarvalhoBusseBrodierViviescasBuchleitner:2005ue},
decoherence \cite{HalliwellZoupas:1995qsddmddham}, non-Markovian
dynamics \cite{DiosiStrunz:1997nmsseos,DiosiGisinStrunz:1998nmqsd,StrunzDiosiGisin:1999osdnmqt,GambettaAskerudWiseman:2004jlunmoqs},
and geometrical phases \cite{BassiIppoliti:2005gpoqssu}.

Similarly, in \cite{EspositoMukamel:2006ftqme}, Esposito and Mukamel
make use of a notion of ``quantum trajectories'' not unlike the
evolving ontic-state trajectories in our interpretation of quantum
theory. Esposito and Mukamel likewise note that defining these quantum
trajectories for a given system $Q$ requires working with the time-dependent
eigenbasis that instantaneously diagonalizes the system's density
matrix $\op{\rho}_{Q}\left(t\right)$, in contrast to the \emph{fixed}
configuration space of a classical system. Moreover, defining a \emph{differential}
linear CPT dynamical mapping $\mathcal{K}_{Q}^{t}\left[\cdot\right]$
in terms of its finite-time counterpart $\mathcal{E}_{Q}^{t^{\prime}\from t}\left[\cdot\right]$
from \eqref{eq:SubsysDensMatrExactCPTTimeEv} according to 
\begin{equation}
\mathcal{K}_{Q}^{t}\left[\cdot\right]\equiv\frac{\mathcal{E}_{Q}^{t+\delta t_{Q}\from t}\left[\cdot\right]-\mathrm{id}\left[\cdot\right]}{\delta t_{Q}},\label{eq:DefDiffLinCPTDynMap}
\end{equation}
 where $\delta t_{Q}$ is the minimal time scale \eqref{eq:QuantCharacWashoutTimescale}
over which our system's dynamics \eqref{eq:ExactOpenSubsysDynMapAfterTimeCoarseGrain}
exists, we can naturally re-express our dynamical quantum conditional
probabilities \eqref{eq:ExactDynQuantCondProbs}, 
\[
p_{Q}\left(j;t^{\prime}\given i;t\right)\equiv\Tr_{Q}\left[\op P_{Q}\left(j;t^{\prime}\right)\mathcal{E}_{Q}^{t^{\prime}\from t}\left[\op P_{Q}\left(i;t\right)\right]\right],
\]
 in terms of a set of quantum transition rates 
\begin{equation}
W_{Q}\left(\left(j\given i\right);t\right)\equiv\frac{p_{Q}\left(j;t+\delta t_{Q}\given i;t\right)-p_{Q}\left(j;t\given i;t\right)}{\delta t_{Q}}=\Tr_{Q}\left[\op P_{Q}\left(j;t^{\prime}\right)\mathcal{K}_{Q}^{t}\left[\op P_{Q}\left(i;t\right)\right]\right]\label{eq:QuantTransRates}
\end{equation}
 that coincide in the formal limit $\delta t_{Q}\to0$ with Esposito
and Mukamel's own definition of quantum transition rates. Esposito,
Mukamel, and others \cite{KawamotoHatano:2011tftnmoqs,LeggioNapoliBreuerMessina:2013ftqme}
use these quantum trajectories and quantum transition rates to study
quantum definitions of work and heat, entropy production, fluctuation
theorems, and other fundamental questions in statistical mechanics
and thermodynamics.

Our quantum conditional probabilities, in their dynamical manifestation
\eqref{eq:ExactDynQuantCondProbs} for a system $Q$ having well-defined
dynamics, are also closely related to the causal quantum conditional states
of Leifer and Spekkens \cite{LeiferSpekkens:2011fqtcntbi}. To introduce
the Leifer-Spekkens construction and to explain this purported connection
to our minimal modal interpretation of quantum theory, we begin by
considering an arbitrary linear CPT dynamical mapping $\mathcal{E}_{Q}^{t^{\prime}\from t}\left[\cdot\right]$
of the form \eqref{eq:SubsysDensMatrExactCPTTimeEv} for a system
$Q$, 
\[
\op{\rho}_{Q}\left(t\right)\mapsto\op{\rho}_{Q}\left(t^{\prime}\right)=\mathcal{E}_{Q}^{t^{\prime}\from t}\left[\op{\rho}_{Q}\left(t\right)\right].
\]
 Then, formally introducing a tensor-product Hilbert space $\mathcal{H}_{Q\left(t^{\prime}\right)}\tensorprod\mathcal{H}_{Q\left(t\right)}$
consisting of two time-separated copies of the Hilbert space of $Q$,
the Choi-Jamio\l{}kowsi isomorphism \cite{Jamiolkowski:1972ltptpso,Choi:1972plmca,Choi:1975cplmcm},
also known as the channel-state duality, implies the existence of
a unique operator $\op{\varrho}_{Q\left(t^{\prime}\right)\given Q\left(t\right)}$
on $\mathcal{H}_{Q\left(t^{\prime}\right)}\tensorprod\mathcal{H}_{Q\left(t\right)}$
that implements precisely the same time evolution for $Q$ through
a quantum generalization of the classical Bayesian propagation rule
$p_{Q}\left(j^{\prime};t\right)=\sum_{i}p_{Q}\left(j;t^{\prime}\given i;t\right)p_{Q}\left(i;t\right)$,
namely, 
\begin{equation}
\op{\rho}_{Q}\left(t^{\prime}\right)=\Tr_{Q\left(t\right)}\left[\op{\varrho}_{Q\left(t^{\prime}\right)\given Q\left(t\right)}\left(\op 1_{Q\left(t^{\prime}\right)}\tensorprod\op{\rho}_{Q}\left(t\right)\right)\right],\label{eq:JamIsoCausalQuantCondStateMapping}
\end{equation}
 where $\op 1_{Q\left(t^{\prime}\right)}$ is the identity operator
on $\mathcal{H}_{Q\left(t^{\prime}\right)}$. Leifer and Spekkens
regard the operator $\op{\varrho}_{Q\left(t^{\prime}\right)\given Q\left(t\right)}$,
which they call a causal quantum conditional state, as being a noncommutative
generalization of classical conditional probabilities, in much the
same way that density matrices $\op{\rho}_{Q}$ themselves serve as
noncommutative generalizations of classical probabilities $p_{Q}$.

Our dynamical quantum conditional probabilities \eqref{eq:ExactDynQuantCondProbs}
are then precisely the \emph{diagonal} matrix elements of the Leifer-Spekkens
quantum conditional state $\op{\varrho}_{Q\left(t^{\prime}\right)\given Q\left(t\right)}$
in the tensor-product basis $\ket{\Psi_{Q}\left(t^{\prime}\right)}\tensorprod\ket{\Psi_{Q}\left(t\right)}$
for $\mathcal{H}_{Q\left(t^{\prime}\right)}\tensorprod\mathcal{H}_{Q\left(t\right)}$
constructed out of the eigenstates $\ket{\Psi_{Q}\left(t^{\prime}\right)}$
of $\op{\rho}_{Q}\left(t^{\prime}\right)$ and the eigenstates $\ket{\Psi_{Q}\left(t\right)}$
of $\op{\rho}_{Q}\left(t\right)$: 
\begin{align}
p_{Q}\left(j;t^{\prime}\given i;t\right) & =\Tr_{Q\left(t^{\prime}\right)}\left[\op P_{Q}\left(j;t^{\prime}\right)\op{\rho}_{Q}\left(t^{\prime}\right)\right]\nonumber \\
 & =\Tr_{Q\left(t^{\prime}\right)}\left[\op P_{Q}\left(j;t^{\prime}\right)\Tr_{Q\left(t\right)}\left[\op{\varrho}_{Q\left(t^{\prime}\right)\given Q\left(t\right)}\left(\op 1_{Q\left(t^{\prime}\right)}\tensorprod\op P_{Q}\left(i;t\right)\right)\right]\right]\nonumber \\
 & =\left(\bra{\Psi_{Q,j}\left(t^{\prime}\right)}\tensorprod\bra{\Psi_{Q,i}\left(t\right)}\right)\op{\varrho}_{Q\left(t^{\prime}\right)\given Q\left(t\right)}\left(\ket{\Psi_{Q,j}\left(t^{\prime}\right)}\tensorprod\ket{\Psi_{Q,i}\left(t\right)}\right).\label{eq:QuantCondProbsFromMatrElemCausalQuantCondStates}
\end{align}
 This result mirrors the way that our interpretation of quantum theory
identifies the diagonal matrix elements of a system's density matrix
$\op{\rho}_{Q}$ in its own eigenbasis as being the system's epistemic
probabilities $p_{Q}$.

These connections go deeper. For a system $Q$ without well-defined
dynamics of its own but belonging to a parent system $W=Q+E$ with
a linear CPT dynamical mapping $\mathcal{E}_{W}^{t^{\prime}\from t}\left[\cdot\right]$
as in \eqref{eq:LinearCPTDynMapParentSys}, 
\[
\op{\rho}_{W}\left(t\right)\mapsto\op{\rho}_{W}\left(t^{\prime}\right)=\mathcal{E}_{W}^{t^{\prime}\from t}\left[\op{\rho}_{W}\left(t\right)\right],
\]
 we introduced the coarse-grained quantum conditional probabilities
$p_{Q\subset W}\left(j;t^{\prime}\given i;t\right)$ according to
\eqref{eq:CoarseGrainedOpenSubsysDynMap}, 
\[
p_{Q\subset W}\left(j;t^{\prime}\given i;t\right)\equiv\Tr_{W}\left[\left(\op P_{Q}\left(j;t^{\prime}\right)\tensorprod\op 1_{E}\right)\mathcal{E}_{W}^{t^{\prime}\from t}\left[\mathcal{A}_{Q\subset W}^{t}\left[\op P_{Q}\left(i;t\right)\right]\right]\right],
\]
 where we defined the nonlinear assignment mapping $\mathcal{A}_{Q\subset W}^{t}\left[\cdot\right]$
in \eqref{eq:DefNonlinearAssgnMap}, 
\[
\mathcal{A}_{Q\subset W}^{t}\left[\cdot\right]\equiv\op{\rho}_{W}^{1/2}\left(t\right)\left(\op{\rho}_{Q}^{-1/2}\left(t\right)\tensorprod\op 1_{E}\right)\left(\left(\cdot\right)\tensorprod\op 1_{E}\right)\left(\op{\rho}_{Q}^{-1/2}\left(t\right)\tensorprod\op 1_{E}\right)\op{\rho}_{W}^{1/2}\left(t\right).
\]
 It follows from a straightforward computation that we can equivalently
express our coarse-grained quantum conditional probabilities $p_{Q\subset W}\left(j;t^{\prime}\given i;t\right)$
as the diagonal matrix elements of a corresponding coarse-grained quantum conditional state
$\op{\varrho}_{Q\left(t^{\prime}\right)\given Q\left(t\right)}^{Q\subset W}$,
\begin{equation}
p_{Q\subset W}\left(j;t^{\prime}\given i;t\right)=\Tr_{Q\left(t^{\prime}\right)}\left[\op P_{Q\left(t^{\prime}\right),j}\Tr_{Q\left(t\right)}\left[\op{\varrho}_{Q\left(t^{\prime}\right)\given Q\left(t\right)}^{Q\subset W}\left(\op 1_{Q\left(t^{\prime}\right)}\tensorprod\op P_{Q\left(t\right),i}\right)\right]\right].\label{eq:CoarseGrainedQuantCondProbsFromCausalQuantCondState}
\end{equation}
 where $\op{\varrho}_{Q\left(t^{\prime}\right)\given Q\left(t\right)}^{Q\subset W}$
is an operator on the formal tensor-product Hilbert space $\mathcal{H}_{Q\left(t^{\prime}\right)}\tensorprod\mathcal{H}_{Q\left(t\right)}$
that we have defined by starting with the \emph{exact} Leifer-Spekkens
quantum conditional state $\op{\varrho}_{W\left(t^{\prime}\right)\given W\left(t\right)}$
of the parent system $W$, multiplying by the density matrix $\op{\rho}_{W}\left(t\right)$
of $W$ at the initial time $t$ to obtain a natural causal quantum joint state
$\op{\varrho}_{W\left(t^{\prime}\right)+W\left(t\right)}$ for $W$,
partial tracing over the environment $E$ at both the initial time
$t$ and the final time $t^{\prime}$ to obtain a coarse-grained causal
quantum joint state $\op{\varrho}_{Q\left(t^{\prime}\right)+Q\left(t\right)}^{Q\subset W}$
for our original subsystem $Q$, and then ``marginalizing'' on $Q$
at the initial time $t$ by multiplying by $\op{\rho}_{Q}^{-1}\left(t\right)$:
\begin{equation}
\begin{aligned} & \op{\varrho}_{Q\left(t^{\prime}\right)\given Q\left(t\right)}^{Q\subset W}\\
 & =\left(\op 1_{Q\left(t^{\prime}\right)}\tensorprod\op{\rho}_{Q}^{-1/2}\left(t\right)\right)\Tr_{E\left(t^{\prime}\right)}\left[\Tr_{E\left(t\right)}\left[\left(\op 1_{W\left(t^{\prime}\right)}\tensorprod\op{\rho}_{W}^{1/2}\left(t\right)\right)\op{\varrho}_{W\left(t^{\prime}\right)\given W\left(t\right)}\left(\op 1_{W\left(t^{\prime}\right)}\tensorprod\op{\rho}_{W}^{1/2}\left(t\right)\right)\right]\right]\left(\op 1_{Q\left(t^{\prime}\right)}\tensorprod\op{\rho}_{Q}^{-1/2}\left(t\right)\right).
\end{aligned}
\label{eq:CoarseGrainedCausalQuantCondStateSubsys}
\end{equation}
 As expected, this coarse-grained quantum conditional state exactly
satisfies a propagation rule of the form \eqref{eq:JamIsoCausalQuantCondStateMapping}:
\begin{equation}
\op{\rho}_{Q}\left(t^{\prime}\right)=\Tr_{Q\left(t\right)}\left[\op{\varrho}_{Q\left(t^{\prime}\right)\given Q\left(t\right)}^{Q\subset W}\left(\op 1_{Q\left(t^{\prime}\right)}\tensorprod\op{\rho}_{Q}\left(t\right)\right)\right].\label{eq:CoarseGrainedCausalQuantCondStateMapping}
\end{equation}

\subsection{Subsystem Spaces\label{sub:Subsystem-Spaces}}

An interesting geometrical structure emerges if we consider the continuously
infinite set of different ways that we can define a particular subsystem
$Q$ with its own Hilbert space $\mathcal{H}_{Q}$ by tensor-product-factorizing
the Hilbert space $\mathcal{H}_{W}=\mathcal{H}_{Q}\tensorprod\mathcal{H}_{E}$
of a given parent system $W=Q+E$ that includes an environment $E$.
This geometrical structure, which we call the subsystem space for
$Q$ and which encompasses all the various possible versions of $\mathcal{H}_{Q}$,
exists even for a parent system $W$ with as few as four mutually
exclusive states ($\dim\mathcal{H}_{W}=4$) and has no obvious counterpart
in classical physics.

\subsubsection{Formal Construction}

We parameterize the smooth family of choices of bipartite tensor-product
factorization of the parent system's Hilbert space $\mathcal{H}_{W}$
using an $n$-tuple of complex numbers $\alpha=\left(\alpha_{1},\dotsc,\alpha_{n}\right)$
for some integer $n$, writing $\mathcal{H}_{W}=\mathcal{H}_{Q\left(\alpha\right)}\tensorprod\mathcal{H}_{E\left(\alpha\right)}$.
We then obtain a complex vector bundle---namely, the subsystem space
for our subsystem $Q$ of interest---that consists of an identical
Hilbert space $\mathcal{H}_{Q\left(\alpha\right)}\isomorphic\mathcal{H}_{Q}$
of some fixed dimension $\dim\mathcal{H}_{Q}$ attached to each point
with coordinates $\alpha$ on a base manifold of complex dimension
$n$. Each \emph{specific} choice of tensor-product factorization
$\mathcal{H}_{W}=\mathcal{H}_{Q\left(\alpha\right)}\tensorprod\mathcal{H}_{E\left(\alpha\right)}$
and partial-trace down to $\mathcal{H}_{Q\left(\alpha\right)}\isomorphic\mathcal{H}_{Q}$
then corresponds to just \emph{one version} of our subsystem $Q$
of interest, or, equivalently, to one ``slice'' of this complex
vector bundle---that is, to one slice of the subsystem space for our
subsystem of interest.%
\footnote{After this work was substantially complete, we noticed that a similar
idea appears in \cite{IizukaKabat:2013omihw}, where the authors parameterize
their subsystem space using a coordinate label $\theta$ and employ
it to study $\sim\exp\left(-S\right)$ breakdowns in locality in the
presence of a black hole of entropy $S$.%
}

The parent system's density matrix $\op{\rho}_{W}$ on $\mathcal{H}_{W}=\mathcal{H}_{Q\left(\alpha\right)}\tensorprod\mathcal{H}_{E\left(\alpha\right)}$
then defines a natural Hermitian inner product between any pair of
vectors $\ket{\psi_{Q\left(\alpha\right)}}\in\mathcal{H}_{Q\left(\alpha\right)}$
and $\ket{\chi_{Q\left(\alpha^{\prime}\right)}}\in\mathcal{H}_{\left(\alpha^{\prime}\right)}$
living in two slices respectively located at $\alpha$ and $\alpha^{\prime}$:
\begin{equation}
\eqsbrace{\begin{aligned}h\left(\psi_{Q\left(\alpha\right)},\chi_{Q\left(\alpha^{\prime}\right)}\right) & \equiv\Tr_{W}\left[\op{\rho}_{W}\left(\ket{\psi_{Q\left(\alpha\right)}}\bra{\psi_{Q\left(\alpha\right)}}\tensorprod\op 1_{E\left(\alpha\right)}\right)\left(\ket{\chi_{Q\left(\alpha^{\prime}\right)}}\bra{\chi_{Q\left(\alpha^{\prime}\right)}}\tensorprod\op 1_{E\left(\alpha^{\prime}\right)}\right)\right]\\
 & =h\left(\chi_{Q\left(\alpha^{\prime}\right)},\psi_{Q\left(\alpha\right)}\right)^{\conj}.
\end{aligned}
}\label{eq:DefSubsysSpaceHermMetric}
\end{equation}
  Here $\op 1_{E\left(\alpha\right)}$ and $\op 1_{E\left(\alpha^{\prime}\right)}$
are respectively the identity operators on the tensor-product factors
$\mathcal{H}_{E\left(\alpha\right)}$ and $\mathcal{H}_{E\left(\alpha^{\prime}\right)}$.

\subsubsection{Actual Ontic Layer of Reality}

The smooth family of reduced density matrices $\op{\rho}_{Q\left(\alpha\right)}$
that we can obtain by partial-tracing $\op{\rho}_{W}$ for different
choices of $\alpha$ defines a set of $N=\dim\mathcal{H}_{Q}$ smooth
sections $\ket{\Psi_{Q\left(\alpha\right),i}}$ ($i=1,\dotsc,N=\dim\mathcal{H}_{Q}$)
on this complex vector bundle. The notion of the subsystem's actual
ontic state therefore extends to a whole ``actual ontic layer of
reality'' as we move through the different values of $\alpha$, but
we \emph{cannot} generically identify this actual ontic layer of reality
as a whole with just a \emph{single} section $\ket{\Psi_{Q\left(\alpha\right),i}}$
for fixed $i$, for the simple reason that the epistemic probabilities
$p_{Q\left(\alpha\right),i}$ across this section may vary considerably
from one value of $\alpha$ to another. However, our kinematical quantum
conditional probabilities \eqref{eq:QuantCondProbKinParentSub} ensure
that if the actual ontic state of $W$ is of the non-entangled form
$\left(\Psi_{Q\left(\alpha\right),i},\Psi_{E\left(\alpha\right),e}\right)$
for a particular choice of $\alpha$, then all versions of our subsystem
$Q$ near this value of $\alpha$ will be nearly certain to be in
ontic states very similar to $\Psi_{Q\left(\alpha\right),i}$ in the
Hilbert-space sense.

\section{The Measurement Process\label{sec:The-Measurement-Process}}

\subsection{Measurements and Decoherence\label{sub:Measurements-and-Decoherence}}

It is not our aim in this paper to investigate the measurement process
and decoherence in any great depth, or to study realistic examples.
For pedagogical treatments, see \cite{Bohm:1951qt,BreuerPetruccione:2002toqs,Joos:2003dacwqt}.
For additional detailed discussions, including a careful study of
the extremely rapid rate of decoherence for typical systems in contact
with macroscopic measurement devices or environments, see, for example,
\cite{Zurek:1981pbqaiwmdwpc,JoosZeh:1985ecptiwe,Zurek:2003dtqc,Schlosshauer:2005dmpiqm,SchlosshauerCamilleri:2008qct}.
For a discussion of how realistic measuring devices with finite spatial
and temporal resolution solve various delocalization \cite{Page:2011qusbgbd}
and instability \cite{AlbertLoewer:1990wdatassp,AlbertLoewer:1991mpss,AlbertLoewer:1993nim}
issues, and, in particular, a treatment of the important role played
both by environmental interactions and by ergodicity breaking in
making sense of realistic measurement processes, see, for example,
\cite{BacciagaluppiHemmo:1996midm,Vermaas:1999puqm,Hollowood:2013cbr,Hollowood:2013eciqm}.%
\footnote{We discuss the modal interpretation of \cite{Hollowood:2013cbr,Hollowood:2013eciqm},
which developed concurrently with our own interpretation of quantum
theory, in Section~\ref{sub:Comparison-with-the-Hollowood-Modal-Interpretation}.%
}

\subsubsection{Von Neumann Measurements\label{subsub:Von-Neumann-Measurements}}

For the purposes of establishing how our minimal modal interpretation
makes sense of measurements, how decoherence turns the environment
into a ``many-dimensional chisel'' that rapidly sculpts the ontic
states of systems into their precise shapes,%
\footnote{The way that decoherence sculpts ontic states into shape is reminiscent
of the way that the external pressure of air molecules above a basin
of water maintains the water in its liquid phase.%
} and how the Born rule naturally emerges to an excellent approximation,
we consider the idealized example of a so-called Von Neumann measurement.
Along the way, we will address the status of both the measurement
problem generally and the notion of wave-function collapse specifically
in the context of our interpretation of quantum theory. Ultimately,
we'll find that our interpretation solves the measurement problem
by replacing instantaneous axiomatic wave-function collapse with an
interpolating ontic-level dynamics, and thereby eliminates any need
for an \emph{ad hoc} Heisenberg cut. 

\begin{figure}
\begin{centering}
\begin{center}
\definecolor{ce1e1e1}{RGB}{225,225,225}
\definecolor{cffffff}{RGB}{255,255,255}

\begin{tikzpicture}[y=0.80pt, x=0.8pt,yscale=-1, inner sep=0pt, outer sep=0pt]
\begin{scope}[shift={(-43.03757,-49.82106)}]
  \path[draw=black,fill=ce1e1e1,miter limit=4.00,line width=0.587pt,rounded
    corners=0.0000cm] (43.4047,50.1881) rectangle (121.0482,118.9161);
  \path[draw=black,fill=cffffff,miter limit=4.00,line width=0.800pt,rounded
    corners=0.0000cm] (49.4975,55.3213) rectangle (116.1675,112.9000);
  \path[fill=black] (77.615807,82.483757) node[above right] (text5199) {$Q$};
  \path[draw=black,fill=ce1e1e1,miter limit=4.00,line width=0.651pt,rounded
    corners=0.0000cm] (221.7193,52.2182) rectangle (296.5049,119.0741);
  \path[draw=black,fill=ce1e1e1,line join=miter,line cap=butt,line width=0.800pt]
    (221.8021,77.5650) -- (199.5889,52.3112) -- (199.5889,118.9812) --
    (221.8123,96.7579) -- cycle;
  \path[draw=black,fill=cffffff,miter limit=4.00,line width=0.800pt,rounded
    corners=0.0000cm] (226.0559,56.8569) rectangle (292.7260,114.4355);
  \path[fill=black] (73.705872,97.846024) node[above right] (text5199-8) {$\left|\Psi\right\rangle$};
  \path[fill=black] (253.5278,82.316536) node[above right] (text5199-5) {$A$};
  \path[fill=black] (235.61786,100.678802) node[above right] (text5199-8-2)
    {$\left|A\left(``\emptyset"\right)\right\rangle$};
  \path[shift={(37.03757,39.82106)},draw=black,miter limit=4.00,line
    width=0.800pt] (354.2605,36.2494)arc(0.000:180.000:211.955262 and
    75.484)arc(-180.000:0.000:211.955262 and 75.484) -- cycle;
  \path[fill=black] (170.96431,20.854202) node[above right] (text5199-5-5) {$E$};
  \path[fill=black] (152.05437,39.216469) node[above right] (text5199-8-2-6)
    {$\left|E\left(``\emptyset"\right)\right\rangle$};
\end{scope}

\end{tikzpicture}
\par\end{center}
\par\end{centering}

\caption{\label{fig:VonNeumannMeasurement}The schematic set-up for a Von Neumann
experiment, consisting of a subject system $Q$ together with a measurement
apparatus $A$ and a larger environment $E$.}
\end{figure}
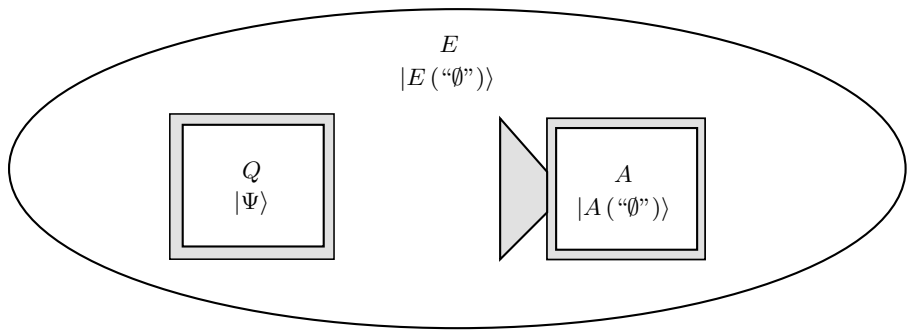

A Von Neumann measurement involves a subject system $Q$ with a complete
basis of mutually exclusive ontic states $Q\left(i\right)$ (with
$\braket{Q\left(i\right)}{Q\left(j\right)}=\delta_{ij}$), a macroscopic
measurement apparatus $A$ in an initial actual ontic state $A\left(\quote{\emptyset}\right)$
with its display dial set to ``empty,'' and an even more macroscopic
environment $E$ in an initial actual ontic state $E\left(\quote{\emptyset}\right)$
in which it sees the dial on the apparatus $A$ as being empty. (See
Figure~\ref{fig:VonNeumannMeasurement}.) We suppose furthermore
that in the special case in which the actual ontic state of the subject
system $Q$ is precisely one of the definite ontic states $Q\left(i\right)$,
the resulting time evolution is unitary (although linear CPT evolution
would be more realistic) and proceeds according to the following idealized
two-step sequence: 
\begin{enumerate}
\item The apparatus $A$ transitions to a new actual ontic state $A\left(\quote i\right)$
in which its dial now displays the value ``the subject system $Q$
appears to be in the state $Q\left(i\right)$,'' and then 
\item the environment $E$ subsequently changes to a new actual ontic state
$E\left(\quote i\right)$ in which its own configuration has been
noticeably perturbed by the change in the apparatus $A$, such as
through unavoidable and irreversible thermal radiation \cite{HackermullerHornbergerBrezgerZeilingerArndt:2004dmwter}.
\end{enumerate}
Written directly in terms of the appropriate state vectors, this two-step
time evolution takes the form 
\begin{equation}
\eqsbrace{\begin{aligned}\ket{Q\left(i\right)}\ket{A\left(\quote{\emptyset}\right)}\ket{E\left(\quote{\emptyset}\right)} & \mapsto\ket{Q\left(i\right)}\ket{A\left(\quote i\right)}\ket{E\left(\quote{\emptyset}\right)}\\
\\
 & \mapsto\ket{Q\left(i\right)}\ket{A\left(\quote i\right)}\ket{E\left(\quote i\right)}.
\end{aligned}
}\label{eq:VonNeumannIndivMeasProc}
\end{equation}

The assumption that the temporal sequence \eqref{eq:VonNeumannIndivMeasProc}
represents \emph{unitary} time evolution presents an obstruction to
assuming that the different final ontic state vectors $\ket{A\left(\quote i\right)}$
of the apparatus are \emph{exactly} mutually orthogonal for different
values of $i$, and similarly for the different final ontic state
vectors $\ket{E\left(\quote i\right)}$ of the environment $E$. However,
our assumption that the apparatus and the environment are macroscopic
systems with large respective Hilbert spaces $\mathcal{H}_{A}$ and
$\mathcal{H}_{E}$ ensures that inner products of the form $\braket{A\left(\quote i\right)}{A\left(\quote j\right)}$
and $\braket{E\left(\quote i\right)}{E\left(\quote j\right)}$ for
$i\ne j$ are both exponentially small in their respective numbers
$\gg10^{23}$ of degrees of freedom: 
\begin{align}
\braket{A\left(\quote i\right)}{A\left(\quote j\right)} & \sim\exp\left(-\left(\#\mathrm{apparatus\ degrees\ of\ freedom}\right)\times\Delta t/\tau_{A}\right),\label{eq:VonNeumannExpSmallAppOverlaps}\\
\nonumber \\
\braket{E\left(\quote i\right)}{E\left(\quote j\right)} & \sim\exp\left(-\left(\#\mathrm{environment\ degrees\ of\ freedom}\right)\times\Delta t/\tau_{E}\right).\label{eq:VonNeumannExpSmallEnvOverlaps}
\end{align}
 Here $\Delta t$ is the time duration of the measurement process
\eqref{eq:VonNeumannIndivMeasProc}, $\tau_{A}$ is the characteristic
time scale for the evolution of each individual degree of freedom
of the apparatus $A$, and $\tau_{E}$ is similarly the characteristic
time scale for the evolution of each individual degree of freedom
of the environment $E$.

If we now instead arrange for the initial actual ontic state $\Psi$
of the subject system $Q$ to be a general superposition of the complete
orthonormal basis of ontic states $Q\left(i\right)$, 
\begin{equation}
\ket{\Psi}=\sum_{i}\alpha_{i}\ket{Q\left(i\right)}=\alpha_{1}\ket{Q\left(1\right)}+\alpha_{2}\ket{Q\left(2\right)}+\dotsb,\qquad\braket{\Psi}{\Psi}=1,\label{eq:VonNeumannPrepSuperposState}
\end{equation}
 then the linearity of the time evolution \eqref{eq:VonNeumannIndivMeasProc}
implies the temporal sequence 
\begin{equation}
\eqsbrace{\begin{aligned}\ket{\Psi}\ket{A\left(\quote{\emptyset}\right)}\ket{E\left(\quote{\emptyset}\right)}=\left(\sum_{i}\alpha_{i}\ket{Q\left(i\right)}\right)\ket{A\left(\quote{\emptyset}\right)}\ket{E\left(\quote{\emptyset}\right)} & \mapsto\left(\sum_{i}\alpha_{i}\ket{Q\left(i\right)}\ket{A\left(\quote i\right)}\right)\ket{E\left(\quote{\emptyset}\right)}\\
\\
 & \mapsto\sum_{i}\alpha_{i}\ket{Q\left(i\right)}\ket{A\left(\quote i\right)}\ket{E\left(\quote i\right)}.
\end{aligned}
\negthickspace\negthickspace\negthickspace\negthickspace\negthickspace\negthickspace\negthickspace\negthickspace\negthickspace}\label{eq:VonNeumannSuperposMeasProc}
\end{equation}
  According to our interpretation of quantum theory, the final density
matrices of the subject system $Q$ and of the measurement apparatus
$A$, obtained by appropriately partial-tracing \eqref{eq:DefPartialTraces}
the composite final state appearing in \eqref{eq:VonNeumannSuperposMeasProc},
are respectively given in their diagonal form by 
\begin{align}
\op{\rho}_{Q} & =\sum_{i}\absval{\alpha_{i}^{\prime}}^{2}\ket{Q\left(i\right)^{\prime}}\bra{Q\left(i\right)^{\prime}},\label{eq:VonNeumannFinalSubjDensMatrix}\\
\nonumber \\
\op{\rho}_{A} & =\sum_{i}\absval{\alpha_{i}^{\prime\prime}}^{2}\ket{A\left(\quote i\right)^{\prime}}\bra{A\left(\quote i\right)^{\prime}}+\left(\mathrm{small}\ \perp\ \mathrm{terms}\right),\label{eq:VonNeumannFinalAppDensMatrix}
\end{align}
 where the primes represent the \emph{very} tiny changes in basis
needed to eliminate exponentially small off-diagonal corrections of
order the product of \eqref{eq:VonNeumannExpSmallAppOverlaps} with
\eqref{eq:VonNeumannExpSmallEnvOverlaps}, 
\begin{equation}
\mathrm{off\hyphen diagonal\ corrections}\sim\exp\left(-\left(\substack{{\displaystyle \#\mathrm{\mathrm{apparatus}+environment}}\\
{\displaystyle \mathrm{degrees\ of\ freedom}}
}
\right)\times\Delta t/\tau\right),\label{eq:VonNeumannEnvOffDiagCorrections}
\end{equation}
 where $\tau\sim\min\left(\tau_{A},\tau_{E}\right)$, and where the
\emph{double} primes appearing on the eigenvalues of $\op{\rho}_{A}$
account for the possibility that they might not precisely agree with
the eigenvalues of $\op{\rho}_{Q}$, again due to discrepancies of
order \eqref{eq:VonNeumannEnvOffDiagCorrections}. No realistic ontic
basis is infinitely sharp anyway in the context of our interpretation
of quantum theory, and certainly no sharper than these sorts of corrections.

The exponential speed with which the corrections \eqref{eq:VonNeumannEnvOffDiagCorrections}
become negligibly small---a phenomenon called decoherence of the
originally coherent superposition \eqref{eq:VonNeumannPrepSuperposState}---is
perfectly in keeping with the notion that a large apparatus conducting
a measurement while sitting in an even larger environment leads to
the appearance of the ``instantaneous'' wave-function collapse
of the traditional Copenhagen interpretation, and indeed \emph{justifies}
the use of wave-function collapse as a heuristic shortcut in elementary
pedagogical treatments of the measurement process in quantum theory.%
\footnote{Typical decoherence time scales for many familiar systems are presented
in \cite{JoosZeh:1985ecptiwe,Zurek:2003dtqc,Schlosshauer:2005dmpiqm,SchlosshauerCamilleri:2008qct}.%
}

In keeping with the basic correspondence \eqref{eq:EpStateDensMatrixCorresp}
between objective epistemic states and density matrices at the heart
of our interpretation of quantum theory, we can then conclude that
the final actual ontic state of the subject system $Q$ is one of
the possibilities $Q\left(i\right)^{\prime}\approx Q\left(i\right)$
with epistemic probability given by the corresponding density-matrix
eigenvalue $p_{Q,i}=\absval{\alpha_{i}^{\prime}}^{2}\approx\absval{\braket{Q\left(i\right)}{\Psi}}^{2}$,
up to exponentially small corrections \eqref{eq:VonNeumannEnvOffDiagCorrections}.
Similarly, the final actual ontic state of the measurement apparatus
$A$ is one of the possibilities $A\left(\quote i\right)^{\prime}\approx A\left(\quote i\right)$
with approximately the \emph{same} epistemic probability $p_{A,\quote i}=\absval{\alpha_{i}^{\prime\prime}}^{2}\approx\absval{\braket{Q\left(i\right)}{\Psi}}^{2}$,
again up to exponentially small corrections \eqref{eq:VonNeumannEnvOffDiagCorrections}.
We therefore see that our interpretation of quantum theory solves
the measurement problem by replacing axiomatic wave-function collapse
with an underlying interpolating evolution for ontic states.

It is widely accepted that measurement-induced decoherence destroys
quantum interference between distinct measurement outcomes and produces
final objective density matrices for both subject systems and apparatuses
that closely resemble the subjective mixtures that result from a so-called
Von Neumann-Lüders projection \cite{vonNeumann:1955mfqm,Luders:1950udzddm}
without post-selection. (See Appendix~\ref{subsub:Wave-Function-Collapse-and-the-Measurement-Problem}.)
Because the latter operation constitutes a linear CPT dynamical mapping,
we see immediately that well-executed measurements by macroscopic
apparatuses can be described to great accuracy by linear CPT dynamical
mappings as well.  Hence, to the extent that the measurement process
experienced by the subject system $Q$ can be captured to an acceptable
level of approximation by a linear CPT dynamical mapping $\mathcal{E}_{Q}^{t+\Delta t\from t}\left[\cdot\right]$
for $Q$ over the relevant time scale $\Delta t$ of the measurement,
we directly find that the stochastic dynamical conditional probability
\eqref{eq:ExactDynQuantCondProbs} for $Q$ to end up in the final
ontic state $Q\left(i\right)^{\prime}\approx Q\left(i\right)$ at
time $t+\Delta t$ given that it was initially in the ontic state
$\Psi$ at time $t$ is $p_{Q}\left(i;t+\Delta t\given\Psi;t\right)\approx\absval{\braket{Q\left(i\right)}{\Psi}}^{2}$,
as expected. Similarly, if we can approximate the evolution of the
apparatus $A$ as a linear CPT dynamical mapping, then $p_{A}\left(\quote i;t+\Delta t\given\quote{\emptyset};t\right)\approx\absval{\braket{Q\left(i\right)}{\Psi}}^{2}$.

By any of these approaches, the famous Born rule 
\begin{equation}
\Probability{\dotsc}=\absval{\dotsc}^{2}\label{eq:BornRule}
\end{equation}
 for computing empirical outcome probabilities therefore emerges \emph{automatically}
to an excellent approximation without having to be assumed \emph{a priori}.

Notice that in line with the central correspondence \eqref{eq:EpStateDensMatrixCorresp}
between objective epistemic states and density matrices, and the boxed
statement \eqref{eq:TautStatementReEpProbs} below it, the quantities
$\absval{\alpha_{i}}^{2}$ do not have an interpretation as \emph{literal}
epistemic probabilities for the underlying actual ontic states until
they (or some approximate versions of them) have become \emph{actualized}
as eigenvalues of density matrices---that is, not until after the
measurement is actually complete. Again, the \emph{only} quantities
that our minimal modal interpretation of quantum theory regards as
being literal epistemic probabilities are those that our interpretation's
axioms explicitly identify as such.

We can also go beyond the Born rule for the \emph{individual} subsystems
and, importantly, say something \emph{jointly} about their actual
ontic states. The post-measurement reduced density matrix $\op{\rho}_{Q+A}$
of the composite system $Q+A$ consisting of the subject system together
with the apparatus, computed according to the standard partial-trace
operation \eqref{eq:DefPartialTraces}, is not generically diagonal
in exactly the ``correct'' tensor-product basis $\ket{Q\left(j\right)^{\prime},A\left(\quote i\right)^{\prime}}\equiv\ket{Q\left(j\right)^{\prime}}\tensorprod\ket{A\left(\quote i\right)^{\prime}}$
consistent with our original definitions of the subject system $Q$
and the apparatus $A$ as subsystems. Nonetheless, the diagonalizing
basis for $\op{\rho}_{Q+A}$ is \emph{exponentially close} to this
tensor-product basis, 
\begin{equation}
\op{\rho}_{Q+A}=\sum_{i,j}\alpha_{i}\alpha_{j}^{\conj}\underbrace{\left(\braket{E\left(\quote i\right)}{E\left(\quote j\right)}\right)}_{\delta_{ij}+\left(\substack{\mathrm{exponentially}\\
\mathrm{small}\\
\mathrm{corrections}
}
\right)}\ket{Q\left(i\right)}\ket{A\left(\quote i\right)}\bra{Q\left(j\right)}\bra{A\left(\quote j\right)},\label{eq:VonNeumannNearTensorProdSubjAppDensMatr}
\end{equation}
 thereby implying that the actual ontic state for $Q+A$ is represented
by a state vector \emph{exponentially close} to a basis vector of
the form $\ket{Q\left(j\right)^{\prime},A\left(\quote i\right)^{\prime}}$
for $j=i$. Hence, given that $Q+A$ is in one of these actual ontic
states, the kinematical parent-subsystem manifestation \eqref{eq:QuantCondProbKinParentSub}
of our quantum conditional probabilities smooths out the discrepancies
and ensures with near-certainty that the final actual ontic states
of the subsystems $Q$ and $A$ have correctly correlated values of
the label $i$. We can therefore safely conclude that if the final
actual ontic state of the apparatus is $A\left(\quote i\right)^{\prime}\approx A\left(\quote i\right)$,
then the final actual ontic state of the subject system must with
near-certainty be $Q\left(i\right)^{\prime}\approx Q\left(i\right)$
for the \emph{same} value of $i$, and vice versa.

Looking back at \eqref{eq:NearDegenOffDiagParamFromError}, we also
see that the off-diagonal corrections \eqref{eq:VonNeumannEnvOffDiagCorrections}
to the final density matrix imply that the Born rule \eqref{eq:BornRule}
is itself only ever accurate up to corrections $\sim\exp\left(-S\times\Delta t/\tau\right)$,
where $S=S_{A+E}$ is the combined entropy of the measurement apparatus
$A$ and environment $E$. As a consequence, all statistical quantities
derived from the Born rule---including all expectation values, final-outcome
state vectors, semiclassical observables, scattering cross sections,
tunneling probabilities, and decay rates computed in quantum-mechanical
theories and quantum field theories---likewise suffer from imprecisions
$\sim\exp\left(-S\times\Delta t/\tau\right)$.

These results are nicely in keeping with the notion of our error-entropy
bound \eqref{eq:MinimumMeasurementErrorEntropyBound}, but are completely
invisible in the traditional Copenhagen interpretation of quantum
theory: In contrast to our own interpretation, the Copenhagen interpretation
\emph{derives} the partial-trace prescription \eqref{eq:DefPartialTraces}
from the \emph{a priori} assumption of the \emph{exact} Born rule
and features a Von Neumann-Lüders projection postulate \cite{vonNeumann:1955mfqm,Luders:1950udzddm}
that \emph{axiomatically} sets the off-diagonal entries of the final
density matrix precisely to zero and converts it into a proper mixture,
as we explain in detail in Appendix~\ref{sub:The-Copenhagen-Interpretation-and-the-Measurement-Problem}.

On the other hand, for ordinary macroscopic measurement devices, our
claimed corrections to the Born rule rapidly become far too small
to notice, so our minimal modal interpretation of quantum theory is
consistent with the predictions of the Copenhagen interpretation in
most practical circumstances. Note that exponentially small corrections
of this kind should, in principle, also generically arise in any other
interpretations of quantum theory (such as the Everett-DeWitt many-worlds
interpretation) that attempt to derive the Born rule from decoherence.%
\footnote{\label{fn:BekensteinBound}These corrections to the Born rule have
interesting consequences for gravitational physics: If every region
of physical space has a bounded maximum entropy---given, say, in terms
of the radius $R$ and energy $E$ of the region by the Bekenstein bound
$S_{\mathrm{max}}=2\pi k_{\mathrm{B}}ER/\hbar c$ \cite{Bekenstein:1981uubeerbs}---then
observable notions of locality in such a region can never be more
precise than $\sim\exp\left(-S\times\Delta t/\tau\right)$ for $S=S_{\mathrm{max}}$.
See also footnote~\ref{fn:BlackHoleInfoParadox}. %
}

Finally, it is easy to show within the scope of our analysis that
if the dial on our apparatus $A$ could display \emph{sequences} of
outcomes arising from \emph{repeated} measurements conducted in rapid
succession, then the final density matrix of the apparatus would have
non-negligible probability eigenvalues \emph{only} for possible ontic
states of the form $A\left(\quote 1\binaryand\quote 1\binaryand\quote 1\dotsc\right)$
and $A\left(\quote 2\binaryand\quote 2\binaryand\quote 2\dotsc\right)$,
but not for, say, $A\left(\quote 1\binaryand\quote 2\binaryand\quote 1\dotsc\right)$.
Thus, realistic measurements, at least of the simplified Von Neumann
type, produce robust, persistent, repeatable outcomes, provided that
we are working over sufficiently short time scales that that uncontrollable
overall dynamics does not have time to alter our subject system $Q$
appreciably. This observation provides yet another reason why our
own interpretation of quantum theory, in contrast to the Copenhagen
interpretation, does not need to assume wave-function collapse as
a distinct axiomatic postulate.

\subsection{Subjective Density Matrices and Proper Mixtures\label{sub:Subjective-Density-Matrices-and-Proper-Mixtures}}

Having derived the Born rule \eqref{eq:BornRule} at last, we can
now safely introduce the conventional practice of employing density
matrices to describe \emph{subjective} epistemic states for quantum
systems: Starting from the Born rule and given a subjective probability
distribution $p_{\alpha}$ over objective epistemic states described
by a collection of density matrices $\op{\rho}_{\alpha}$ for a given
system, where we allow that $p_{\alpha}$ may be only a \emph{formal}
probability distribution in the spirit of Section~\ref{subsub:Formal-Epistemic-States}
and thus do not insist on mutual exclusivity, standard textbook arguments
show that we can compute empirical expectation values in terms of
the subjective density matrix 
\begin{equation}
\op{\rho}=\sum_{\alpha}p_{\alpha}\op{\rho}_{\alpha}.\label{eq:SubjDensMatr}
\end{equation}
 Note, of course, that the mapping from formal subjective epistemic
states to subjective density matrices is many-to-one because the mapping
does not invariantly encode the original choice of objective density
matrices $\op{\rho}_{\alpha}$---the same subjective density matrix
$\op{\rho}$ can be expressed in terms of infinitely many different
choices of coefficients $p_{\alpha}$ and their corresponding objective
density matrices $\op{\rho}_{\alpha}$.

In the special case of a proper mixture, meaning an epistemic state
that is \emph{wholly} subjective in nature, each objective density
matrix $\op{\rho}_{\alpha}$ describes a single pure state $\Psi_{\alpha}$
of the given system and thus the subjective density matrix \eqref{eq:SubjDensMatr}
of the system takes the simpler form 
\begin{equation}
\op{\rho}=\sum_{\alpha}p_{\alpha}\ket{\Psi_{\alpha}}\bra{\Psi_{\alpha}}.\label{eq:SubjDensMatrProperMix}
\end{equation}
 Note, however, that the Von Neumann entropy formula $S=-\Tr\left[\op{\rho}\log\op{\rho}\right]$
from \eqref{eq:DefVonNeumannEntropy} does not generically yield $-\sum_{\alpha}p_{\alpha}\log p_{\alpha}$
for the proper mixture if $p_{\alpha}$ is merely a \emph{formal}
probability distribution over non-exclusive possible ontic states;
the Von Neumann entropy formula gives $-\sum_{\alpha}p_{\alpha}\log p_{\alpha}$
only if $p_{\alpha}$ is a \emph{logically rigorous} probability distribution
involving \emph{mutually exclusive} possibilities $\Psi_{\alpha}$
in the sense that the associated state vectors $\ket{\Psi_{\alpha}}$
are mutually orthogonal---see \eqref{eq:MutualExclOrthog}---in which
case the quantities $p_{\alpha}$ are the eigenvalues of $\op{\rho}$.

It is important to note that a proper mixture \eqref{eq:SubjDensMatrProperMix}
fundamentally describes classical uncertainty over the state vector
$\ket{\Psi_{\alpha}}$ of a \emph{single} system. For example, if
a source generates a sequence of $N\gg1$ physical copies of an elementary
subsystem, with classical frequency ratios $p_{\alpha}\in\left[0,1\right]$
for the emitted elementary subsystems to be described by definite
corresponding state vectors $\ket{\Psi_{\alpha}}$ in the elementary-subsystem
Hilbert space $\mathcal{H}$, then we can employ a subjective density
matrix of the form \eqref{eq:SubjDensMatrProperMix} to provide an
approximate description of the resulting proper mixture for a \emph{single}
randomly chosen copy of the elementary subsystem. However, in the
idealized limit in which we can neglect any entanglements with the
larger environment, the state of the parent system consisting of the
\emph{entire physical sequence} of copies is not described by a nontrivial
density matrix at all, but instead by a tensor-product state vector
of the form 
\begin{equation}
\ket{\Psi}=\ket{\Psi_{\alpha_{1}}}\tensorprod\ket{\Psi_{\alpha_{2}}}\tensorprod\dotsm\tensorprod\ket{\Psi_{\alpha_{N}}}\label{eq:LongSeqStateVec}
\end{equation}
 belonging to the tensor-product Hilbert space $\mathcal{H}_{\mathrm{sequence}}=\mathcal{H}\tensorprod\dotsm\tensorprod\mathcal{H}$
($N$ factors) of the parent system. In essence, formally replacing
the state vector \eqref{eq:LongSeqStateVec} of the full parent system
with the subjective density matrix \eqref{eq:SubjDensMatrProperMix}
of an arbitrary elementary subsystem---as is often done merely for
practical convenience in order to simplify calculations---implicitly
assumes that we have erased all information about the original ordering
of the physical sequence. By contrast, if we \emph{do} wish to retain
and make use of information about the original ordering of the sequence---such
as if we wish to examine or select particular subsequences of interest---then
we must be careful not to confuse the full sequence's own state vector
\eqref{eq:LongSeqStateVec} with the subjective density matrix \eqref{eq:SubjDensMatrProperMix}
of an arbitrarily chosen member of the sequence, lest we run into
apparent contradictions of the sort uncovered in \cite{MillerFarr:2014opomqm}.%
\footnote{The authors of \cite{MillerFarr:2014opomqm} elide these subtle distinctions
in order to generate a paradox in which the proper mixture describing
a carefully post-selected subsequence of non-entangled two-qubit systems
can also seemingly be represented as an entangled pure state, with
the ultimate goal of casting doubt on interpretations of quantum theory
(such as our own) that attempt to ascribe an ontological meaning to
state vectors.%
}

\subsection{Paradoxes of Quantum Theory Revisited\label{sub:Paradoxes-of-Quantum-Theory-Revisited}}

\subsubsection{Schrödinger's Cat, Wigner's Friend, and the Local Nature of Ontology}

In his famous 1935 thought experiment arguing against the Copenhagen
interpretation \cite{Schrodinger:1935dgsdq},%
\footnote{Schrödinger himself coined the term ``entanglement'' (\emph{Verschränkung})
in this 1935 paper.%
} Schrödinger considered the hypothetical case of an unfortunate cat
trapped in a perfectly sealed box together with a killing mechanism
triggered by the decay of an unstable atom. If the experiment begins
at time $t=0$, and if the atom has a 50-50 empirical outcome probability
of quantum-mechanically decaying over a nonzero time interval $\Delta t$,
then the atom's state vector at exactly the time $t=\Delta t$ is
the quantum superposition 
\begin{equation}
\ket{\mathrm{atom}}=\frac{1}{\sqrt{2}}\left(\ket{\mathrm{not\ decayed}}+\ket{\mathrm{decayed}}\right).\label{eq:SchroCatAtomFinalStateVec}
\end{equation}
 Immediately thereafter, the linearity of time evolution would seem
to imply that the final state vector of the composite parent system
\begin{equation}
\left(\mathrm{parent\ system}\right)\equiv\left(\mathrm{atom}\right)+\left(\mathrm{killing\ mechanism}\right)+\left(\mathrm{cat}\right)+\left(\mathrm{air\ in\ box}\right)+\left(\mathrm{box}\right)\label{eq:SchroCatParentSystemOriginal}
\end{equation}
 then has the highly entangled, quantum-superposed form 
\begin{equation}
\begin{aligned}\ket{\mathrm{parent\ system}} & =\frac{1}{\sqrt{2}}\left(\ket{\mathrm{not\ decayed}}\ket{\mathrm{not\ triggered}}\ket{\mathrm{alive}}\ket{\mathrm{air\ in\ box}}\ket{\mathrm{box}}\right.\\
 & \qquad\qquad+\left.\ket{\mathrm{decayed}}\ket{\mathrm{triggered}}\ket{\mathrm{dead}}\ket{\left(\mathrm{air\ in\ box}\right)^{\prime}}\ket{\left(\mathrm{box}\right)^{\prime}}\right).
\end{aligned}
\label{eq:SchroCatFinalStateVec}
\end{equation}
 Does this expression imply that the cat's state of existence is smeared
out into a quantum superposition of $\mathrm{alive}$ and $\mathrm{dead}$,
and remains in that superposed condition until a human experimenter
opens the box to perform a measurement on the cat's final state and
thereby collapses the cat's wave function?

The Everett-DeWitt many-worlds interpretation claims to resolve this
paradoxical state of affairs by dropping the assumption that the final
measurement carried out by the human experimenter has a definite outcome
in any familiar sense. Instead, according to the many-worlds interpretation,
the human experimenter merely ``joins'' the superposition upon opening
the box to look inside, thereby splitting into two clones that each
observe a different outcome. As always, beyond whatever metaphysical
discomfort accompanies the notion that the world ``unzips'' into
multiple copies with each passing moment in time (and, after all,
such discomfort may be nothing more than human philosophical prejudice),
there remains the tricky issue of making sense of probability itself
when all possible outcomes of an experiment simultaneously occur,
as well as understanding why the numerical coefficients $1/\sqrt{2}$
appearing in the superposed state vector \eqref{eq:SchroCatFinalStateVec}
have the physical meaning of probabilities in the first place.%
\footnote{We will discuss possible connections between the many-worlds interpretation
and our own interpretation in greater depth in Section~\ref{sub:Quantum-Theory-and-Classical-Gauge-Theories}.%
}

By contrast, our minimal modal interpretation of quantum theory resolves
the problem much less extravagantly: We begin by noting that the moment
the cat interacts with the killing mechanism (and inevitably also
with the air in the box and with the walls of the box itself), the
cat's own density matrix rapidly decoheres to the approximate form
\[
\op{\rho}_{\mathrm{cat}}=\frac{1}{2}\ket{\mathrm{alive}}\bra{\mathrm{alive}}+\frac{1}{2}\ket{\mathrm{dead}}\bra{\mathrm{dead}},
\]
 up to small but unavoidable degeneracy-breaking corrections similar
to those appearing in \eqref{eq:EPRPairStateVecDegenBreaking}.%
\footnote{Contrary to some accounts \cite{Griffiths:2005iqm}, at no time during
the experiment is the cat alone ever represented by a pure-state superposition
of the form $\left(1/\sqrt{2}\right)\left(\ket{\mathrm{alive}}+\ket{\mathrm{dead}}\right)$.%
} According to our interpretation, the cat's actual ontic state is
therefore definitely alive or dead almost immediately after the killing
mechanism acts---each possibility with corresponding epistemic probability
$1/2$---and well before the human experimenter opens the box.

Of course, the composite parent system \eqref{eq:SchroCatParentSystemOriginal}
has a strangely quantum-superposed pure state \eqref{eq:SchroCatFinalStateVec}
until the human experimenter opens the box,  but this parent system
is simply \emph{not} the system that we identify as the cat alone.
The relationship between a parent system and any one of its subsystems
must ultimately be postulated as part of the interpretation of \emph{any}
physical theory, especially quantum theory; in our own interpretation,
this relationship---and thus the cat's ontology---is determined \emph{locally}
(in the sense of being system-centric) through the cat's \emph{own}
reduced density matrix.

After the human experimenter has opened the box, the original parent
system \eqref{eq:SchroCatParentSystemOriginal} undergoes rapid decoherence
and no longer remains in a quantum-superposed pure state. However,
the \emph{larger} parent system 
\begin{equation}
\begin{aligned}\left(\mathrm{larger\ parent\ system}\right) & \equiv\left(\mathrm{atom}\right)+\left(\mathrm{killing\ mechanism}\right)+\left(\mathrm{cat}\right)+\left(\mathrm{air\ in\ box}\right)+\left(\mathrm{box}\right)\\
 & \qquad+\left(\mathrm{human\ experimenter}\right)+\left(\mathrm{environment}\right)
\end{aligned}
\label{eq:SchroCatParentSystemEnlarged}
\end{equation}
 \emph{does} remain in a quantum-superposed pure state, provided that
we allow our definition of the environment to grow rapidly in physical
size with the unavoidable and irreversible outward emission of thermal
radiation at the speed of light. If a second human experimenter---``Wigner''---now
enters the story, then even before Wigner makes causal contact with
this larger parent system \eqref{eq:SchroCatParentSystemEnlarged}
to find out the result of the measurement by the first experimenter---whom
we'll call ``Wigner's friend'' \cite{Wigner:1967sr}---decoherence
ensures that Wigner could still say that the cat and Wigner's friend
($\mathrm{WF}$) each have classical-looking reduced density matrices
given respectively by 
\[
\op{\rho}_{\mathrm{cat}}=\frac{1}{2}\ket{\mathrm{alive}}\bra{\mathrm{alive}}+\frac{1}{2}\ket{\mathrm{dead}}\bra{\mathrm{dead}}
\]
 and 
\[
\op{\rho}_{\mathrm{WF}}=\frac{1}{2}\ket{\mathrm{WF}\left(\quote{\mathrm{alive}}\right)}\bra{\mathrm{WF}\left(\quote{\mathrm{alive}}\right)}+\frac{1}{2}\ket{\mathrm{WF}\left(\quote{\mathrm{dead}}\right)}\bra{\mathrm{WF}\left(\quote{\mathrm{dead}}\right)},
\]
 all with classical-looking possible ontic states $\mathrm{alive}$,
$\mathrm{dead}$, $\mathrm{WF}\left(\quote{\mathrm{alive}}\right)$,
and $\mathrm{WF}\left(\quote{\mathrm{dead}}\right)$, and again up
to tiny degeneracy-breaking corrections. In fact, thanks once more
to decoherence, Wigner can even say that the composite system consisting
of just (cat) + (Wigner's friend) has a classical-looking reduced
density matrix 
\[
\op{\rho}_{\mathrm{cat}+\mathrm{WF}}=\frac{1}{2}\ket{\mathrm{alive}}\ket{\mathrm{WF}\left(\quote{\mathrm{alive}}\right)}\bra{\mathrm{alive}}\bra{\mathrm{WF}\left(\quote{\mathrm{alive}}\right)}+\frac{1}{2}\ket{\mathrm{dead}}\ket{\mathrm{WF}\left(\quote{\mathrm{dead}}\right)}\bra{\mathrm{dead}}\bra{\mathrm{WF}\left(\quote{\mathrm{dead}}\right)}.
\]
 Hence, our instantaneous kinematical parent-subsystem quantum conditional
probabilities \eqref{eq:QuantCondProbKinParentSub} ensure that if
the composite system (cat) + (Wigner's friend) has, say, the ontic
state $\left(\mathrm{alive},\mathrm{WF}\left(\quote{\mathrm{alive}}\right)\right)$,
then, with essentially unit conditional probability, the ontic state
of the cat is $\mathrm{alive}$ and the ontic state of Wigner's friend
is $\mathrm{WF}\left(\quote{\mathrm{alive}}\right)$, as expected.

The preceding discussion makes clear that in our interpretation of
quantum theory, ontology is a \emph{local} phenomenon, where by ``local''
we simply mean system-centric: In order to establish the ontology
of a given system, one must examine that system's \emph{own} (reduced)
density matrix, and not the density matrices of any other systems---not
even those of parent systems, except to the extent that a parent system's
density matrix determines the subsystem's density matrix through the
partial-trace operation \eqref{eq:DefPartialTraces}.%
\footnote{As we explained in Section~\ref{sub:Comparison-with-Other-Interpretations-of-Quantum-Theory},
there exists a very real possibility that there is no well-defined
maximal closed system in a pure state and enclosing all other systems.
In that case, we are \emph{forced} to approach notions of ontology
in a system-centric way in any \emph{realist} interpretation of quantum
theory, including in the de Broglie-Bohm and Everett-DeWitt interpretations.
Indeed, if we eventually have to stop going up the never-ending hierarchy
of open parent systems at some particular system's reduced density
matrix, then we have no choice but to try to make sense of that reduced
density matrix on its own terms.%
}

The way that our interpretation localizes notions of ontology is analogous
to the way that general relativity localizes the notions of inertial
references frames and observables like relative velocities that are
familiar from special relativity: Just as there's no well-defined
sense in which widely separated galaxies in an expanding universe
have a sensible relative velocity (contrary to colloquial assertions
that galaxies sufficiently far from our own are ``receding from our
own galaxy faster than the speed of light''), there's no well-defined
sense in which we can generically read off the ontologies of highly
entangled subsystems directly from the density matrix of a larger
parent system.

In particular, given a system whose ontic state involves a highly
quantum superposition, naïvely assuming (as one does in the Everett-DeWitt
many-worlds interpretation) that each of the system's subsystems likewise
has a highly quantum-superposed existence  would mean committing
the same kind of fallacy of division that we first described in the
context of characterizing the precise kinematical relationship \eqref{eq:QuantCondProbKinParentSub}
between the ontic states of parent systems and their subsystems in
Section~\ref{subsub:Kinematical-Relationships-Between-Ontic-States-of-Parent-Systems-and-Subsystems}.

Schrödinger intended his thought experiment to be a proof of principle
against the Copenhagen interpretation, but it is worth asking whether
his experiment could ever practically be realized (barring obvious
ethical questions). Microscopic versions of his experiment involving
small assemblages of particles, known as ``Schrödinger kittens,''
are performed all the time \cite{SongCavesYurke:1989skobae,OurjoumtsevTualleBrouriLauratGrangier:2006goscfpns,GerlichEibenbergerTomandlNimmrichterHornbergerFaganTuxenMayorArndt:2011qilom,ArndtBassiGiuliniHeidmannRaimond:2011ffqst,ArndtNairzVosAndreaeKellerVanDerZouwZeilinger:1999wpdcm},
and these experiments have even been scaled up beyond microscopic
size in recent years \cite{FriedmanPatelChenTolpygoLukens:2000qsdms,VanDerWalTerHaarWilhelmSchoutenHarmansOrlandoLloydMooij:2000qsmpcs,WilhelmVanDerWalTerHaarSchoutenHarmansMooijOrlandoLloyd:2001mqscsjjl,OurjoumtsevJeongTualleBrouriGrangier:2007goscfpns}.

However, the systems involved in these experiments are far from what
we would consider complex living organisms. The trouble is that maintaining
a large system in a non-negligible superposition of macroscopically
distinct states requires completely decoupling the system of interest
from the sort of warm, hospitable environment that life as we know
it seems to require. When we recognize moreover that all large living
creatures constantly give off thermal radiation themselves, it becomes
untenable to imagine that partial-tracing down to just the degrees
of freedom that we associate with a living cat could ever yield an
actual ontic state for the cat that remains in such a distinctly non-classical
superposition for more than an infinitesimal instant in time. One
can certainly imagine some day having the technological ability to
construct a quantum superposition involving a frozen-solid cat at
a temperature near absolute zero in an evacuated chamber, but such
an experiment would hardly be in keeping with the original spirit
of Schrödinger's ``paradox.''

\subsubsection{The Quantum Zeno Paradox}

The quantum Zeno paradox \cite{DegasperisFondaChirardi:1974dlusdma,Misra:1977zpqt,ChiuSudarshanMisra:1977teuqsrzp,Ballentine:1998qmmd,Griffiths:2005iqm}
demonstrates yet another limitation of the traditional Copenhagen
interpretation, and so it is important to explain why the problem
is avoided in decoherence-based interpretations of quantum theory
such as the one that we present in this paper.

The paradox concerns the behavior of systems that experience rapid
sequences of measurements but that otherwise undergo unitary time
evolution according to some Hamiltonian $\op H$. For time intervals
$T$ that are sufficiently large compared to the inverse-energy-differences
between the system's energy eigenstates, but small enough that time-dependent
perturbation theory is valid, the probability of observing a transition
away from an unstable state $\Psi_{\mathrm{initial}}$ is approximately
\emph{linear} in $T$, 
\begin{equation}
p_{\mathrm{initial}\to\mathrm{anything}\ne\mathrm{initial}}\left(T\right)\simeq\alpha T,\label{eq:ZenoLinearProb}
\end{equation}
 where $\alpha\simeq\const$ is just the total transition rate away
from $\Psi_{\mathrm{initial}}$ and where we assume for consistency
with perturbation theory that $\alpha$ is much smaller than $1/T$.
Hence, the probability that a measurement after a time $T$ will find
the system still in its \emph{original} state $\Psi_{\mathrm{initial}}$
is 
\begin{equation}
p_{\mathrm{initial}}\left(T\right)\simeq1-\alpha T.\label{eq:ZenoLinRemainProb}
\end{equation}
 It follows that if we perform $N\gg1$ sequential measurements separated
by time intervals $T=t/N$ over which \eqref{eq:ZenoLinRemainProb}
holds, then the probability of finding the system \emph{still in the
initial state} $\Psi_{\mathrm{initial}}$ at the final time $t=N\times\left(t/N\right)$
is given by the familiar exponential decay law 
\begin{equation}
p_{\mathrm{initial}}\left(t\right)\simeq e^{-\alpha t}.\label{eq:ZenoExpDecayLaw}
\end{equation}
 This argument helps explain why exponential laws are so ubiquitous
among systems exhibiting quantum decay.

By contrast, for \emph{extremely tiny} time intervals $\Delta t\to0$,
the Born rule \eqref{eq:BornRule} implies that the probability that
a measurement after a time $\Delta t$ will find the system still
in its original state $\Psi_{\mathrm{initial}}$ depends \emph{quadratically}
on $\Delta t$, 
\begin{equation}
p_{\mathrm{initial}}\left(\Delta t\right)=\absval{\bra{\Psi_{\mathrm{initial}}}e^{-i\op H\Delta t/\hbar}\ket{\Psi_{\mathrm{initial}}}}^{2}=1-\beta^{2}\Delta t^{2},\label{eq:ZenoQuadProb}
\end{equation}
 where 
\begin{equation}
\beta=\frac{\sqrt{\expectval{\left(H-\expectval H\right)^{2}}}}{\hbar}=\frac{\Delta E}{\hbar}\sim\frac{1}{\tau}\label{eq:ZenoQuadCoeff}
\end{equation}
 is just the variance of the system's energy in the state $\Psi_{\mathrm{initial}}$
in units of $\hbar$, and is related to the system's characteristic
time scale $\tau$ through the energy-time uncertainty principle $\Delta E\times\tau\sim\hbar$.
If we carry out $N\gg1$ sequential measurements separated by time
intervals $\Delta t=t/N\ll\beta$, then the probability that we will
see the system still in its initial state $\Psi_{\mathrm{initial}}$
at the final time $t=N\times\left(t/N\right)$ seemingly asymptotes
to unity, 
\begin{equation}
p_{\mathrm{initial}}\left(t\right)\to1,\label{eq:ZenoProbRemainUnity}
\end{equation}
 apparently implying that the quantum state never transitions. This
result, known as the quantum Zeno paradox (or the watched-pot paradox),
would appear to suggest that a \emph{continuously observed} unstable
system can never decay.

As with many of the supposed ``paradoxes'' of quantum physics, the
quantum Zeno paradox can be resolved by examining the measurement
process more carefully. To begin, any interpretation of quantum theory
(unlike the Copenhagen interpretation) that regards measurements as
physical processes implies that they occur over a \emph{nonzero} time
interval, and therefore implies that $\Delta t$ cannot, in fact,
be taken literally to zero. The notion of a \emph{continuously observed}
system is therefore a fiction, and so we can safely rule out the\emph{
}quantum Zeno paradox in its strongest sense, namely, in its suggestion
that frequently observed systems should \emph{never} decay.

But we can go further and explain why even a weaker version of the
quantum Zeno effect---that is, a noticeable delay in the system's
decay rate or a deviation in the exponential decay law \eqref{eq:ZenoExpDecayLaw}---is
extremely difficult to produce and observe. By momentarily connecting
our unstable system to a measurement apparatus $A$ with its own temporal
resolution scale $\Delta t_{A}\sim\hbar/\Delta E_{A}$---meaning that
we replace our initial state vector $\ket{\Psi_{\mathrm{initial}}}$
with $\ket{\Psi_{\mathrm{initial}}}\ket A$---the resulting interactions
risk enlarging the system's total energy variance and thereby increasing
the coefficient $\beta$ appearing in the decay probability \eqref{eq:ZenoQuadProb}
\cite{GhirardiOmeroWeberRimini:1979stbqndzpqm}.  The quantum Zeno
effect is therefore far from a generic process for systems exposed
to environmental interactions, and successfully producing the effect
requires very careful experiments \cite{ItanoHeinzenBollingerWineland:1990qze,WoltersStraussSchoenfeldBenson:2013oqzessss}.

\subsubsection{Particle-Field Duality}

Disputes crop up occasionally over whether particles or fields represent
the more fundamental feature of reality.%
\footnote{For a recent instance, see \cite{Hobson:2013tanpof}.%
} Such questions are perfectly valid to ask in the context of certain
interpretations of quantum theory, including the so-called ``fixed''
modal interpretations---the de Broglie-Bohm pilot-wave interpretation
being an example---in which a particular Hilbert-space basis is singled
out permanently as defining the system's ontology.

From the standpoint of our own interpretation, however, these questions
are much less meaningful. A quantum field is a quantum system having
some specific dynamics and a Hilbert space that is usefully described
at low energies and weak coupling as a Fock space.%
\footnote{Note that we use the term ``quantum field'' here to refer to the
physical system itself, and not to mathematical field operators.%
} Some of the states in this Hilbert space look particle-like, in the
sense of being eigenstates of a particle-number operator, whereas---at
least in the case of a bosonic system---other states in the Hilbert
space look more like classical fields, in the sense of being coherent
states \cite{Schrodinger:1926dsuvdmzm,Glauber:1963cisrf} with a well-defined
classical field amplitude.%
\footnote{As we explained in Section~\ref{subsub:The-de-Broglie-Bohm-Pilot-Wave-Interpretation-of-Quantum-Theory},
the reliance of the de Broglie-Bohm pilot-wave interpretation on a
universal, permanent ontic basis leads to serious trouble dealing
with relativistic systems: States of particles with definite positions
are unavailable in general, because relativistic systems do not admit
a basis of sharp orthonormal position eigenstates, and neither coherent
states nor field eigenstates exist for fermionic systems \cite{Struvye:2012ozzpwaf}.%
}

According to our interpretation of quantum theory, the ontic basis
of the system is contextual: Experiments that count quanta decohere
the system to an ontic basis of particle-like states, and experiments
that measure forces on test bodies decohere the system to an ontic
basis of field-like states. Hence, neither of these two possible classes
of states represents the ``fundamental'' basis for the system in
any permanent or universal sense. Indeed, to say one of these two
classes of states is more fundamental than the other would be precisely
like saying that the energy eigenstates of a harmonic oscillator are
more fundamental than its coherent states or vice versa, or like saying
that a harmonic oscillator ``is really'' made up of energy eigenstates
or that it ``is really'' made up of coherent states.

\section{Lorentz Invariance and Locality\label{sec:Lorentz-Invariance-and-Locality}}

\subsection{Special Relativity and the No-Communication Theorem\label{sub:Special-Relativity-and-the-No-Communication-Theorem}}

Special relativity links locality with causality, but only to the
extent of forbidding \emph{observable signals} from propagating superluminally.
Hidden variables that exhibit nonlocal dynamics are therefore not
necessarily a problem: The no-communication theorem \cite{Hall:1987imnlqm,PeresTerno:2004qirt}
ensures that quantum systems with local density-matrix dynamics do
not transmit superluminal observable signals, which could otherwise
spell trouble for causality.

Indeed, consider the simplified example of a dynamically closed system
with a Hamiltonian $\op H$ having a separable decomposition of the
form $\op H=\op H_{A}\tensorprod\op 1+\op 1\tensorprod\op H_{B}$
for a pair of subsystems $A$ and $B$---as is necessarily the case
in practice if the subsystems $A$ and $B$ are well-separated from
each other in space. Then it is straightforward to show \cite{vonNeumann:1932mgdq,BasdevantDalibard:2002qm}
directly by taking the partial trace \eqref{eq:DefPartialTraces}
of the unitary Liouville-Von-Neumann equation \eqref{eq:QuantumLiouvilleEq},
\[
\frac{\partial\op{\rho}\left(t\right)}{\partial t}=-\frac{i}{\hbar}\comm{\op H}{\op{\rho}\left(t\right)},
\]
 that the dynamics of the reduced density matrix $\op{\rho}_{A}\left(t\right)$
of subsystem $A$ has no influence whatsoever on the reduced density
matrix $\op{\rho}_{B}\left(t\right)$ of subsystem $B$, and vice
versa.

But a more subtle issue arises when we consider whether an interpretation's
hidden variables are well-defined under arbitrary choices of inertial
reference frame, or, equivalently, under arbitrary choices of foliation
of four-dimensional spacetime into the three-dimensional spacelike
hyperplanes that define slices of constant time. We explore this and
related questions in subsequent sections, including the well-known
EPR-Bohm and GHZ-Mermin thought experiments, and ultimately conclude
that our minimal modal interpretation of quantum theory is indeed
consistent with Lorentz invariance and that its nonlocality is no
more severe than is the case for classical gauge theories.

\subsection{The EPR-Bohm Thought Experiment and Bell's Theorem\label{sub:The-EPR-Bohm-Thought-Experiment-and-Bell's-Theorem}}

In 1964, Bell \cite{Bell:1964oeprp} employed a thought experiment
due to Bohm \cite{Bohm:1951qt,BohmAharonov:1957depperp} and based
originally on a 1935 paper by Einstein, Podolsky, and Rosen \cite{EinsteinPodolskyRosen:1935cqmdprbcc}
to prove that no realist interpretation of quantum theory in which
experiments have definite outcomes could be based on \emph{dynamically
local} hidden variables.%
\footnote{It is important to note again that, despite Bell's theorem, any interpretation
of quantum theory consistent with the theory's standard observable
rules does not permit superluminal \emph{observable signals}, as ensured
by the no-communication theorem \cite{Hall:1987imnlqm,PeresTerno:2004qirt}.%
}

Bell's theorem has exerted a powerful influence on the foundations
of quantum theory in all the years since, to the extent that all interpretations
today drop at least one of Bell's stated assumptions---either the
existence of hidden variables, realism, dynamical locality, or the
assertion that experiments have definite outcomes. For example, the
de Broglie-Bohm pilot-wave interpretation \cite{deBroglie:1930iswm,Bohm:1952siqtthvi,Bohm:1952siqtthvii,BohmHiley:1993uu}
involves a hidden level of \emph{manifestly nonlocal} dynamics, as
do all of the modal interpretations \cite{Krips:1969tpqm,Krips:1975siqt,vanFraassen:1972faps,Cartwright:1974vfmmqm,Krips:1987mqt,vanFraassen:1991qmmv,Bub:1992qmwpp,VermaasDieks:1995miqmgdo,BacciagaluppiDickson:1997gqtm,BacciagaluppiDickson:1999dmi,Vermaas:1999puqm}
including our own. The Everett-DeWitt many-worlds interpretation \cite{Everett:1957rsfqm,Wheeler:1957aersfqt,DeWitt:1970qmr,EverettDeWittGraham:1973mwiqm,Everett:1973tuwf,Deutsch:1985qtupt,Deutsch:1999qtpd,Wallace:2002wei,Wallace:2003erddapei,BrownWallace:2004smpdbble}
does away with the assumption that experiments have definite outcomes,
and instead asserts that all outcomes occur simultaneously and are
thus equally real. (We argue in Section~\ref{subsub:Nonlocality in the Everett-DeWitt Many-Worlds Interpretation}
that the many-worlds interpretation, as traditionally formulated,
still suffers from some residual nonlocality.)

It is interesting to recapitulate Bell's arguments carefully in order
to understand precisely how they interface with our own interpretation
of quantum theory. Along the way, we will spot a subtle \emph{additional}
loophole in his assumptions, although our interpretation does not
ultimately exploit this loophole and thus does not void Bell's conclusions
about nonlocality.%
\footnote{As we will discuss shortly, the same loophole also exists in the assumptions
of similar no-go theorems, such as the CHSH theorem of \cite{ClauserHorneShimonyHold:1969petlhvt,ClauserHorne:1974ecolt,Bell:1971fqm,Leggett:2003nhvtqmit},
although not in the GHZ-Mermin arguments to be described in Section~\ref{sub:The-GHZ-Mermin-Thought-Experiment}.
}

\subsubsection{The Basic Set-Up of the EPR-Bohm Thought Experiment}

\begin{figure}
\begin{centering}
\begin{center}
\definecolor{ce1e1e1}{RGB}{225,225,225}
\definecolor{cffffff}{RGB}{255,255,255}

\begin{tikzpicture}[y=0.80pt,x=0.80pt,yscale=-1, inner sep=0pt, outer sep=0pt]
\begin{scope}[shift={(-75.36362,-46.1165)}]
  \path[fill=black] (306.47684,78.512917) node[above right] (text3005) {$1$};
  \path[fill=black] (345.35025,78.513031) node[above right] (text3005-3) {$2$};
  \path[fill=black] (269.18335,60.629883) node[above right] (text3005-0) {$\left|\Psi\right\rangle=\frac{1}{\sqrt{2}}\left(\left|\uparrow\downarrow\right\rangle-\left|\downarrow\uparrow\right\rangle\right)$};
  \path[->,>=latex,draw=black,line join=miter,line cap=butt,line width=0.800pt]
    (290.4619,74.0315) -- (182.8017,74.0315);
  \path[<-,>=latex,draw=black,line join=miter,line cap=butt,line width=0.800pt]
    (479.8062,74.0312) -- (372.1460,74.0312);
  \path[draw=black,fill=ce1e1e1,miter limit=4.00,line width=1.600pt,rounded
    corners=0.0000cm] (76.3636,47.1165) rectangle (143.9642,99.6947);
  \path[draw=black,fill=ce1e1e1,line join=miter,line cap=butt,line width=0.800pt]
    (167.0816,48.5448) -- (143.8451,63.3719) -- (143.8451,84.8380) --
    (167.3029,98.1160) -- cycle;
  \path[draw=black,fill=cffffff,miter limit=4.00,line width=1.600pt,rounded
    corners=0.0000cm] (92.5035,62.9293) rectangle (127.4689,84.1741);
  \path[fill=black] (103.69986,78.756378) node[above right] (text3075) {$A$};
  \path[xscale=-1.000,yscale=1.000,draw=black,fill=ce1e1e1,miter limit=4.00,line
    width=1.600pt,rounded corners=0.0000cm] (-582.2515,47.1170) rectangle
    (-514.6509,99.6952);
  \path[draw=black,fill=ce1e1e1,line join=miter,line cap=butt,line width=0.800pt]
    (491.5335,48.5452) -- (514.7700,63.3723) -- (514.7700,84.8384) --
    (491.3122,98.1164) -- cycle;
  \path[xscale=-1.000,yscale=1.000,draw=black,fill=cffffff,miter limit=4.00,line
    width=1.600pt,rounded corners=0.0000cm] (-566.1116,62.9297) rectangle
    (-531.1461,84.1745);
  \path[fill=black] (542.37311,78.756851) node[above right] (text3075-4) {$B$};
  \path[shift={(298.8748,135.49057)},draw=black,fill=cffffff,miter limit=4.00,line
    width=1.600pt]
    (25.6632,-61.5316)arc(0.000:180.000:2.817)arc(-180.000:0.000:2.817) -- cycle;
  \path[shift={(315.01228,135.49107)},draw=black,fill=cffffff,miter
    limit=4.00,line width=1.600pt]
    (25.6632,-61.5316)arc(0.000:180.000:2.817)arc(-180.000:0.000:2.817) -- cycle;
\end{scope}

\end{tikzpicture}
\par\end{center}
\par\end{centering}

\caption{\label{fig:EPRBohmExperiment}The schematic set-up for the EPR-Bohm
thought experiment, consisting of particles $1$ and $2$ together
with spin detectors $A$ and $B$.}
\end{figure}
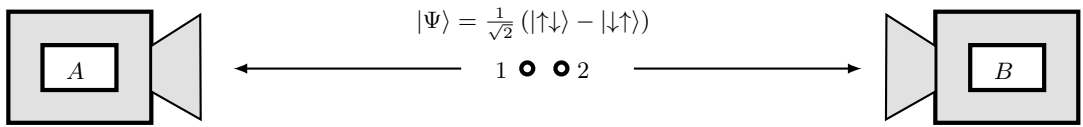

The EPR-Bohm set-up begins with a pair of distinguishable spin-1/2
particles---denoted particle $1$ and particle $2$---that are initially
prepared in an entangled pure state of total spin zero, so that the
ontic state $\Psi$ of the composite two-particle system $1+2$ is
described by the state vector \eqref{eq:EPRPairStateVecPerfDegen}:
\begin{equation}
\ket{\Psi}=\frac{1}{\sqrt{2}}\left(\ket{\uparrow\downarrow}-\ket{\downarrow\uparrow}\right),\qquad\op S_{1+2,z}\ket{\Psi}=0.\label{eq:EPRPairStateVec}
\end{equation}
 We then consider two spatially separated spin detectors $A$ and
$B$, where $A$ locally measures the spin of particle $1$ and $B$
locally measures the spin of particle $2$. We suppose that $A$ measures
the component $S_{1,\svec a}=\svec a\cdot\svec S_{1}$ of the spin
$\svec S_{1}$ of particle $1$ along the direction of a three-dimensional
spatial unit vector $\svec a$ and that $B$ measures the component
$S_{2,\svec b}=\svec b\cdot\svec S_{2}$ of the spin $\svec S_{2}$
of particle $2$ along the direction of a three-dimensional spatial
unit vector $\svec b$. (See Figure~\ref{fig:EPRBohmExperiment}.)

\subsubsection{Spin Correlation Predicted by Quantum Theory}

After many repetitions of the entire experiment---each time preparing
a composite two-particle system $1+2$ in the ontic state \eqref{eq:EPRPairStateVec}
and then allowing the spin detectors $A$ and $B$ to perform local
measurements on the spins of their respective individual particles---we
can ask for the average value of the product $S_{1,\svec a}S_{2,\svec b}$,
which is just the correlation function for the pair of measured spins.
Working in units of $\hbar/2$ for clarity (that is, setting $\hbar/2\equiv1$),
quantum theory predicts that this expectation value should be given
by 
\begin{equation}
\expectval{S_{1,\svec a}S_{2,\svec b}}_{\mathrm{QM}}=-\svec a\cdot\svec b,\label{eq:EPRCorrFromQM}
\end{equation}
 a result that follows directly from linearity and rotation invariance.%
\footnote{Proof: First, observe that $\expectval{S_{1,\svec a}S_{2,\svec b}}=\sum_{i}\sum_{j}a_{i}b_{j}\expectval{S_{1,i}S_{2,j}}=\sum_{i}\sum_{j}a_{i}b_{j}F_{ij}$,
where $F_{ij}=\expectval{S_{1,i}S_{2,j}}$ is a $3\times3$ matrix
that is manifestly independent of $\svec a$ and $\svec b$. Because
our final result for $\expectval{S_{1,\svec a}S_{2,\svec b}}$ must
be real and invariant under global rotations, and should obviously
have the specific value $-1$ if $\svec a=\svec b$, we must have
$F_{ij}=-\delta_{ij}$. Hence, $\expectval{S_{1,\svec a}S_{2,\svec b}}=\sum_{i}\sum_{j}a_{i}b_{j}\left(-\delta_{ij}\right)=-\svec a\cdot\svec b.$
$\QED$%
}

\subsubsection{The Bell Inequality}

Bell argued that no \emph{dynamically local} hidden variables could
account for the result \eqref{eq:EPRCorrFromQM}. To prove this claim,
he derived a ``triangle inequality'' 
\begin{equation}
\absval{\expectval{S_{1,\svec a}S_{2,\svec b}}_{\mathrm{LHV}}-\expectval{S_{1,\svec a}S_{2,\svec c}}_{\mathrm{LHV}}}\leq1+\expectval{S_{1,\svec b}S_{2,\svec c}}_{\mathrm{LHV}}\label{eq:BellIneq}
\end{equation}
 that all interpretations based on local hidden variables (LHV) would
necessarily have to satisfy, where $\svec c$ is a unit vector specifying
the alignment of a third detector $C$. The correct quantum result
\eqref{eq:EPRCorrFromQM} can easily violate this inequality as well
as various generalizations, as has been confirmed repeatedly in experiments
\cite{FreedmanClauser:1972etlhvt,AspectGrangierRoger:1981etrltvbt,AspectDalibardRoger:1982etbiutva,TittelBrendelZbindenGisin:1998vbibpmt10kma,TittelBrendelGisinHerzogZbindenGisin:1998edqcom10k,WeihsJenneweinSimonWeinfurterZeilinger:1998vbiuselc,RoweKielpinskiMeyerSackettItanoMonroeWineland:2001evbied,GroblacherPaterekKaltenbaekBruknerZukowskiAspelmeyerZeilinger:2007etnlr,AnsmannWangBialczakHofheinzLuceroNeeleyOConnellSankWeidesWennerClelandMartinis:2009vbijpq,GiustinaMechRamelowWittmannKoflerBeyerLitaCalkinsGerritsNamUrsinZeilinger:2013bvuepwfsa}.%
\footnote{For example, if we take $\svec a$ and $\svec b$ to be orthogonal
and for $\svec c$ to be 45 degrees between both of them, then the
quantum formula \eqref{eq:EPRCorrFromQM} for the spin correlation
function yields $\expectval{S_{1,\svec a}S_{2,\svec b}}=0,$ $\expectval{S_{1,\svec a}S_{2,\svec c}}\simeq-0.707,$
and $\expectval{S_{1,\svec b}S_{2,\svec c}}\simeq-0.707$, in which
case the Bell inequality \eqref{eq:BellIneq} is expressly violated:
$\absval{\expectval{S_{1,\svec a}S_{2,\svec b}}-\expectval{S_{1,\svec a}S_{2,\svec c}}}\simeq0.707\not\leq0.293\simeq1+\expectval{S_{1,\svec b}S_{2,\svec c}}$.
For a discussion of potential shortcomings and loopholes in these
Bell-test experiments, see, for example, \cite{GargMermin:1987diepre,Larsson:1998bidi,GerhardtLiuLamasLinaresSkaarScaraniMakarovKurtsiefer:2011efvbi,GallicchioFriedmanKaiser:2013tbicpcsil}.%
}

To derive the inequality \eqref{eq:BellIneq}, Bell assumed that the
hidden variables $\lambda$ obeyed a probability distribution $p\left(\lambda\right)$
satisfying the standard Kolmogorov conditions \cite{Kolmogorov:1933ftp}:
\begin{equation}
0\leq p\left(\lambda\right)\leq1,\qquad\integration{\int}{d\lambda}p\left(\lambda\right)=1.\label{eq:BellProbDistribConds}
\end{equation}
 The proof of the Bell inequality \eqref{eq:BellIneq}, which we present
in Appendix~\ref{subsub:Proof-of-the-Bell-Theorem}, then rests entirely
on manipulations of averages weighted by $p\left(\lambda\right)$.

\subsubsection{The Additional Loophole\label{subsub:The-Additional-Loophole}}

Bell's theorem and all the various generalizations that have arisen
in subsequent papers by Bell and others, including the CHSH inequalities
\cite{ClauserHorneShimonyHold:1969petlhvt,ClauserHorne:1974ecolt,Bell:1971fqm}
and Legget's inequality for ``crypto-nonlocal'' interpretations
\cite{Leggett:2003nhvtqmit}, implicitly assume that the hidden variables
$\lambda$ admit a sensible \emph{joint} epistemic probability distribution
$p\left(\lambda\right)$ as described in \eqref{eq:BellProbDistribConds}.
 As we made clear in Section~\ref{subsub:Probabilistic-and-Non-Probabilistic-Uncertainty},
this assertion is nontrivial, but none of these papers attempt to
justify its validity. However, this assumption represents a significant
loophole, one that appears to have been noticed first by Bene \cite{Bene:1997qrsnfqm,Bene:1997qpdnvpl,Bene:1997onqsms,Bene:2001qrsrlqm,Bene:2001rmirqft,BeneDieks:2002pvmiqm}
in the context of his ``perspectival'' interpretation of quantum
theory, but then also discovered independently and examined in a small
number of additional papers \cite{Valdenebro:2002aubi,HessPhilipp:2004btcpwwi,Zhao:2009fpbrdta}.%
\footnote{Fine \cite{Fine:1982hvjpbi,Fine:1982jdqcco} appears to have worked
on similar ideas in 1982. He studied whether one could assume the
existence of the particular joint probability $p\left(A,B,1,2\right)$,
but he did not regard the states of $A$, $B$, $1$, and $2$ as
hidden variables themselves. Indeed, he assumed a deeper level of
hidden variables $\lambda$ with their own probability distribution
$\rho\left(\lambda\right)$, and he never considered that $\rho\left(\lambda\right)$
might not exist. Some interpretations of quantum theory, such as Bene's,
essentially identify the states of $A,B,1,2$ as hidden variables
and question the existence of $\rho\left(\lambda\right)$ itself.%
} Interesting though this loophole may be, we do not exploit it in
our own interpretation.

\subsubsection{Analysis of the EPR-Bohm Thought Experiment}

To the extent that our interpretation of quantum theory involves hidden
variables (see Section~\ref{subsub:Hidden-Variables-and-the-Irreducibility-of-Ontic-States}),
they are just the actual ontic states $\lambda=\set{\Psi,\Psi_{1},\Psi_{2}}$
of the two-particle parent system and of the individual one-particle
subsystems, where $\Psi$ is the entangled state defined in \eqref{eq:EPRPairStateVec}
and where $\Psi_{i}=\uparrow$ or $\downarrow$. Before any measurements
take place, the epistemic probabilities for the composite two-particle
parent system $1+2$ are 
\begin{equation}
\mathrm{Before}\colon\qquad p_{1+2}\left(\Psi\right)=1,\qquad p_{1+2}\left(\mathrm{states}\perp\Psi\right)=0,\label{eq:EPRPairEpProbsBefore}
\end{equation}
 the epistemic probabilities for the individual one-particle subsystems
$1$ and $2$ are%
\footnote{As we explained in our discussion surrounding \eqref{eq:EPRPairStateVecDegenBreaking},
there will unavoidably exist tiny degeneracy-breaking parameters in
the entangled two-particle parent system's ontic state $\Psi$ that---by
an appropriate choice of our coordinate system---pick out the spin-$z$
basis for each one-particle subsystem.%
} 
\begin{equation}
\mathrm{Before}\colon\qquad p_{1}\left(\uparrow\right)=p_{1}\left(\downarrow\right)=\frac{1}{2},\qquad p_{2}\left(\uparrow\right)=p_{2}\left(\downarrow\right)=\frac{1}{2},\label{eq:EPRIndivPartEpProbsBefore}
\end{equation}
 and the epistemic probabilities for the individual spin detectors
$A$ and $B$ are 
\begin{equation}
\mathrm{Before}\colon\qquad p_{A}\left(\quote{\emptyset}\right)=1,\qquad p_{B}\left(\quote{\emptyset}\right)=1,\label{eq:EPRAliceBobEpProbsBefore}
\end{equation}
 with all epistemic probabilities for orthogonal possible ontic states
being zero. Importantly, in accordance with our instantaneous kinematical
conditional probabilities \eqref{eq:QuantCondProbKinParentSub} (but
contrary to the hypothetical loophole that we described in Section~\ref{subsub:The-Additional-Loophole}),
the two particles $1$ and $2$ also have the initial \emph{joint}
epistemic probabilities 
\begin{equation}
\eqsbrace{\begin{aligned}\mathrm{Before}\colon\qquad & p_{1,2\given1+2}\left(\uparrow,\downarrow\given\Psi\right)=\frac{1}{2},\qquad p_{1,2\given1+2}\left(\downarrow,\uparrow\given\Psi\right)=\frac{1}{2},\\
 & p_{1,2\given1+2}\left(\uparrow,\uparrow\given\Psi\right)=0,\qquad p_{1,2\given1+2}\left(\downarrow,\downarrow\given\Psi\right)=0,
\end{aligned}
}\label{eq:EPRPairEpJointProbsBefore}
\end{equation}
 which ensure that the individual ontic states of the two particles
are initially anti-correlated. If we introduce a macroscopic measurement
apparatus $M$ (unavoidably coupled to an even larger environment
$E$, at the very least through irreversible thermal radiation \cite{HackermullerHornbergerBrezgerZeilingerArndt:2004dmwter})
that will locally record and compare the two readings on the spin
detectors after their pair of measurements, then before any measurements
have taken place, its only nonzero epistemic probability is 
\begin{equation}
\mathrm{Before}\colon\qquad p_{M}\left(\quote{\emptyset}\right)=1.\label{eq:EPRAppEpProbsBefore}
\end{equation}

It is easy to check that after a local spin measurement by either
one of the spin detectors $A$ or $B$ aligned along the $z$ direction,
and the consequent rapid decoherence, the objective epistemic state
of the composite two-particle system $1+2$ becomes 
\begin{equation}
\mathrm{After\ }A\binaryor B\colon\qquad p_{1+2}\left(\left(\uparrow\downarrow\right)^{\prime}\right)=p_{1+2}\left(\left(\downarrow\uparrow\right)^{\prime}\right)=\frac{1}{2},\qquad p_{1+2}\left(\left(\uparrow\uparrow\right)^{\prime}\right)=p_{1+2}\left(\left(\downarrow\downarrow\right)^{\prime}\right)=0,\label{eq:EPRPairEpProbsAfter}
\end{equation}
 where the primes denote corrections \eqref{eq:VonNeumannEnvOffDiagCorrections}---exponentially
small in the number of (spin-detector) + (environment) degrees of
freedom---to the two-particle basis states that are necessary to diagonalize
the final density matrix $\op{\rho}_{1+2}$ exactly. Given that the
ontic state of the composite system $1+2$ after the measurement is
$\left(\uparrow\downarrow\right)^{\prime}$, a simple application
of the kinematical parent-subsystem formula \eqref{eq:QuantCondProbKinParentSub}
shows that there is nearly unit probability that particle $1$ is
in the ontic state $\uparrow$ and particle $2$ is in the ontic state
$\downarrow$, whereas given that the ontic state of $1+2$ is $\left(\downarrow\uparrow\right)^{\prime}$
after the measurement, there is nearly unit probability that particle
$1$ is in the ontic state $\downarrow$ and particle $2$ is in the
ontic state $\uparrow$. These results are nicely consistent with
our initial joint epistemic probabilities \eqref{eq:EPRPairEpJointProbsBefore}
for the two particles.

Taking partial traces \eqref{eq:DefPartialTraces} of the density
matrix of the full system after both spin detectors $A$ and $B$
have performed their measurements, we find that the final epistemic
probabilities for the individual one-particle subsystems $1$ and
$2$ are given to very high numerical accuracy by 
\begin{equation}
\mathrm{After}\colon\qquad p_{1}\left(\uparrow^{\prime}\right)=p_{1}\left(\downarrow^{\prime}\right)=\frac{1}{2},\qquad p_{2}\left(\uparrow^{\prime}\right)=p_{2}\left(\downarrow^{\prime}\right)=\frac{1}{2},\label{eq:EPRIndivPartEpProbsAfter}
\end{equation}
 in approximate agreement with \eqref{eq:EPRIndivPartEpProbsBefore}
before the measurements took place, where, again, the primes denote
exponentially tiny corrections \eqref{eq:VonNeumannEnvOffDiagCorrections}
to the definitions of the ontic basis states.  Likewise, the final
\emph{joint} epistemic probabilities for the two particles are given
by 
\begin{equation}
\eqsbrace{\begin{aligned}\mathrm{After}\colon\qquad & p_{1,2\given1+2}\left(\uparrow^{\prime},\downarrow^{\prime}\given\left(\uparrow\downarrow\right)^{\prime}\right)=1,\qquad p_{1,2\given1+2}\left(\downarrow^{\prime},\uparrow^{\prime}\given\left(\uparrow\downarrow\right)^{\prime}\right)=0,\\
 & p_{1,2\given1+2}\left(\uparrow^{\prime},\uparrow^{\prime}\given\left(\uparrow\downarrow\right)^{\prime}\right)=0,\qquad p_{1,2\given1+2}\left(\downarrow^{\prime},\downarrow^{\prime}\given\left(\uparrow\downarrow\right)^{\prime}\right)=0,\\
 & p_{1,2\given1+2}\left(\uparrow^{\prime},\downarrow^{\prime}\given\left(\downarrow\uparrow\right)^{\prime}\right)=0,\qquad p_{1,2\given1+2}\left(\downarrow^{\prime},\uparrow^{\prime}\given\left(\downarrow\uparrow\right)^{\prime}\right)=1,\\
 & p_{1,2\given1+2}\left(\uparrow^{\prime},\uparrow^{\prime}\given\left(\downarrow\uparrow\right)^{\prime}\right)=0,\qquad p_{1,2\given1+2}\left(\downarrow^{\prime},\downarrow^{\prime}\given\left(\downarrow\uparrow\right)^{\prime}\right)=0.
\end{aligned}
}\label{eq:EPRPairEpJointProbsAfter}
\end{equation}
As for the individual spin detectors $A$ and $B$, we find the final
epistemic probabilities 
\begin{equation}
\eqsbrace{\begin{aligned}\mathrm{After}\colon\qquad & p_{A}\left(\left(\quote{\uparrow}\right)^{\prime}\right)=\frac{1}{2},\qquad p_{A}\left(\left(\quote{\downarrow}\right)^{\prime}\right)=\frac{1}{2},\\
 & p_{B}\left(\left(\quote{\downarrow}\right)^{\prime}\right)=\frac{1}{2},\qquad p_{B}\left(\left(\quote{\uparrow}\right)^{\prime}\right)=\frac{1}{2},
\end{aligned}
}\label{eq:EPRAliceBobEpProbsAfter}
\end{equation}
 and for the apparatus $M$ that locally reads off the pair of results,
\begin{equation}
\eqsbrace{\begin{aligned}\mathrm{After}\colon\qquad & p_{M}\left(\left(\quote{\mathrm{found\ }A\left(\quote{\uparrow}\right)\binaryand B\left(\quote{\downarrow}\right)}\right)^{\prime}\right)=\frac{1}{2},\\
 & p_{M}\left(\left(\quote{\mathrm{found\ }A\left(\quote{\downarrow}\right)\binaryand B\left(\quote{\uparrow}\right)}\right)^{\prime}\right)=\frac{1}{2}.
\end{aligned}
}\label{eq:EPRAppEpProbsAfter}
\end{equation}
 Notice that if our apparatus $M$ could immediately check the spin
detectors again and record their dials a second time, then its final
epistemic probabilities would be 
\begin{equation}
\eqsbrace{\begin{aligned}\mathrm{After}\colon\qquad & p_{M}\left(\left(\quote{\mathrm{found\ }A\left(\quote{\uparrow}\right)\binaryand B\left(\quote{\downarrow}\right)\ \emph{twice}}\right)^{\prime}\right)=\frac{1}{2},\\
 & p_{M}\left(\left(\quote{\mathrm{found\ }A\left(\quote{\downarrow}\right)\binaryand B\left(\quote{\uparrow}\right)\ \emph{twice}}\right)^{\prime}\right)=\frac{1}{2},
\end{aligned}
}\label{eq:EPRAppEpProbsAfterRecheck}
\end{equation}
 thereby implying that the measurement results are robust and persistent,
as expected.

The spin detectors $A$ and $B$ play the role of macroscopic, highly
classical intermediaries between the apparatus $M$ and the individual
particles $1$ and $2$. As a consequence, we would face troubling
metaphysical issues if we were unable to assert a strong connection
between the final actual ontic states of the spin detectors $A$ and
$B$ and the final actual ontic state of the apparatus $M$ that looks
at them. Fortunately, one can easily show that after environmental
decoherence, the composite system $W=1+2+A+B+M$ has the final epistemic
state 
\begin{equation}
\eqsbrace{\begin{aligned}\mathrm{After}\colon\qquad & p_{W}\left(\left(\quote{\mathrm{found\ }A\left(\quote{\uparrow}\right)\binaryand B\left(\quote{\downarrow}\right)\ \emph{twice}},A\left(\quote{\uparrow}\right),B\left(\quote{\downarrow}\right),\uparrow,\downarrow\right)^{\prime}\right)=\frac{1}{2},\\
 & p_{W}\left(\left(\quote{\mathrm{found\ }A\left(\quote{\downarrow}\right)\binaryand B\left(\quote{\uparrow}\right)\ \emph{twice}},A\left(\quote{\downarrow}\right),B\left(\quote{\uparrow}\right),\downarrow,\uparrow\right)^{\prime}\right)=\frac{1}{2},
\end{aligned}
}\label{eq:EPRCompositeEpProbsAfter}
\end{equation}
 where, as usual, primes denote exponentially suppressed corrections
\eqref{eq:VonNeumannEnvOffDiagCorrections} in the definitions of
the ontic states that are necessary to diagonalize the density matrix
of $W$ exactly. Hence, invoking again our kinematical parent-subsystem
probabilistic smoothness condition \eqref{eq:QuantCondProbKinParentSub},
we can say with near-certainty at the level of \emph{conditional}
probabilities that if, say, the final actual ontic state of the apparatus
$M$ happens to be $\left(\quote{\mathrm{found\ }A\left(\quote{\uparrow}\right)\binaryand B\left(\quote{\downarrow}\right)\ \emph{twice}}\right)^{\prime}$,
then the final actual ontic states of the two spin detectors are really
respectively $A\left(\quote{\uparrow}\right)$ and $B\left(\quote{\downarrow}\right)$
and, furthermore, that the final actual ontic states of the individual
particles are really $\uparrow$ for $1$ and $\downarrow$ for $2$.
These results are obviously all consistent with our expectations for
the post-measurement state of affairs.

\subsubsection{More General Alignments and Nonlocality}

Generating a violation of the Bell inequality \eqref{eq:BellIneq}
necessitates choosing \emph{nonparallel} detector alignments, so it
is no surprise that our investigation so far has not turned up any
clear evidence of nonlocality. Hence, let us now suppose that the
alignment $\svec a$ of spin detector $A$ is along the $x$ direction.
For clarity, we will suppress the primes on ontic states, remembering
always that decoherence is an imperfect process and that we should
expect exponentially small corrections to all our results.

If spin detector $A$ performs its measurement on particle $1$ first,
then after the resulting measurement-induced decoherence, a simple
calculation shows that particle $1$ now has possible ontic spin states
$\leftarrow$ (spin $x$ left) and $\rightarrow$ (spin $x$ right),
each with epistemic probability $1/2$. But the local unitary dynamics
of particle $2$ implies that its own possible ontic states remain
$\uparrow$ and $\downarrow$, each with epistemic probability $1/2$.
As expected, a subsequent local spin measurement by spin detector
$B$ aligned along the $z$ direction would yield the outcome $\quote{\uparrow}$
or $\quote{\downarrow}$, each with corresponding empirical outcome
probability $1/2$.

On the other hand, suppose that before performing its own local spin
measurement, spin detector $B$ changes its alignment $\svec b$ to
point along the $x$ direction just like the alignment $\svec a$
of spin detector $A$. In that case, if we use the local dynamics
of particle $2$ and spin detector $B$ in our dynamical conditional
probabilities \eqref{eq:ExactDynQuantCondProbs}, then we find that
regardless of whether the initial actual ontic state of particle $2$
was $\uparrow$ or $\downarrow$, the conditional probability for
$B$ to obtain either $\quote{\leftarrow}$ or $\quote{\rightarrow}$
for particle $2$ is $1/2$, independent of the result $\quote{\leftarrow}$
or $\quote{\rightarrow}$ obtained by spin detector $A$ for particle
$1$; this result is unsurprising because by tracing out spin detector
$A$ and particle $1$, we have chosen to ignore crucial information
about the necessary final anti-correlation between the spins of the
two particles. But if we instead condition on either the initial or
final actual ontic state of the composite parent two-particle system
$1+2$, which has evolved nonlocally during the experiment, then a
simple calculation yields with unit probability that the final measurement
outcome for the spin of particle $2$ by spin detector $B$ will indeed
show the correct anti-correlation. This nonlocal influence on the
final measurement result obtained by $B$ is in line with our claim
in Section~\ref{subsub:Hidden-Ontic-Level-Nonlocality} that our
quantum conditional probabilities safely allow for a hidden level
of nonlocality.

\subsection{The GHZ-Mermin Thought Experiment\label{sub:The-GHZ-Mermin-Thought-Experiment}}

In a 1989 paper \cite{GreenbergerHorneZeilinger:1989gbbt}, Greenberger,
Horne, and Zeilinger (GHZ) showed that certain generalizations of
the EPR-Bohm thought experiment involving three or more particles
with spin could not be described in terms of dynamically local hidden
variables. Unlike the arguments employed in Bell's theorem, the GHZ
result does not depend on the assumption of joint probability distributions
for the hidden variables, and can be captured by talking about just
a single measurement outcome.%
\footnote{Hardy \cite{Hardy:1992qmlrtlirt,Hardy:1993ntpwiaaes} has found certain
two-particle variants of the GHZ construction that likewise do not
depend on assumptions regarding joint probability distributions. %
}

Following the simplest version of the GHZ argument, described in detail
by Mermin in 1990 \cite{Mermin:1990qmr}, we consider a system of
three distinguishable spin-1/2 particles $1$, $2$, and $3$ in an
initial pure state represented by the entangled state vector 
\begin{equation}
\ket{\Psi_{\mathrm{GHZ}}}=\frac{1}{\sqrt{2}}\left(\ket{\uparrow\uparrow\uparrow}-\ket{\downarrow\downarrow\downarrow}\right),\label{eq:GHZStateVec}
\end{equation}
 where $\ket{\uparrow}$ and $\ket{\downarrow}$ are one-particle
eigenstates of the spin-$z$ operator $\op S_{z}$ and where again
we work in units for which $\hbar/2\equiv1$: 
\begin{equation}
\op S_{z}\ket{\uparrow}=+\ket{\uparrow},\qquad\op S_{z}\ket{\downarrow}=-\ket{\downarrow}.\label{eq:SpinzEigenstates}
\end{equation}
 It is then straightforward to show that the GHZ state vector described
by \eqref{eq:GHZStateVec} is a definite eigenstate of the three operators
$\op S_{1,x}\op S_{2,y}\op S_{3,y}$, $\op S_{1,y}\op S_{2,x}\op S_{3,y}$,
and $\op S_{1,y}\op S_{2,y}\op S_{3,x}$ with eigenvalue $+1$, 
\begin{equation}
\eqsbrace{\begin{aligned}\op S_{1,x}\op S_{2,y}\op S_{3,y}\ket{\Psi_{\mathrm{GHZ}}} & =+\ket{\Psi_{\mathrm{GHZ}}},\\
\op S_{1,y}\op S_{2,x}\op S_{3,y}\ket{\Psi_{\mathrm{GHZ}}} & =+\ket{\Psi_{\mathrm{GHZ}}},\\
\op S_{1,y}\op S_{2,y}\op S_{3,x}\ket{\Psi_{\mathrm{GHZ}}} & =+\ket{\Psi_{\mathrm{GHZ}}},
\end{aligned}
}\label{eq:GHZxyyEigenstate}
\end{equation}
 but is \emph{also} a definite eigenstate of the operator 
\[
\op S_{1,x}\op S_{2,x}\op S_{3,x}=-\left(\op S_{1,x}\op S_{2,y}\op S_{3,y}\right)\left(\op S_{1,y}\op S_{2,x}\op S_{3,y}\right)\left(\op S_{1,y}\op S_{2,y}\op S_{3,x}\right)
\]
 with eigenvalue $-1$, 
\begin{equation}
\op S_{1,x}\op S_{2,x}\op S_{3,x}\ket{\Psi_{\mathrm{GHZ}}}=-\ket{\Psi_{\mathrm{GHZ}}}.\label{eq:GHZxxxEigenstate}
\end{equation}
 One can carefully list all the possible ``local instructions''
that we could imagine somehow packaging along with each of the individual
particles to ensure agreement with the outcomes required by the first
three eigenvalue equations \eqref{eq:GHZxyyEigenstate}, but not one
of those sets of local instructions could then accommodate the fourth
eigenvalue equation \eqref{eq:GHZxxxEigenstate}. The consequence
is that any hidden variables accounting for all these possible measurement
results must change nonlocally during the course of the experiment
in order to ensure agreement with the necessary eigenvalue equations.

\begin{figure}
\begin{centering}
\begin{center}
\definecolor{ce1e1e1}{RGB}{225,225,225}
\definecolor{cffffff}{RGB}{255,255,255}

\begin{tikzpicture}[y=0.80pt,x=0.80pt,yscale=-1, inner sep=0pt, outer sep=0pt]
\begin{scope}[shift={(-75.36362,-46.1165)}]
  \path[fill=black] (306.47684,78.512917) node[above right] (text3005) {$1$};
  \path[fill=black] (345.35025,78.513031) node[above right] (text3005-3) {$2$};
  \path[fill=black] (259.05057,60.82629) node[above right] (text3005-0) {$\left|\Psi\right\rangle=\frac{1}{\sqrt{2}}\left(\left|\uparrow\uparrow\uparrow\right\rangle-\left|\downarrow\downarrow\downarrow\right\rangle\right)$};
  \path[->,>=latex,draw=black,line join=miter,line cap=butt,line width=0.800pt]
    (290.4619,74.0315) -- (182.8017,74.0315);
  \path[<-,>=latex,draw=black,line join=miter,line cap=butt,line width=0.800pt]
    (479.8062,74.0312) -- (372.1460,74.0312);
  \path[draw=black,fill=ce1e1e1,miter limit=4.00,line width=1.600pt,rounded
    corners=0.0000cm] (76.3636,47.1165) rectangle (143.9642,99.6947);
  \path[draw=black,fill=ce1e1e1,line join=miter,line cap=butt,line width=0.800pt]
    (167.0816,48.5448) -- (143.8451,63.3719) -- (143.8451,84.8380) --
    (167.3029,98.1160) -- cycle;
  \path[draw=black,fill=cffffff,miter limit=4.00,line width=1.600pt,rounded
    corners=0.0000cm] (92.5035,62.9293) rectangle (127.4689,84.1741);
  \path[fill=black] (103.69986,78.756378) node[above right] (text3075) {$A$};
  \path[xscale=-1.000,yscale=1.000,draw=black,fill=ce1e1e1,miter limit=4.00,line
    width=1.600pt,rounded corners=0.0000cm] (-582.2515,47.1170) rectangle
    (-514.6509,99.6952);
  \path[draw=black,fill=ce1e1e1,line join=miter,line cap=butt,line width=0.800pt]
    (491.5335,48.5452) -- (514.7700,63.3723) -- (514.7700,84.8384) --
    (491.3122,98.1164) -- cycle;
  \path[xscale=-1.000,yscale=1.000,draw=black,fill=cffffff,miter limit=4.00,line
    width=1.600pt,rounded corners=0.0000cm] (-566.1116,62.9297) rectangle
    (-531.1461,84.1745);
  \path[fill=black] (542.37311,78.756851) node[above right] (text3075-4) {$B$};
  \path[shift={(298.8748,135.49057)},draw=black,fill=cffffff,miter limit=4.00,line
    width=1.600pt]
    (25.6632,-61.5316)arc(0.000:180.000:2.817)arc(-180.000:0.000:2.817) -- cycle;
  \path[shift={(315.01228,135.49107)},draw=black,fill=cffffff,miter
    limit=4.00,line width=1.600pt]
    (25.6632,-61.5316)arc(0.000:180.000:2.817)arc(-180.000:0.000:2.817) -- cycle;
  \path[shift={(307.04548,149.42454)},draw=black,fill=cffffff,miter
    limit=4.00,line width=1.600pt]
    (25.6632,-61.5316)arc(0.000:180.000:2.817)arc(-180.000:0.000:2.817) -- cycle;
  \path[fill=black] (314.20554,99.685837) node[above right] (text3005-01) {$3$};
  \path[cm={{0.0,-1.0,1.0,0.0,(0.0,0.0)}},draw=black,fill=ce1e1e1,miter
    limit=4.00,line width=1.600pt,rounded corners=0.0000cm] (-245.0155,301.9373)
    rectangle (-177.4149,354.5156);
  \path[draw=black,fill=ce1e1e1,line join=miter,line cap=butt,line width=0.800pt]
    (303.7057,154.2576) -- (318.5328,177.4941) -- (339.9989,177.4941) --
    (353.2769,154.0363) -- cycle;
  \path[draw=black,fill=cffffff,miter limit=4.00,line width=1.600pt,rounded
    corners=0.0000cm] (310.5660,200.7389) rectangle (345.5314,221.9837);
  \path[fill=black] (322.76239,216.56601) node[above right] (text3075-6) {$C$};
  \path[->,>=latex,shift={(75.36362,46.1165)},draw=black,line join=miter,line cap=butt,line
    width=0.800pt] (254.0526,57.2343) -- (254.0526,98.8387);
\end{scope}

\end{tikzpicture}
\par\end{center}
\par\end{centering}

\caption{\label{fig:GHZMerminExperiment}The schematic set-up for the GHZ-Mermin
thought experiment, consisting of particles $1$, $2$, and $3$ together
with spin detectors $A$, $B$, and $C$.}
\end{figure}
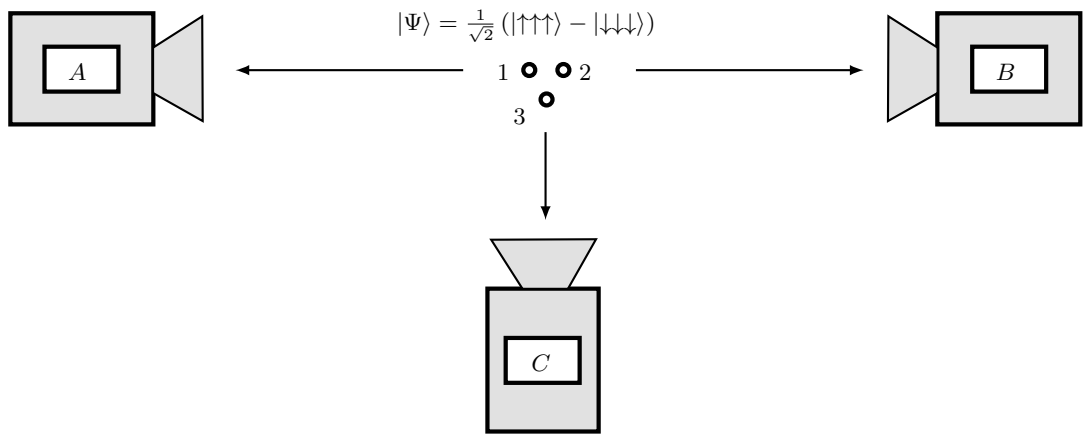

Lest one suppose that this issue involves only subatomic particles,
we could consider placing macroscopic spin detectors $A$, $B$, and
$C$ respectively near each of the three particles (see Figure~\ref{fig:GHZMerminExperiment}),
and include also a larger measurement apparatus $M$ that will ultimately
visit each spin detector at the end of the full experiment and then
compare their final readings, so that the initial state vector of
the full parent system $W=1+2+3+A+B+M$ is 
\begin{equation}
\ket{\Psi_{W}}=\frac{1}{\sqrt{2}}\left(\ket{\uparrow\uparrow\uparrow}-\ket{\downarrow\downarrow\downarrow}\right)\ket{A\left(\quote{\emptyset}\right)}\ket{B\left(\quote{\emptyset}\right)}\ket{C\left(\quote{\emptyset}\right)}\ket{M\left(\quote{\emptyset}\right)}.\label{eq:GHZStateVecWithSpinDetectors}
\end{equation}
If the spin detectors $A$, $B$, and $C$ perform their various local
measurements on each particle's spin-$x$ or spin-$y$ component,
then the hidden ontic states of these \emph{macroscopic} spin detectors
would need to interact nonlocally in order to ensure that when the
apparatus $M$ finally comes along to look at them, it would find
agreement with the four eigenvalue equations in \eqref{eq:GHZxyyEigenstate}
and \eqref{eq:GHZxxxEigenstate}.

\subsection{The Myrvold No-Go Theorem\label{sub:The-Myrvold-No-Go-Theorem}}

Building on a paper by Dickson and Clifton \cite{DicksonClifton:1998limi}
and employing arguments similar to Hardy \cite{Hardy:1992qmlrtlirt},
Myrvold \cite{Myrvold:2002mir} argued that modal interpretations
are fundamentally inconsistent with Lorentz invariance at a deeper
level than mere unobservable nonlocality, leading to much additional
work \cite{EarmanRuetsche:2005rimi,BerkovitzHemmo:2005miqmrr,Myrvold:2009cc}
in subsequent years to determine the implications of his result. Specifically,
Myrvold argued that one could not safely assume that a quantum system,
regardless of its size or complexity, always possesses a specific
actual ontic state beneath its epistemic state, because any such actual
ontic state could seemingly change under Lorentz transformations more
radically than, say, a four-vector does---for example, binary digits
$1$ and $0$ could switch, words written on a paper could change,
and true could become false. The only way to avoid this conclusion
would then be to break Lorentz symmetry in a fundamental way by asserting
the existence of a ``preferred'' Lorentz frame in which all ontic-state
assignments must be made.

Myrvold's claims, and those of Dickson and Clifton as well as Hardy,
rest on several invalid assumptions. Dickson and Clifton \cite{DicksonClifton:1998limi}
assume that the hidden ontic states admit certain joint epistemic
probabilities that are conditioned on \emph{multiple} disjoint systems
at an initial time, and such probabilities are not a part of our interpretation
of quantum theory, as we explained in Section~\ref{subsub:Probabilistic-and-Non-Probabilistic-Uncertainty}.%
\footnote{The same implicit assumption occurs in Section 9.2 of \cite{Vermaas:1999puqm}.%
} Similarly, one of Myrvold's arguments hinges on the assumed existence
of joint epistemic probabilities for two disjoint systems at \emph{two
separate times}, and again such probabilities are not present in our
interpretation. As Dickson and Clifton point out in Appendix B of
their paper, Hardy's argument also relies on several faulty assumptions
about ontic property assignments in modal interpretations.

A final argument of Myrvold---and repeated by Berkovitz and Hemmo
\cite{BerkovitzHemmo:2005miqmrr}---is based on inadmissible assumptions
about the proper way to implement Lorentz transformations in quantum
theory and the relationship between density matrices and ontic states
in our minimal modal interpretation. Myrvold's mistake is subtle,
so we go through his argument in detail.

\subsubsection{Measurements Performed at Different Times and in Different Inertial
Reference Frames}

Myrvold begins by considering the general case of a composite parent
system $W=1+A+2+B$ consisting of a pair of well-localized, isolated
subsystems $1$ and $2$ at spacelike separation and with respective
observables $\op X$ and $\op Y$, together with a macroscopic detector
$A$ localized near subsystem $1$ and a macroscopic detector $B$
localized near subsystem $2$. Letting $\op P_{X}\left(x\right)$
for outcome $x$ denote an arbitrary eigenprojector of $\op X$, and,
similarly, letting $\op P_{Y}\left(y\right)$ for outcome $y$ denote
an arbitrary eigenprojector for $\op Y$, Myrvold argues that the
probability that $B$ obtains $y$ in a measurement of $\op Y$, and
then, at a later time, $A$ obtains $x$ in a measurement of $\op X$,
is given by 
\begin{equation}
\Tr_{W}\left[\left(\op U_{1}^{\adj}\op P_{X}\left(x\right)\op U_{1}\right)\tensorprod\op P_{Y}\left(y\right)\op{\rho}_{W}\right],\label{eq:MyrvoldDiffTimeBornRule}
\end{equation}
 where $\op U_{1}$ is the unitary time-evolution operator for subsystem
$1$ over the time interval between the measurements. This formula
follows directly from a double application of the Born rule---once
for $\op Y$, and then again for $\op X$---provided that we perform
a Von Neumann-Lüders projection (see Appendix~\ref{subsub:The-Copenhagen-Interpretation})
immediately after the measurement of $\op Y$.

Because the entire experiment could have been described from a safe
distance by an observer $M$ who was moving at relativistic speed
relative to $W$ and for whom the measurements were simultaneous,
Myrvold then asserts on empirical grounds that \eqref{eq:MyrvoldDiffTimeBornRule}
must agree with 
\begin{equation}
\Tr_{W}\left[\op P_{X}^{\prime}\left(x\right)\tensorprod\op P_{Y}^{\prime}\left(y\right)\op{\rho}_{W}^{\prime}\right],\label{eq:MyrvoldSameTimeLorFrameBornRule}
\end{equation}
 where the primes refer to the fact that the operators inside the
trace have been appropriately Lorentz-boosted compared to their unprimed
counterparts. Extracting the unitary operators $\op U_{\Lambda,1}$
and $\op U_{\Lambda,2}$ that respectively implement the Lorentz boost
on the two eigenprojectors, 
\begin{equation}
\op P_{X}^{\prime}\left(x\right)=\op U_{\Lambda,1}\op P_{X}\left(x\right)\op U_{\Lambda,1}^{\adj},\qquad\op P_{Y}^{\prime}\left(y\right)=\op U_{\Lambda,2}\op P_{Y}\left(y\right)\op U_{\Lambda,2}^{\adj},\label{eq:MyrvoldLorTransfObsEigenproj}
\end{equation}
 Myrvold then uses the arbitrariness of the choice of eigenprojectors
to show that 
\begin{equation}
\op{\rho}_{W}^{\prime}=\left(\op U_{\Lambda,1}\tensorprod\op U_{\Lambda,2}\right)\left(\op U_{1}\tensorprod\op 1_{2}\right)\op{\rho}_{W}\left(\op U_{1}^{\adj}\tensorprod\op 1_{2}\right)\left(\op U_{\Lambda,1}^{\adj}\tensorprod\op U_{\Lambda,2}^{\adj}\right).\label{eq:MyrvoldLorTransfLawDensMatr}
\end{equation}
 In other words, Lorentz transformations act on density matrices in
a more subtle manner than other kinds of unitary transformations,
because they also involve changes in time---hence the factor of $\left(\op U_{1}\tensorprod\op 1_{2}\right)$
and its adjoint.

Indeed, a density matrix is not a local object, but is more properly
associated with an entire three-dimensional spacelike slice of constant
time. And so, when we perform a Lorentz transformation, we have to
be careful to account for the attendant change in time-slice, with
possibly different consequences for different well-localized subsystems.
In the present case, we account for this effect using the time-evolution
operator $\op U_{1}$. An important corollary for our interpretation
of quantum theory is that the ontic states of a spatially extended
composite system can change nontrivially under Lorentz transformations
that shift some of their subsystems across large temporal displacements,
although the same is obviously true even in classical physics; indeed,
under a sufficiently large Lorentz transformation, we can significantly
alter the ``current'' state of far-away galaxies.

For ``fixed'' modal interpretations like \cite{Bub:1992qmwpp},
for which there exists a \emph{universal} ontic basis independent
of density matrices or arbitrary choices of orthonormal basis, we
are free to drop the Lorentz-transformation operator $\left(\op U_{\Lambda,1}\tensorprod\op U_{\Lambda,2}\right)$
and its adjoint from \eqref{eq:MyrvoldLorTransfLawDensMatr}, because
it merely represents a physically meaningless unitary change of basis.%
\footnote{As Myrvold writes on p. 1776 of \cite{Myrvold:2002mir}: ``Now, the
Lorentz boost operators $\Lambda,\Lambda^{\prime}$ merely effect
a transformation from a state given with respect to one reference
frame's coordinates to one given with respect to another reference
frame's coordinates. In what follows, it will be more convenient to
utilize the coordinate basis of one reference frame $\Sigma$ for
all states, even those on hypersurfaces that are not equal-time hyperplanes
for $\Sigma$.''%
} However, in \emph{our} modal interpretation, the full right-hand-side
of \eqref{eq:MyrvoldLorTransfLawDensMatr} is necessary for yielding
the correct density matrix $\op{\rho}_{W}^{\prime}$---and the correct
ontic basis---for the parent system $W$ in the Lorentz-boosted frame.
That is, without including the Lorentz-transformation operator $\left(\op U_{\Lambda,1}\tensorprod\op U_{\Lambda,2}\right)$,
the left-hand-side simply wouldn't be the density matrix of any relevant
physical system, and so its eigenstates would have no ontological
meaning. Myrvold neglects this important fact completely, and we will
see that it invalidates the applicability of his no-go theorem to
our interpretation of quantum theory.

\subsubsection{Myrvold's Thought Experiment}

Following Myrvold \cite{Myrvold:2002mir}, we next examine a scenario
inspired by the EPR-Bohm thought experiment of Section~\ref{sub:The-EPR-Bohm-Thought-Experiment-and-Bell's-Theorem}
but seen from several different inertial reference frames, each corresponding
to a different foliation of spacetime into three-dimensional slices
of constant time.

Consider a composite four-state system $W=1+A+2+B$ consisting of
two localized qubits $1$ and $2$ that each have orthonormal basis
states $\ket +$ and $\ket -$ and that lie very far apart in space
at sharply defined positions. As in the later part of Myrvold's papers,
as well as in Berkovitz and Hemmo's treatment \cite{BerkovitzHemmo:2005miqmrr},
consider also a macroscopic qubit detector $A$ located near qubit
$1$ and a similar macroscopic qubit detector $B$ located near qubit
$2$, where the qubit detector $A$ has ontic states $A\left(\quote{\emptyset}\right)$,
$A\left(\quote +\right)$, and $A\left(\quote -\right)$ respectively
describing its initial state and its allowed measurement outcomes
for qubit $1$, and, similarly, where the qubit detector $B$ has
ontic states $B\left(\quote{\emptyset}\right)$, $B\left(\quote +\right)$,
and $B\left(\quote -\right)$ respectively describing its initial
state and its allowed measurement outcomes for qubit $2$. We imagine
that the local detectors measure their respective qubits approximately
continuously, so that they always show the correct readings on their
dials.

As with the spin detectors in the EPR-Bohm thought experiment analyzed
in Section~\ref{sub:The-EPR-Bohm-Thought-Experiment-and-Bell's-Theorem},
and in parallel with our discussion surrounding \eqref{eq:EPRCompositeEpProbsAfter},
the qubit detectors $A$ and $B$ will play an important role as macroscopic,
highly classical intermediaries between observers and the quantum
qubits themselves. In particular, the presence of $A$ and $B$ makes
clear our eventual metaphysical need for joint statements about the
final actual ontic states of the various subsystems at the end of
the experiment. At the very least, the final results recorded by observers
should be correlated with the final actual ontic states of the detectors
$A$ and $B$.

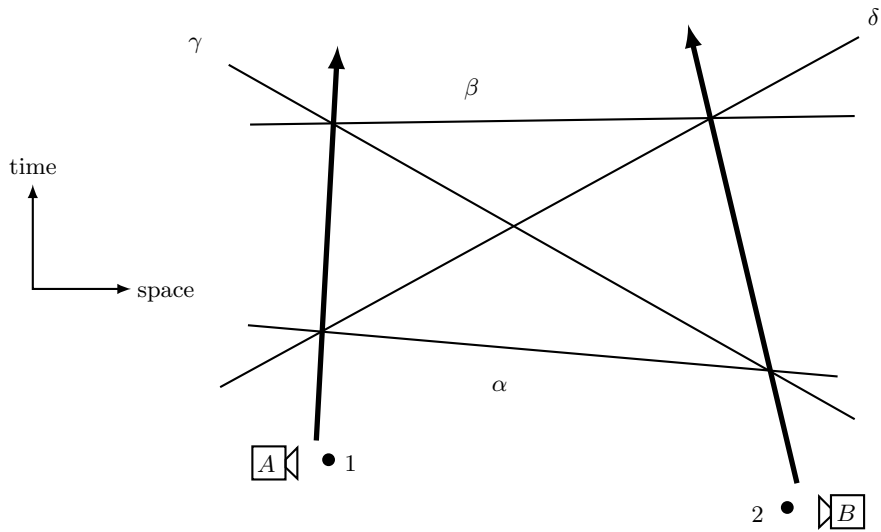
\begin{figure}
\begin{centering}
\begin{center}
\begin{tikzpicture}[y=0.80pt,x=0.80pt,yscale=-1, inner sep=0pt, outer sep=0pt]
\begin{scope}[shift={(-128.41829,-105.21349)}]
  \path[draw=black,line join=miter,line cap=butt,line width=0.800pt]
    (217.1828,256.3620) -- (495.9849,280.6056);
  \path[draw=black,line join=miter,line cap=butt,line width=0.800pt]
    (218.1929,161.4076) -- (504.0661,157.3670);
  \path[draw=black,line join=miter,line cap=butt,line width=0.800pt]
    (208.0914,133.1234) -- (504.0661,300.8188);
  \path[draw=black,line join=miter,line cap=butt,line width=0.800pt]
    (506.0966,119.9914) -- (204.0508,285.6564);
  \path[<-,>=latex,draw=black,line join=miter,line cap=butt,miter limit=4.00,line
    width=2.000pt] (259.6092,124.0320) -- (249.5077,310.9102);
  \path[<-,>=latex,draw=black,line join=miter,line cap=butt,miter limit=4.00,line
    width=2.000pt] (425.2742,113.9305) -- (476.7920,331.1133);
  \path[<->,>=latex,draw=black,line join=miter,line cap=butt,line width=0.800pt]
    (115.4620,189.6919) -- (115.4620,239.1894) -- (161.9290,239.1894);
  \path[fill=black] (104.28938,184.62085) node[above right] (text7243) {time};
  \path[fill=black] (164.9001,245.05234) node[above right] (text7243-6) {space};
  \path[fill=black] (332.53467,287.6615) node[above right] (text7243-9)
    {$\alpha$};
  \path[fill=black] (319.43784,149.45163) node[above right] (text7243-9-3)
    {$\beta$};
  \path[fill=black] (189.01645,127.22828) node[above right] (text7243-9-1)
    {$\gamma$};
  \path[fill=black] (510.2348,114.0963) node[above right] (text7243-9-34)
    {$\delta$};
  \path[draw=black,fill=black,miter limit=4.00,line width=0.800pt]
    (257.8414,320.0016) .. controls (257.8414,321.3963) and (256.7108,322.5270) ..
    (255.3161,322.5270) .. controls (253.9213,322.5270) and (252.7907,321.3963) ..
    (252.7907,320.0016) .. controls (252.7907,318.6069) and (253.9213,317.4762) ..
    (255.3161,317.4762) .. controls (256.7108,317.4762) and (257.8414,318.6069) ..
    (257.8414,320.0016) -- cycle;
  \path[draw=black,miter limit=4.00,line width=0.800pt,rounded corners=0.0000cm]
    (219.2843,313.6881) rectangle (234.9417,329.3455);
  \path[fill=black] (262.83881,325.2182) node[above right] (text7243-9-6) {$1$};
  \path[shift={(216.87981,22.60576)},draw=black,fill=black,miter limit=4.00,line
    width=0.800pt] (257.8414,320.0016) .. controls (257.8414,321.3963) and
    (256.7108,322.5270) .. (255.3161,322.5270) .. controls (253.9213,322.5270) and
    (252.7907,321.3963) .. (252.7907,320.0016) .. controls (252.7907,318.6069) and
    (253.9213,317.4762) .. (255.3161,317.4762) .. controls (256.7108,317.4762) and
    (257.8414,318.6069) .. (257.8414,320.0016) -- cycle;
  \path[fill=black] (455.20212,349.3392) node[above right] (text7243-9-6-1) {$2$};
  \path[draw=black,line join=miter,line cap=butt,line width=0.800pt]
    (234.9894,319.8666) -- (240.5452,314.3108) -- (240.5452,328.9580) --
    (234.9894,323.6547) -- cycle;
  \path[fill=black] (221.48963,326.02249) node[above right] (text3154) {$A$};
  \path[xscale=-1.000,yscale=1.000,draw=black,miter limit=4.00,line
    width=0.800pt,rounded corners=0.0000cm] (-508.5120,336.7867) rectangle
    (-492.8547,352.4440);
  \path[draw=black,line join=miter,line cap=butt,line width=0.800pt]
    (492.8070,342.9652) -- (487.2511,337.4094) -- (487.2511,352.0566) --
    (492.8070,346.7533) -- cycle;
  \path[fill=black] (495.48337,349.20344) node[above right] (text3179) {$B$};
\end{scope}
\end{tikzpicture}
\par\end{center}
\par\end{centering}

\caption{\label{fig:MyrvoldSpacetimeDiagram}A spacetime diagram representing
the Myrvold thought experiment, consisting of particles $1$ and $2$
together with qubit detectors $A$ and $B$. The constant-time slices
$\alpha$, $\beta$, $\gamma$, and $\delta$ are also shown.}
\end{figure}

Suppose that on one particular three-dimensional constant-time slice,
which we will denote by $\alpha$,%
\footnote{This label $\alpha$ should not be confused with the coordinate parameterization
we used for subsystem spaces in Section~\ref{sub:Subsystem-Spaces}.%
} the parent system $W=1+A+2+B$ is in a pure state represented by
the post-measurement state vector 
\begin{equation}
\eqsbrace{\begin{aligned}\ket{\Psi_{W}\left(\alpha\right)}= & +\frac{1}{\sqrt{12}}\ket{+_{1}}\ket{A\left(\quote +\right)}\ket{+_{2}}\ket{B\left(\quote +\right)}\\
 & -\frac{1}{\sqrt{12}}\ket{+_{1}}\ket{A\left(\quote +\right)}\ket{-_{2}}\ket{B\left(\quote -\right)}\\
 & -\frac{1}{\sqrt{12}}\ket{-_{1}}\ket{A\left(\quote -\right)}\ket{+_{2}}\ket{B\left(\quote +\right)}\\
 & -\sqrt{\frac{9}{12}}\ket{-_{1}}\ket{A\left(\quote -\right)}\ket{-_{2}}\ket{B\left(\quote -\right)}.
\end{aligned}
}\label{eq:MyrvoldStateVecAlpha}
\end{equation}
 (See Figure~\ref{fig:MyrvoldSpacetimeDiagram}.) Taking the partial
trace \eqref{eq:DefPartialTraces} to obtain the corresponding reduced
density matrix $\op{\rho}_{1+2}\left(\alpha\right)$ for the composite
two-qubit subsystem $1+2$, we find that it describes the possible
ontic states $\left(+_{1}+_{2}\right)$, $\left(+_{1}-_{2}\right)$,
$\left(-_{1}+_{2}\right)$, and $\left(-_{1}-_{2}\right)$, each with
significant nonzero epistemic probabilities given respectively by
$1/12$, $1/12$, $1/12$, $9/12$; that is, 
\begin{equation}
\op{\rho}_{1+2}\left(\alpha\right)=\frac{1}{12}\ket{+_{1}}\ket{+_{2}}\bra{+_{1}}\bra{+_{2}}+\frac{1}{12}\ket{+_{1}}\ket{-_{2}}\bra{+_{1}}\bra{-_{2}}+\frac{1}{12}\ket{-_{1}}\ket{+_{2}}\bra{-_{1}}\bra{+_{2}}+\frac{9}{12}\ket{-_{1}}\ket{-_{2}}\bra{-_{1}}\bra{-_{2}},\label{eq:MyrvoldDensMatrixAlpha}
\end{equation}
 where, as explained in our discussion surrounding \eqref{eq:EPRPairStateVecDegenBreaking},
we implicitly assume small degeneracy-breaking corrections to avoid
the measure-zero case of any \emph{exactly} equal probability eigenvalues.
According to our interpretation of quantum theory, the actual ontic
state of the composite two-qubit subsystem $1+2$ on the constant-time
slice $\alpha$ could be any one of the four possible ontic states
$\left(+_{1}+_{2}\right)$, $\left(+_{1}-_{2}\right)$, $\left(-_{1}+_{2}\right)$,
and $\left(-_{1}-_{2}\right)$.

Next, Myrvold supposes that the system undergoes unitary time evolution
governed by a local time-evolution operator $\op U_{1+A}\tensorprod\op U_{2+B}$
given by the following Hadamard transformations: 
\begin{equation}
\eqsbrace{\begin{aligned}\op U_{1+A}\ket{+_{1}}\ket{A\left(\quote +\right)} & =\frac{1}{\sqrt{2}}\left(\ket{+_{1}}\ket{A\left(\quote +\right)}+\ket{-_{1}}\ket{A\left(\quote -\right)}\right),\\
\op U_{1+A}\ket{-_{1}}\ket{A\left(\quote -\right)} & =\frac{1}{\sqrt{2}}\left(\ket{+_{1}}\ket{A\left(\quote +\right)}-\ket{-_{1}}\ket{A\left(\quote -\right)}\right),\\
\op U_{2+B}\ket{+_{2}}\ket{B\left(\quote +\right)} & =\frac{1}{\sqrt{2}}\left(\ket{+_{2}}\ket{B\left(\quote +\right)}+\ket{-_{2}}\ket{B\left(\quote -\right)}\right),\\
\op U_{2+B}\ket{-_{2}}\ket{B\left(\quote -\right)} & =\frac{1}{\sqrt{2}}\left(\ket{+_{2}}\ket{B\left(\quote +\right)}+\ket{-_{2}}\ket{B\left(\quote -\right)}\right).
\end{aligned}
}\label{eq:MyrvoldHadamard}
\end{equation}
 (Notice that the entanglement between each qubit and its corresponding
detector implies that the evolution of the qubit alone \emph{is not
unitary}, although it can be captured by a suitable linear CPT dynamical
mapping.) The pure state of the parent system $W=1+A+2+B$ then ends
up represented by the following state vector on the later constant-time
slice $\beta$: 
\begin{equation}
\eqsbrace{\begin{aligned}\ket{\Psi_{W}\left(\beta\right)}= & -\frac{1}{\sqrt{3}}\ket{+_{1}}\ket{A\left(\quote +\right)}\ket{+_{2}}\ket{B\left(\quote +\right)}\\
 & +\frac{1}{\sqrt{3}}\ket{+_{1}}\ket{A\left(\quote +\right)}\ket{-_{2}}\ket{B\left(\quote -\right)}\\
 & +\frac{1}{\sqrt{3}}\ket{-_{1}}\ket{A\left(\quote -\right)}\ket{+_{2}}\ket{B\left(\quote +\right)}.
\end{aligned}
}\label{eq:MyrvoldStateVecBeta}
\end{equation}
 The possible ontic states of the composite two-qubit subsystem $1+2$
are the same as before, but now with epistemic probability zero for
$\left(-_{1}-_{2}\right)$: 
\begin{equation}
\op{\rho}_{1+2}\left(\beta\right)=\frac{1}{3}\ket{+_{1}}\ket{+_{2}}\bra{+_{1}}\bra{+_{2}}+\frac{1}{3}\ket{+_{1}}\ket{-_{2}}\bra{+_{1}}\bra{-_{2}}+\frac{1}{3}\ket{-_{1}}\ket{+_{2}}\bra{-_{1}}\bra{+_{2}}.\label{eq:MyrvoldDensMatrixBeta}
\end{equation}

Now Myrvold considers a Lorentz-boosted (and therefore ``tilted'')
spacelike constant-time slice $\gamma$ intersecting the localized
composite detector-qubit subsystem $1+A$ at the \emph{final} time
and intersecting the detector-qubit subsystem $2+B$ at the \emph{initial}
time. Then, looking back at the unitary transformation law \eqref{eq:MyrvoldLorTransfLawDensMatr},
we see that we can obtain the state vector representing the parent
system's pure state on the constant-time slice $\gamma$ by starting
with the pure state \eqref{eq:MyrvoldStateVecAlpha} on the constant-time
slice $\alpha$, carrying out the local unitary time evolution just
on $1+A$, and then acting with the Lorentz-boost operator $\left(\op U_{\Lambda,1+A}\tensorprod\op U_{\Lambda,2+B}\right)$,
with the result being 
\begin{equation}
\eqsbrace{\begin{aligned}\ket{\Psi_{W}^{\prime}\left(\gamma\right)}= & -\sqrt{\frac{2}{3}}\op U_{\Lambda,1+A}\left(\ket{+_{1}}\ket{A\left(\quote +\right)}\right)\op U_{\Lambda,2+B}\left(\ket{-_{2}}\ket{B\left(\quote -\right)}\right)\\
 & +\frac{1}{\sqrt{6}}\op U_{\Lambda,1+A}\left(\ket{-_{1}}\ket{A\left(\quote -\right)}\right)\op U_{\Lambda,2+B}\left(\ket{+_{2}}\ket{B\left(\quote +\right)}\right)\\
 & +\frac{1}{\sqrt{6}}\op U_{\Lambda,1+A}\left(\ket{-_{1}}\ket{A\left(\quote -\right)}\right)\op U_{\Lambda,2+B}\left(\ket{-_{2}}\ket{B\left(\quote -\right)}\right).
\end{aligned}
}\label{eq:MyrvoldStateVecGamma}
\end{equation}

\emph{Crucially}, Myrvold \emph{drops} the Lorentz-transformation
operators $\op U_{\Lambda,1+A}$ and $\op U_{\Lambda,2+B}$ from his
corresponding expression for \eqref{eq:MyrvoldStateVecGamma}, despite
that fact that, at the very least, $\op U_{\Lambda,1+A}$ has a significant
temporal effect on the composite detector-qubit subsystem $1+A$,
which is located far from the spacetime-center of the Lorentz boost.
Moreover, because $\op U_{\Lambda,1+A}$ implements a significant
time shift on $1+A$, and in light of the entanglement-inducing Hadamard
time evolution \eqref{eq:MyrvoldHadamard}, $\op U_{\Lambda,1+A}$
is certainly not factorizable into separate transformation operators
for qubit $1$ and detector $A$, meaning that qubit $1$ will necessarily
change according to a \emph{non-unitary} dynamical mapping. It is
therefore \emph{impossible} to arrive at Myrvold's subsequent conclusion
that the reduced density matrix of the composite two-qubit system
$1+2$ on this new constant-time slice $\gamma$ is given by 
\begin{equation}
\op{\rho}_{1+2}\left(\gamma\right)=\frac{2}{3}\ket{+-}\bra{+-}+\frac{1}{6}\ket{-+}\bra{-+}+\frac{1}{6}\ket{--}\bra{--},\label{eq:MyrvoldDensMatrixGamma}
\end{equation}
 which has zero epistemic probability for $\left(+_{1}+_{2}\right)$.
The same erroneous reasoning applies to Myrvold's second choice of
Lorentz-boosted constant-time slice $\delta$ intersecting the localized
composite detector-qubit subsystem $1+A$ at the \emph{initial} time
and the detector-qubit subsystem $2+B$ at the \emph{final} time.

Had these results held up, Myrvold would have been able to compare
ontic states on each of the constant-time slices $\alpha$, $\delta$,
$\gamma$, and $\beta$ and thereby arrive at the conclusion that
the composite two-qubit system must be in the final ontic state $\left(-_{1}-_{2}\right)$
on the constant-time slice $\beta$, in seeming contradiction with
the epistemic probabilities encoded in the density matrix \eqref{eq:MyrvoldDensMatrixBeta}.
However, without these central results, Myrvold's no-go theorem breaks
down.

\subsection{Quantum Theory and Classical Gauge Theories\label{sub:Quantum-Theory-and-Classical-Gauge-Theories}}

Suppose that we were to imagine reifying all of the \emph{possible}
ontic states defined by each system's density matrix as simultaneous
\emph{actual} ontic states in the sense of the many-worlds interpretation.%
\footnote{Observe that the eigenbasis of each system's density matrix therefore
defines a preferred basis \emph{for that system alone}. We \emph{do
not} assume the sort of universe-spanning preferred basis shared by
all systems that is featured in the traditional many-worlds interpretation;
such a universe-spanning preferred basis would lead to new forms of
nonlocality, as we describe in Section~\ref{subsub:Nonlocality in the Everett-DeWitt Many-Worlds Interpretation}.%
} Then because every density matrix \emph{as a whole} evolves locally,
no nonlocal dynamics between the actual ontic states is necessary
and our interpretation of quantum theory becomes manifestly dynamically
local: For example, each spin detector in the EPR-Bohm or GHZ-Mermin
thought experiments possesses all its possible results in actuality,
and the larger measurement apparatus locally ``splits'' into all
the various possibilities when it visits each spin detector and looks
at the detector's final reading.

To help make sense of this step of adding unphysical actual ontic
states into our interpretation of quantum theory, recall the story
of classical gauge theories, and specifically the example of the
Maxwell theory of electromagnetism. The physical states of the theory
involve only \emph{two} possible polarizations, corresponding fundamentally
to the two physical spin states of the underlying species of massless
spin-1 gauge boson that we call the photon.

In a unitary gauge, meaning a choice of gauge in which we describe
the theory using just two polarization states for the gauge field
$A^{\mu}$, the theory appears to be nonlocal and not Lorentz covariant.
For instance, in Weyl-Coulomb gauge, we impose the two manifestly
non-covariant gauge conditions $A^{0}=0$ and $\divergence\svec A=0$,
which together eliminate the timelike and longitudinal polarizations
of the gauge field $A^{\mu}$ and thereby leave intact just the two
physical transverse states orthogonal to the direction of wave propagation.
In this ``ontologically correct'' choice of gauge, the vector potential
$\svec A$ becomes a nonlocal action-at-a-distance function of the
distribution of currents over all of three-dimensional space \cite{Jackson:1998ce},
and is no longer part of a true Lorentz four-vector \cite{Weinberg:1996qtfi}.

However, if we \emph{formally} add an additional unphysical polarization
state by replacing the two Weyl-Coulomb gauge conditions $A^{0}=0$
and $\divergence\svec A=0$ with the weaker but Lorentz-covariant
single condition $\partial_{\mu}A^{\mu}=\left(1/c\right)\partial A^{0}/\partial t-\divergence\svec A=0$
that defines Lorenz gauge, then we obtain a manifestly local, Lorentz-covariant
description of the Maxwell theory in which all the gauge potentials
become true Lorentz four-vectors given by \emph{causal} integrals
over appropriately \emph{time-delayed} distributions of charges and
currents. Because the Maxwell theory therefore has a mathematical
description in which the nonlocality of the ontologically correct
unitary gauge disappears, we see that apparent nonlocality of the
Maxwell theory is totally harmless.

Nonetheless, switching to Lorenz gauge does not imply that the extra
unphysical polarization state that we have formally added to the description
achieves a \emph{true} ontological status, and so we take precisely
the same view toward the addition of extra unphysical ``actual''
ontic states that would provide a manifestly local mathematical description
of our own interpretation of quantum theory. Just as different choices
of gauge for a given classical gauge theory make different calculations
or properties of the theory more or less manifest---each choice of
gauge inevitably involves trade-offs---we see that switching from
the ``unitary gauge'' corresponding to our interpretation of quantum
theory to the ``Lorenz gauge'' in which it looks more like a density-matrix-centered
version of the many-worlds interpretation makes the locality and Lorentz
covariance of the interpretation more manifest at the cost of obscuring
the interpretation's underlying ontology and the meaning of probability.

In this analogy, gauge potentials $A^{\mu}$ correspond to ontic states
$\Psi_{i}$, which can similarly undergo nonlocal changes. Gauge transformations
$A^{\mu}\mapsto A^{\mu}+\partial^{\mu}\lambda$ describe an unobservable
change in the gauge theory's ontology, and are analogous in our interpretation
of quantum theory to carrying out a simultaneous but fundamentally
unobservable shift in the hidden actual ontic states (and thus also
the memories) of all our systems, such as by reassigning each system's
hidden (``private'') actual ontic state to one of the other possible
ontic states defined by the system's density matrix. Just as all observable
predictions of electromagnetism can be expressed in terms of gauge-invariant,
dynamically local quantities like electric fields $\svec E$ and
magnetic fields $\svec B$, all outwardly observable statistical
predictions of quantum theory ultimately derive from density matrices
$\op{\rho}$, which are insensitive to our identification of each
system's hidden actual ontic state from among its possible ontic states
and which always evolve locally in accordance with the no-communication
theorem \cite{Hall:1987imnlqm,PeresTerno:2004qirt}.

Seen from this perspective, we can also better understand why it is
so challenging \cite{Hsu:2012oopqm} to make sense of a many-worlds-type
interpretation as an ontologically and epistemologically reasonable
interpretation of quantum theory: Attempting to do so leads to as
much metaphysical difficulty as trying to make sense of the Lorenz
gauge of Maxwell electromagnetism as an ``ontologically correct interpretation''
of the Maxwell theory.%
\footnote{Indeed, in large part for this reason, some textbooks \cite{Weinberg:1996qtfi}
develop quantum electrodynamics fundamentally from the perspective
of Weyl-Coulomb gauge.%
} Hence, taking a lesson from classical gauge theories, we propose
instead regarding many-worlds-type interpretations as merely a convenient
mathematical tool---a particular ``gauge choice''---for establishing
definitively that a given ``unitary-gauge'' interpretation of quantum
theory like our own is ultimately consistent with locality and Lorentz
invariance.

\section{Conclusion\label{sec:Conclusion}}

\subsection{Summary\label{sub:Summary}}

In this paper, we have introduced what we call the minimal modal interpretation of quantum theory.
Our interpretation consists of several parsimonious ingredients:
\begin{enumerate}
\item In \eqref{eq:OntStateVecCorresp}, we define ontic states $\Psi_{i}$,
meaning the states of the given system as it could actually exist
in reality, in terms of arbitrary (unit-norm) state vectors $\ket{\Psi_{i}}$
in the system's Hilbert space $\mathcal{H}$, 
\[
\Psi_{i}\exchange\ket{\Psi_{i}}\in\mathcal{H}\ \left(\mathrm{up\ to\ overall\ phase}\right),
\]
 and we define epistemic states $\setindexed{\left(p_{i},\Psi_{i}\right)}i$
as probability distributions over sets of possible ontic states, 
\[
\setindexed{\left(p_{i},\Psi_{i}\right)}i,\qquad p_{i}\in\left[0,1\right],
\]
 where these definitions parallel the corresponding notions from classical
physics. We translate logical mutual exclusivity of ontic states $\Psi_{i}$
as mutual orthogonality of state vectors $\ket{\Psi_{i}}$, and we
make a distinction between subjective epistemic states (proper mixtures)
and objective epistemic states (improper mixtures): The former arise
from classical ignorance and are uncontroversial, whereas the latter
arise from quantum entanglements to other systems and do not have
a widely accepted \emph{a priori} meaning outside of our interpretation
of quantum theory. Indeed, the problem of interpreting objective epistemic
states may well be unavoidable: Essentially all realistic systems
are entangled to other systems to a nonzero degree and thus cannot
be described exactly by pure states or by purely subjective epistemic
states.%
\footnote{As we explain in Section~\ref{sub:Comparison-with-Other-Interpretations-of-Quantum-Theory},
there are reasons to be skeptical of the common assumption that one
can always assign an exactly pure state or purely subjective epistemic
state to ``the universe as a whole.''%
}
\item We posit a correspondence \eqref{eq:EpStateDensMatrixCorresp} between
objective epistemic states $\setindexed{\left(p_{i},\Psi_{i}\right)}i$
and density matrices $\op{\rho}$: 
\[
\setindexed{\left(p_{i},\Psi_{i}\right)}i\exchange\op{\rho}=\sum_{i}p_{i}\ket{\Psi_{i}}\bra{\Psi_{i}}.
\]
 The relationship between subjective epistemic states and density
matrices is not as strict, as we explain in Section~\ref{sub:Subjective-Density-Matrices-and-Proper-Mixtures}.
\item We invoke the partial-trace operation $\op{\rho}_{Q}\equiv\Tr_{E}\left[\op{\rho}_{Q+E}\right]$,
motivated and defined in \eqref{eq:DefPartialTraces} without appeals
to the Born rule or Born-rule-based averages, to relate the density
matrix (and thus the epistemic state) of any subsystem $Q$ to that
of any parent system $W=Q+E$.
\item We introduce a general class of quantum conditional probabilities
\eqref{eq:DefQuantCondProbGenFromTrParent}, 
\begin{align*}
p_{Q_{1},\dotsc,Q_{n}\given W}\left(i_{1},\dotsc,i_{n};t^{\prime}\given w;t\right) & \equiv\Tr_{W}\left[\left(\op P_{Q_{1}}\left(i_{1};t^{\prime}\right)\tensorprod\dotsm\tensorprod\op P_{Q_{n}}\left(i_{n};t^{\prime}\right)\right)\mathcal{E}_{W}^{t^{\prime}\from t}\left[\op P_{W}\left(w;t\right)\right]\right]\\
 & \sim\Tr\left[\op P_{i_{1}}\left(t^{\prime}\right)\dotsm\op P_{i_{n}}\left(t^{\prime}\right)\mathcal{E}\left[\op P_{w}\left(t\right)\right]\right],
\end{align*}
 relating the possible ontic states of any partitioning collection
of mutually disjoint subsystems $Q_{1},\dotsc,Q_{n}$ to the possible
ontic states of a corresponding parent system $W=Q_{1}+\dotsb+Q_{n}$
whose own dynamics is governed by a linear completely-positive-trace-preserving
(``CPT'') dynamical mapping $\mathcal{E}_{W}^{t^{\prime}\from t}\left[\cdot\right]$
over the given time interval $t^{\prime}-t$. Here $\op P_{W}\left(w;t\right)$
denotes the projection operator onto the eigenstate $\ket{\Psi_{W}\left(w;t\right)}$
of the density matrix $\op{\rho}_{W}\left(t\right)$ of the parent
system $W$ at the initial time $t$, and, similarly, $\op P_{Q_{\alpha}}\left(i;t^{\prime}\right)$
denotes the projection operator onto the eigenstate $\ket{\Psi_{Q_{\alpha}}\left(i;t^{\prime}\right)}$
of the density matrix $\op{\rho}_{Q_{\alpha}}\left(t^{\prime}\right)$
of the subsystem $Q_{\alpha}$ at the final time $t^{\prime}$ for
$\alpha=1,\dotsc,n$. In a rough sense, the dynamical mapping $\mathcal{E}_{W}^{t^{\prime}\from t}\left[\cdot\right]$
acts as a parallel-transport superoperator that moves the parent-system
projection operator $\op P_{W}\left(w;t\right)$ from $t$ to $t^{\prime}$
before we compare it with the subsystem projection operators $\op P_{Q_{\alpha}}\left(i;t^{\prime}\right)$.
As special cases, these quantum conditional probabilities provide
a \emph{kinematical} smoothing relationship \eqref{eq:QuantCondProbKinParentSub},
\begin{align*}
p_{Q_{1},\dotsc,Q_{n}\given W}\left(i_{1},\dotsc,i_{n}\given w\right) & \equiv\Tr_{W}\left[\left(\op P_{Q_{1}}\left(i_{1}\right)\tensorprod\dotsm\tensorprod\op P_{Q_{n}}\left(i_{n}\right)\right)\op P_{W}\left(w\right)\right]\\
 & =\bra{\Psi_{W,w}}\left(\ket{\Psi_{Q_{1},i_{1}}}\bra{\Psi_{Q_{1},i_{1}}}\tensorprod\dotsm\tensorprod\ket{\Psi_{Q_{n},i_{n}}}\bra{\Psi_{Q_{n},i_{n}}}\right)\ket{\Psi_{W,w}},
\end{align*}
 between the possible ontic states of any partitioning collection
of mutually disjoint subsystems $Q_{1},\dotsc,Q_{n}$ and the possible
ontic states of the corresponding parent system $W=Q_{1}+\dotsb+Q_{n}$
at any \emph{single} moment in time, and, taking $Q\equiv Q_{1}=W$,
also provide a \emph{dynamical} smoothing relationship \eqref{eq:ExactDynQuantCondProbs},
\[
p_{Q}\left(j;t^{\prime}\given i;t\right)\equiv\Tr_{Q}\left[\op P_{Q}\left(j;t^{\prime}\right)\mathcal{E}_{Q}^{t^{\prime}\from t}\left[\op P_{Q}\left(i;t\right)\right]\right]\sim\Tr\left[\op P_{j}\left(t^{\prime}\right)\mathcal{E}\left[\op P_{i}\left(t\right)\right]\right],
\]
 between the possible ontic states of a system $Q$ over time and
also between the objective epistemic states of a system $Q$ over
time. 
\end{enumerate}
Essentially, 1 establishes a linkage between ontic states and epistemic
states, 2 establishes a linkage between (objective) epistemic states
and density matrices, 3 establishes a linkage between parent-system
density matrices and subsystem density matrices, and 4 establishes
a linkage between parent-system ontic states and subsystem ontic states
as well as between parent-system epistemic states and subsystem epistemic
states, either at the same time or at different times.

After verifying that our quantum conditional probabilities satisfy
a number of consistency requirements \eqref{eq:QuantCondProbsNonNeg}-\eqref{eq:QuantCondProbsTrivUnitary},
we showed that they allow us to avoid ontological instabilities that
have presented problems for other modal interpretations, analyzed
the measurement process, studied various familiar ``paradoxes''
and thought experiments, and examined the status of Lorentz invariance
and locality in our interpretation of quantum theory. In particular,
our interpretation accommodates the nonlocality implied by the EPR-Bohm
and GHZ-Mermin thought experiments without leading to superluminal
signaling, and evades claims by Myrvold purporting to show that interpretations
like our own lead to unacceptable contradictions with Lorentz invariance.
As a consequence of its compatibility with Lorentz invariance, we
claim that our interpretation is capable of encompassing all the familiar
quantum models of physical systems widely in use today, from nonrelativistic
point particles to quantum field theories and even string theory.
We also compared our interpretation to some of the other prominent
interpretations of quantum theory, such as the de Broglie-Bohm pilot-wave
interpretation and the Everett-DeWitt many-worlds interpretation,
and concluded that we could view the latter interpretation as being
a local ``Lorenz gauge'' of our own interpretation.

\subsection{Falsifiability and the Role of Decoherence\label{sub:Falsifiability-and-the-Role-of-Decoherence}}

As some other interpretations do, our own interpretation puts decoherence
in the central role of transforming the Born rule from an axiomatic
postulate into a derived consequence and thereby solving the measurement
problem of quantum theory. We regard it as a positive feature of our
interpretation of quantum theory that falsification of the capacity
for decoherence to manage this responsibility would mean falsification
of our interpretation. We therefore also take great interest in the
ongoing arms race between proponents and critics of decoherence, in
which critics offer up examples of decoherence coming up short \cite{AlbertLoewer:1990wdatassp,AlbertLoewer:1991mpss,AlbertLoewer:1993nim,Albert:1994qme,Page:2011qusbgbd}
and thereby push proponents to argue that increasingly realistic measurement
set-ups involving non-negligible environmental interactions resolve
the claimed inconsistencies \cite{BacciagaluppiHemmo:1996midm,Hollowood:2013cbr,Hollowood:2013eciqm,Vermaas:1999puqm}.

\subsection{Future Directions\label{sub:Future-Directions}}

\subsubsection{Understanding and Generalizing the Hilbert-Space Structure of Quantum
Theory\label{subsub:Understanding-and-Generalizing-the-Hilbert-Space-Structure-of-Quantum-Theory}}

The Hilbert-space structure underlying quantum theory is closely related
to the principle of linear superposition. Interesting recent work
has examined whether this Hilbert-space structure can be motivated
from more primitive ideas \cite{Hardy:2001jk}, perhaps by a ``purification
postulate'' that all mixed states should have a correspondence to
pure states in some formally defined larger system, together with
the known linear structure of classical spaces of mixed epistemic
states.

Note that requiring the vector space $\mathcal{H}$ to be \emph{complex}
is necessary for the existence of energy as an observable, as well
as for the existence of states having definite energy, at least for
systems that are dynamically closed and thus possess a well-defined
Hamiltonian in the first place; however there exist subtle ways to
get around these requirements \cite{Stueckelberg:1959fqtrrhs,Stueckelberg:1960qtrhs,Wootters:2013oitrvsqt}.
And because the Born rule \eqref{eq:BornRule} involves \emph{absolute-value-squares}
of inner products, one could also, in principle, explore dropping
the requirement that the Hilbert space's inner product must be positive
definite, although then avoiding the dynamical appearance of ``null''
state vectors having vanishing norm (and thus vanishing probability)
requires a delicate choice of Hamiltonian.%
\footnote{Unphysical null and negative-norm (``ghost'') states also arise
when \emph{formally} enlarging the Hilbert space of gauge theories
in order to make their symmetries more manifest, but in the present
context we are imagining that we treat positive- and negative-norm
states as both being physical.%
} One could also try to alter the definition of the inner product to
involve a PT transformation \cite{BenderBoettcher1998:rsnhhhpts,BenderBoettcherMeisinger:1998ptsqm,Bender:2005iptsqt}.
These and other approaches \cite{Aaronson:2004qmit} to modifying
the Hilbert-space structure of quantum theory may have interesting
implications for our interpretation of quantum theory that would be
worth exploring.

\subsubsection{Coherent States\label{subsub:Coherent-States}}

For a system with continuously valued degrees of freedom, the similarity
between the classical Liouville equation \eqref{eq:ClassicalLiouvilleEq}
and the quantum Liouville equation \eqref{eq:QuantumLiouvilleEq}
becomes much closer if we re-express the $2N$-dimensional classical
phase space $\left(q,p\right)$ and the Poisson brackets \eqref{eq:PBracks}
in terms of the $N$ dimensionless complex variables 
\begin{equation}
z_{\alpha}\equiv\frac{1}{\sqrt{2}}\left(\frac{1}{\ell_{\alpha}}q_{\alpha}+i\frac{\ell_{\alpha}}{\hbar}p_{\alpha}\right)\label{eq:ComplexPhaseSpCoords}
\end{equation}
 and their complex conjugates $z_{\alpha}^{\conj}$, where $\ell_{\alpha}$
are characteristic length scales in the system \cite{Strocchi:1966ccqm}.
In that case, introducing the complexified Poisson brackets 
\begin{equation}
\left\{ f,g\right\} _{z,z^{\conj}}\equiv\sum_{\alpha}\left[\frac{\partial f}{\partial z_{\alpha}}\frac{\partial g}{\partial z_{\alpha}^{\conj}}-\frac{\partial g}{\partial z_{\alpha}}\frac{\partial f}{\partial z_{\alpha}^{\conj}}\right],\label{eq:ComplexPBracks}
\end{equation}
 the classical Liouville equation takes a much more quantum-looking
form, complete with the familiar prefactor of $-i/\hbar$ on the right-hand
side: 
\begin{equation}
\frac{\partial\rho}{\partial t}=-\frac{i}{\hbar}\left\{ H,\rho\right\} _{z,z^{\conj}}\label{eq:ComplexClassicalLiouvilleEq}
\end{equation}

The complex variables $z_{\alpha}$ are natural for another reason,
namely, because they label the corresponding quantum system's coherent states
$\ket z\equiv\ket{\setindexed{z_{\alpha}}{\alpha}}$ \cite{Schrodinger:1926dsuvdmzm,Glauber:1963cisrf},
which are defined to be the solutions to the eigenvalue equations
$\op z_{\alpha}\ket z=z_{\alpha}\ket z,$ with $q_{\alpha}\equiv\sqrt{2}\ell_{\alpha}\re z_{\alpha}$
and $p_{\alpha}\equiv\sqrt{2}\hbar\im z_{\alpha}/\ell_{\alpha}$ the
expectation values of the operators $\op q_{\alpha}$ and $\op p_{\alpha}$.
Coherent states are the closest quantum analogues to classical states:
\begin{itemize}
\item They saturate the Heisenberg uncertainty bounds $\Delta q_{\alpha}\Delta p_{\alpha}\geq\hbar/2$;
\item they have Gaussian wave functions in both coordinate space and momentum
space that both approach delta functions in the limit $\hbar\to0$;
\item they become orthogonal in the limit of large coordinate separation
$\sum_{\alpha}\absval{z_{\alpha}-z_{\alpha}^{\prime}}^{2}\gg1$;
\item they each occupy an $N$-dimensional disc  of approximate volume
$\left(2\pi\hbar\right)^{N}=h^{N}$ in $N$-dimensional phase space
$\left(q,p\right)$ and thereby nicely account for phase-space quantization;
\item they satisfy the overcompleteness relation $\integration{\int}{\left(1/\pi^{N}\right)d^{2}z^{N}}\ket z\bra z$,
with a measure $d^{2}z^{N}/\pi^{N}=dq^{N}dp^{N}/h^{N}$ that exactly
replicates the familiar phase-space measure from semiclassical statistical
mechanics;
\item and, for coupled systems with small interactions between them, the
rate at which coherent states become mutually entangled is very low,
so that they remain uncorrelated even in the macroscopic limit.%
\footnote{As explained in \cite{JoosZeh:1985ecptiwe}, ``This argument may
explain the dominance of the field aspect over the particle aspect
for boson{[}ic{]} fields.''%
}
\end{itemize}
Moreover, we can regard delta-function coordinate-basis eigenstates
as coherent states in the limit $\ell_{\alpha}\to0$, and delta-function
momentum-basis eigenstates as coherent states in the limit $\ell_{\alpha}\to\infty$.
If we instead choose the length scales $\ell_{\alpha}$ so that the
terms in the Hamiltonian quadratic in coordinates and momenta are
proportional to $\sum_{\alpha}\op z_{\alpha}^{\adj}\op z_{\alpha}$,
then the corresponding coherent states \emph{remain} coherent states
under unitary time evolution over short time intervals, in which case
the expectation values $q_{\alpha}\left(t\right)$ and $p_{\alpha}\left(t\right)$
approximately follow the same dynamical equations as would be expected
if the system were classical and governed by second-order dynamics;
this last fact, in particular, helps explain the ubiquity of second-order
dynamics among classical systems with continuous degrees of freedom.

Because there is no real observable whose Hermitian operator's eigenstates
are coherent states, systems don't end up in coherent states as a
consequence of a Von Neumann measurement. Instead, sufficiently large
systems with continuously valued degrees of freedom end up approximately
in coherent states due to messy environmental perturbations because
coherent states are so robust \cite{ZurekHabibPaz:1993csvd,TegmarkShapiro:1994dpcsephc}.
A set of coherent states farther apart in phase space than the Planck
constant $h$ form an approximately orthogonal set, and we can imagine
extending such a set to an orthonormal basis suitable for the spectrum
of a macroscopic system's density matrix by including additional state
vectors as needed; for small $h$, coherent states are very, very
close to being orthogonal, so this approximation becomes better and
better in the formal classical limit $h\to0$. It would be worthwhile
to investigate this story in greater detail as it relates to our interpretation
of quantum theory.

\subsubsection{The First-Quantized Formalism for Quantum Theories, Quantum Gravity,
and Cosmology}

In describing first-quantized closed systems of particles or strings
\cite{Polyakov:1987gfs,Polchinski:2005st1,BeckerBeckerSchwarz:2007stmtmi},
as well as in canonical methods for describing quantum gravity \cite{Misner:1957fqgr,DeWitt:1967qtgict},
it is often useful to work with a generalized Hilbert space consisting
of infinitely many copies of the given system's physical Hilbert space,
each copy corresponding to one instant along a suitable temporal parameterization.
Unitary dynamics is then expressed as a Hamiltonian constraint equation
of the Wheeler-DeWitt form $\op{\mathcal{H}}\ket{\Psi}=0$. It would
be an intriguing exercise to study how to re-express the formalism
of our interpretation of quantum theory in this alternative framework,
and, indeed, how to accommodate open systems exhibiting more general
linear CPT dynamics. More broadly, we look forward to exploring quantum
features of black holes and cosmology within the context of our interpretation,
including the measure problem of eternal inflation \cite{GarrigaSchwartzPerlovVilenkinWinitzki:2006,Guth:2007eii,Freivogel:2011mpm,Vilenkin:2013gsmmp,GarrigaVilenkin:2013wm}.

\subsection{Relevant Metaphysical Speculations\label{sub:Relevant-Metaphysical-Speculations}}

\subsubsection{The Status of Superdeterminism\label{subsub:The-Status-of-Superdeterminism}}

Throughout this paper, we have repeatedly emphasized the importance
and nontriviality of the existence of dynamics for a classical or
quantum system. In particular, the existence of dynamics for a system
is a much stronger property than the mere possession of a \emph{particular}
kinematical trajectory by the system. Indeed, one could easily imagine
a ``superdeterministic'' universe in which every system has some
specific trajectory---written down at the beginning of time on some
mystical ``cosmic ledger,'' say---but has no dynamics (not even
deterministic dynamics) in the sense of the existence of a mapping
\eqref{eq:ClassicalOnticLevelDynMapGeneral} for \emph{arbitrary}
values of initial ontic states and that allows us to compute hypothetical
\emph{alternative} trajectories.%
\footnote{It's important to realize that the existence of dynamics---even deterministic
dynamics---still leaves open a great deal of flexibility via initial
conditions. Building on this idea, Aaronson \cite{Aaronson:2013gqtm}
suggests that if our universe is not superdeterministic but is instead
governed by ``merely'' deterministic or probabilistically stochastic
dynamical laws, then un-cloneable quantum details (``freebits'')
of our observable universe's initial conditions could allow for the
kind of non-probabilistic uncertainty that we discussed in Section~\ref{subsub:Probabilistic-and-Non-Probabilistic-Uncertainty}
and thus make room for notions of free will, much as the low entropy
of our observable universe's initial conditions makes room for a thermodynamic
arrow of time.%
}

It is therefore a remarkable fact that so many systems in Nature---the
Standard Model of particle physics being an example \emph{par excellence}---are
well described by dynamics simple enough  that we can write them
down on a sheet of paper. The time evolution of most systems apparently
encodes \emph{far} less information than their complicated trajectories
might naïvely suggest---that is, the information encoded in their
trajectories is highly compressible---and certainly far less information
than would be expected for systems belonging to a superdeterministic
universe.%
\footnote{Following Aaronson's classification of philosophical problems in \cite{Aaronson:2013gqtm},
the $Q$ question ``Is the universe superdeterministic?'' may be
fundamentally unanswerable, but our present discussion suggests the
more meaningful and well-defined $Q^{\prime}$ question ``How compressible
is trajectory information for the systems that make up the universe?''%
}

\subsubsection{Implications for Presentism and Block-Time Universes}

A perennial debate in metaphysics concerns the fundamental nature
of time itself: Presentism is the philosophical proposition that
the present moment in time is an ontologically real concept, whereas
an obvious alternative is that our notion of ``the moving now''
is merely an illusion experience by beings who actually live in a
``block-time'' universe whose extent into the past and future are
equally real. In the classical deterministic case, we can always depict
spacetime in block form, and we can even accommodate nontrivial stochastic
dynamics by considering appropriate ensembles of block-time universes
\cite{Aaronson:2013gqtm}.

However, according to our interpretation of quantum theory, naïve
notions of reductionism generically break down at the microscale,
as we explained immediately following \eqref{eq:QuantCondProbKinParentSub},
and thus epistemic states for microscopic quantum systems each have
their own time evolution and don't sew together in a classically intuitive
manner; indeed, the same may well be true on length scales that exceed
the size of our cosmic horizon \cite{BoussoSusskind:2012miqm}. Hence,
the notion of a block-time universe, which is, in a certain sense,
a ``many-times'' interpretation of physics, may turn out to be just
as untenable as the many-worlds interpretation of quantum theory.

\subsubsection{Sentient Quantum Computers}

In Section~\ref{subsub:Hidden-Variables-and-the-Irreducibility-of-Ontic-States}
and elsewhere in this paper, we have repeatedly emphasized that our
interpretation of quantum theory regards ontic states as being irreducible
objects, rather than as being epistemic probability distributions
over a more basic set of preferred basis states or hidden variables.
Because human brains are warm systems in constant contact with a messy
larger environment, their reduced density matrices and thus their
possible ontic states are guaranteed by decoherence to look classical
and not to involve macroscopic quantum superpositions \cite{Tegmark:2000iqdbp};
putting a human brain into an overall quantum superposition would
therefore seem to require completely isolating the brain from its
environment (including its blood supply) and lowering its temperature
to nearly absolute zero, in which case no human awareness is conceivable.
We can therefore sidestep metaphysical questions about the subjective
first-person experiences of human observers who temporarily exist
in quantum superpositions of mutually exclusive, classical-looking
state vectors.

However, one might imagine someday building a sentient quantum computer
capable of human-level intelligence. In principle, and in contrast
to human beings, such a machine would be perfectly functional even
at a temperature close to absolute zero and without any need for continual
interactions with a larger environment. Could such a machine achieve
something like subjective first-person experiences, and, if so, how
would the machine experience existing in a quantum superposition of
mutually exclusive state vectors? Or is there something about the
existence of subjective first-person experiences that fundamentally
requires continual decoherence and information exchange with a larger
environment?%
\footnote{We thank Scott Aaronson (private communication) for suggesting these
points.%
}

A recurring problem in the history of interpretations of quantum theory
is the tendency for the subject to become mixed up with other thorny
problems in philosophy. Just as our interpretation deliberately aims
to be model independent and thereby attempts to avoid getting tangled
up in the philosophical debate over the rigorous meaning of probability
and its many schools of interpretation (from Laplacianism to frequentism
to Bayesianism to decision theory), we also intentionally avoid making
any definitive statements about a preferred philosophy of mind and
the important metaphysical problem of understanding the connection
between the physical third-person reality of atoms, planets, and galaxies
and the subjective first-person reality of colors, thoughts, and emotions.
By design, our interpretation concerns itself solely with physical
third-person reality, and does not favor any of the schools of thought
on the reality of first-person experiences (from dualism to eliminativism
to functionalism to panpsychism).

\subsubsection{The Wigner Representation Theorem}

As we mentioned in Section~\ref{sub:Comparison-with-Other-Interpretations-of-Quantum-Theory},
we have no evidence for the existence of a closed maximal parent system
described by a cosmic pure state or universal wave function---there
may only be a succession of increasingly large parent systems that
are all in nontrivial improper mixtures---so there may be ``no place
to stand'' to say that we wish to perform a global physical transformation.

But looking again at the Wigner representation theorem \cite{Wigner:1931guiaadqda,Weinberg:1996qtfi},
we can now provide a new definition of physical transformations that
does not depend on the existence of such cosmic pure-state systems:
We can simply say that our system admits a particular class of physical
transformations if there exists some sort of \emph{active} way of
altering the physical state of the external observer (which we now
understand is equivalent to using the phrase ``passive transformation'')
that leaves the external observer's own calculated empirical outcome
probabilities invariant. That statement is still perfectly true, and
one can express it at the level of mathematics in terms of unitary
or anti-unitary transformations \emph{formally} acting on the subject
system, provided that we keep in mind that the transformation of the
external observer is not truly changing the subject system.

This general line of reasoning makes possible a very concise proof
of the Wigner representation theorem. Consider a subject system $Q$
together with an external observer $O$ in the presence of a larger
environment $E$, and suppose that the composite system $Q+O$ goes
through a process in which $O$ performs a measurement on $Q$ and
then the environment $E$ immediately causes decoherence of the final
density matrix of $Q+O$ to a diagonalizing eigenbasis whose individual
eigenstates correspond to the expected measurement outcomes. We require
that the resulting probability eigenvalues of the final post-measurement
density matrix of the composite system $Q+O$ must be the same regardless
of whether or not we actively transformed the observer $O$ before
the experiment. This condition is equivalent to the usual definition
of a physical transformation for the purposes of the Wigner representation
theorem, as it translates directly into an invariance constraint on
Born-rule probabilities. Note that the physical transformation is
to be performed \emph{before} the experiment, and thus \emph{before}
we have a density matrix with our final probability eigenvalues.

We can now employ the following trick: If we know in advance what
observable the observer $O$ is planning to measure, then, even before
the experiment has been performed, we can \emph{pretend} that the
subject system $Q$ was \emph{already} in a mixed state whose density
matrix is diagonal in the eigenbasis of that observable's Hermitian
operator, with the Born-rule probabilities appearing as the eigenvalues
of this mixed-state density matrix. Taking this approach ends up producing
the same final density matrix for the composite system $Q+O$ at the
end of the experiment and with the correct final probability eigenvalues,
and so we can safely substitute our mixed-state density matrix for
the subject system $Q$ into the Born-rule formulas in advance. Then,
to the extent that we can always pretend inside our Born-rule formulas
that a passive transformation---really an \emph{active} transformation
of just the external observer $O$---is \emph{mathematically} equivalent
to an (inverse) transformation of the subject system $Q$, we can
demand that the probability eigenvalues of the mixed-state density
matrix of $Q$ should remain unchanged under the physical transformation.
Because the only transformations that leave the eigenvalues of a general
Hermitian matrix unchanged are unitary or anti-unitary transformations,
the theorem is proved. $\QED$

\subsubsection{Dynamical Symmetries and the Absence of Preferred Perspectives}

As an aside, it is worth noting that a physical transformation in
the present context coincides with the notion of an (unbroken) \emph{dynamical}
symmetry when it maps solutions of the dynamics to new solutions of
the \emph{same} dynamics---for example, at the classical level, because
the physical transformation leaves the given system's action functional
invariant up to possible boundary terms. Not all physical transformations
are dynamical symmetries in this sense; for example, although a change
in the length of all our rulers may be a physical transformation in
the sense of Wigner's representation theorem, it is not a \emph{dynamical}
symmetry of Nature because there exist plenty of well-known physical
systems whose dynamical equations involve dimensionful length scales
and are therefore not scale-invariant. By contrast, translations in
space, as well as Lorentz transformations, appear to be dynamical
symmetries of Nature.

In particular, if a given physical transformation happens to correspond
to a dynamical symmetry, then there cannot exist a fundamentally ``preferred''
perspective with respect to that symmetry transformation. For example,
because spatial translations are a dynamical symmetry of Nature, there
cannot exist a fundamentally preferred location in space, and because
Lorentz transformations are also a dynamical symmetry of Nature, there
cannot exist a fundamentally preferred inertial reference frame, at
least when considering sufficiently small spacetime intervals that
general relativity reduces to special relativity.

\subsubsection{Nonlocality in the Everett-DeWitt Many-Worlds Interpretation\label{subsub:Nonlocality in the Everett-DeWitt Many-Worlds Interpretation}}

Although often claimed to be manifestly local, the Everett-DeWitt
many-worlds interpretation of quantum theory can only maintain this
manifest locality by abandoning any sharply defined probabilistic
branching structure that spans all systems or assertions that it can
solve the preferred-basis problem, as we mentioned in Section~\ref{subsub:The-Everett-DeWitt-Many-Worlds-Interpretation-of-Quantum-Theory}.

To see explicitly how this problem arises, consider again the standard
EPR-Bohm experiment that we originally examined in Section~\ref{sub:The-EPR-Bohm-Thought-Experiment-and-Bell's-Theorem}.
The initial state vector of the composite system $1+2+A+B$ is 
\[
\ket{\Psi_{1+2+A+B}}=\frac{1}{\sqrt{2}}\left(\ket{\uparrow\downarrow}-\ket{\downarrow\uparrow}\right)\ket{A\left(\quote{\emptyset}\right)}\ket{B\left(\quote{\emptyset}\right)}.
\]
 How should one regard this state vector in terms of branches (``worlds'')
and their associated probabilities according to the many-worlds interpretation?

If the claim is that there exists just one branch with unit probability
and on which the state vector of the two-particle system $1+2$ is
really $\ket{\Psi_{1+2}}=\left(1/\sqrt{2}\right)\left(\ket{\uparrow\downarrow}-\ket{\downarrow\uparrow}\right)$,
then one runs into trouble with nonlocality, because after the spin
detector $A$ performs its local measurement on particle $1$, two
branches instantaneously emerge on which particles 1 and 2 are suddenly
classically correlated ($\uparrow\downarrow$ in one branch and $\downarrow\uparrow$
in the other branch) despite not having been classically correlated
before.

An alternative approach is to argue that there were actually two branches
with respective probabilities of $1/2$ all along, meaning that one
should regard the initial state vector of the composite system $1+2+A+B$
as really being 
\[
\ket{\Psi_{1+2+A+B}}=\frac{1}{\sqrt{2}}\ket{\uparrow\downarrow}\ket{A\left(\quote{\emptyset}\right)}\ket{B\left(\quote{\emptyset}\right)}-\frac{1}{\sqrt{2}}\ket{\downarrow\uparrow}\ket{A\left(\quote{\emptyset}\right)}\ket{B\left(\quote{\emptyset}\right)}.
\]
 In that case, the two particles were really classically correlated
on each branch even before the experiment and thus no ``new'' classical
correlation suddenly appears nonlocally after the spin detector $A$
carries out its local measurement on particle $1$. But then one runs
into a serious problem making the notion of the branches and their
probabilities well-defined, because there is nothing that privileges
splitting up the branches in the spin-$z$ basis for the two-particle
system $1+2$; indeed, one could just as well have split up the branches
in the spin-$x$ basis ($\ket{\leftarrow}$ and $\ket{\rightarrow}$)
instead, in which case the initial state vector of the composite system
$1+2+A+B$ would be 
\[
\frac{1}{\sqrt{2}}\ket{\leftarrow\rightarrow}\ket{A\left(\quote{\emptyset}\right)}\ket{B\left(\quote{\emptyset}\right)}-\frac{1}{\sqrt{2}}\ket{\rightarrow\leftarrow}\ket{A\left(\quote{\emptyset}\right)}\ket{B\left(\quote{\emptyset}\right)}.
\]
 One is therefore confronted directly with the preferred-basis problem,
and decoherence cannot help resolve the paradox because the spin detectors
$A$ and $B$ haven't performed their measurements yet and thus all
the various possible bases are on an equal footing.%
\footnote{Note that these conclusions cannot be evaded by the assumption of
small degeneracy-breaking effects of the form \eqref{eq:EPRPairStateVecDegenBreaking},
which would merely have the effect of increasing the total number
of branches.%
}

So what are the branches? Which basis is the ``correct'' one for
deciding? As we have seen in this section, if we pick one preferred
definition for the branches that span the systems under consideration---and
their associated probabilities---then we immediately run into the
nonlocality issue again; for example, if we pick, say, the spin-$x$
branches, and then $A$ performs a spin-$z$ measurement, then again
the branches need to change in a nonlocal way in order to ensure the
correct final correlation. If we declare that we must be noncommittal
about assigning entangled systems to branches, then we run into the
trouble that entanglement is a ubiquitous phenomenon afflicting \emph{all}
systems to a nonzero degree, and thus we don't obtain sharp definitions
of what we mean by branches. Because there is no fixed choice of branch-set
compatible with manifest locality, the many-worlds interpretation
isn't manifestly local unless we give up any notion of a preferred
branch set that span the systems, but then we lose any hope of making
sense of probabilities in the interpretation.

\subsection{Comparison with the Hollowood Modal Interpretation\label{sub:Comparison-with-the-Hollowood-Modal-Interpretation}}

Our minimal modal interpretation of quantum theory differs in several
key aspects from the recently introduced ``emergent Copenhagen interpretation''
by Hollowood \cite{Hollowood:2013cbr,Hollowood:2013nmqm,Hollowood:2013eciqm},
with whom we have collaborated over the past year. Specifically, we
employ a different, manifestly non-negative, more general formula
for our quantum conditional probabilities that doesn't depend \emph{fundamentally}
on a temporal cut-off time scale; we use these quantum conditional
probabilities not only for dynamical purposes but also to interpolate
between the ontologies of parent systems and their subsystems; our
quantum conditional probabilities allow ontic states of disjoint systems
to influence each other; joint ontic-state and epistemic-state assignments
exist for mutually disjoint systems; and we accept the inevitability
of ontic-level nonlocality implied by the EPR-Bell and GHZ-Mermin
thought experiments.

\begin{acknowledgments}
J.\,A.\,B. has benefited tremendously from personal communications
with Matthew Leifer, Timothy Hollowood, and Francesco Buscemi. D.\,K.
thanks Gaurav Khanna, Pontus Ahlqvist, Adam Brown, Darya Krym, and
Paul Cadden-Zimansky for many useful discussions on related topics,
and is supported in part by FQXi minigrant \#FQXi-MGB-1322. Both authors
have greatly appreciated their interactions with the Harvard Philosophy
of Science group and its organizers Andrew Friedman and Elizabeth
Petrik, as well as with Scott Aaronson and Ned Hall. Both authors
are indebted to Steven Weinberg, who has generously exchanged relevant
ideas and whose own recent paper \cite{Weinberg:2014qmwsv} also explores
the idea of reformulating quantum theory in terms of density matrices.
The work of Pieter Vermaas has also been a continuing source of inspiration
for both authors, as have many conversations with Allan Blaer.
\end{acknowledgments}
\appendix

\hypertarget{hypertarget:appendix}{}

\section*{Appendix\label{sec:Appendix}}

In this appendix, we summarize the traditional Copenhagen interpretation,
in part just to establish our notation and terminology. With this
background established, we then define the measurement problem and
systematically analyze attempts to solve it according to the various
prominent interpretations of quantum theory, including the instrumentalist
approach. Finally, we describe several important theorems that have
been developed over the years to constrain candidate interpretations
of quantum theory.

\subsection{The Copenhagen Interpretation and the Measurement Problem\label{sub:The-Copenhagen-Interpretation-and-the-Measurement-Problem}}

\subsubsection{A Review of the Copenhagen Interpretation}

The Copenhagen interpretation asserts that every isolated quantum
system is completely described by a particular unit-norm state vector
$\ket{\Psi}$ in an associated Hilbert space $\mathcal{H}$. Furthermore,
according to the Copenhagen interpretation, every observable property
$\Lambda$ of the system corresponds to a Hermitian operator $\op{\Lambda}=\op{\Lambda}^{\adj}$
necessarily having a complete orthonormal basis of eigenstates $\ket a$
with corresponding real eigenvalues $\lambda_{a}$: 
\begin{equation}
\op{\Lambda}\ket a=\lambda_{a}\ket a,\qquad\lambda_{a}\in\mathbb{R},\qquad\braket a{a^{\prime}}=\delta_{aa^{\prime}},\qquad\sum_{a}\ket a\bra a=\op 1.\label{eq:CopInterpObservables}
\end{equation}
 The eigenvalues $\lambda_{a}$ represent the possible measurement
outcomes for the random variable $\Lambda$, and the empirical outcome
probability $p\left(\lambda\right)$ with which a particular eigenvalue
$\lambda=\lambda_{a}$ is obtained is given by the Born rule, 
\begin{equation}
p\left(\lambda\right)=\sum_{\substack{a\\
\left(\lambda_{a}=\lambda\right)
}
}\absval{\braket a{\Psi}}^{2},\label{eq:CopInterpBornRuleDeg}
\end{equation}
 where, in order to accommodate the case of degeneracies in the eigenvalue
spectrum of $\op{\Lambda}$, the sum is over all values of the label
$a$ for which the eigenvalue $\lambda_{a}$ of the eigenstate $\ket a$
is equal to the outcome eigenvalue $\lambda$. When degeneracy is
absent, so that we can uniquely label the eigenstates of $\op{\Lambda}$
by $\lambda$, the Born rule \eqref{eq:CopInterpBornRuleDeg} reduces
to the simpler expression 
\begin{equation}
p\left(\lambda\right)=\absval{\braket{\lambda}{\Psi}}^{2}.\label{eq:CopInterpBornRuleNondeg}
\end{equation}

Either way, we can then express expectation values of observables
$\Lambda$ in terms of the system's state vector $\ket{\Psi}$ in
the following way: 
\begin{equation}
\expectval{\Lambda}=\bra{\Psi}\op{\Lambda}\ket{\Psi}.\label{eq:CopInterpExpectVal}
\end{equation}
 In particular, letting $\op P_{\lambda}$ denote the Hermitian projection
operator onto eigenstates of $\op{\Lambda}$ having the eigenvalue
$\lambda$, 
\begin{equation}
\op P_{\lambda}\equiv\sum_{\substack{a\\
\left(\lambda_{a}=\lambda\right)
}
}\ket a\bra a,\label{eq:CopInterpProjOp}
\end{equation}
 we can use \eqref{eq:CopInterpExpectVal} to express the Born rule
\eqref{eq:CopInterpBornRuleDeg} in the alternative form 
\begin{equation}
p\left(\lambda\right)=\expectval{P_{\lambda}}=\bra{\Psi}\op P_{\lambda}\ket{\Psi}.\label{eq:CopInterpBornRuleFromExpVal}
\end{equation}
 These statements all naturally extend to density matrices $\op{\rho}$
according to the formulas 
\begin{equation}
p\left(\lambda\right)=\Tr\left[\op{\rho}\op P_{\lambda}\right],\qquad\expectval{\Lambda}=\Tr\left[\op{\rho}\op{\Lambda}\right],\label{eq:CopInterpBornRuleExpectValFromDensMatrix}
\end{equation}
 which we can identify as noncommutative generalizations of the respective
classical formulas 
\begin{equation}
p\left(\lambda\right)=\sum_{a}p_{a}P_{\lambda,a},\qquad\expectval{\Lambda}=\sum_{a}p_{a}\lambda_{a},\label{eq:CopInterpClassicalStatisticalFormulas}
\end{equation}
 where the quantities $p_{a}\in\left[0,1\right]$ constitute a classical
probability distribution over elementary outcomes $a$ and 
\begin{equation}
P_{\lambda,a}=\begin{cases}
1 & \whichfor\lambda_{a}=\lambda,\\
0 & \whichfor\lambda_{a}\ne\lambda
\end{cases}\label{eq:DefClassCharactFunc}
\end{equation}
 is the characteristic (or indicator) function for the set of elementary
outcomes $a$ corresponding to the generalized outcome $\lambda$.

Systems that are dynamically closed undergo smooth, linear time evolution
according to a unitary time-evolution operator $\op U\left(t\right)$,
\begin{equation}
\ket{\Psi\left(t\right)}=\op U\left(t\right)\ket{\Psi\left(0\right)}\qquad\left[\op U\left(t\right)^{\adj}=\op U\left(t\right)^{-1},\qquad\braket{\Psi\left(t\right)}{\Psi\left(t\right)}=\braket{\Psi\left(0\right)}{\Psi\left(0\right)}=1\right].\label{eq:CopInterpUnitaryTimeEv}
\end{equation}
 If we can express the time-evolution operator in terms of a Hermitian,
time-independent Hamiltonian operator $\op H$ according to 
\begin{equation}
\op U\left(t\right)=e^{-i\op Ht/\hbar},\label{eq:CopInterpTimeEvOpFromHam}
\end{equation}
 then the system's state vector obeys the famous Schrödinger equation:
\begin{equation}
i\hbar\frac{\partial}{\partial t}\ket{\Psi\left(t\right)}=\op H\ket{\Psi\left(t\right)}.\label{eq:CopInterpSchroEq}
\end{equation}
 At the level of the system's density matrix $\op{\rho}\left(t\right)$,
these integral and differential dynamical equations respectively become
\begin{equation}
\op{\rho}\left(t\right)=\op U\left(t\right)\op{\rho}\left(0\right)\op U\left(t\right)^{\adj},\qquad\frac{\partial}{\partial t}\op{\rho}\left(t\right)=-\frac{i}{\hbar}\comm{\op H}{\op{\rho}\left(t\right)}.\label{eq:CopInterpLiouvilleEq}
\end{equation}

The Copenhagen interpretation axiomatically regards the Born rule
\eqref{eq:CopInterpBornRuleDeg} as an exact postulate, and defines
the partial-trace prescription together with reduced density matrices
for subsystems precisely to ensure that the Born-rule-based expectation
value \eqref{eq:CopInterpBornRuleExpectValFromDensMatrix} obtained
for any observable $\op{\Lambda}$ of any subsystem agrees with the
expectation value of the corresponding observable $\op{\Lambda}\tensorprod\op 1$
of any parent system: 
\[
\Tr_{\mathrm{subsystem}}\left[\op{\rho}_{\mathrm{subsystem}}\op{\Lambda}\right]=\negthickspace\negthickspace\negthickspace\negthickspace\negthickspace\underbrace{\Tr_{\mathrm{parent}}}_{\Tr_{\mathrm{subsystem}}\Tr_{\mathrm{other}}}\negthickspace\negthickspace\negthickspace\negthickspace\negthickspace\left[\op{\rho}_{\mathrm{parent}}\op{\Lambda}\tensorprod\op 1\right]
\]
\begin{equation}
\implies\op{\rho}_{\mathrm{subsystem}}=\Tr_{\mathrm{other}}\left[\op{\rho}_{\mathrm{parent}}\right].\label{eq:CopInterpPartialTrace}
\end{equation}
 Consequently, the dynamics governing a general (open) subsystem of
a dynamically closed parent system is determined by an equation that
is generically non-unitary: 
\begin{equation}
\frac{\partial}{\partial t}\op{\rho}\left(t\right)=-\frac{i}{\hbar}\Tr_{\mathrm{parent}}\left[\left[\op H_{\mathrm{parent}},\op{\rho}_{\mathrm{parent}}\right]\right].\label{eq:CopInterpNonUnitaryOpenSubsys}
\end{equation}

The final axiom of the Copenhagen interpretation stipulates that if
an external observer measures an observable $\Lambda$ of the system
and obtains an outcome $\lambda$ corresponding to the projection
operator $\op P_{\lambda}$ defined in \eqref{eq:CopInterpProjOp},
then the system instantaneously ``collapses'' according to a non-unitary
rule called the Von Neumann-Lüders projection postulate \cite{vonNeumann:1955mfqm,Luders:1950udzddm}:
Regardless of the system's initial density matrix $\op{\rho}_{\mathrm{initial}}$,
which may well be pure $\ket{\Psi_{\mathrm{initial}}}\bra{\Psi_{\mathrm{initial}}}$
or may instead describe an improper mixture---meaning that the nontriviality
of the density matrix arises at least in part from quantum entanglements
with other systems---the system's \emph{final} density matrix $\op{\rho}_{\mathrm{final},\lambda}$
is given by 
\begin{equation}
\op{\rho}_{\mathrm{final},\lambda}=\frac{\op P_{\lambda}\op{\rho}_{\mathrm{initial}}\op P_{\lambda}}{\Tr\left[\op P_{\lambda}\op{\rho}_{\mathrm{initial}}\right]}.\label{eq:CopInterpVonNeumannLudersProjPost}
\end{equation}
 In keeping with \eqref{eq:CopInterpBornRuleExpectValFromDensMatrix},
the denominator is the probability $p\left(\lambda\right)$ of obtaining
the outcome $\lambda$ and ensures that the final density matrix $\op{\rho}_{\mathrm{final},\lambda}$
has unit trace. In the special case in which the eigenvalue spectrum
of the operator $\op{\Lambda}$ representing the observable $\Lambda$
has no degeneracies, so that the projection operator $\op P_{\lambda}$
reduces to the simple form $\ket{\lambda}\bra{\lambda}$ for a single
eigenstate $\ket{\lambda}$ of $\op{\Lambda}$, the Von Neumann-Lüders
projection postulate \eqref{eq:CopInterpVonNeumannLudersProjPost}
reduces to 
\begin{equation}
\op{\rho}_{\mathrm{final},\lambda}=\ket{\lambda}\bra{\lambda},\label{eq:CopInterpVonNeumannLudersProjPostNondeg}
\end{equation}
 so that the system ends up in a pure state represented by $\ket{\lambda}$.

\subsubsection{Wave-Function Collapse and the Measurement Problem\label{subsub:Wave-Function-Collapse-and-the-Measurement-Problem}}

The Von Neumann-Lüders projection postulate \eqref{eq:CopInterpVonNeumannLudersProjPost}---known
informally as wave-function collapse---both accounts for the statistical
features of the post-measurement state of affairs and also ensures
that definite measurement outcomes \emph{persist} under identical
repeated experiments performed over sufficiently short time intervals.
However, \eqref{eq:CopInterpVonNeumannLudersProjPost} represents
a discontinuous departure from the smooth time evolution determined
by the Schrödinger equation \eqref{eq:CopInterpSchroEq}, a discrepancy
known as the measurement problem of quantum theory.

As a first step toward better characterizing the measurement problem,
notice that we can gather together the different possible post-measurement
density matrices $\op{\rho}_{\mathrm{final},\lambda}$ appearing in
\eqref{eq:CopInterpVonNeumannLudersProjPost} into a subjective probability
distribution of the form 
\begin{equation}
\setindexed{p\left(\lambda\right)\implies\op{\rho}_{\mathrm{final},\lambda}}{\lambda}.\label{eq:CopInterFinalClassicalMixt}
\end{equation}
 We can obtain the same statistical predictions from the block-diagonal
``subjective'' density matrix 
\begin{equation}
\op{\rho}_{\mathrm{final}}=\sum_{\lambda}p\left(\lambda\right)\op{\rho}_{\mathrm{final},\lambda}=\sum_{\lambda}\op P_{\lambda}\op{\rho}_{\mathrm{initial}}\op P_{\lambda},\label{eq:CopInterpFinalLargeDensMatr}
\end{equation}
 which describes the post-measurement system in the absence of post-selecting
or conditioning on the actually observed value $\lambda$. In the
special case \eqref{eq:CopInterpVonNeumannLudersProjPostNondeg} in
which the set of possible measurement outcomes exhibit no degeneracies,
\eqref{eq:CopInterpFinalLargeDensMatr} reduces to a proper mixture,
meaning that the nontriviality of the density matrix arises \emph{solely}
from subjective uncertainty over the true underlying state vector
of the system: 
\begin{equation}
\op{\rho}_{\mathrm{final}}=\sum_{\lambda}p\left(\lambda\right)\ket{\lambda}\bra{\lambda}.\label{eq:CopInterpFinalLargeDensMatrProperMixt}
\end{equation}
 More generally, note that the specific decomposition appearing on
the right-hand-side of \eqref{eq:CopInterpFinalLargeDensMatr} is
preferred among all possible decompositions of $\op{\rho}_{\mathrm{final}}$
because we know that the system's \emph{true} density matrix is \emph{really}
one of the possibilities $\op{\rho}_{\mathrm{final},\lambda}$ obtained
from the Von Neumann-Lüders projection postulate \eqref{eq:CopInterpVonNeumannLudersProjPost}.
That is, our system is fundamentally described by the subjective probability
distribution \eqref{eq:CopInterFinalClassicalMixt}, and we have introduced
the subjective density matrix \eqref{eq:CopInterpFinalLargeDensMatrProperMixt}
merely for mathematical convenience.

Remarkably, provided that the measurement in question is performed
by a sufficiently macroscopic observer or measuring device, the quantum
phenomenon of decoherence \cite{Bohm:1951qt,JoosZeh:1985ecptiwe,Joos:2003dacwqt,Zurek:2003dtqc,Schlosshauer:2005dmpiqm,SchlosshauerCamilleri:2008qct,BreuerPetruccione:2002toqs}
naturally produces \emph{reduced} (``objective'') density matrices
describing \emph{improper} mixtures that look formally just like the
subjective density matrix \eqref{eq:CopInterpFinalLargeDensMatrProperMixt}
up to tiny corrections, with eigenstates that exhibit negligible quantum
interference with one another under further time evolution. However,
the subjective density matrix \eqref{eq:CopInterpFinalLargeDensMatrProperMixt},
unlike a decoherence-generated density matrix, is not a \emph{reduced}
density matrix arising from external quantum entanglements---it is
ultimately just a formal stand-in for the subjective probability distribution
\eqref{eq:CopInterFinalClassicalMixt} over \emph{persistent} measurement
outcomes, and, in the simplest case \eqref{eq:CopInterpFinalLargeDensMatrProperMixt},
is a proper mixture. Moreover, the Copenhagen interpretation provides
no canonical recipe for assigning preferred decompositions to generic
reduced density matrices. Hence, solving the measurement problem requires
either additional ingredients or a new interpretation altogether.

\subsubsection{A Variety of Approaches}

Within the framework of the Copenhagen interpretation, one declares
that observers or measurement devices that are ``sufficiently classical''---meaning
that they are on the classical side of the so-called Heisenberg cut---cause
decoherence-generated density matrices to \emph{cease} being \emph{reduced}
density matrices by \emph{breaking} their quantum entanglements with
any external systems, and, furthermore, cause them to develop the
necessary preferred decomposition appearing on the right-hand-side
of \eqref{eq:CopInterpFinalLargeDensMatrProperMixt}; the overall
effect is therefore to convert decoherence-generated density matrices
into subjective probability distributions of the form \eqref{eq:CopInterFinalClassicalMixt}.
Equivalently, one can phrase the Heisenberg cut as a threshold on
the amount of quantum interference between the eigenstates of a decoherence-generated
density matrix: If the amount of quantum interference falls below
that threshold, then we can treat those eigenstates as describing
classical possibilities in a subjective probability distribution \eqref{eq:CopInterFinalClassicalMixt}.
Unfortunately, according to either formulation, the Heisenberg cut
remains ill-defined and has not be identified in any experiment so
far.

An alternative approach is to postulate that \emph{all} reduced density
matrices gradually evolve into corresponding subjective probability
distributions \eqref{eq:CopInterFinalClassicalMixt}, with macroscopic
systems evolving in this way more rapidly than microscopic systems.
Achieving these effects requires altering the basic dynamical structure
of quantum theory along the lines of GRW dynamical-collapse or spontaneous-localization
constructions \cite{GhirardiRiminiWeber:1986udmms,Penrose:1989enm,Pearle:1989csdsvrwsl,BassiGhirardi:2003drm,Weinberg:2011csv,AdlerBassi:2009qte}.

The Everett-DeWitt many-worlds interpretation \cite{Everett:1957rsfqm,Wheeler:1957aersfqt,DeWitt:1970qmr,EverettDeWittGraham:1973mwiqm,Everett:1973tuwf,Deutsch:1985qtupt,Deutsch:1999qtpd,Wallace:2002wei,Wallace:2003erddapei,BrownWallace:2004smpdbble}
attempts to solve the measurement problem by reifying all the members
of a suitably chosen basis as simultaneous ``worlds'' or ``branches''
while somehow also regarding them as the elements of an appropriate
subjective probability distribution. Apart from making sense of probabilities
when all outcomes are simultaneously realized, one key trouble with
this interpretation is deciding which basis to choose, a quandary
known as the preferred-basis problem and that we describe in Sections~\ref{subsub:The-Everett-DeWitt-Many-Worlds-Interpretation-of-Quantum-Theory}
and \ref{subsub:Nonlocality in the Everett-DeWitt Many-Worlds Interpretation}.

The modal interpretations instead reify \emph{just one} member of
a suitable basis. In ``fixed'' modal interpretations---of which
the de Broglie-Bohm interpretation \cite{deBroglie:1930iswm,Bohm:1952siqtthvi,Bohm:1952siqtthvii,BohmHiley:1993uu}
is the most well-known example---this preferred basis is fixed for
all systems, leading to problems that we detail in Section~\ref{subsub:The-de-Broglie-Bohm-Pilot-Wave-Interpretation-of-Quantum-Theory}.
By contrast, in density-matrix-centered modal interpretations such
as the one that we introduce in this paper, one chooses the basis
to be the diagonalizing eigenbasis of the density matrix of whatever
system is currently under consideration. Essentially, the central
idea of our own minimal modal interpretation is a conservative one:
We identify improper density matrices as closely as possible with
subjective probability distributions, and add the minimal axiomatic
ingredients that are necessary to make this identification viable.

\subsubsection{The Instrumentalist Approach\label{subsub:The-Instrumentalist-Approach}}

A final prominent option is the instrumentalist approach, in which
one formally accepts the basic axioms of the Copenhagen interpretation
without taking a definitive stand on the ontological meaning of state
vectors or density matrices, or on any reality that underlies the
formalism of quantum theory more generally. In that case, we can regard
the Von Neumann-Lüders projection postulate \eqref{eq:CopInterpVonNeumannLudersProjPost}
as being the natural noncommutative generalization of the classical
post-measurement probability-update formula 
\begin{equation}
p_{\mathrm{initial},a}\mapsto p_{\mathrm{final},\lambda,a}=\frac{P_{\lambda,a}p_{a}}{\sum_{a^{\prime}}P_{\lambda,a^{\prime}}p_{a^{\prime}}},\label{eq:CopInterpClassicalProbUpdateFormula}
\end{equation}
 where we defined $P_{\lambda,a}$ in \eqref{eq:DefClassCharactFunc}
as the characteristic function for the generalized outcome $\lambda$.

We can then replace the Heisenberg cut with a self-consistency condition
on the definition of valid observers, called agents in this context.
Specifically, in order for a system such as a creature or a measurement
device to count as an agent, we require that whenever it performs
generic measurements, the observable $Q$ representing the true-or-false
question ``Did the agent obtain a set of frequency ratios in agreement
with the Born formula?'' and the observable $Q^{\prime}$ representing
the true-or-false question ``Did the agent find a persistent measurement
outcome that appears to be in keeping with the Von Neumann-Lüders
projection postulate?'' each have respective Born-rule probabilities
\eqref{eq:CopInterpBornRuleExpectValFromDensMatrix} $p\left(Q=\quote{\mathrm{true}}\right)$
and $p\left(Q^{\prime}=\quote{\mathrm{true}}\right)$ acceptably close
to unity---say, $99.9\%$. Provided that the agent is sufficiently
macroscopic, decoherence will generally guarantee that these conditions
hold.

It is important to keep in mind, however, that the Born rule \eqref{eq:CopInterpBornRuleExpectValFromDensMatrix}
and the Von Neumann-Lüders projection postulate \eqref{eq:CopInterpVonNeumannLudersProjPost}
are still nontrivial axioms and cannot be dropped within the instrumentalist
approach: Without these axioms, we \emph{cannot} conclude \emph{solely}
from $p\left(Q=\quote{\mathrm{true}}\right)\approx1$ and $p\left(Q^{\prime}=\quote{\mathrm{true}}\right)\approx1$
that sufficiently macroscopic measurement devices (such as human beings)
will experience the Born rule and the Von Neumann-Lüders projection
postulate, the reason being that neither of the associated Hermitian
operators $\op Q$ and $\op Q^{\prime}$ actually correspond to \emph{unique}
questions; indeed, each of these operators generically has two \emph{highly}
degenerate eigenvalues $1$ (``true'') and $0$ (``false''), meaning
that neither $\op Q$ nor $\op Q^{\prime}$ can uniquely pick out
an orthonormal basis of states describing classically sensible realities.
Thus, at best, the conditions $p\left(Q=\quote{\mathrm{true}}\right)\approx1$
and $p\left(Q^{\prime}=\quote{\mathrm{true}}\right)\approx1$ merely
supply us with a self-consistency check---a necessary but not sufficient
condition---on our definition of agents and a quantitative criterion
for determining how macroscopic a valid agent must be, rather than
making possible an \emph{ab initio} derivation of the Born rule or
the Von Neumann-Lüders projection postulate.

\subsection{Foundational Theorems\label{sub:Foundational-Theorems}}

We discuss the Bell theorem in Section~\ref{sub:The-EPR-Bohm-Thought-Experiment-and-Bell's-Theorem}
of the main text, and the Myrvold theorem in Section~\ref{sub:The-Myrvold-No-Go-Theorem}.
Here we present a brief proof of the Bell theorem and describe several
other important theorems that put strong constraints on candidate
interpretations of quantum theory.

\subsubsection{Proof of the Bell Theorem\label{subsub:Proof-of-the-Bell-Theorem}}

As we explained in Section~\ref{sub:The-EPR-Bohm-Thought-Experiment-and-Bell's-Theorem},
the Bell theorem involves a set of three spin detectors $A$, $B$,
and $C$ aligned respectively along three unit vectors $\svec a$,
$\svec b$, and $\svec c$ and that make measurements on pairs of
spin-1/2 particles governed by local hidden variables $\lambda$.
Granting the fact that spin is quantized in units of $\hbar/2$ (one
can account for this condition in classical language by insisting
that the particles automatically line themselves up along the local
detector alignments as they are being measured), Bell assumed that
the respective results $A\left(\svec a,\lambda\right)=\pm1$, $B\left(\svec b,\lambda\right)=\pm1$,
and $C\left(\svec c,\lambda\right)=\pm1$ (in units of $\hbar/2$)
of the three detectors depend only on data local to each detector.
In keeping with the supposed anti-correlated state \eqref{eq:EPRPairStateVec}
of each pair of spin-1/2 particles, Bell also required that if any
two detectors are aligned, then they must always measure opposite
spins: 
\begin{align*}
A\left(\svec a,\lambda\right) & =-B\left(\svec b=\svec a,\lambda\right)=-C\left(\svec c=\svec a,\lambda\right),\\
B\left(\svec b,\lambda\right) & =-A\left(\svec a=\svec b,\lambda\right)=-C\left(\svec c=\svec b,\lambda\right),\\
C\left(\svec c,\lambda\right) & =-A\left(\svec a=\svec c,\lambda\right)=-B\left(\svec b=\svec c,\lambda\right).
\end{align*}
 Finally, Bell posited the existence of a probability distribution
\eqref{eq:BellProbDistribConds} for the hidden variables $\lambda$
themselves: 
\[
0\leq p\left(\lambda\right)\leq1,\qquad\integration{\int}{d\lambda}p\left(\lambda\right)=1.
\]

The Bell theorem then asserts that the three average spin correlations
\begin{align*}
\expectval{S_{1,\svec a}S_{2,\svec b}}_{\mathrm{LHV}} & =\integration{\int}{d\lambda}p\left(\lambda\right)A\left(\svec a,\lambda\right)B\left(\svec b,\lambda\right),\\
\expectval{S_{1,\svec a}S_{2,\svec b}}_{\mathrm{LHV}} & =\integration{\int}{d\lambda}p\left(\lambda\right)A\left(\svec a,\lambda\right)B\left(\svec b,\lambda\right),\\
\expectval{S_{1,\svec a}S_{2,\svec b}}_{\mathrm{LHV}} & =\integration{\int}{d\lambda}p\left(\lambda\right)A\left(\svec a,\lambda\right)B\left(\svec b,\lambda\right)
\end{align*}
must satisfy the inequality \eqref{eq:BellIneq}, 
\[
\absval{\expectval{S_{1,\svec a}S_{2,\svec b}}_{\mathrm{LHV}}-\expectval{S_{1,\svec a}S_{2,\svec c}}_{\mathrm{LHV}}}\leq1+\expectval{S_{1,\svec b}S_{2,\svec c}}_{\mathrm{LHV}},
\]
 as follows from a straightforward computation: 
\begin{align*}
\absval{\expectval{S_{1,\svec a}S_{2,\svec b}}_{\mathrm{LHV}}-\expectval{S_{1,\svec a}S_{2,\svec c}}_{\mathrm{LHV}}} & =\absval{\integration{\int}{d\lambda}p\left(\lambda\right)\left(A\left(\svec a,\lambda\right)\underbrace{B\left(\svec b,\lambda\right)}_{-A\left(\svec b,\lambda\right)}-\underbrace{1}_{\left(A\left(\svec b,\lambda\right)\right)^{2}}A\left(\svec a,\lambda\right)\underbrace{C\left(\svec c,\lambda\right)}_{-A\left(\svec c,\lambda\right)}\right)}
\end{align*}
\begin{align*}
 & \leq\integration{\int}{d\lambda}p\left(\lambda\right)\underbrace{\absval{\overbrace{A\left(\svec a,\lambda\right)}^{\pm1}\overbrace{A\left(\svec b,\lambda\right)}^{\pm1}}}_{1}\absval{\underbrace{1-A\left(\svec b,\lambda\right)\overbrace{A\left(\svec c,\lambda\right)}^{-C\left(\svec c,\lambda\right)}}_{>0}}\\
 & =\integration{\int}{d\lambda}p\left(\lambda\right)\left(1+A\left(\svec b,\lambda\right)C\left(\svec c,\lambda\right)\right)=1+\expectval{S_{1,\svec b}S_{2,\svec c}}_{\mathrm{LHV}}.\qquad\QED
\end{align*}

\subsubsection{Gleason's Theorem}

Gleason's theorem \cite{Gleason:1957mcshs} asserts that the only
consistent probability measures for closed subspaces of a Hilbert
space of dimension $\geq3$  must be given by $p_{i}=\Tr\left[\op{\rho}\op P_{i}\right]$,
which generalizes the state-vector version of the Born rule. Here
$\op{\rho}$ is a unit-trace, positive semi-definite operator that
we interpret as the system's density matrix and $\op P_{i}=\ket{\Psi_{i}}\bra{\Psi_{i}}$
is a projection operator onto some state vector $\ket{\Psi_{i}}$.

All the well-known interpretations of quantum theory satisfy Gleason's
theorem. In particular, the basic correspondence \eqref{eq:EpStateDensMatrixCorresp}
at the heart of our own minimal modal interpretation is consistent
with the theorem.

\subsubsection{The Kochen-Specker Theorem}

It might seem plausible for an interpretation of quantum theory to
claim that when we measure observables $\Lambda_{1},\Lambda_{2},\dotsc$
belonging to a quantum system and obtain some set of outcome values
$\lambda_{1},\lambda_{2},\dotsc$, we are merely revealing that those
observables secretly possessed all those values $\lambda_{1},\lambda_{2},\dotsc$
simultaneously even before the measurement took place. The Kochen-Specker theorem
\cite{KochenSpecker:1967phvqm} presents an obstruction to such claims.
More precisely, the theorem rules out hidden-variables interpretations
that assert that all observables that \emph{could} be measured have
simultaneous, sharply defined values \emph{before} they are measured---values
that depend \emph{only} on the observables themselves and on the system
to which they belong---and that measurements merely reveal those supposedly
preexisting values. For viable hidden-variables interpretations of
quantum theory, an immediate consequence of the theorem is that the
properties of a quantum system must generically be contextual, in
the sense that they can depend on the kinds of measurements performed
on the system. Here we present one of the simplest versions of the
theorem, due to Peres \cite{Peres:1996gkst}.

We begin by considering a four-state quantum system, which we can
regard as consisting of a pair of two-state spin-1/2 subsystems and
therefore having a Hilbert space $\mathcal{H}$ of the form 
\begin{equation}
\mathcal{H}=\mathcal{H}_{1/2}\tensorprod\mathcal{H}_{1/2},\qquad\dim\mathcal{H}=2\times2=4.\label{eq:KSTheoremHilbSpace}
\end{equation}
 We consider also the following nine Hermitian matrices on this four-dimensional
Hilbert space: 
\begin{equation}
\eqsbrace{\begin{aligned}P_{2z} & \equiv\left(1-\sigma_{2z}\right)/2,\\
P_{1x} & \equiv\left(1-\sigma_{1x}\right)/2,\\
P_{1x,2z} & \equiv\left(1-\sigma_{1x}\sigma_{2z}\right)/2,
\end{aligned}
\qquad\begin{aligned}P_{1z} & \equiv\left(1-\sigma_{1z}\right)/2,\\
P_{2x} & \equiv\left(1-\sigma_{2x}\right)/2,\\
P_{1z,2x} & \equiv\left(1-\sigma_{1z}\sigma_{2x}\right)/2,
\end{aligned}
\qquad\begin{aligned}P_{1z,2z} & \equiv\left(1-\sigma_{1z}\sigma_{2z}\right)/2,\\
P_{1x,2x} & \equiv\left(1-\sigma_{1x}\sigma_{2x}\right)/2,\\
P_{1y,2y} & \equiv\left(1-\sigma_{1y}\sigma_{2y}\right)/2.
\end{aligned}
}\label{eq:KSTheoremOperators}
\end{equation}
 In our notation here, $1\equiv1_{2\times2}\tensorprod1_{2\times2}$
is the $4\times4$ identity matrix, and $\sigma_{1i}\equiv\sigma_{i}\tensorprod1_{2\times2}$
and $\sigma_{2i}\equiv1_{2\times2}\tensorprod\sigma_{i}$ are respectively
the Pauli sigma matrices for the first and second spin-1/2 subsystems.

Each of the $4\times4$ matrices in \eqref{eq:KSTheoremOperators}
has two $+1$ eigenvalues and two $-1$ eigenvalues; in the usual
language of quantum theory, the associated observables therefore have
possible measured values $+1$ and $-1$. Furthermore, all three matrices
in each column of \eqref{eq:KSTheoremOperators} commute with each
other, and, likewise, all three matrices in each row commute with
each other; hence, for each column or row, the three corresponding
observables are mutually compatible and thus can be made simultaneously
sharply defined by an appropriate preparation of the state of our
system.

We next define three new $4\times4$ matrices $A_{1},A_{2},A_{3}$
by respectively summing each column of \eqref{eq:KSTheoremOperators},
\begin{equation}
\eqsbrace{\begin{aligned}A_{1} & \equiv P_{2z}+P_{1x}+P_{1x,2z} &  & \qquad\left(\mathrm{eigenvalues\ }+2,\ +2,\ 0,\ 0\right),\\
A_{2} & \equiv P_{1z}+P_{2x}+P_{1z,2x} &  & \qquad\left(\mathrm{eigenvalues\ }+2,\ +2,\ 0,\ 0\right),\\
A_{3} & \equiv P_{1z,2z}+P_{1x,2x}+P_{1y,2y} &  & \qquad\left(\mathrm{eigenvalues\ }+3,\ +3,\ +1,\ +1\right)
\end{aligned}
}\label{eq:KSTheoremAMatrices}
\end{equation}
 and three $4\times4$ matrices $B_{1},B_{2},B_{3}$ by respectively
summing each row of \eqref{eq:KSTheoremOperators}, 
\begin{equation}
\eqsbrace{\begin{aligned}B_{1} & \equiv P_{2z}+P_{1z}+P_{1z,2z} &  & \qquad\left(\mathrm{eigenvalues\ }+2,\ +2,\ 0,\ 0\right),\\
B_{2} & \equiv P_{1x}+P_{2x}+P_{1x,2x} &  & \qquad\left(\mathrm{eigenvalues\ }+2,\ +2,\ 0,\ 0\right),\\
B_{3} & \equiv P_{1x,2z}+P_{1z,2x}+P_{1y,2y} &  & \qquad\left(\mathrm{eigenvalues\ }+2,\ +2,\ 0,\ 0\right).
\end{aligned}
}\label{eq:KSTheoremBMatrices}
\end{equation}
 Finally, we introduce an observable $\Sigma$ defined to be the sum
(times two) of the nine original observables defined in \eqref{eq:KSTheoremOperators},
or, equivalently, defined to be the sum of the six observables $A_{1},A_{2},A_{3},B_{1},B_{2},B_{3}$:
\begin{equation}
\begin{aligned}\Sigma\equiv~ & 2P_{2z}+2P_{1x}+2P_{1x,2z}\\
 & +2P_{1z}+2P_{2x}+2P_{1z,2x}\\
 & +2P_{1z,2z}+2P_{1x,2x}+2P_{1y,2y}\\
= & ~A_{1}+A_{2}+A_{3}+B_{1}+B_{2}+B_{3}.
\end{aligned}
\label{eq:KSTheoremDefSigma}
\end{equation}

Each of the nine observables defined in \eqref{eq:KSTheoremOperators}
has permissible values $+1$ or $-1$, and so, from the first expression
for $\Sigma$ in \eqref{eq:KSTheoremDefSigma}, we see that the existence
of simultaneous pre-measurement values for these nine observables
would imply that $\Sigma$ has an \emph{even} pre-measurement value:
\begin{equation}
\eqsbrace{\begin{aligned}\Sigma=~ & 2\times\left(+1\binaryor-1\right)+2\times\left(+1\binaryor-1\right)+2\times\left(+1\binaryor-1\right)\\
 & +2\times\left(+1\binaryor-1\right)+2\times\left(+1\binaryor-1\right)+2\times\left(+1\binaryor-1\right)\\
= & ~\mathrm{even}.
\end{aligned}
}\label{eq:KSTheoremSigmaEven}
\end{equation}
 However, if we assume that the six observables $A_{1},A_{2},A_{3},B_{1},B_{2},B_{3}$
likewise have simultaneous pre-measurement values, then we find instead
that $\Sigma$ has an \emph{odd} pre-measurement value, a contradiction:
\begin{equation}
\eqsbrace{\begin{aligned}\Sigma & =\left(0\binaryor2\right)+\left(0\binaryor2\right)+\left(1\binaryor3\right)+\left(0\binaryor2\right)+\left(0\binaryor2\right)+\left(0\binaryor2\right)\\
 & =\mathrm{odd}.
\end{aligned}
}\label{eq:KSTheoremSigmaOdd}
\end{equation}
We are therefore forced to give up our assumption that a quantum system
can always have simultaneous sharply defined pre-measurement values
for all of its observables. $\QED$

Neither the traditional Copenhagen interpretation nor the minimal
modal interpretation we introduce in this paper asserts that all observables
have well-defined values before measurements---indeed, our own interpretation
makes manifest that a system's set of possible ontic states can change
contextually from one orthonormal basis of the system's Hilbert space
to another orthonormal basis in the course of interactions with other
systems---and so are both consistent with the Kochen-Specker theorem.
(For macroscopic systems, however, decoherence ensures that all observables
simultaneously develop \emph{approximate} pre-measurement values.)
The de Broglie-Bohm pilot-wave interpretation evades this theorem
as well, because although the interpretation assumes systems have
hidden well-defined values of both canonical coordinates and canonical
momenta at all times, the canonical momenta are not identified with
the \emph{observable} momenta that actually show up in measurements.

\subsubsection{The Pusey-Barrett-Rudolph (PBR) Theorem}

The Pusey-Barret-Rudolph (PBR) theorem \cite{PuseyBarrettRudolph:2012rqs,BarrettCavalcantiLalMaroney:2013rqs,ColbeckRenner:2013swfudups}
rules out so-called psi-epistemic interpretations of quantum theory
that directly regard \emph{state vectors} (as opposed to density matrices)
as merely being epistemic probability distributions for hidden variables
whose configurations determine the outcomes of measurements. Essentially,
the theorem derives a one-to-one correspondence between each specific
configuration of hidden variables and each state vector (up to trivial
overall phase factor), thus implying that each configuration singles
out a unique state vector. It is therefore impossible to regard each
state vector as being an epistemic probability distribution over a
nontrivial collection of \emph{different} configurations of hidden
variables.

To illustrate the theorem in a simple case, the authors consider a
two-state system with an orthonormal basis $\ket 0,\ket 1$ and a
second orthonormal basis defined by 
\begin{equation}
\ket +\equiv\frac{1}{\sqrt{2}}\left(\ket 0+\ket 1\right),\qquad\ket -\equiv\frac{1}{\sqrt{2}}\left(\ket 0-\ket 1\right).\label{eq:PBRTheoremPlusMinusBasis}
\end{equation}
 If the two state vectors $\ket 0$ and $\ket +$ merely describe
probability distributions over configurations of hidden variables,
and those probability distributions are allowed to overlap on the
sample space of configurations of hidden variables, then there exists
some nonzero probability $q>0$ that a particular configuration $\lambda$
of hidden variables will reside in both probability distributions.
Hence, if the system's hidden variables have the configuration $\lambda$,
then the corresponding probability distribution could be either $\ket 0$
or $\ket +$.

If we now set up a pair of independent such systems as a composite
system, then, with probability $q^{2}$, the configurations $\lambda_{1}$
and $\lambda_{2}$ of the two subsystems would permit being jointly
described by the probability distributions arising from any of the
possible tensor-product state vectors $\ket 0\ket 0$, $\ket 0\ket +$,
$\ket +\ket 0$, $\ket +\ket +$. But then a measurement of an observable
whose corresponding basis of orthonormal eigenstates is 
\begin{equation}
\eqsbrace{\begin{aligned}\ket{\xi_{1}} & =\frac{1}{\sqrt{2}}\left(\ket 0\ket 1+\ket 1\ket 0\right),\\
\ket{\xi_{2}} & =\frac{1}{\sqrt{2}}\left(\ket 0\ket -+\ket 1\ket +\right),\\
\ket{\xi_{3}} & =\frac{1}{\sqrt{2}}\left(\ket +\ket 1+\ket -\ket 0\right),\\
\ket{\xi_{4}} & =\frac{1}{\sqrt{2}}\left(\ket +\ket -+\ket -\ket +\right)
\end{aligned}
}\label{eq:PBRTheoremMeasurementBasis}
\end{equation}
 would have zero probability of yielding $\ket{\xi_{1}}$ if the composite
system's state vector happened to be $\ket 0\ket 0$, zero probability
of yielding $\ket{\xi_{2}}$ if the composite system's state vector
happened to be $\ket 0\ket +$, zero probability of yielding $\ket{\xi_{3}}$
if the composite system's state vector happened to be $\ket +\ket 0$,
and zero probability of yielding $\ket{\xi_{4}}$ if the composite
system's state vector happened to be $\ket +\ket +$. Hence, whichever
result is found by the measurement, there exists a contradiction with
the notion that the original simultaneous configurations $\lambda_{1}$
and $\lambda_{2}$ of the conjoined subsystems were really compatible
with \emph{all four} possible state vectors $\ket 0\ket 0$, $\ket 0\ket +$,
$\ket +\ket 0$, $\ket +\ket +$. The authors then extend this general
argument to a much larger class of possibilities beyond the particular
pair $\ket 0$ and $\ket +$ to argue that no two state vectors could
ever describe overlapping probability distributions over hidden variables.

The Copenhagen interpretation and our own interpretation satisfy the
PBR theorem, because they both regard state vectors as irreducible
features of reality rather than as mere epistemic probability distributions
over a deeper layer of hidden variables. The de Broglie-Bohm pilot-wave
interpretation also evades the theorem because, in that interpretation,
the state vector plays both the role of an epistemic probability distribution
as well as a physical pilot wave that (nonlocally if necessary) guides
the hidden variables during measurements to values that are always
consistent with final measurement outcomes.

\subsubsection{The Fine and Vermaas No-Go Theorems}

Finally, there exist no-go theorems due to Fine \cite{Fine:1982hvjpbi,Fine:1982jdqcco}
and Vermaas \cite{Vermaas:1997ngtjpamiqm} suggesting that axiomatically
imposing joint probability distributions for properties of \emph{non-disjoint}
subsystems of a larger parent system leads to contradictions with
axiomatic impositions of joint probability distributions for properties
of \emph{disjoint} subsystems. In our minimal modal interpretation
of quantum theory, we do not postulate joint probability distributions
for non-disjoint subsystems of a given parent system, and so our interpretation
evades both theorems.

\bibliographystyle{plainurl}
\nocite{*}
\bibliography{Bibliography,BibliographyEntryLetter}

\end{document}